\documentclass[aps,amsmath,amssymb,nofootinbib,preprint numbers]{revtex4} 
\usepackage{graphicx} 
\usepackage{amsmath}
\usepackage{amsfonts,amsbsy}
\usepackage{amssymb}
\allowdisplaybreaks

\usepackage{xcolor}
\definecolor{lcolor}{rgb}{0.5,0,0}
\definecolor{citcolor}{rgb}{0,0.3,0.0}
\usepackage[breaklinks,colorlinks,urlcolor=blue,citecolor=citcolor,linkcolor=lcolor]{hyperref}

\usepackage{mciteplus}

\def\gsim{ \,\, \vcenter{\hbox{$\buildrel{\displaystyle >}\over\sim$}}
 \,\,}

\def\be{\begin{equation}}
\def\ee{\end{equation}}
\def\bea{\begin{eqnarray}}
\def\eea{\end{eqnarray}}

\newcommand{\dd}{{\rm d}}
\newcommand{\nn}{\nonumber}

\begin{document}

\title{\bf Cubic color charge correlator in a proton made of
  three quarks and a gluon}

\preprint{HIP-2021-21/TH}

\author{Adrian Dumitru}
\email{adrian.dumitru@baruch.cuny.edu}
\affiliation{Department of Natural Sciences, Baruch College, CUNY,
17 Lexington Avenue, New York, NY 10010, USA}
\affiliation{The Graduate School and University Center, The City University
  of New York, 365 Fifth Avenue, New York, NY 10016, USA}

\author{Heikki Mäntysaari}
\email{heikki.mantysaari@jyu.fi}
\affiliation{
Department of Physics, University of Jyväskylä,  P.O. Box 35, 40014 University of Jyväskylä, Finland
}
\affiliation{
Helsinki Institute of Physics, P.O. Box 64, 00014 University of Helsinki, Finland
}

\author{Risto Paatelainen} \email{risto.sakari.paatelainen@cern.ch}
\affiliation{Helsinki Institute of Physics and Department of Physics, FI-00014 University of Helsinki, Finland}

\begin{abstract}
The three point correlation function of color charge densities is
evaluated explicitly in light cone gauge for a proton on the light
cone. This includes both $C$-conjugation even and odd contributions.
We account for perturbative corrections to the three-quark light cone
wave function due to the emission of an internal gluon which is not
required to be soft. We verify the Ward identity as well as the
cancellation of UV divergences in the sum of all diagrams so that the
correlator is independent of the renormalization scale. It does,
however, exhibit the well known soft and collinear singularities. The
expressions derived here provide the $C$-odd contribution to the
initial conditions for high-energy evolution of the dipole scattering
amplitude to small $x$.  Finally, we also present a numerical model
estimate of the impact parameter dependence of quantum color charge
three-point correlations in the proton at moderately small $x$.
\end{abstract}

\maketitle

\tableofcontents
\newpage

\newcommand{\Ptp}{{\mathbf{P}'}}
\newcommand{\Pt}{{\mathbf{P}}}
\newcommand{\qtp}{{\mathbf{q}'}}
\newcommand{\ktp}{{\mathbf{k}'}}
\newcommand{\ktpp}{{\mathbf{k}''}}
\newcommand{\ptp}{{\mathbf{p}'}}
\newcommand{\ptpp}{{\mathbf{p}''}}
\newcommand{\xtp}{{\mathbf{x}'}}
\newcommand{\ytp}{{\mathbf{y}'}}
\newcommand{\ztp}{{\mathbf{z}'}}
\newcommand{\rtp}{{\mathbf{r}'}}
\newcommand{\rt}{{\mathbf{r}}}
\newcommand{\xt}{{\mathbf{x}}}
\newcommand{\bt}{{\mathbf{b}}}
\newcommand{\yt}{{\mathbf{y}}}
\newcommand{\zt}{{\mathbf{z}}}
\newcommand{\pt}{{\mathbf{p}}}
\newcommand{\qt}{{\mathbf{q}}}
\newcommand{\kt}{{\mathbf{k}}}
\newcommand{\ktpzero}{{\mathbf{k}_{0'}}}
\newcommand{\nt}{{\mathbf{n}}}
\newcommand{\mt}{{\mathbf{m}}}
\newcommand{\Rt}{{\mathbf{R}}}
\newcommand{\ot}{\mathbf{0}}
\newcommand{\itt}{\mathbf{i}}
\newcommand{\jt}{\mathbf{j}}
\newcommand{\hht}{\mathbf{h}}
\newcommand{\Kt}{\mathbf{K}}
\newcommand{\Lt}{\mathbf{L}}
\newcommand{\at}{\mathbf{a}}
\newcommand{\ptt}{p_T} 
\newcommand{\ktt}{k_T} 
\newcommand{\qtt}{q_T} 
\newcommand{\lt}{\mathbf{l}}
\newcommand{\nabt}{\boldsymbol{\nabla}}
\newcommand{\epst}{\boldsymbol{\varepsilon}}
\newcommand{\At}{\mathbf{A}}
\newcommand{\Deltat}{{\boldsymbol{\Delta}}}
\newcommand{\Deltatt}{{\Delta_T}}
\newcommand{\kvec}{{\vec{k}}}
\newcommand{\pvec}{{\vec{p}}}
\newcommand{\lvec}{{\vec{l}}}
\newcommand{\ppvec}{{{\vec{p}}{\, '}}}
\newcommand{\pppvec}{{{\vec{p}}{\, ''}}}
\newcommand{\kpvec}{{{\vec{k}}{\, '}}}
\newcommand{\kpzerovec}{{{\vec{k}_{0'}}}}
\newcommand{\kppzerovec}{{{\vec{k}_{0''}}}}
\newcommand{\kponevec}{{{\vec{k}_{1'}}}}
\newcommand{\kpponevec}{{{\vec{k}_{1''}}}}
\newcommand{\kppvec}{{{\vec{k}}{\, ''}}}
\newcommand{\qvec}{{\vec{q}}}
\newcommand{\qpvec}{{{\vec{q}}{\, '}}}
\newcommand{\xvec}{{\vec{x}}}
\newcommand{\yvec}{{\vec{y}}}

\newcommand{\lo}{{\textnormal{LO}}}
\newcommand{\nlo}{{\textnormal{NLO}}}

\newcommand{\epsl}{{\varepsilon\!\!\!/}}
\newcommand{\pasl}{{p\!\!\!/_{a}}}
\newcommand{\pbsl}{{p\!\!\!/_{b}}}
\newcommand{\psl}{{p\!\!\!/}}
\newcommand{\ppsl}{{p\!\!\!/}'}
\newcommand{\ksl}{{k\!\!\!/}}


\newcommand{\dk}{{\widetilde{\mathrm{d} k}}}
\newcommand{\dkp}{{\widetilde{\mathrm{d} k}{'}}}
\newcommand{\dkpzero}{{\widetilde{\mathrm{d} k_{0'}}}}
\newcommand{\dkppzero}{{\widetilde{\mathrm{d} k_{0''}}}}
\newcommand{\dkpone}{{\widetilde{\mathrm{d} k_{1'}}}}
\newcommand{\dkpp}{{\widetilde{\mathrm{d} k}{''}}}
\newcommand{\ddp}{{\widetilde{\mathrm{d} p}}}
\newcommand{\dpp}{{\widetilde{\mathrm{d} p}{'}}}
\newcommand{\dppp}{{\widetilde{\mathrm{d} p}{''}}}
\newcommand{\dqp}{{\widetilde{\mathrm{d} q}{'}}}
\newcommand{\dq}{{\widetilde{\mathrm{d} q}}}

\newcommand{\dPp}{{\widetilde{\mathrm{d} P}{'}}}

\newcommand{\qemit}{V}
\newcommand{\qemitas}{\mathcal{V}}
\newcommand{\qbemit}{\overline{V}}
\newcommand{\qbemitas}{\overline{\mathcal{V}}}
\newcommand{\paircr}{A}
\newcommand{\paircras}{\mathcal{A}}
\newcommand{\annih}{\overline{A}}
\newcommand{\annihas}{\overline{\mathcal{A}}}

\newcommand{\ud}{\, \mathrm{d}}
\newcommand{\uc}{{\mathrm{c}}}
\newcommand{\ul}{{\mathrm{L}}}
\newcommand{\intd}{\int \!}
\newcommand{\tr}{\, \mathrm{tr} \, }
\newcommand{\R}{\mathrm{Re}}
\newcommand{\nc}{{N_\mathrm{c}}}
\newcommand{\nf}{{N_\mathrm{F}}}
\newcommand{\half}{\frac{1}{2}}
\newcommand{\hc}{\mathrm{\ h.c.\ }}
\newcommand{\nosum}[1]{\textrm{ (no sum over } #1 )}
\newcommand{\na}{\, :\!}
\newcommand{\nb}{\!: \,}
\newcommand{\cf}{C_\mathrm{F}}
\newcommand{\ca}{C_\mathrm{A}}
\newcommand{\df}{d_\mathrm{F}}
\newcommand{\da}{d_\mathrm{A}}
\newcommand{\nr}[1]{(\ref{#1})}
\newcommand{\dadj}{D_{\mathrm{adj}}}
\newcommand{\ra}{R_A}
\newcommand{\rp}{R_p}

\newcommand{\mev}{\ \textrm{MeV}}
\newcommand{\tev}{\ \textrm{TeV}}
\newcommand{\gev}{\ \textrm{GeV}}
\newcommand{\fm}{\ \textrm{fm}}
\newcommand{\mb}{\ \textrm{mb}}
\newcommand{\ls}{\Lambda_\mathrm{s}}
\newcommand{\qs}{Q_\mathrm{s}}
\newcommand{\qsprime}{Q_\mathrm{s}'}
\newcommand{\qsprimep}{Q_{\mathrm{s}p}'}
\newcommand{\qsprimea}{Q_\mathrm{sA}'}
\newcommand{\qsa}{Q_{\mathrm{s}A}}
\newcommand{\qsp}{Q_{\mathrm{s}p}}
\newcommand{\qsadj}{\widetilde{Q}_{\mathrm{s}}}
\newcommand{\qsoadj}{\widetilde{Q}_{\mathrm{s0}}}
\newcommand{\qso}{Q_\mathrm{s0}}

\newcommand{\lqcd}{\Lambda_{\mathrm{QCD}}}
\newcommand{\as}{\alpha_{\mathrm{s}}}

\newcommand{\subA}{A}
\newcommand{\subB}{B}
\newcommand{\B}{A_{\subB}}

\newcommand{\fig}{Fig.~}
\newcommand{\figs}{Figs.~}
\newcommand{\eq}{Eq.~}
\newcommand{\se}{Sec.~}
\newcommand{\eqs}{Eqs.~}

\newcommand{\npart}{{N_\mathrm{part}}}
\newcommand{\nch}{{N_\mathrm{ch}}}

\newcommand{\xbj}{{x}}
\newcommand{\sigmaa}{{ \sigma^A_\textrm{dip} }}
\newcommand{\sigmap}{{ \sigma^\textrm{p}_\textrm{dip} }}
\newcommand{\sigmadip}{{ \sigma_\textrm{dip} }}
\newcommand{\dsigmap}{{\frac{\ud \sigma^\textrm{p}_\textrm{dip}}{\ud^2 \bt}}}
\newcommand{\dsigmaa}{{\frac{\ud \sigma^A_\textrm{dip}}{\ud^2 \bt}}}
\newcommand{\dsigmab}{{\frac{\ud \sigma^B_\textrm{dip}}{\ud^2 \bt}}}
\newcommand{\dsigma}{{\frac{\ud \sigma_\textrm{dip}}{\ud^2 \bt}}}
\newcommand{\dsigmaadj}{{\frac{\ud \tilde{\sigma}_\textrm{dip}}{\ud^2 \bt}}}
\newcommand{\xpom}{{x_\mathbb{P}}}

\newcommand{\llsim}{{\underset{\sim}{\ll}}}
\newcommand{\ggsim}{{\underset{\sim}{\gg}}}

\newcounter{diag}
\setcounter{diag}{0}

\newcommand{\namediag}[1]{\refstepcounter{diag} \thediag \label{#1}}
\renewcommand{\thediag}{(\alph{diag})}

\newcommand{\ps}{\textnormal{PS}}
\newcommand{\epsmsbar}{\varepsilon_{\overline{\textnormal{MS}}}}

\section{Introduction}

In this paper we present explicit expressions for all diagrams which
determine the cubic light cone gauge color charge correlator
$\langle\rho^a(\vec q_1)\, \rho^b(\vec q_2)\,\rho^c(\vec q_3)\rangle$
in a proton. The proton is approximated by a non-perturbative
three-quark Fock state, plus a perturbative gluon. This is in
continuation of ref.~\cite{Dumitru:2020gla} where we derived analogous
expressions for $\langle\rho^a(\vec q_1)\, \rho^b(\vec q_2) \rangle$,
and of ref.~\cite{Dumitru:2021tvw} where we presented numerical
results for the quadratic correlator (see also Ref.~\cite{Dumitru:2021mab} for a first phenomenological
application).
 The cubic correlator quantifies
corrections to Gaussian color charge fluctuations in the proton, and
it provides a contribution to the scattering matrix of a dipole which
is odd under $C$-conjugation.\\

The $S$-matrix for eikonal scattering of a quark - antiquark dipole
off the fields in the target proton can be expressed
as\footnote{$\langle\cdots\rangle$ denotes the matrix element between
  incoming and outgoing proton states. It is defined in
  eq.~(\ref{eq:proton-matrix-element}) below.}~\cite{Mueller:2001fv,
  Kovchegov:2012mbw}
\be \label{eq:S_dipole_b}
   {\cal S} (\vec r,\vec b) =
   \frac{1}{N_c}\,{\rm tr} \,\left< U\left(\vec b +
     \frac{\vec r}{2}\right)\,
     U^\dagger\left( \vec b - \frac{\vec r}{2}\right)\right> \, .
\ee
Here, $\vec b$ is the impact parameter vector while $\vec r$ denotes
the transverse separation of quark and anti-quark. The operators $U$
($U^\dagger$) are (anti-)path ordered Wilson lines of the field in
covariant gauge, representing the eikonal scattering of the quarks at
transverse coordinate $\vec x$:
\be \label{eq:WilsonLines}
U(\vec x) = {\cal P} e^{ig \int \ud x^- A^{+a}(x^-,\vec x)\, t^a} ~~~~~,~~~~~
U^\dagger(\vec x) = \overline{\cal P} e^{-ig \int \ud x^- A^{+a}(x^-,\vec x)\, t^a} ~.
\ee
$C$-conjugation transforms the generators of the fundamental representation
$t^a \to - (t^a)^T$.

The $S$-matrix can be separated into its real part which at high
energy is dominated by $C$-conjugation even two-gluon exchange, and
its imaginary part which starts out as $C$-odd three gluon exchange:
\bea
D(\vec r,\vec b) \equiv \mathrm{Re}~{\cal S} (\vec r,\vec b) &=&
\frac{1}{2N_c}\,{\rm tr} \,\left<
U\left(\vec b + \frac{\vec r}{2}\right)\,
U^\dagger\left( \vec b - \frac{\vec r}{2}\right)
+
U\left(\vec b - \frac{\vec r}{2}\right)\,
U^\dagger\left( \vec b + \frac{\vec r}{2}\right)
\right>~,\\
O(\vec r,\vec b) \equiv \mathrm{Im}~{\cal S} (\vec r,\vec b) &=&
\frac{1}{2iN_c}\,{\rm tr} \,\left<
U\left(\vec b + \frac{\vec r}{2}\right)\,
U^\dagger\left( \vec b - \frac{\vec r}{2}\right)
-
U\left(\vec b - \frac{\vec r}{2}\right)\,
U^\dagger\left( \vec b + \frac{\vec r}{2}\right)
\right>~.
\eea

Thus, the fact that the imaginary part of ${\cal S} (\vec r,\vec b)$
is non-zero is due to the existence of a color singlet three gluon
($t$-channel) exchange with negative $C$-parity in
QCD~\cite{Bartels:1980pe,Jaroszewicz:1980mq,
  Kwiecinski:1980wb,Braun:1998fs,Ewerz:2003xi,Kovner:2005qj}. Recently,
the TOTEM and D0 collaborations have presented evidence for a
difference in $p-p$ vs.\ $p-\bar{p}$ elastic scattering cross sections
at a CM energy of $\sqrt s \simeq 2$~TeV, and low momentum transfer
$|t| < 1$~GeV$^2$~\cite{Abazov:2020rus, Antchev:2017yns} (also see
ref.~\cite{Martynov:2018sga}).  
However, our focus here is on cubic color charge correlations in the 
semi-hard regime, which is related to the $C$ and $P$ odd contribution 
to the dipole scattering amplitude.\\

The evolution of the dipole $S$-matrix in the high-energy regime is
described by the JIMWLK renormalization group
equations~\cite{Balitsky:1995ub, Balitsky:1998ya, Balitsky:2001re,
  Jalilian-Marian:1997jx, Jalilian-Marian:1997gr,
  JalilianMarian:1997dw, Iancu:2001ad, Iancu:2000hn, Ferreiro:2001qy,
  Weigert:2000gi}. These reduce in the large-$N_c$ limit to the
Balitsky-Kovchegov (BK) equation~\cite{Balitsky:1995ub,
  Kovchegov:1999yj}. High-energy resummation may modify the intercept of
the ``hard Odderon'' from its value of unity; see the
review~\cite{Ewerz:2003xi} and references therein. The evolution with energy specifically of the (hard) Odderon $O(\vec
r,\vec b)$ has been studied in refs.~\cite{Kovchegov:2003dm,
  Hatta:2005as, Lappi:2016gqe}. Its knowledge is important for various
spin dependent Transverse Momentum Dependent (TMD) distributions such
as the (dipole) gluon Sivers function of a transversely polarized
proton~\cite{Zhou:2013gsa,Boer:2015pni,Yao:2018vcg}.  Furthermore,
this amplitude is responsible for charge asymmetries in diffractive
electroproduction of a $\pi^+\, \pi^-$
pair~\cite{Hagler:2002nh,Hagler:2002nf}, and exclusive production of a
pseudo-scalar meson~\cite{Dumitru:2019qec, Czyzewski:1996bv,
  Engel:1997cga, Kilian:1997ew, Rueter:1998gj}. Lastly, the odderon is
related to cubic color charge density fluctuations $\langle\rho^a(\vec
q_1)\, \rho^b(\vec q_2) \, \rho^c(\vec q_3)\rangle$ (see below) and
therefore provides insight into three-body correlations in the proton.
This may guide phenomenological models of correlated ``hot spots''
which have been applied to proton-proton scattering at high
energies~\cite{Albacete:2016pmp, Albacete:2016gxu,
  Albacete:2017ajt,Csorgo:2019egs} (see also Ref.~\cite{Mantysaari:2020axf}). 
   The existence of the cubic
correlator also implies that color charge density fluctuations in the
proton are not Gaussian.\\

A key limitation for quantitative predictions in the energy regime of
the Electron-Ion Collider
(EIC)~\cite{Boer:2011fh,Accardi:2012qut,Aschenauer:2017jsk,AbdulKhalek:2021gbh}
is the
crude knowledge of the initial condition for the evolution equations
at moderately small $x$. Deriving the next-to-leading order (NLO) expressions for $O(\vec
r,\vec b)$ due to one gluon emission corrections in a proton target at
$x\sim 0.01 - 0.1$ is the main purpose of this paper. The
corresponding expressions at leading order (LO) have been published in
refs.~\cite{Dumitru:2018vpr, Dumitru:2019qec, Dumitru:2020fdh}. The
latter paper also provides numerical estimates of cubic color charge
correlators and of $O(\vec r,\vec b)$ at LO, i.e.\ in the valence
quark regime. Bartels and Motyka~\cite{Bartels:2007aa} have also
calculated the proton impact factor for $t$-channel three gluon
exchange, which agrees with the LO expressions for $\langle\rho^a(\vec
q_1)\, \rho^b(\vec q_2) \, \rho^c(\vec q_3)\rangle$ given in
refs.~\cite{Dumitru:2018vpr, Dumitru:2019qec, Dumitru:2020fdh}, and
soft gluon emission corrections to proton-proton scattering at high
energy.\\

The initial condition for the small-$x$ evolution~\cite{Kovchegov:2003dm,
  Hatta:2005as, Lappi:2016gqe} 
  of $\mathrm{Im}~{\cal S}$, which we derive here,
depends not only on the impact parameter and the dipole vectors but
also on their relative angle, and on the light-cone momentum fraction
$x$ in the {\em target}.  In fact, the BK equation in its standard
formulation evolves the wave function of the dipole projectile, and
the evolution ``time'' is then related to the minus component of the
momentum of the gluon in the proton target~\cite{Beuf:2014uia,
  Ducloue:2019ezk, Boussarie:2020fpb}. Duclou\'e {\em et al.} have
reformulated~\cite{Ducloue:2019ezk} BK evolution at NLO in terms of
the target rapidity (or Bjorken-$x$).  They obtained an evolution
equation which is non-local in rapidity and which depends explicitly
on the gluon's plus momentum fraction $x=k_g^+/P^+$.  Therefore, it is
important to determine the dependence of the initial scattering
amplitude not only on impact parameter $\vec b$ and dipole size $\vec
r$ but also its dependence on $x$.

The amplitude for $C$-odd three gluon exchange is related to the
correlator of $C$-odd color charge fluctuations~\cite{Dumitru:2018vpr,
  Dumitru:2019qec, Dumitru:2020fdh}\footnote{The sign of ${\cal
    T}_{ggg}(\vec r,\vec b)$ in eq.~(\ref{eq:Odderon-operator})
  differs from ref.~\cite{Dumitru:2019qec} because here we follow the
  convention of Kovchegov and Sievert~\cite{Kovchegov:2012ga} with
  $+ig$ in the exponent of the Wilson line $U(\vec x)$ in
  eq.~(\ref{eq:WilsonLines}), and with the covariant gauge operator
  relation $-\nabla_\perp^2 \int \dd x^- A^{+a}(x^-,\vec x) =
  \rho^a(\vec x)$.},
\bea
-i O(\vec r, \vec b) &=&  
- \frac{5}{18}\, g^6
\int\limits_{q_1, q_2, q_3}
\frac{1}{q_1^2}\frac{1}{q_2^2}\frac{1}{q_3^2}\,
e^{-i \vec b \cdot \vec K}\,
G_3^-(\vec q_1,\vec q_2,\vec q_3)\,\left[
\sin\left(\vec r\cdot \vec q_1 + \frac{1}{2} \vec r\cdot \vec
    K\right)
    - \frac{1}{3}\sin\left(\frac{{1}}{2}\vec r\cdot
  \vec K\right)\right]~.
\label{eq:Odderon-operator}
\eea
Here, $\vec K = - (\vec q_1 + \vec q_2 + \vec q_3)$ is the
(transverse) momentum transfer given $\vec P=0$ for the incoming
proton, and $\int_q$ is shorthand for $\int \dd^2q/(2\pi)^2$. We
denote the $C$-odd part of the correlator of three color charges as
\be
\left<
\rho^a(\vec q_1) \, \rho^b(\vec q_2)\, \rho^c(\vec q_3)\right>_{C=-} \equiv
\frac{1}{4} d^{abc}\, g^3\, G_3^-(\vec q_1,\vec q_2,\vec q_3)
\ee
Note that $G_3^-(\vec q_1,\vec q_2,\vec q_3)$ from
eq.~(\ref{eq:Odderon-operator}) is given by the correlator of three
covariant-gauge color charge densities. However, in the weak field
limit, a computation in light-cone gauge is applicable.

The above correlator is symmetric under a simultaneous
sign flip of all three gluon momenta, and so $-i\mathrm{Im}~{\cal S}(\vec r,\vec b) = -iO(\vec r,\vec
b)$ is imaginary\footnote{In mixed representation $iO(\vec r,\vec K)$
  is real, however.}.  Also, it
vanishes quadratically in any of the transverse momentum arguments so
that $-iO(\vec r,\vec b)$ is free of infrared
divergences. The light-cone gauge color charge density operator in the
eikonal ``shock wave limit'' is given by $\rho^a(\vec k) =
\rho^a_{\mathrm {qu}}(\vec k) + \rho^a_{\mathrm {gl}}(\vec k)$
with~\cite{Dumitru:2020gla}
\bea
\rho^a_\mathrm{qu}(\vec k) = g \sum_{i,j,\sigma} (t^a)_{ij}
\int\frac{\dd x_q \dd^2q}{16\pi^3\, x_q}\,
 b^\dagger_{i\sigma}(x_q,\vec q)\, b_{j\sigma}(x_q,\vec k+\vec q) ~,
 \label{eq:rho_q} \\
\rho_\mathrm{gl}^a(\vec k) = g  \sum_{\lambda b c} (T^a)_{bc}
\int \frac{\dd x_g \dd^2q}{16\pi^3\, x_g}\,
a^\dagger_{b\lambda}(x_g,\vec q) \, a_{c\lambda}(x_g,\vec q + \vec k)~.
 \label{eq:rho_gl}
\eea
Here $a^\dagger, a$ and $b^\dagger, b$ denote creation and
annihilation operators for gluons and quarks, respectively.  \\

In the next section~\ref{sec:rho^3} we compute all contributions to
$\left< \rho^a(\vec q_1) \, \rho^b(\vec q_2)\, \rho^c(\vec
q_3)\right>$ in a proton on the light cone in light-cone gauge;
specifically we consider the NLO correction due to the emission or
exchange of a gluon which is not required to be soft. In
sec.~\ref{eq:sec_G3-_b-space} we describe the Fourier transform of the
correlator to impact parameter space, and present a numerical model
estimate. A brief summary is presented in sec.~\ref{sec:summary}.
Appendix~\ref{sec:P_LF} summarizes the Fock state description of the
proton on the light front used throughout this paper,
appendix~\ref{sec:UVcancel} shows the cancellation of UV divergences
in the sum of all diagrams for $\left< \rho^a(\vec q_1) \, \rho^b(\vec
q_2)\, \rho^c(\vec q_3)\right>$, and in appendix~\ref{sec:Ward} we
check the vanishing of this correlator when $\vec q_1\to 0$ or $\vec
q_3\to 0$.


\section{Correlator of three color charge operators, $\langle
  \rho^a(\vec q_1)\,\rho^b(\vec q_2) \,\rho^c(\vec q_3)\rangle$}
\label{sec:rho^3}

In this section we compute the correlator of three color charge
operators $\langle \rho^a(\vec q_1)\,\rho^b(\vec q_2) \,\rho^c(\vec
q_3)\rangle$ where $\rho^a(\vec q) = \rho^a_\mathrm{gl}(\vec q) +
\rho^a_\mathrm{qu}(\vec q)$.
This expectation value is defined as the
matrix element of the product of three color charge operators between
the incoming ($|P\rangle$) and outgoing ($\langle K|$) proton states,
stripped of the delta functions expressing conservation of L.C.\ and
transverse momentum:
\be
16\pi^3\, P^+ \delta(P^+- K^+)\,\delta(\vec P-\vec K - \sum_i\vec q_i)\,
\langle \rho^a(\vec q_1)\,\rho^b(\vec q_2) \,\rho^c(\vec
q_3)\rangle \equiv \langle K|\, \rho^a(\vec q_1)\,\rho^b(\vec q_2)
\,\rho^c(\vec q_3)\, |P\rangle~.
\label{eq:proton-matrix-element}
\ee
The structure of the proton state assumed in this work is explained
briefly in appendix~\ref{sec:P_LF}.

In general, this correlator has both even and odd components under
$C$-parity which transforms $(t^a)_{ij} \to -
(t^a)_{ji}$~\footnote{Hence, $\tr t^a t^b t^c = \frac{1}{4}(d^{abc} +i
  f^{abc}) \to - \tr t^c t^b t^a = - \frac{1}{4}(d^{abc} -i
  f^{abc})$. Therefore, terms proportional to $d^{abc}$ are odd under
  $C$ conjugation while terms proportional to $if^{abc}$ are even.}
and $(T^a)_{bc} \to -(T^a)_{cb} = (T^a)_{bc}$.~\footnote{One may also
  classify according to ``signature'', i.e.\ the sign under exchange
  of the colors and momenta of any two external gluons (charge
  operators).} Note that the following expressions apply when the number
of colors $N_c=3$.

We shall use the shorthand notation $\vec q = \vec q_1 + \vec q_2 +
\vec q_3= \vec P - \vec K$ and $\vec q_{ij} = \vec q_i + \vec q_j$ in
the following expressions\footnote{Thus, the notation here is
  different from refs.~\cite{Dumitru:2020gla,Dumitru:2021tvw} where
  $\vec q_{12}$ stood for $\vec q_1 - \vec q_2$~!}.  Their
corresponding diagrams are shown in the figures. We label them as
fig.~1$(q_3 q_2 g)$, for example, corresponding to a diagram of the
type shown in fig.~1 (i.e., a gluon exchange across the operator
insertion by a quark with itself, with at least one of the probes
attached to that internal gluon)  where the first probe gluon
  (momentum $\vec q_1$, color $a$) couples to the internal gluon, the
  second probe gluon (momentum $\vec q_2$, color $b$) couples to the
  second quark, and the third probe (momentum $\vec q_3$, color $c$)
  couples to the third quark.

\subsection{UV divergent diagrams}

\begin{figure}[htb]
  \centering
  \begin{minipage}[hb]{\linewidth}
  \includegraphics[width=0.28\linewidth]{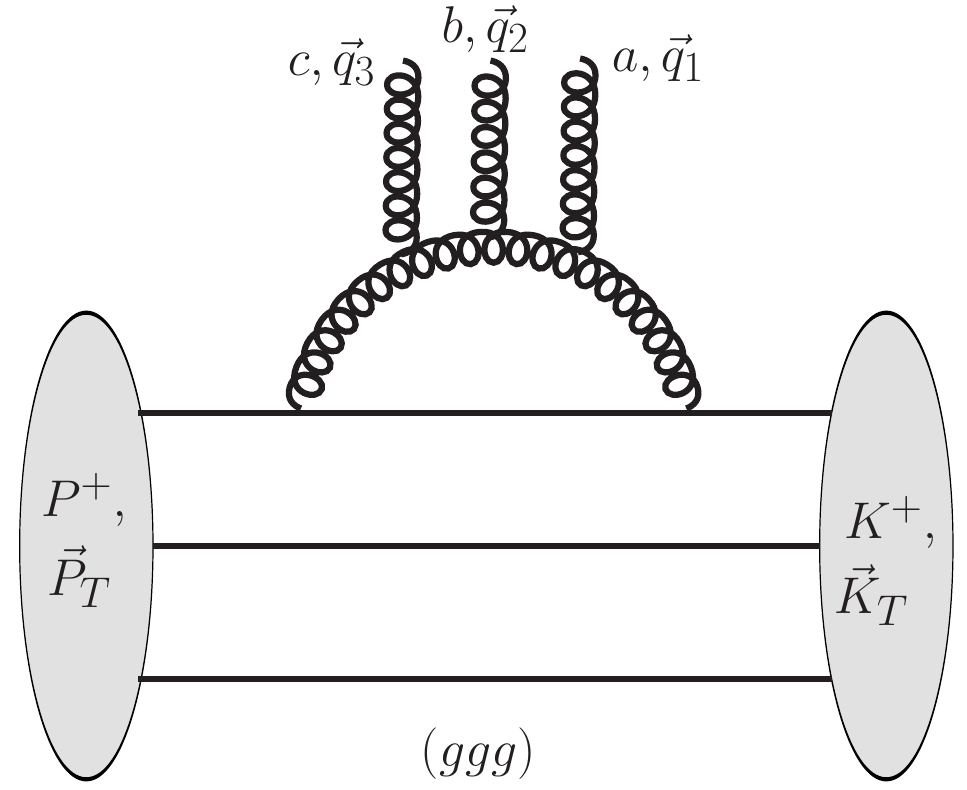}
    \hspace*{.4cm}
    \includegraphics[width=0.28\linewidth]{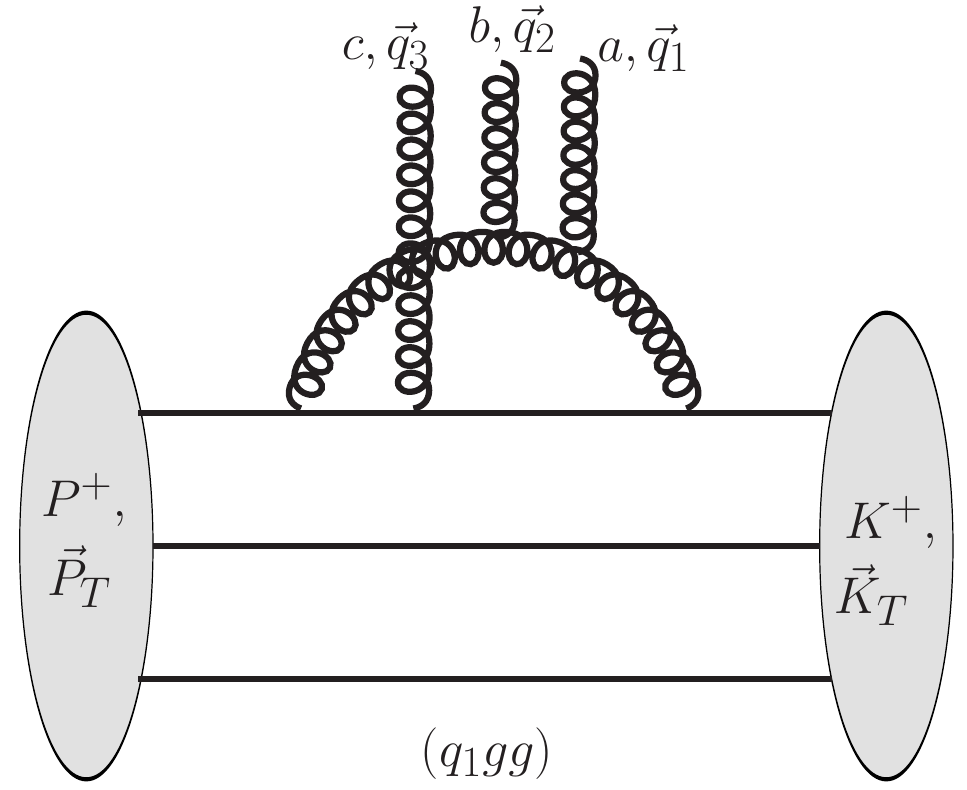}
    \hspace*{.4cm}
    \includegraphics[width=0.28\linewidth]{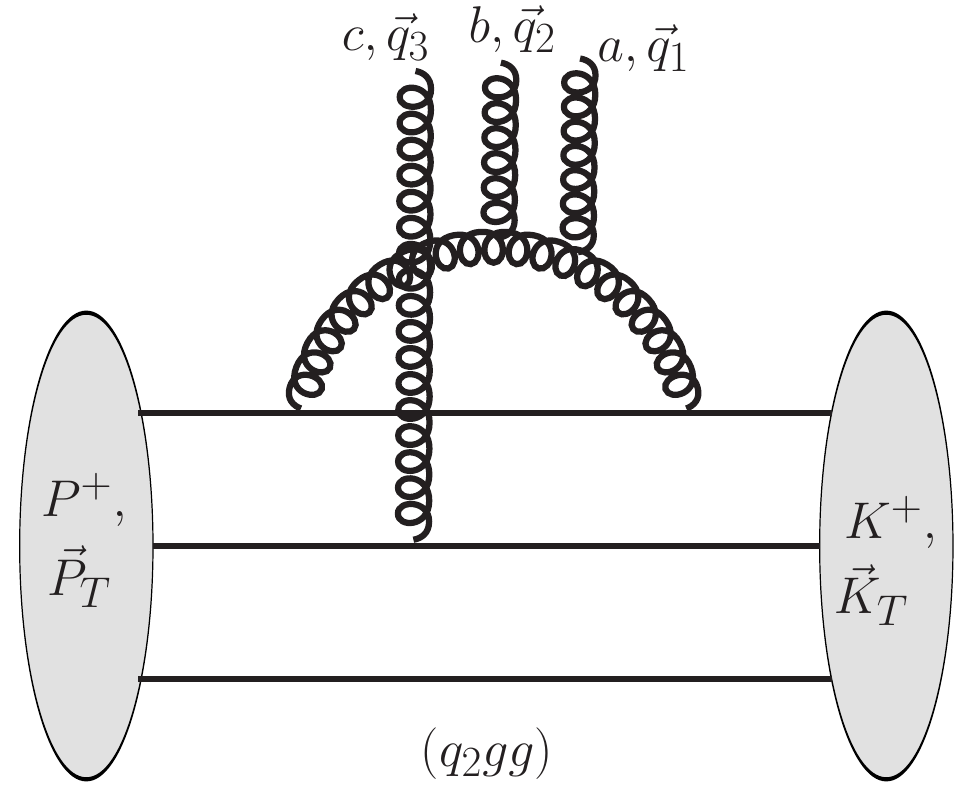}
  \end{minipage}
  \begin{minipage}[hb]{\linewidth}
  \includegraphics[width=0.23\linewidth]{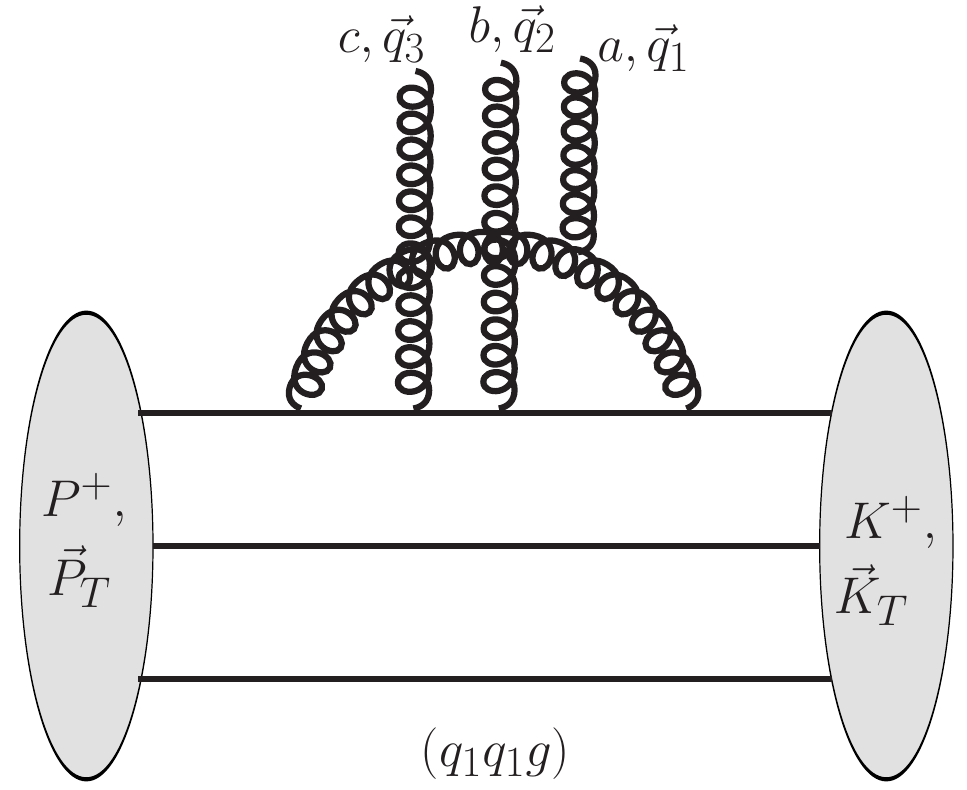}
    \hspace*{.1cm}
    \includegraphics[width=0.23\linewidth]{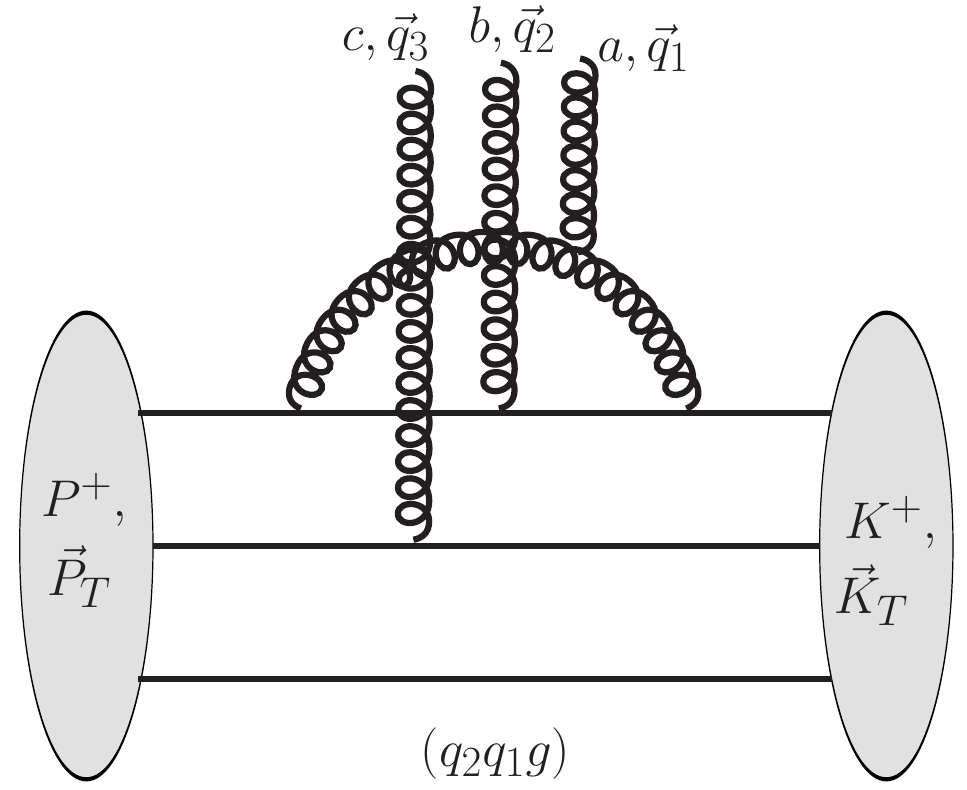}
    \hspace*{.1cm}
    \includegraphics[width=0.23\linewidth]{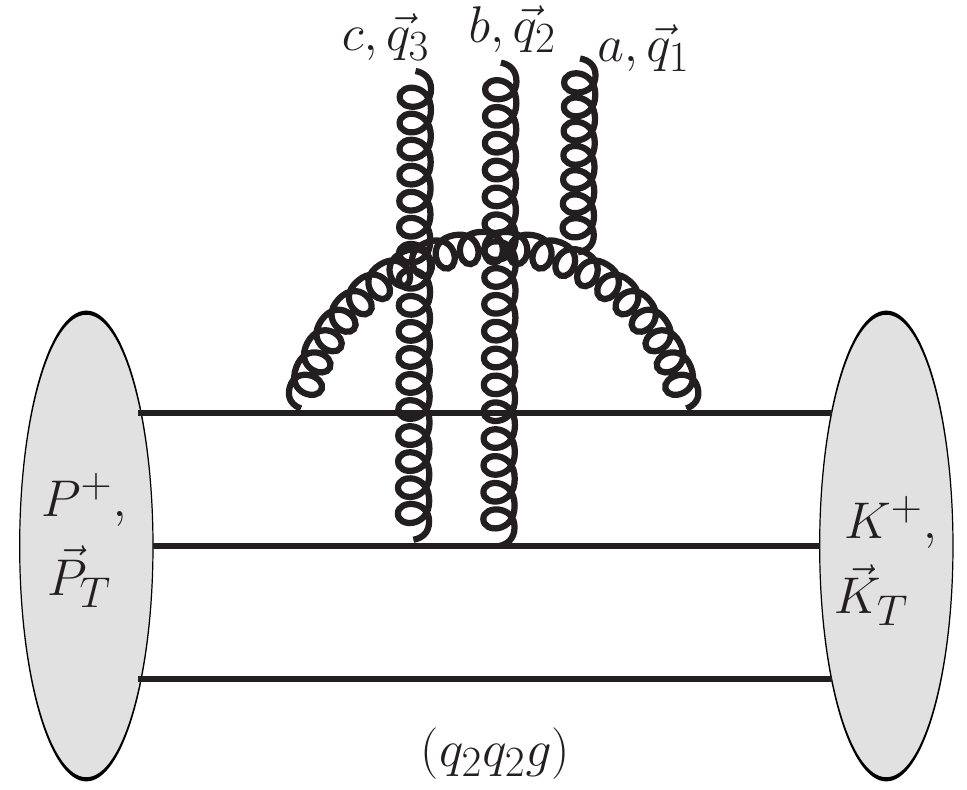}
    \hspace*{.1cm}
    \includegraphics[width=0.23\linewidth]{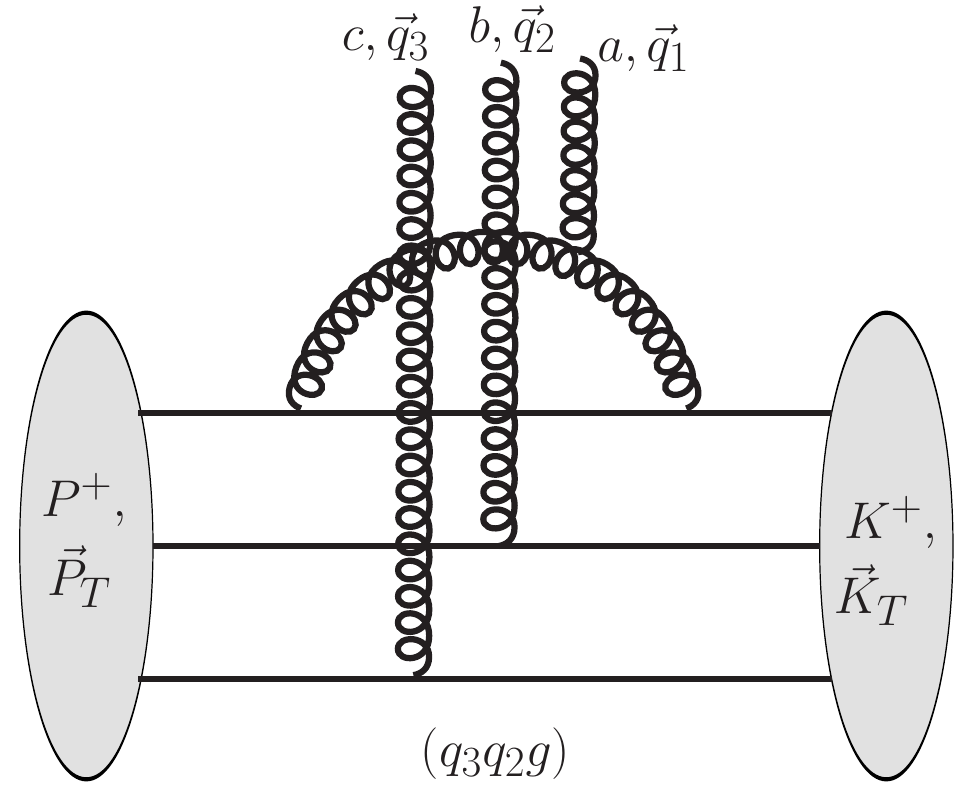}
  \end{minipage}
\caption{UV divergent diagrams (propagators of external probes to be
  amputated) for $\langle \rho^a(\vec q_1)\, \rho^b(\vec
  q_2)\,\rho^c(\vec q_3)\rangle$ where at least one of the probes
  attaches to the gluon in the proton.  The cut is located at the
  insertion of the three color charge operators.}
\label{fig:rho-rho-rho_UVdiv}
\end{figure}
We begin with the UV divergent diagrams where a quark exchanges a
gluon with itself. The diagrams where one or more of the probes attach
to the gluon are shown in fig.~\ref{fig:rho-rho-rho_UVdiv}.\\

To prepare, we first list the matrix elements of one, two, and three
$\rho_\mathrm{gl}(\vec q)$ between one-gluon states:
\bea \label{eq:<g|rho_gl-123|g>}
\left< \ell, \rho, d \vert\, \rho^a_\mathrm{gl}(\vec q_1)\,
\vert k_g, \sigma, c\right> &=& g
\left(T^a\right)_{d c}\, \delta_{\rho\sigma}\,
(2\pi)^{D-1}\, 2k_g^+ \delta(k_g^+ -
\ell^+)\, \delta(\vec k_g - \vec \ell -\vec q_1) \\
\left< \ell, \rho, d \vert\, \rho^a_\mathrm{gl}(\vec q_1)\,
\rho^b_\mathrm{gl}(\vec q_2)\, \vert k_g, \sigma, c\right> &=& g^2
\left(T^a T^b\right)_{d c}\, \delta_{\rho\sigma}\,
(2\pi)^{D-1}\, 2k_g^+ \delta(k_g^+ -
\ell^+)\, \delta(\vec k_g - \vec \ell -\vec q_1 - \vec q_2) \\
\left< \ell, \rho, d \vert\, \rho^a_\mathrm{gl}(\vec q_1)\,
\rho^b_\mathrm{gl}(\vec q_2)\, \rho^c_\mathrm{gl}(\vec q_3)\,
\vert k_g, \sigma, e\right> &=& g^3
\left(T^a T^b T^c\right)_{d e}\, \delta_{\rho\sigma}\,
(2\pi)^{D-1}\, 2k_g^+ \delta(k_g^+ -
\ell^+)\, \delta(\vec k_g - \vec \ell -\vec q_1 - \vec q_2 - \vec q_3)~.
\eea
The matrix elements of $\rho_\mathrm{qu}(\vec q)$ between one-quark states
are similar, with $T^a \to t^a$.
\\

With this we obtain
\bea
\mathrm{fig.}\, \ref{fig:rho-rho-rho_UVdiv}(ggg) &=&
\frac{2g^5}{3\cdot 16\pi^3} \,\tr T^a T^b T^c
\int \left[\dd x_i\right]
  \int \left[\dd^2 k_i\right]\,
  \Psi_{qqq}(x_1,\vec k_1; x_2,\vec k_2; x_3,\vec k_3)\,
\nn\\
& &     \Psi_{qqq}^*(x_1,\vec k_1-(1-x_1)\vec q;
x_2,\vec k_2 +x_2\vec q;
x_3,\vec k_3 +x_3\vec q) \nn\\
& &
\frac{(2\pi)^{D-1}}{2p_1^+} \int \frac{\dk_g}{2(p_1^+ - k_g^+)}
\langle S| \hat\psi_{q\to qg}(\pvec_1; \pvec_1-\kvec_g, \kvec_g)\,
\hat\psi^*_{q\to qg}(\pvec_1-\vec q; \pvec_1-\kvec_g,
\kvec_g-\vec q) |S\rangle~,
     \label{eq:rho-rho-rho_UV_ggg}
\eea
with a symmetry factor of 3.
Here the integration measures $ [\dd k_i]$ and $[\dd x_i]$ and the proton valence quark
wave function $\Psi_{qqq}$ are defined in Appendix~\ref{sec:LFwf}. The Lorentz
invariant gluon phase space measure $\dk_g$ is given in Appendix.~\ref{sec:LFwf_qqqg},  the
phase space integral is calculated in Ref.~\cite{Dumitru:2020gla}, and the result
is also included in Appendix.~\ref{sec:UVcancel}.
Since $\tr T^a T^b T^c = \frac{i}{2}N_c
f^{abc} = - \frac{3}{2} (T^a)_{bc}$ it follows that this contribution
is even under $C$ conjugation. The remaining integral over the
longitudinal and transverse momenta of the emitted gluon can be decomposed
into a finite function and a UV divergent part~\cite{Dumitru:2020gla}.
We verify the cancellation of the UV divergences in appendix~\ref{sec:UVcancel}.

\bea
\mathrm{fig.}\,  \ref{fig:rho-rho-rho_UVdiv}(q_1 g g) &=&
\frac{2g^5}{3\cdot 16\pi^3} \,
\frac{1}{2}\tr\left(T^a T^b D^c - T^a T^b T^c\right)
\int \left[\dd x_i\right]
  \int \left[\dd^2 k_i\right]\,
  \Psi_{qqq}(x_1,\vec k_1; x_2,\vec k_2; x_3,\vec k_3)\,
  \nn\\
     & &     \Psi_{qqq}^*(x_1,\vec k_1-(1-x_1)\vec q;
x_2,\vec k_2 +x_2\vec q;
x_3,\vec k_3 +x_3\vec q) \nn\\  
& &
\frac{(2\pi)^{D-1}}{2p_1^+} \int \frac{\dk_g}{2(p_1^+ - k_g^+)}
\langle S| \hat\psi_{q\to qg}(\pvec_1; \pvec_1-\kvec_g, \kvec_g)\,
\hat\psi^*_{q\to qg}(\pvec_1-\vec q; \pvec_1-\kvec_g-\vec q_3,
\kvec_g-\vec q_{12}) |S\rangle~,
\label{eq:rho-rho-rho_UV_q1gg}
\eea
where $(D^c)_{ab} = d^{cab}$. Performing the traces, the color factor
becomes $\frac{1}{2}\tr\left(D^c T^a T^b - \tr T^b T^c T^a\right) =
\frac{N_c}{4} (d^{abc} - i f^{abc})$. The first term corresponds to a
$C$-odd contribution while the second one is even under $C$. The
symmetry factor for this diagram is 3.

The diagram where instead the second gluon attaches to quark 1 (not shown)
is given by
\bea
\mathrm{fig.}\,  \ref{fig:rho-rho-rho_UVdiv}(g q_1 g) &=&
\frac{2g^5}{3\cdot 16\pi^3} \,
\frac{1}{2}\tr\left(T^a T^c D^b - T^a T^c T^b\right)
\int \left[\dd x_i\right]
  \int \left[\dd^2 k_i\right]\,
  \Psi_{qqq}(x_1,\vec k_1; x_2,\vec k_2; x_3,\vec k_3)\,
  \nn\\
     & &     \Psi_{qqq}^*(x_1,\vec k_1-(1-x_1)\vec q;
x_2,\vec k_2 +x_2\vec q;
x_3,\vec k_3 +x_3\vec q) \nn\\  
& &
\frac{(2\pi)^{D-1}}{2p_1^+} \int \frac{\dk_g}{2(p_1^+ - k_g^+)}
\langle S| \hat\psi_{q\to qg}(\pvec_1; \pvec_1-\kvec_g, \kvec_g)\,
\hat\psi^*_{q\to qg}(\pvec_1-\vec q; \pvec_1-\kvec_g-\vec q_2,
\kvec_g-\vec q_{13}) |S\rangle~,
\label{eq:rho-rho-rho_UV_gq1g}
\eea
and the one where the first gluon attaches to quark 1:
\bea
\mathrm{fig.}\,  \ref{fig:rho-rho-rho_UVdiv}(g g q_1) &=&
\frac{2g^5}{3\cdot 16\pi^3} \,
\frac{1}{2}\tr\left(T^b T^c D^a - T^a T^b T^c\right)
\int \left[\dd x_i\right]
  \int \left[\dd^2 k_i\right]\,
  \Psi_{qqq}(x_1,\vec k_1; x_2,\vec k_2; x_3,\vec k_3)\,
  \nn\\
     & &     \Psi_{qqq}^*(x_1,\vec k_1-(1-x_1)\vec q;
x_2,\vec k_2 +x_2\vec q;
x_3,\vec k_3 +x_3\vec q) \nn\\  
& &
\frac{(2\pi)^{D-1}}{2p_1^+} \int \frac{\dk_g}{2(p_1^+ - k_g^+)}
\langle S| \hat\psi_{q\to qg}(\pvec_1; \pvec_1-\kvec_g, \kvec_g)\,
\hat\psi^*_{q\to qg}(\pvec_1-\vec q; \pvec_1-\kvec_g-\vec q_1,
\kvec_g-\vec q_{23}) |S\rangle~.
\label{eq:rho-rho-rho_UV_ggq1}
\eea
These diagrams come with a symmetry factor of 3 since the ``active'' quark
may just as well be quark 2 or quark 3.\\

Continuing with the diagram where the third probe attaches to quark 2:
\bea
\mathrm{fig.}\, \ref{fig:rho-rho-rho_UVdiv}(q_2 g g) &=&
-\frac{2g^5}{3\cdot 16\pi^3} \,
\frac{1}{4}\tr\left(T^a T^b D^c + T^a T^b T^c\right)
\int \left[\dd x_i\right]
  \int \left[\dd^2 k_i\right]\,
  \Psi_{qqq}(x_1,\vec k_1; x_2,\vec k_2; x_3,\vec k_3)\,
  \nn\\
     & &     \Psi_{qqq}^*(x_1,\vec k_1+x_1\vec q- \vec q_{12};
x_2,\vec k_2 +x_2\vec q - \vec q_3;
x_3,\vec k_3 +x_3\vec q) \nn\\  
& &
\frac{(2\pi)^{D-1}}{2p_1^+} \int \frac{\dk_g}{2(p_1^+ - k_g^+)}
\langle S| \hat\psi_{q\to qg}(\pvec_1; \pvec_1-\kvec_g, \kvec_g)\,
\hat\psi^*_{q\to qg}(\pvec_1-\vec q_{12}; \pvec_1-\kvec_g,
\kvec_g-\vec q_{12}) |S\rangle~.
\label{eq:rho-rho-rho_UV_q2gg}
\eea
The symmetry factor is 6 because the third gluon probe may also attach
to quark 3.

Once again there are analogous diagrams (not shown) where the second or the
first probe attaches to quark 2 (or to quark 3):
\bea
\mathrm{fig.}\, \ref{fig:rho-rho-rho_UVdiv}(g q_2 g) &=&
-\frac{2g^5}{3\cdot 16\pi^3} \,
\frac{1}{4}\tr\left(T^a T^c D^b + T^a T^c T^b\right)
\int \left[\dd x_i\right]
  \int \left[\dd^2 k_i\right]\,
  \Psi_{qqq}(x_1,\vec k_1; x_2,\vec k_2; x_3,\vec k_3)\,
  \nn\\
     & &     \Psi_{qqq}^*(x_1,\vec k_1+x_1\vec q- \vec q_{13};
x_2,\vec k_2 +x_2\vec q - \vec q_2;
x_3,\vec k_3 +x_3\vec q) \nn\\  
& &
\frac{(2\pi)^{D-1}}{2p_1^+} \int \frac{\dk_g}{2(p_1^+ - k_g^+)}
\langle S| \hat\psi_{q\to qg}(\pvec_1; \pvec_1-\kvec_g, \kvec_g)\,
\hat\psi^*_{q\to qg}(\pvec_1-\vec q_{13}; \pvec_1-\kvec_g,
\kvec_g-\vec q_{13}) |S\rangle~,
\label{eq:rho-rho-rho_UV_gq2g}
\eea
\bea
\mathrm{fig.}\, \ref{fig:rho-rho-rho_UVdiv}(g g q_2) &=&
-\frac{2g^5}{3\cdot 16\pi^3} \,
\frac{1}{4}\tr\left(T^b T^c D^a + T^a T^b T^c\right)
\int \left[\dd x_i\right]
  \int \left[\dd^2 k_i\right]\,
  \Psi_{qqq}(x_1,\vec k_1; x_2,\vec k_2; x_3,\vec k_3)\,
  \nn\\
     & &     \Psi_{qqq}^*(x_1,\vec k_1+x_1\vec q- \vec q_{23};
x_2,\vec k_2 +x_2\vec q - \vec q_1;
x_3,\vec k_3 +x_3\vec q) \nn\\  
& &
\frac{(2\pi)^{D-1}}{2p_1^+} \int \frac{\dk_g}{2(p_1^+ - k_g^+)}
\langle S| \hat\psi_{q\to qg}(\pvec_1; \pvec_1-\kvec_g, \kvec_g)\,
\hat\psi^*_{q\to qg}(\pvec_1-\vec q_{23}; \pvec_1-\kvec_g,
\kvec_g-\vec q_{23}) |S\rangle~.
\label{eq:rho-rho-rho_UV_ggq2}
\eea
Their symmetry factors are 6.\\

We continue with the diagrams where two of the probes attach to quarks.
\bea
\mathrm{fig.}\, \ref{fig:rho-rho-rho_UVdiv}(q_1 q_1 g) &=&
-\frac{2g^5}{3\cdot 16\pi^3} \,
N_c \tr t^a t^b t^c
\int \left[\dd x_i\right]
  \int \left[\dd^2 k_i\right]\,
  \Psi_{qqq}(x_1,\vec k_1; x_2,\vec k_2; x_3,\vec k_3)\,
  \nn\\
     & &     \Psi_{qqq}^*(x_1,\vec k_1+x_1\vec q- \vec q;
x_2,\vec k_2 +x_2\vec q;
x_3,\vec k_3 +x_3\vec q) \nn\\  
& &
\frac{(2\pi)^{D-1}}{2p_1^+} \int \frac{\dk_g}{2(p_1^+ - k_g^+)}
\langle S| \hat\psi_{q\to qg}(\pvec_1; \pvec_1-\kvec_g, \kvec_g)\,
\hat\psi^*_{q\to qg}(\pvec_1-\vec q; \pvec_1-\kvec_g-\vec q_{23},
\kvec_g-\vec q_{1}) |S\rangle~.
\label{eq:rho-rho-rho_UV_q1q1g}
\eea
The symmetry factor is 3. The SU($N_c$) relation $f^{abc} t^a t^b =
\frac{i}{2} N_c \, t^c$ is useful for evaluating the color factor for
this diagram. Performing the trace, $\tr t^a t^b t^c =
\frac{1}{4}(d^{abc} + i f^{abc})$, separates the $C$-odd contribution
proportional to $d^{abc}$ from the $C$-even contribution proportional
to $if^{abc}$.

\bea
\mathrm{fig.}\, \ref{fig:rho-rho-rho_UVdiv}(q_1 g q_1 ) &=&
-\frac{2g^5}{3\cdot 16\pi^3} \,
N_c \tr t^b t^a t^c
\int \left[\dd x_i\right]
  \int \left[\dd^2 k_i\right]\,
  \Psi_{qqq}(x_1,\vec k_1; x_2,\vec k_2; x_3,\vec k_3)\,
  \nn\\
     & &     \Psi_{qqq}^*(x_1,\vec k_1+x_1\vec q- \vec q;
x_2,\vec k_2 +x_2\vec q;
x_3,\vec k_3 +x_3\vec q) \nn\\  
& &
\frac{(2\pi)^{D-1}}{2p_1^+} \int \frac{\dk_g}{2(p_1^+ - k_g^+)}
\langle S| \hat\psi_{q\to qg}(\pvec_1; \pvec_1-\kvec_g, \kvec_g)\,
\hat\psi^*_{q\to qg}(\pvec_1-\vec q; \pvec_1-\kvec_g-\vec q_{13},
\kvec_g-\vec q_{2}) |S\rangle~.
\label{eq:rho-rho-rho_UV_q1gq1}
\eea
The symmetry factor is 3.
\bea
\mathrm{fig.}\, \ref{fig:rho-rho-rho_UVdiv}(g q_1 q_1 ) &=&
-\frac{2g^5}{3\cdot 16\pi^3} \,
N_c \tr t^a t^b t^c
\int \left[\dd x_i\right]
  \int \left[\dd^2 k_i\right]\,
  \Psi_{qqq}(x_1,\vec k_1; x_2,\vec k_2; x_3,\vec k_3)\,
  \nn\\
     & &     \Psi_{qqq}^*(x_1,\vec k_1+x_1\vec q- \vec q;
x_2,\vec k_2 +x_2\vec q;
x_3,\vec k_3 +x_3\vec q) \nn\\  
& &
\frac{(2\pi)^{D-1}}{2p_1^+} \int \frac{\dk_g}{2(p_1^+ - k_g^+)}
\langle S| \hat\psi_{q\to qg}(\pvec_1; \pvec_1-\kvec_g, \kvec_g)\,
\hat\psi^*_{q\to qg}(\pvec_1-\vec q; \pvec_1-\kvec_g-\vec q_{12},
\kvec_g-\vec q_{3}) |S\rangle~.
\label{eq:rho-rho-rho_UV_gq1q1}
\eea
The symmetry factor is 3.
\bea
\mathrm{fig.}\, \ref{fig:rho-rho-rho_UVdiv}(q_2 q_1 g) &=& 0~.
\label{eq:rho-rho-rho_UV_q2q1g}
\eea
The symmetry factor is 6, to include the contribution where the third
gluon attaches to quark 3.
\bea
\mathrm{fig.}\, \ref{fig:rho-rho-rho_UVdiv}(q_1 q_2 g) &=&
0~.
\label{eq:rho-rho-rho_UV_q1q2g}
\eea
The symmetry factor is 6.
\bea
\mathrm{fig.}\, \ref{fig:rho-rho-rho_UVdiv}(q_2 g q_1) &=&
0~.
\label{eq:rho-rho-rho_UV_q2gq1}
\eea
The symmetry factor is 6.
\bea
\mathrm{fig.}\, \ref{fig:rho-rho-rho_UVdiv}(q_1 g q_2) &=&
0~.
\label{eq:rho-rho-rho_UV_q1gq2}
\eea
The symmetry factor is 6.
\bea
\mathrm{fig.}\, \ref{fig:rho-rho-rho_UVdiv}(g q_2 q_1) &=&
0~.
\label{eq:rho-rho-rho_UV_gq2q1}
\eea
The symmetry factor is 6.
\bea
\mathrm{fig.}\, \ref{fig:rho-rho-rho_UVdiv}(g q_1 q_2) &=&
0~.
\label{eq:rho-rho-rho_UV_gq1q2}
\eea
The symmetry factor is 6.
\bea
\mathrm{fig.}\, \ref{fig:rho-rho-rho_UVdiv}(q_2 q_2 g) &=&
-\frac{2g^5}{3\cdot 16\pi^3} \,
\frac{1}{2} N_c \, \tr t^a t^b t^c\,
\int \left[\dd x_i\right]
  \int \left[\dd^2 k_i\right]\,
  \Psi_{qqq}(x_1,\vec k_1; x_2,\vec k_2; x_3,\vec k_3)\,
  \nn\\
     & &     \Psi_{qqq}^*(x_1,\vec k_1+x_1\vec q- \vec q_{1};
x_2,\vec k_2 +x_2\vec q - \vec q_{23};
x_3,\vec k_3 +x_3\vec q) \nn\\  
& &
\frac{(2\pi)^{D-1}}{2p_1^+} \int \frac{\dk_g}{2(p_1^+ - k_g^+)}
\langle S| \hat\psi_{q\to qg}(\pvec_1; \pvec_1-\kvec_g, \kvec_g)\,
\hat\psi^*_{q\to qg}(\pvec_1-\vec q_{1}; \pvec_1-\kvec_g,
\kvec_g-\vec q_{1}) |S\rangle~.
\label{eq:rho-rho-rho_UV_q2q2g}
\eea
The symmetry factor is 6.
\bea
\mathrm{fig.}\, \ref{fig:rho-rho-rho_UVdiv}(q_2 g q_2) &=&
-\frac{2g^5}{3\cdot 16\pi^3} \,
\frac{1}{2} N_c \, \tr t^a t^c t^b\,
\int \left[\dd x_i\right]
  \int \left[\dd^2 k_i\right]\,
  \Psi_{qqq}(x_1,\vec k_1; x_2,\vec k_2; x_3,\vec k_3)\,
  \nn\\
     & &     \Psi_{qqq}^*(x_1,\vec k_1+x_1\vec q- \vec q_{2};
x_2,\vec k_2 +x_2\vec q - \vec q_{13};
x_3,\vec k_3 +x_3\vec q) \nn\\  
& &
\frac{(2\pi)^{D-1}}{2p_1^+} \int \frac{\dk_g}{2(p_1^+ - k_g^+)}
\langle S| \hat\psi_{q\to qg}(\pvec_1; \pvec_1-\kvec_g, \kvec_g)\,
\hat\psi^*_{q\to qg}(\pvec_1-\vec q_{2}; \pvec_1-\kvec_g,
\kvec_g-\vec q_{2}) |S\rangle~.
\label{eq:rho-rho-rho_UV_q2gq2}
\eea
The symmetry factor is 6.
\bea
\mathrm{fig.}\, \ref{fig:rho-rho-rho_UVdiv}(gq_2 q_2) &=&
-\frac{2g^5}{3\cdot 16\pi^3} \,
\frac{1}{2} N_c \, \tr t^a t^b t^c\,
\int \left[\dd x_i\right]
  \int \left[\dd^2 k_i\right]\,
  \Psi_{qqq}(x_1,\vec k_1; x_2,\vec k_2; x_3,\vec k_3)\,
  \nn\\
     & &     \Psi_{qqq}^*(x_1,\vec k_1+x_1\vec q- \vec q_{3};
x_2,\vec k_2 +x_2\vec q - \vec q_{12};
x_3,\vec k_3 +x_3\vec q) \nn\\  
& &
\frac{(2\pi)^{D-1}}{2p_1^+} \int \frac{\dk_g}{2(p_1^+ - k_g^+)}
\langle S| \hat\psi_{q\to qg}(\pvec_1; \pvec_1-\kvec_g, \kvec_g)\,
\hat\psi^*_{q\to qg}(\pvec_1-\vec q_{3}; \pvec_1-\kvec_g,
\kvec_g-\vec q_{3}) |S\rangle~.
\label{eq:rho-rho-rho_UV_gq2q2}
\eea
The symmetry factor is 6.
\bea
\mathrm{fig.}\, \ref{fig:rho-rho-rho_UVdiv}(q_3 q_2g) &=&
\frac{2g^5}{3\cdot 16\pi^3} \,
\frac{1}{2} N_c \tr (t^a t^b t^c + t^a t^c t^b)
\int \left[\dd x_i\right]
  \int \left[\dd^2 k_i\right]\,
  \Psi_{qqq}(x_1,\vec k_1; x_2,\vec k_2; x_3,\vec k_3)\,
  \nn\\
     & &     \Psi_{qqq}^*(x_1,\vec k_1+x_1\vec q- \vec q_{1};
x_2,\vec k_2 +x_2\vec q - \vec q_{2};
x_3,\vec k_3 +x_3\vec q - \vec q_{3}) \nn\\  
& &
\frac{(2\pi)^{D-1}}{2p_1^+} \int \frac{\dk_g}{2(p_1^+ - k_g^+)}
\langle S| \hat\psi_{q\to qg}(\pvec_1; \pvec_1-\kvec_g, \kvec_g)\,
\hat\psi^*_{q\to qg}(\pvec_1-\vec q_{1}; \pvec_1-\kvec_g,
\kvec_g-\vec q_{1}) |S\rangle~.
\label{eq:rho-rho-rho_UV_q3q2g}
\eea
The symmetry factor is 6.
\bea
\mathrm{fig.}\, \ref{fig:rho-rho-rho_UVdiv}(q_3 g q_2) &=&
\frac{2g^5}{3\cdot 16\pi^3} \,
\frac{1}{2} N_c \tr (t^a t^b t^c + t^a t^c t^b)
\int \left[\dd x_i\right]
  \int \left[\dd^2 k_i\right]\,
  \Psi_{qqq}(x_1,\vec k_1; x_2,\vec k_2; x_3,\vec k_3)\,
  \nn\\
     & &     \Psi_{qqq}^*(x_1,\vec k_1+x_1\vec q- \vec q_{2};
x_2,\vec k_2 +x_2\vec q - \vec q_{1};
x_3,\vec k_3 +x_3\vec q - \vec q_{3}) \nn\\  
& &
\frac{(2\pi)^{D-1}}{2p_1^+} \int \frac{\dk_g}{2(p_1^+ - k_g^+)}
\langle S| \hat\psi_{q\to qg}(\pvec_1; \pvec_1-\kvec_g, \kvec_g)\,
\hat\psi^*_{q\to qg}(\pvec_1-\vec q_{2}; \pvec_1-\kvec_g,
\kvec_g-\vec q_{2}) |S\rangle~.
\label{eq:rho-rho-rho_UV_q3gq2}
\eea
The symmetry factor is 6.
\bea
\mathrm{fig.}\, \ref{fig:rho-rho-rho_UVdiv}(g q_3 q_2) &=&
\frac{2g^5}{3\cdot 16\pi^3} \,
\frac{1}{2} N_c \tr (t^a t^b t^c + t^a t^c t^b)
\int \left[\dd x_i\right]
  \int \left[\dd^2 k_i\right]\,
  \Psi_{qqq}(x_1,\vec k_1; x_2,\vec k_2; x_3,\vec k_3)\,
  \nn\\
     & &     \Psi_{qqq}^*(x_1,\vec k_1+x_1\vec q- \vec q_{3};
x_2,\vec k_2 +x_2\vec q - \vec q_{1};
x_3,\vec k_3 +x_3\vec q - \vec q_{2}) \nn\\  
& &
\frac{(2\pi)^{D-1}}{2p_1^+} \int \frac{\dk_g}{2(p_1^+ - k_g^+)}
\langle S| \hat\psi_{q\to qg}(\pvec_1; \pvec_1-\kvec_g, \kvec_g)\,
\hat\psi^*_{q\to qg}(\pvec_1-\vec q_{3}; \pvec_1-\kvec_g,
\kvec_g-\vec q_{3}) |S\rangle~.
\label{eq:rho-rho-rho_UV_gq3q2}
\eea
The symmetry factor is 6.
\\

\begin{figure}[htb]
\centerline{
    \includegraphics[width=0.3\linewidth]{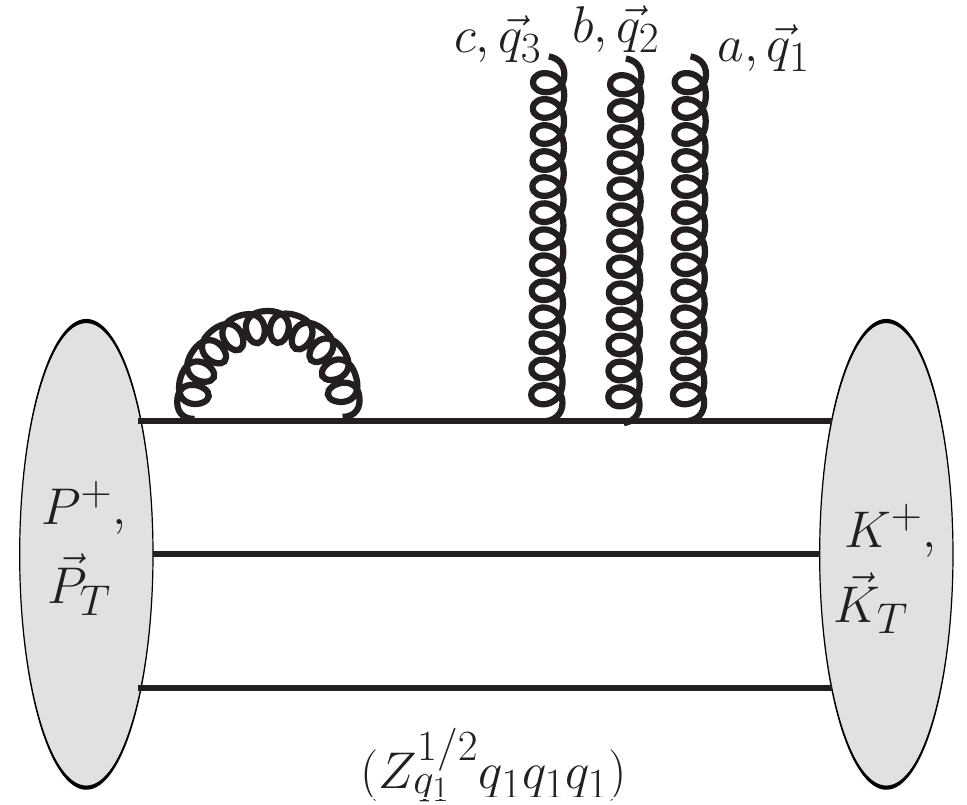}
    \hspace*{.4cm}
  \includegraphics[width=0.3\linewidth]{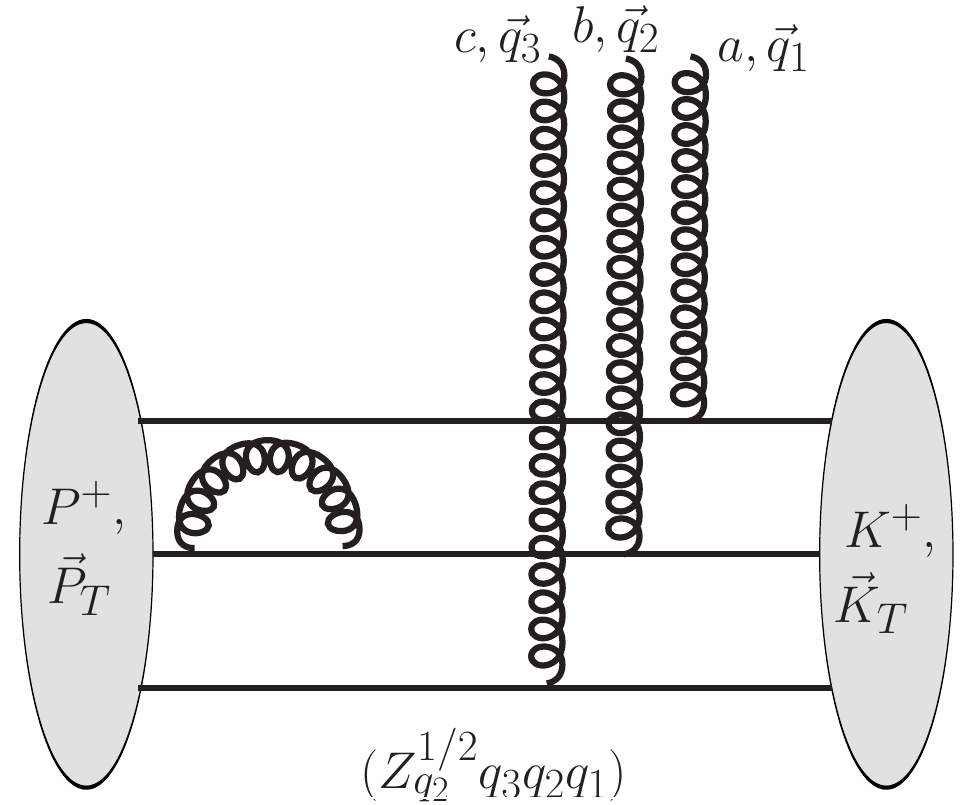}
  }
\caption{Two of the diagrams for $\langle \rho^a_\mathrm{qu}(\vec
  q_1)\, \rho^b_\mathrm{qu}(\vec q_2)\, \rho^c_\mathrm{qu}(\vec
  q_3)\rangle$ in the three-quark Fock state which involve the quark
  wave function renormalization factor $Z_q^{1/2}(x)$ (``virtual
  corrections'').  The cut is located at the insertion of the three
  color charge operators.}
\label{fig:rho-rho-rho_qu_Z}
\end{figure}
We now proceed to the diagrams where all three gluon probes couple to quarks.
First, there is the expectation value of $\rho_\mathrm{qu}(\vec q_1)\,
\rho_\mathrm{qu}(\vec q_2)\,\rho_\mathrm{qu}(\vec q_3)$ between $|qqq\rangle$
three quark states:
\bea
\left< \rho^a(\vec q_1) \, \rho^b(\vec q_2) \, \rho^c(\vec
q_3) \,\right>
&=& \frac{g^3}{6} \,
\int [\dd x_i]
\int [\dd^2 k_i]\,
\Bigl[ \nn\\
  & &
  \tr t^a t^b t^c\, \Psi^*(x_1,\vec k_1-\vec q+x_1q;  x_2, \vec{k}_2+x_2\vec q
; x_3, \vec k_3+x_3\vec q) \nonumber\\
& & - \tr t^a t^b t^c\, 
  \Psi^*(x_1,\vec k_1-\vec q_1 +x_1\vec q ;  x_2, \vec k_2-\vec q_{23}
  + x_2\vec q; x_3, \vec k_3+x_3\vec q) \nonumber\\
& &  - \tr t^a t^c t^b\, 
 \Psi^*(x_1,\vec k_1-\vec q_{13} +x_1\vec q; x_2,\vec k_2-\vec q_2
 + x_2\vec q; x_3, \vec k_3+x_3\vec q) \nonumber\\
& & - \tr t^a t^b t^c\, 
  \Psi^*(x_1,\vec k_1-\vec q_{12} +x_1 \vec q;  x_2,
  \vec k_2-\vec q_3+ x_2\vec q; x_3,
  \vec k_3+x_3\vec q) \nonumber\\
& & + (\tr t^a t^b t^c + \tr t^a t^c t^b)\,
  \Psi^*(x_1,\vec k_1-\vec q_1 +x_1\vec q; x_2,
  \vec k_2-\vec q_2 +x_2 \vec q; x_3,
  \vec k_3-\vec q_3 + x_3\vec q) \nn\\
  & & \Bigr]\, \psi(x_1,\vec k_1; x_2, \vec k_2 ; x_3,\vec k_3)
~.
\label{eq:rho3_qqq}
\eea
The first to fifth term, respectively, correspond to
fig.~\ref{fig:rho-rho-rho_qu_Z}($q_1 q_1 q_1$),
fig.~\ref{fig:rho-rho-rho_qu_Z}($q_2 q_2 q_1$) +
fig.~\ref{fig:rho-rho-rho_qu_Z}($q_3 q_3 q_1$),
fig.~\ref{fig:rho-rho-rho_qu_Z}($q_1 q_2 q_1$) +
fig.~\ref{fig:rho-rho-rho_qu_Z}($q_1 q_3 q_1$),
fig.~\ref{fig:rho-rho-rho_qu_Z}($q_2 q_1 q_1$) +
fig.~\ref{fig:rho-rho-rho_qu_Z}($q_3 q_1 q_1$),
fig.~\ref{fig:rho-rho-rho_qu_Z}($q_3 q_2 q_1$) +
fig.~\ref{fig:rho-rho-rho_qu_Z}($q_2 q_3 q_1$).  The entire expression
comes with a symmetry factor of 3.  To account for the quark wave
function renormalization factor, we multiply it by $Z_q(x_1)\,
Z_q(x_2)\, Z_q(x_3) = 1 - C_q(x_1) - C_q(x_2) - C_q(x_3)$ where
$C_q(x) = {\cal O}(g^2)$. The explicit expression for $C_q(x)$ in the
$\overline{MS}$ scheme is given in ref.~\cite{Dumitru:2020gla} but is
not needed here because we will verify that all UV divergences cancel.\\

\begin{figure}[htb]
  \centering
  \begin{minipage}[hb]{\linewidth}
  \includegraphics[width=0.18\linewidth]{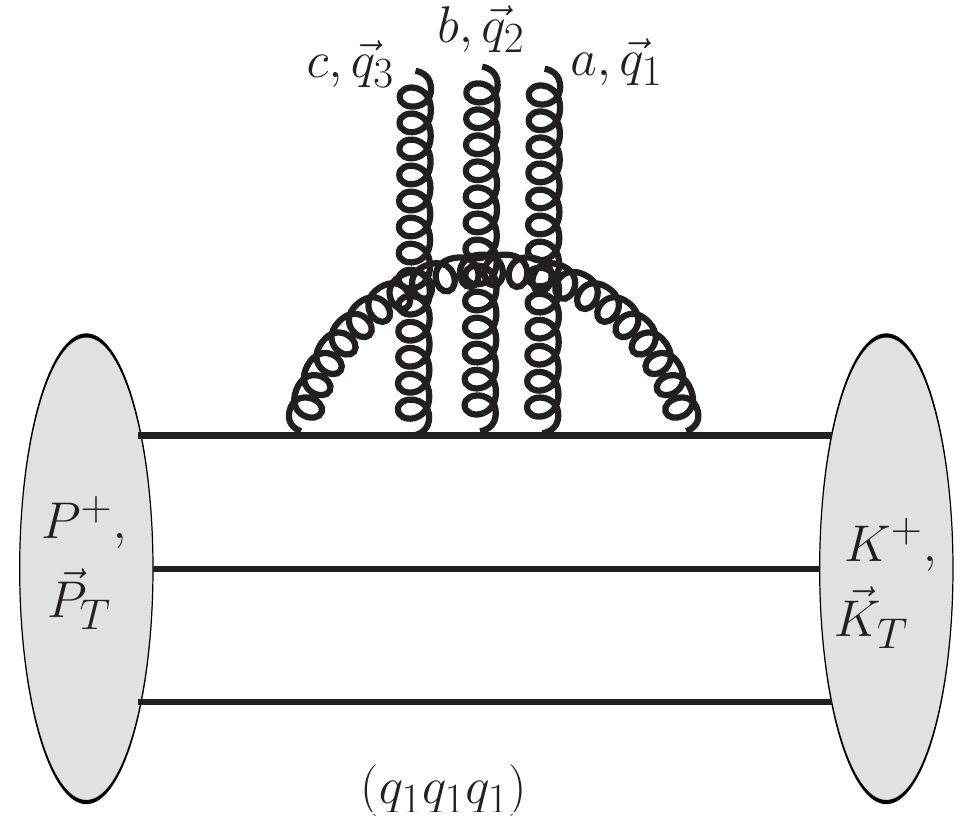}
  \hspace*{.1cm}
  \includegraphics[width=0.18\linewidth]{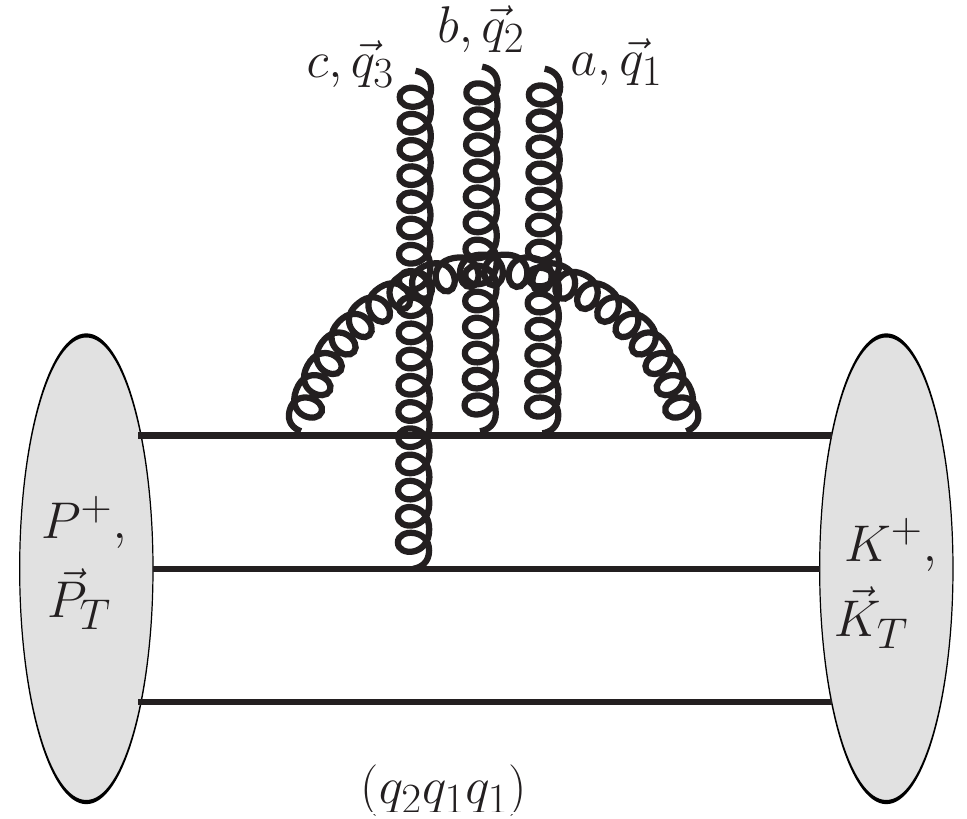}
  \hspace*{.1cm}
  \includegraphics[width=0.18\linewidth]{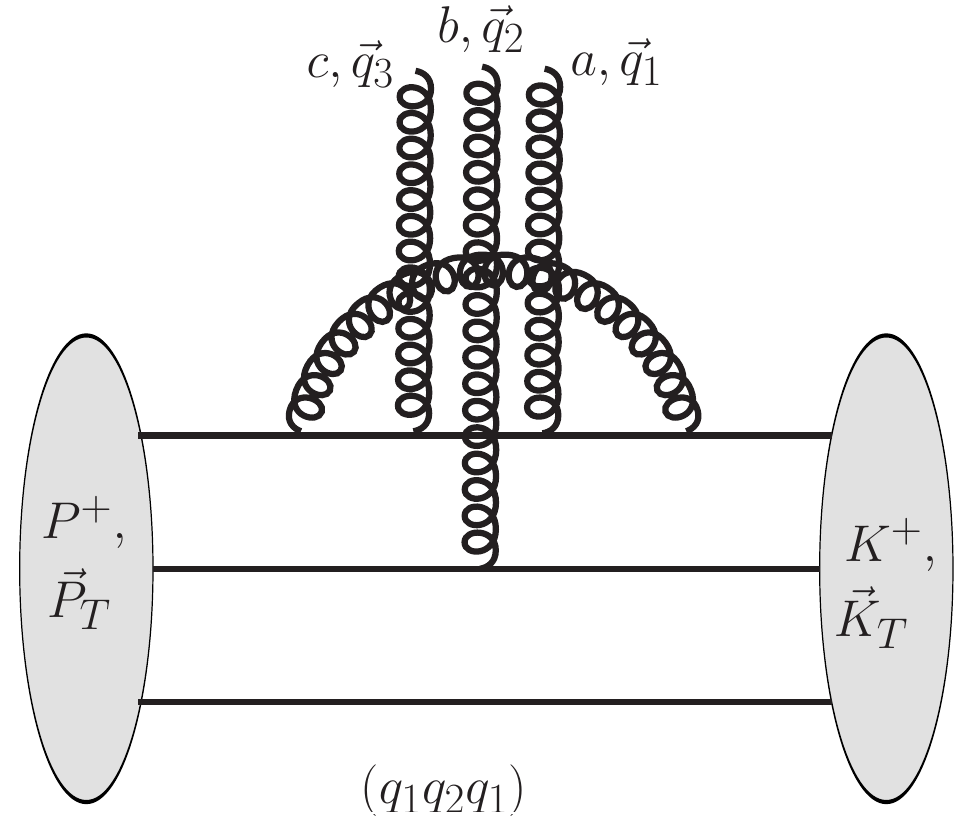}
  \hspace*{.1cm}
  \includegraphics[width=0.18\linewidth]{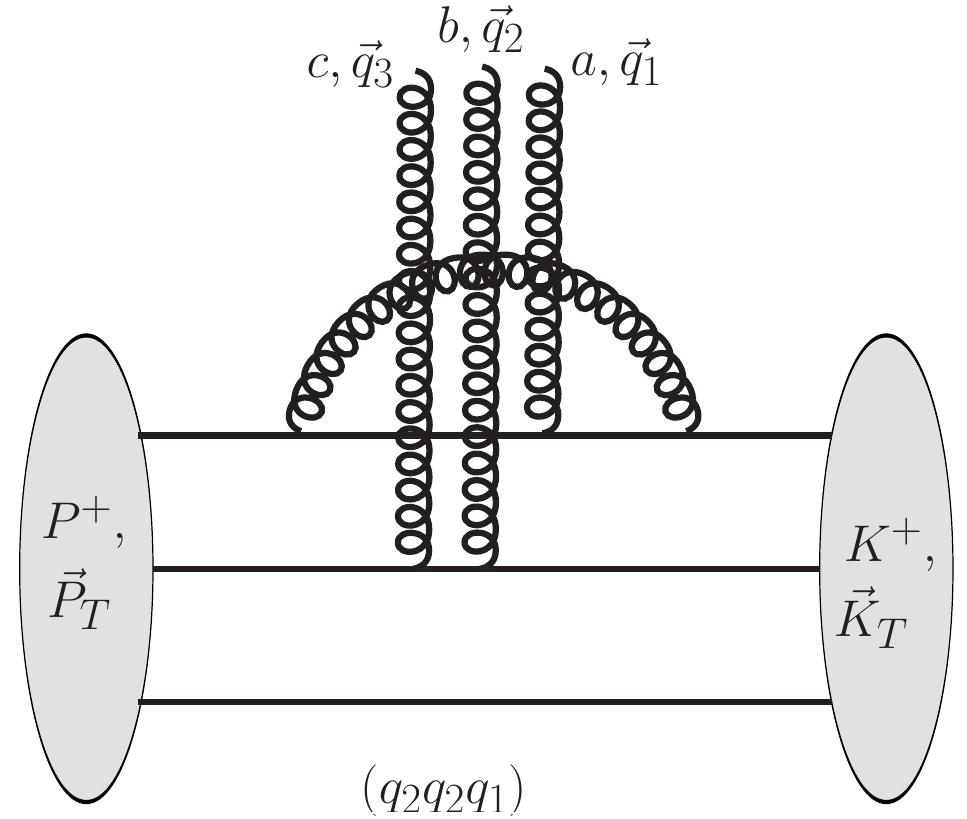}
  \hspace*{.1cm}
  \includegraphics[width=0.18\linewidth]{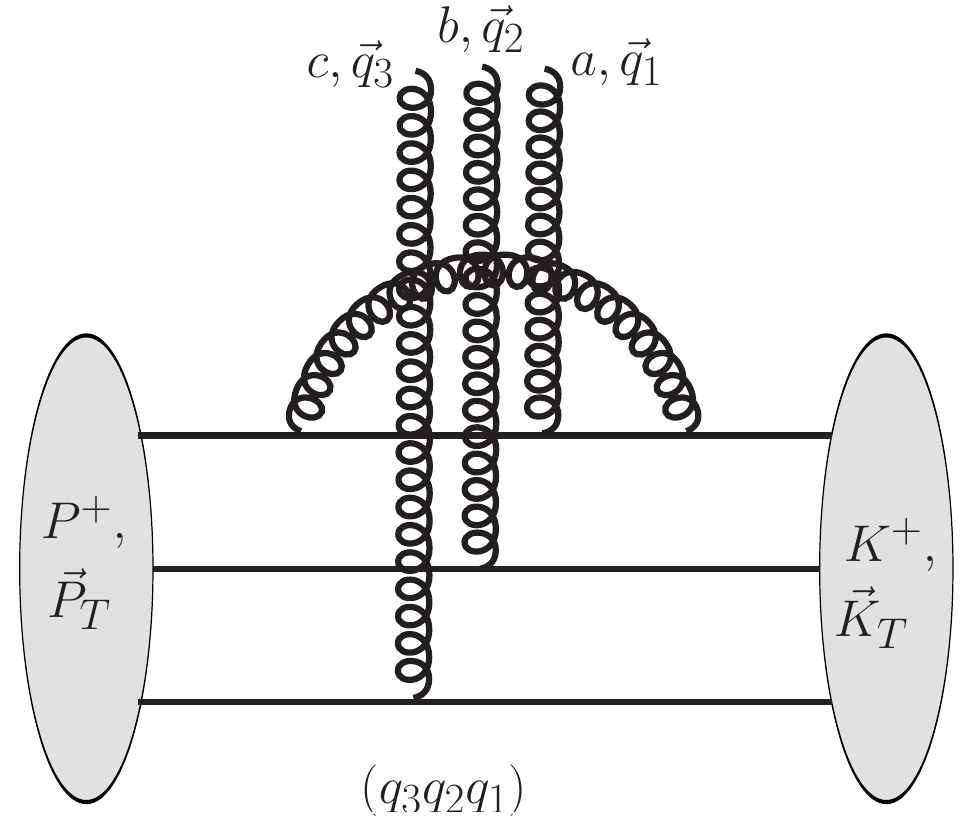}
  \end{minipage}
  \begin{minipage}[hb]{\linewidth}
  \includegraphics[width=0.18\linewidth]{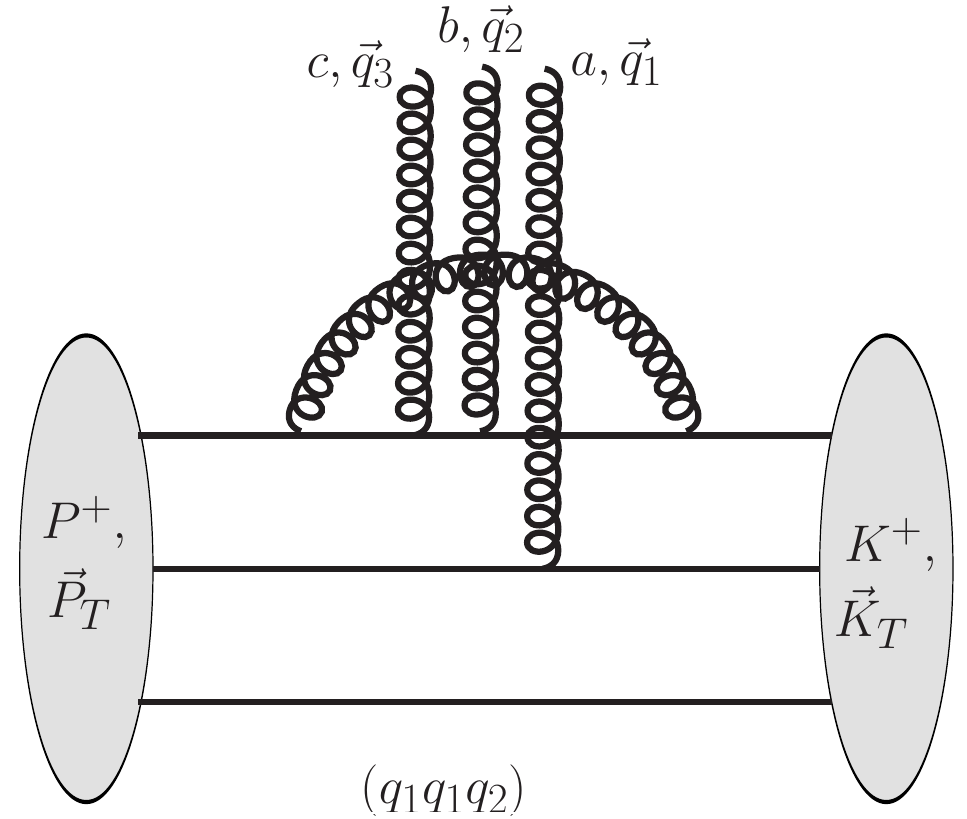}
  \hspace*{.1cm}
  \includegraphics[width=0.18\linewidth]{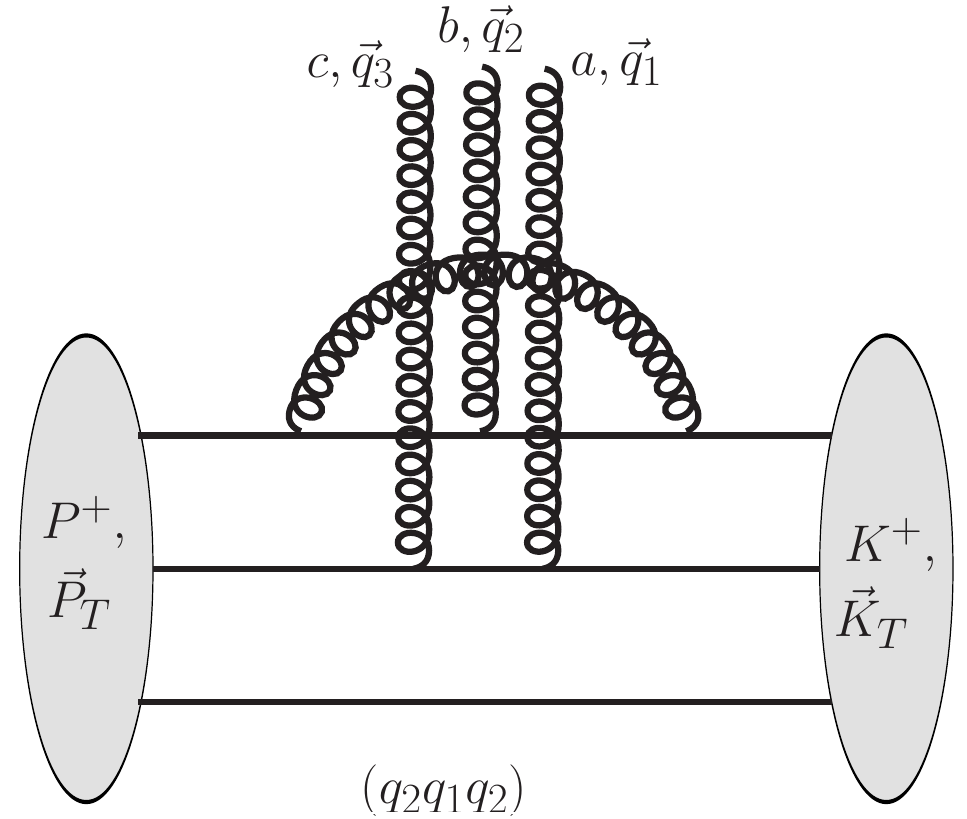}
  \hspace*{.1cm}
  \includegraphics[width=0.18\linewidth]{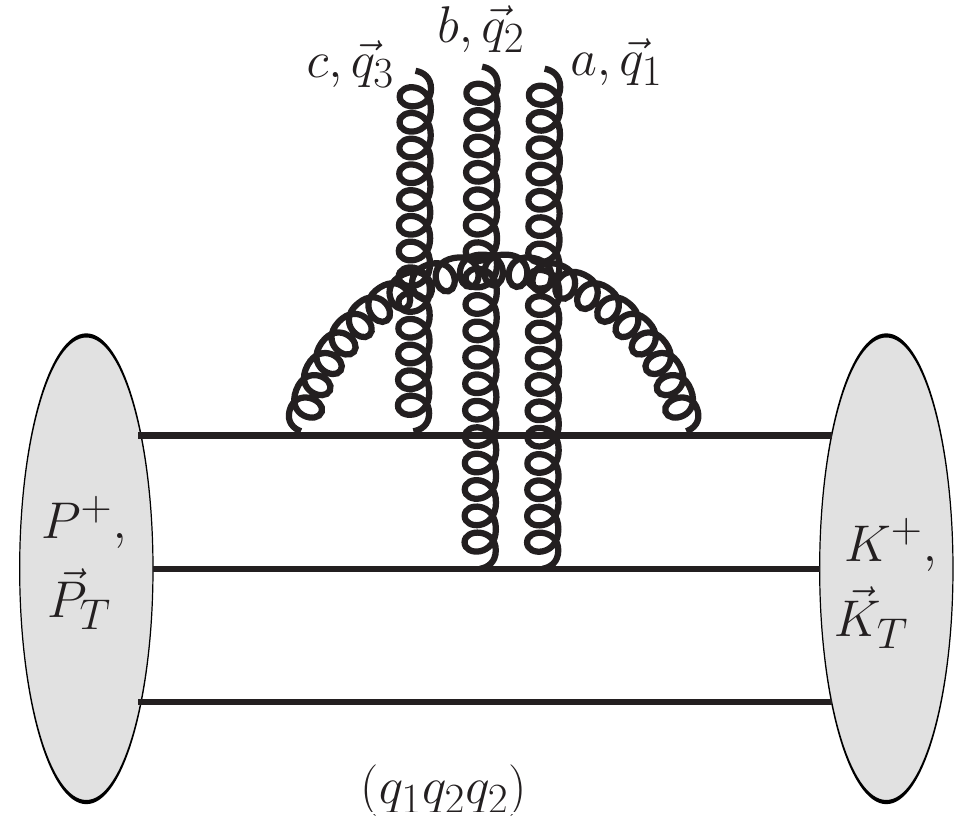}
  \hspace*{.1cm}
  \includegraphics[width=0.18\linewidth]{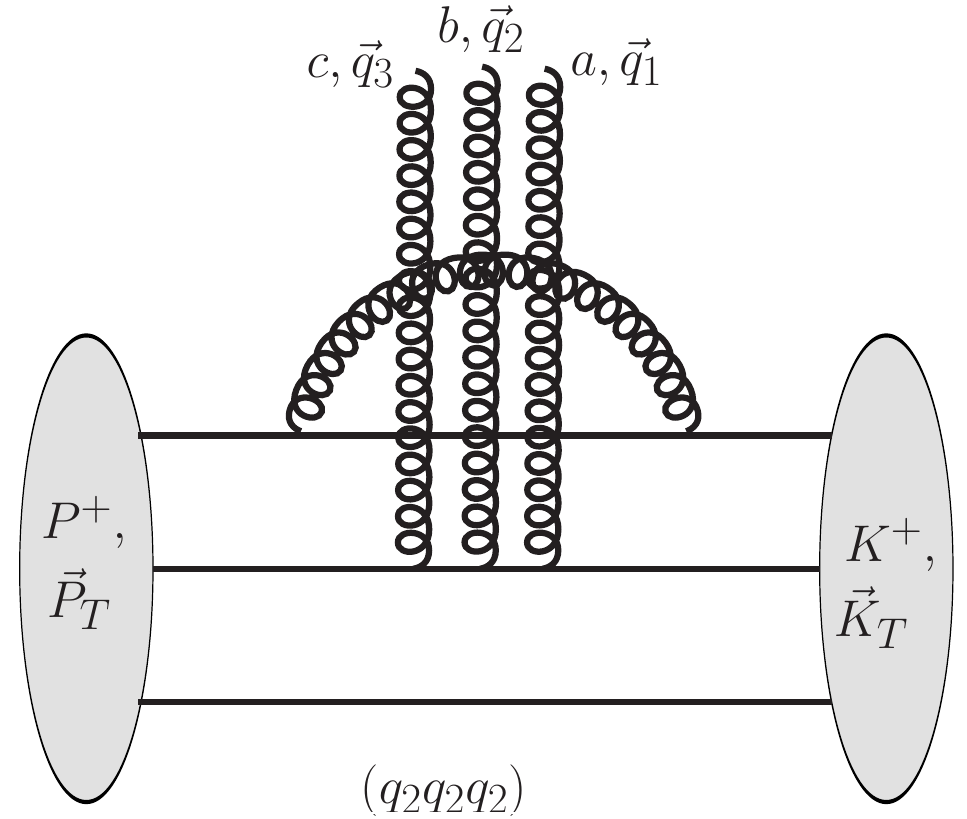}
  \end{minipage}
  \begin{minipage}[hb]{\linewidth}
  \includegraphics[width=0.18\linewidth]{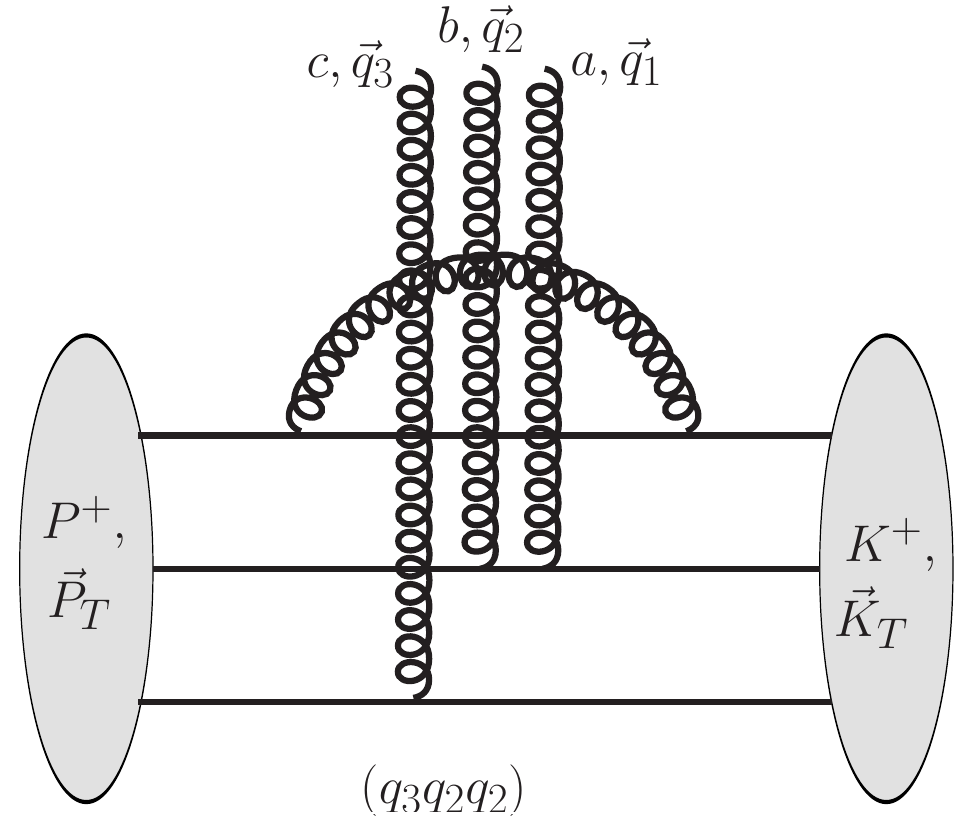}
  \hspace*{.1cm}
  \includegraphics[width=0.18\linewidth]{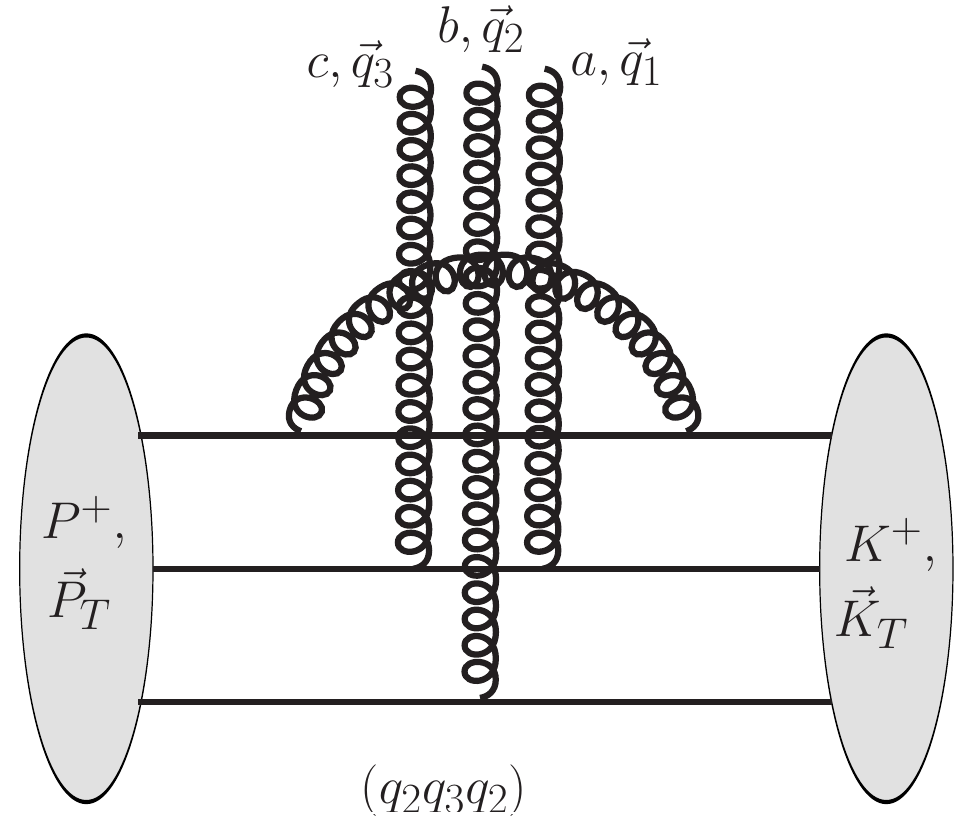}
  \hspace*{.1cm}
  \includegraphics[width=0.18\linewidth]{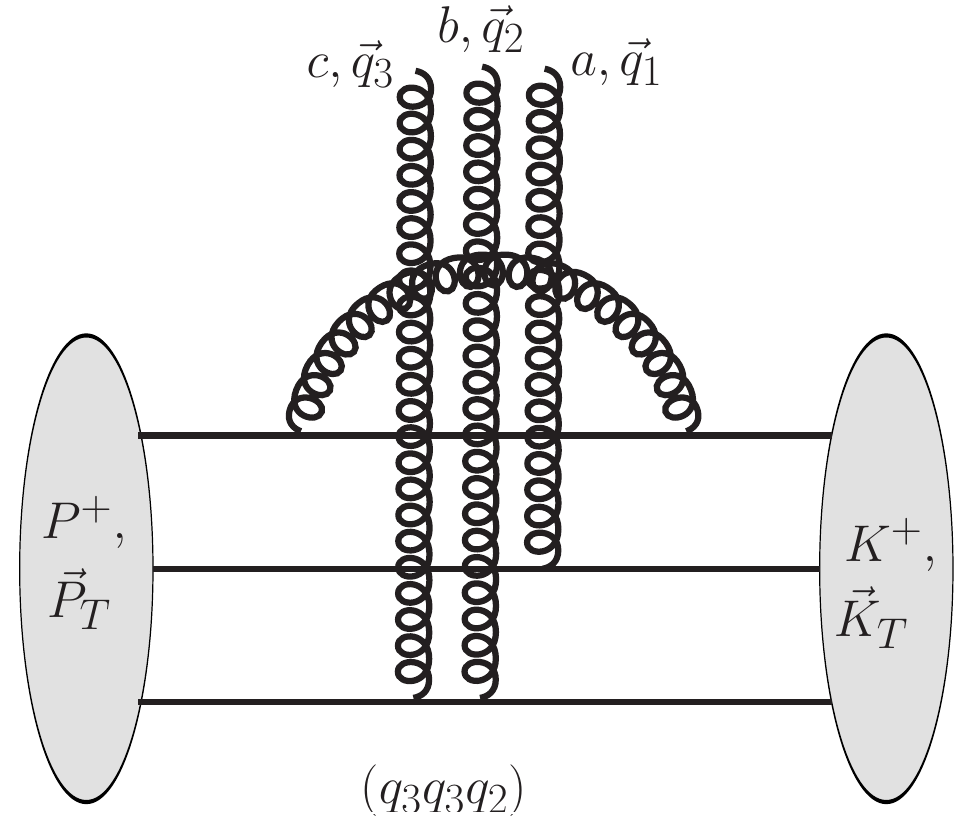}
  \hspace*{.1cm}
  \includegraphics[width=0.18\linewidth]{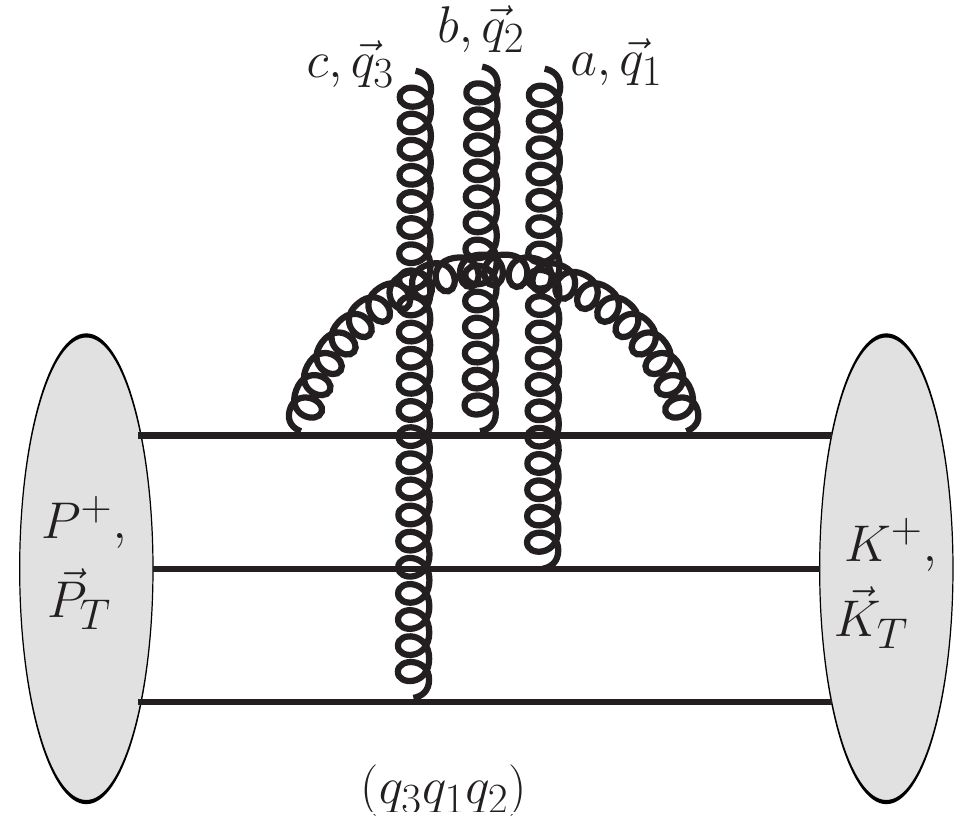}
  \hspace*{.1cm}
  \includegraphics[width=0.18\linewidth]{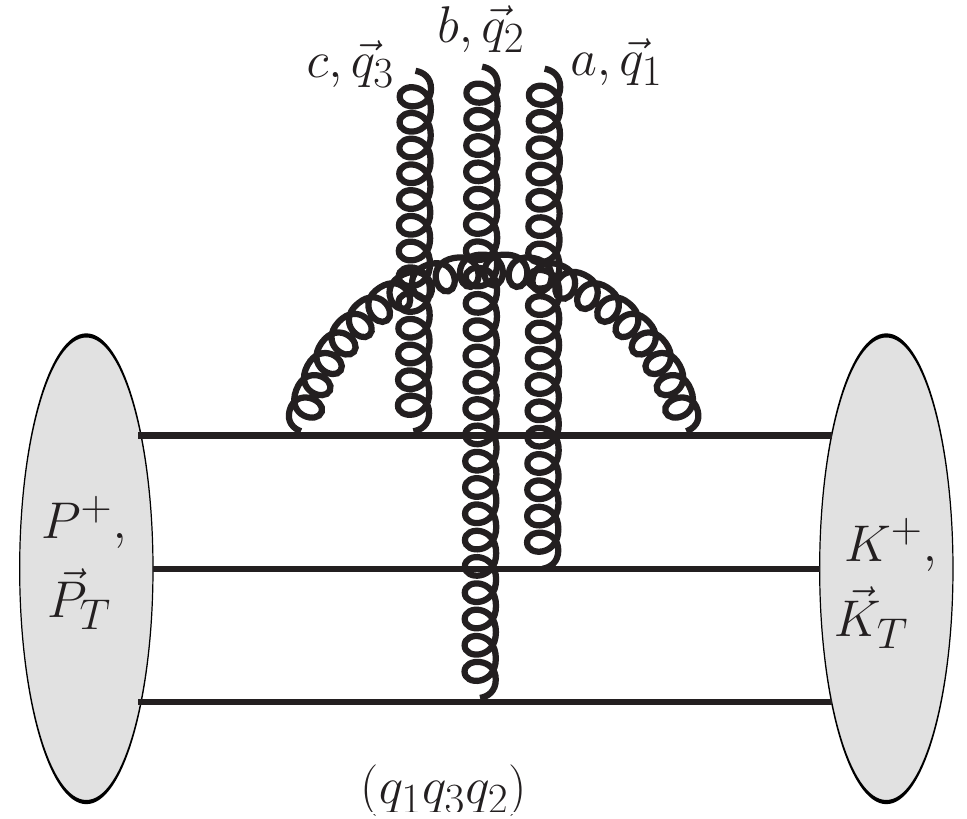}
  \end{minipage}
\caption{Final set of UV divergent ``real emission'' diagrams
  for $\langle \rho^a(\vec q_1)\, \rho^b(\vec
  q_2)\,\rho^c(\vec q_3)\rangle$ where all three gluon probes
  attach to quarks in the proton.  The cut is located at the
  insertion of the three color charge operators.}
\label{fig:rho-rho-rho_UVdiv-qqq}
\end{figure}
The final set of UV divergent diagrams is shown in
fig.~\ref{fig:rho-rho-rho_UVdiv-qqq}. Here, quark 1 exchanges a gluon
with itself across the insertion of the three charge operators while
the three gluon probes attach in all possible ways to the three
quarks.
\bea
\mathrm{fig.}\, \ref{fig:rho-rho-rho_UVdiv-qqq}(q_1q_1q_1) &=&
\frac{2g^5}{3\cdot 16\pi^3} \,2C_F\, \tr t^a t^b t^c
\int \left[\dd x_i\right]
  \int \left[\dd^2 k_i\right]\,
  \Psi_{qqq}(x_1,\vec k_1; x_2,\vec k_2; x_3,\vec k_3)\,
\nn\\
& &     \Psi_{qqq}^*(x_1,\vec k_1-(1-x_1)\vec q;
x_2,\vec k_2 +x_2\vec q;
x_3,\vec k_3 +x_3\vec q) \nn\\
& &
\frac{(2\pi)^{D-1}}{2p_1^+} \int \frac{\dk_g}{2(p_1^+ - k_g^+)}
\langle S| \hat\psi_{q\to qg}(\pvec_1; \pvec_1-\kvec_g, \kvec_g)\,
\hat\psi^*_{q\to qg}(\pvec_1-\vec q; \pvec_1-\kvec_g-\vec q,
\kvec_g) |S\rangle~,
     \label{eq:rho-rho-rho_UV_q1q1q1}
\eea
with a symmetry factor of 3.
\bea
\mathrm{fig.}\, \ref{fig:rho-rho-rho_UVdiv-qqq}(q_2q_1q_1) &=&
\frac{2g^5}{3\cdot 16\pi^3} \,\frac{1}{2N_c}\, \tr t^a t^b t^c
\int \left[\dd x_i\right]
  \int \left[\dd^2 k_i\right]\,
  \Psi_{qqq}(x_1,\vec k_1; x_2,\vec k_2; x_3,\vec k_3)\,
\nn\\
& &     \Psi_{qqq}^*(x_1,\vec k_1+x_1\vec q-\vec q_{12};
x_2,\vec k_2 +x_2\vec q-\vec q_{3};
x_3,\vec k_3 +x_3\vec q) \nn\\
& &
\frac{(2\pi)^{D-1}}{2p_1^+} \int \frac{\dk_g}{2(p_1^+ - k_g^+)}
\langle S| \hat\psi_{q\to qg}(\pvec_1; \pvec_1-\kvec_g, \kvec_g)\,
\hat\psi^*_{q\to qg}(\pvec_1-\vec q_{12}; \pvec_1-\kvec_g-\vec q_{12},
\kvec_g) |S\rangle~,
     \label{eq:rho-rho-rho_UV_q2q1q1}
\eea
with a symmetry factor of 6.
\bea
\mathrm{fig.}\, \ref{fig:rho-rho-rho_UVdiv-qqq}(q_1q_2q_1) &=&
\frac{2g^5}{3\cdot 16\pi^3} \,\frac{1}{2N_c}\, \tr t^a t^c t^b
\int \left[\dd x_i\right]
  \int \left[\dd^2 k_i\right]\,
  \Psi_{qqq}(x_1,\vec k_1; x_2,\vec k_2; x_3,\vec k_3)\,
\nn\\
& &     \Psi_{qqq}^*(x_1,\vec k_1+x_1\vec q-\vec q_{13};
x_2,\vec k_2 +x_2\vec q-\vec q_{2};
x_3,\vec k_3 +x_3\vec q) \nn\\
& &
\frac{(2\pi)^{D-1}}{2p_1^+} \int \frac{\dk_g}{2(p_1^+ - k_g^+)}
\langle S| \hat\psi_{q\to qg}(\pvec_1; \pvec_1-\kvec_g, \kvec_g)\,
\hat\psi^*_{q\to qg}(\pvec_1-\vec q_{13}; \pvec_1-\kvec_g-\vec q_{13},
\kvec_g) |S\rangle~,
     \label{eq:rho-rho-rho_UV_q1q2q1}
\eea
with a symmetry factor of 6.
\bea
\mathrm{fig.}\, \ref{fig:rho-rho-rho_UVdiv-qqq}(q_2q_2q_1) &=&
\frac{2g^5}{3\cdot 16\pi^3} \,\frac{1}{2N_c}\, \tr t^a t^b t^c
\int \left[\dd x_i\right]
  \int \left[\dd^2 k_i\right]\,
  \Psi_{qqq}(x_1,\vec k_1; x_2,\vec k_2; x_3,\vec k_3)\,
\nn\\
& &     \Psi_{qqq}^*(x_1,\vec k_1+x_1\vec q-\vec q_{1};
x_2,\vec k_2 +x_2\vec q-\vec q_{23};
x_3,\vec k_3 +x_3\vec q) \nn\\
& &
\frac{(2\pi)^{D-1}}{2p_1^+} \int \frac{\dk_g}{2(p_1^+ - k_g^+)}
\langle S| \hat\psi_{q\to qg}(\pvec_1; \pvec_1-\kvec_g, \kvec_g)\,
\hat\psi^*_{q\to qg}(\pvec_1-\vec q_{1}; \pvec_1-\kvec_g-\vec q_{1},
\kvec_g) |S\rangle~,
     \label{eq:rho-rho-rho_UV_q2q2q1}
\eea
with a symmetry factor of 6.
\bea
\mathrm{fig.}\, \ref{fig:rho-rho-rho_UVdiv-qqq}(q_3q_2q_1) &=&
-\frac{2g^5}{3\cdot 16\pi^3} \,\frac{1}{2N_c}\,
(\tr t^a t^b t^c + \tr t^a t^c t^b)\,
\int \left[\dd x_i\right]
  \int \left[\dd^2 k_i\right]\,
  \Psi_{qqq}(x_1,\vec k_1; x_2,\vec k_2; x_3,\vec k_3)\,
\nn\\
& &     \Psi_{qqq}^*(x_1,\vec k_1+x_1\vec q-\vec q_{1};
x_2,\vec k_2 +x_2\vec q-\vec q_{2};
x_3,\vec k_3 +x_3\vec q-\vec q_{3}) \nn\\
& &
\frac{(2\pi)^{D-1}}{2p_1^+} \int \frac{\dk_g}{2(p_1^+ - k_g^+)}
\langle S| \hat\psi_{q\to qg}(\pvec_1; \pvec_1-\kvec_g, \kvec_g)\,
\hat\psi^*_{q\to qg}(\pvec_1-\vec q_{1}; \pvec_1-\kvec_g-\vec q_{1},
\kvec_g) |S\rangle~,
     \label{eq:rho-rho-rho_UV_q3q2q1}
\eea
with a symmetry factor of 6.
\bea
\mathrm{fig.}\, \ref{fig:rho-rho-rho_UVdiv-qqq}(q_1q_1q_2) &=&
\frac{2g^5}{3\cdot 16\pi^3} \,\frac{1}{2N_c}\,
\tr t^a t^b t^c\,
\int \left[\dd x_i\right]
  \int \left[\dd^2 k_i\right]\,
  \Psi_{qqq}(x_1,\vec k_1; x_2,\vec k_2; x_3,\vec k_3)\,
\nn\\
& &     \Psi_{qqq}^*(x_1,\vec k_1+x_1\vec q-\vec q_{23};
x_2,\vec k_2 +x_2\vec q-\vec q_{1};
x_3,\vec k_3 +x_3\vec q) \nn\\
& &
\frac{(2\pi)^{D-1}}{2p_1^+} \int \frac{\dk_g}{2(p_1^+ - k_g^+)}
\langle S| \hat\psi_{q\to qg}(\pvec_1; \pvec_1-\kvec_g, \kvec_g)\,
\hat\psi^*_{q\to qg}(\pvec_1-\vec q_{23}; \pvec_1-\kvec_g-\vec q_{23},
\kvec_g) |S\rangle~,
     \label{eq:rho-rho-rho_UV_q1q1q2}
\eea
with a symmetry factor of 6.
\bea
\mathrm{fig.}\, \ref{fig:rho-rho-rho_UVdiv-qqq}(q_2q_1q_2) &=&
\frac{2g^5}{3\cdot 16\pi^3} \,\frac{1}{2N_c}\,
\tr t^a t^c t^b\,
\int \left[\dd x_i\right]
  \int \left[\dd^2 k_i\right]\,
  \Psi_{qqq}(x_1,\vec k_1; x_2,\vec k_2; x_3,\vec k_3)\,
\nn\\
& &     \Psi_{qqq}^*(x_1,\vec k_1+x_1\vec q-\vec q_{2};
x_2,\vec k_2 +x_2\vec q-\vec q_{13};
x_3,\vec k_3 +x_3\vec q) \nn\\
& &
\frac{(2\pi)^{D-1}}{2p_1^+} \int \frac{\dk_g}{2(p_1^+ - k_g^+)}
\langle S| \hat\psi_{q\to qg}(\pvec_1; \pvec_1-\kvec_g, \kvec_g)\,
\hat\psi^*_{q\to qg}(\pvec_1-\vec q_{2}; \pvec_1-\kvec_g-\vec q_{2},
\kvec_g) |S\rangle~,
     \label{eq:rho-rho-rho_UV_q2q1q2}
\eea
with a symmetry factor of 6.
\bea
\mathrm{fig.}\, \ref{fig:rho-rho-rho_UVdiv-qqq}(q_1q_2q_2) &=&
\frac{2g^5}{3\cdot 16\pi^3} \,\frac{1}{2N_c}\,
\tr t^a t^b t^c\,
\int \left[\dd x_i\right]
  \int \left[\dd^2 k_i\right]\,
  \Psi_{qqq}(x_1,\vec k_1; x_2,\vec k_2; x_3,\vec k_3)\,
\nn\\
& &     \Psi_{qqq}^*(x_1,\vec k_1+x_1\vec q-\vec q_{3};
x_2,\vec k_2 +x_2\vec q-\vec q_{12};
x_3,\vec k_3 +x_3\vec q) \nn\\
& &
\frac{(2\pi)^{D-1}}{2p_1^+} \int \frac{\dk_g}{2(p_1^+ - k_g^+)}
\langle S| \hat\psi_{q\to qg}(\pvec_1; \pvec_1-\kvec_g, \kvec_g)\,
\hat\psi^*_{q\to qg}(\pvec_1-\vec q_{3}; \pvec_1-\kvec_g-\vec q_{3},
\kvec_g) |S\rangle~,
     \label{eq:rho-rho-rho_UV_q1q2q2}
\eea
with a symmetry factor of 6.
\bea
\mathrm{fig.}\, \ref{fig:rho-rho-rho_UVdiv-qqq}(q_2q_2q_2) &=&
\frac{2g^5}{3\cdot 16\pi^3} \,C_F (N_c-1)\, \tr t^a t^b t^c\,
\int \left[\dd x_i\right]
  \int \left[\dd^2 k_i\right]\,
  \Psi_{qqq}(x_1,\vec k_1; x_2,\vec k_2; x_3,\vec k_3)\,
\nn\\
& &     \Psi_{qqq}^*(x_1,\vec k_1+x_1\vec q;
x_2,\vec k_2 +x_2\vec q-\vec q;
x_3,\vec k_3 +x_3\vec q) \nn\\
& &
\frac{(2\pi)^{D-1}}{2p_1^+} \int \frac{\dk_g}{2(p_1^+ - k_g^+)}
\langle S| \hat\psi_{q\to qg}(\pvec_1; \pvec_1-\kvec_g, \kvec_g)\,
\hat\psi^*_{q\to qg}(\pvec_1; \pvec_1-\kvec_g,
\kvec_g) |S\rangle~,
     \label{eq:rho-rho-rho_UV_q2q2q2}
\eea
with a symmetry factor of 6.
\bea
\mathrm{fig.}\, \ref{fig:rho-rho-rho_UVdiv-qqq}(q_3q_2q_2) &=&
- \frac{2g^5}{3\cdot 16\pi^3} \,(N_c-2)C_F\, \tr t^a t^b t^c\,
\int \left[\dd x_i\right]
  \int \left[\dd^2 k_i\right]\,
  \Psi_{qqq}(x_1,\vec k_1; x_2,\vec k_2; x_3,\vec k_3)\,
\nn\\
& &     \Psi_{qqq}^*(x_1,\vec k_1+x_1\vec q;
x_2,\vec k_2 +x_2\vec q-\vec q_{12};
x_3,\vec k_3 +x_3\vec q-\vec q_{3}) \nn\\
& &
\frac{(2\pi)^{D-1}}{2p_1^+} \int \frac{\dk_g}{2(p_1^+ - k_g^+)}
\langle S| \hat\psi_{q\to qg}(\pvec_1; \pvec_1-\kvec_g, \kvec_g)\,
\hat\psi^*_{q\to qg}(\pvec_1; \pvec_1-\kvec_g, \kvec_g) |S\rangle~,
     \label{eq:rho-rho-rho_UV_q3q2q2}
\eea
with a symmetry factor of 6.
\bea
\mathrm{fig.}\, \ref{fig:rho-rho-rho_UVdiv-qqq}(q_2q_3q_2) &=&
- \frac{2g^5}{3\cdot 16\pi^3} \,(N_c-2)C_F\, \tr t^a t^c t^b\,
\int \left[\dd x_i\right]
  \int \left[\dd^2 k_i\right]\,
  \Psi_{qqq}(x_1,\vec k_1; x_2,\vec k_2; x_3,\vec k_3)\,
\nn\\
& &     \Psi_{qqq}^*(x_1,\vec k_1+x_1\vec q;
x_2,\vec k_2 +x_2\vec q-\vec q_{13};
x_3,\vec k_3 +x_3\vec q-\vec q_{2}) \nn\\
& &
\frac{(2\pi)^{D-1}}{2p_1^+} \int \frac{\dk_g}{2(p_1^+ - k_g^+)}
\langle S| \hat\psi_{q\to qg}(\pvec_1; \pvec_1-\kvec_g, \kvec_g)\,
\hat\psi^*_{q\to qg}(\pvec_1; \pvec_1-\kvec_g, \kvec_g) |S\rangle~,
     \label{eq:rho-rho-rho_UV_q2q3q2}
\eea
with a symmetry factor of 6.
\bea
\mathrm{fig.}\, \ref{fig:rho-rho-rho_UVdiv-qqq}(q_3q_3q_2) &=&
- \frac{2g^5}{3\cdot 16\pi^3} \,(N_c-2)C_F\, \tr t^a t^b t^c\,
\int \left[\dd x_i\right]
  \int \left[\dd^2 k_i\right]\,
  \Psi_{qqq}(x_1,\vec k_1; x_2,\vec k_2; x_3,\vec k_3)\,
\nn\\
& &     \Psi_{qqq}^*(x_1,\vec k_1+x_1\vec q;
x_2,\vec k_2 +x_2\vec q-\vec q_{1};
x_3,\vec k_3 +x_3\vec q-\vec q_{23}) \nn\\
& &
\frac{(2\pi)^{D-1}}{2p_1^+} \int \frac{\dk_g}{2(p_1^+ - k_g^+)}
\langle S| \hat\psi_{q\to qg}(\pvec_1; \pvec_1-\kvec_g, \kvec_g)\,
\hat\psi^*_{q\to qg}(\pvec_1; \pvec_1-\kvec_g, \kvec_g) |S\rangle~,
     \label{eq:rho-rho-rho_UV_q3q3q2}
\eea
with a symmetry factor of 6.
\bea
\mathrm{fig.}\, \ref{fig:rho-rho-rho_UVdiv-qqq}(q_3q_1q_2) &=&
- \frac{2g^5}{3\cdot 16\pi^3} \,\frac{1}{2N_c}\, (\tr t^a t^b t^c +
\tr t^a t^c t^b)\,
\int \left[\dd x_i\right]
  \int \left[\dd^2 k_i\right]\,
  \Psi_{qqq}(x_1,\vec k_1; x_2,\vec k_2; x_3,\vec k_3)\,
\nn\\
& &     \Psi_{qqq}^*(x_1,\vec k_1+x_1\vec q-\vec q_{2};
x_2,\vec k_2 +x_2\vec q-\vec q_{1};
x_3,\vec k_3 +x_3\vec q-\vec q_{3}) \nn\\
& &
\frac{(2\pi)^{D-1}}{2p_1^+} \int \frac{\dk_g}{2(p_1^+ - k_g^+)}
\langle S| \hat\psi_{q\to qg}(\pvec_1; \pvec_1-\kvec_g, \kvec_g)\,
\hat\psi^*_{q\to qg}(\pvec_1-\vec q_{2}; \pvec_1-\kvec_g-\vec q_{2},
\kvec_g) |S\rangle~,
     \label{eq:rho-rho-rho_UV_q3q1q2}
\eea
with a symmetry factor of 6.
\bea
\mathrm{fig.}\, \ref{fig:rho-rho-rho_UVdiv-qqq}(q_1q_3q_2) &=&
- \frac{2g^5}{3\cdot 16\pi^3} \,\frac{1}{2N_c}\, (\tr t^a t^b t^c +
\tr t^a t^c t^b)\,
\int \left[\dd x_i\right]
  \int \left[\dd^2 k_i\right]\,
  \Psi_{qqq}(x_1,\vec k_1; x_2,\vec k_2; x_3,\vec k_3)\,
\nn\\
& &     \Psi_{qqq}^*(x_1,\vec k_1+x_1\vec q-\vec q_{3};
x_2,\vec k_2 +x_2\vec q-\vec q_{1};
x_3,\vec k_3 +x_3\vec q-\vec q_{2}) \nn\\
& &
\frac{(2\pi)^{D-1}}{2p_1^+} \int \frac{\dk_g}{2(p_1^+ - k_g^+)}
\langle S| \hat\psi_{q\to qg}(\pvec_1; \pvec_1-\kvec_g, \kvec_g)\,
\hat\psi^*_{q\to qg}(\pvec_1-\vec q_{3}; \pvec_1-\kvec_g-\vec q_{3},
\kvec_g) |S\rangle~,
     \label{eq:rho-rho-rho_UV_q1q3q2}
\eea
with a symmetry factor of 6.

\subsection{Finite diagrams}

\subsubsection{Coupling at least once to the gluon}

\begin{figure}[htb]
  \centering
  \begin{minipage}[hb]{\linewidth}
    \includegraphics[width=0.23\linewidth]{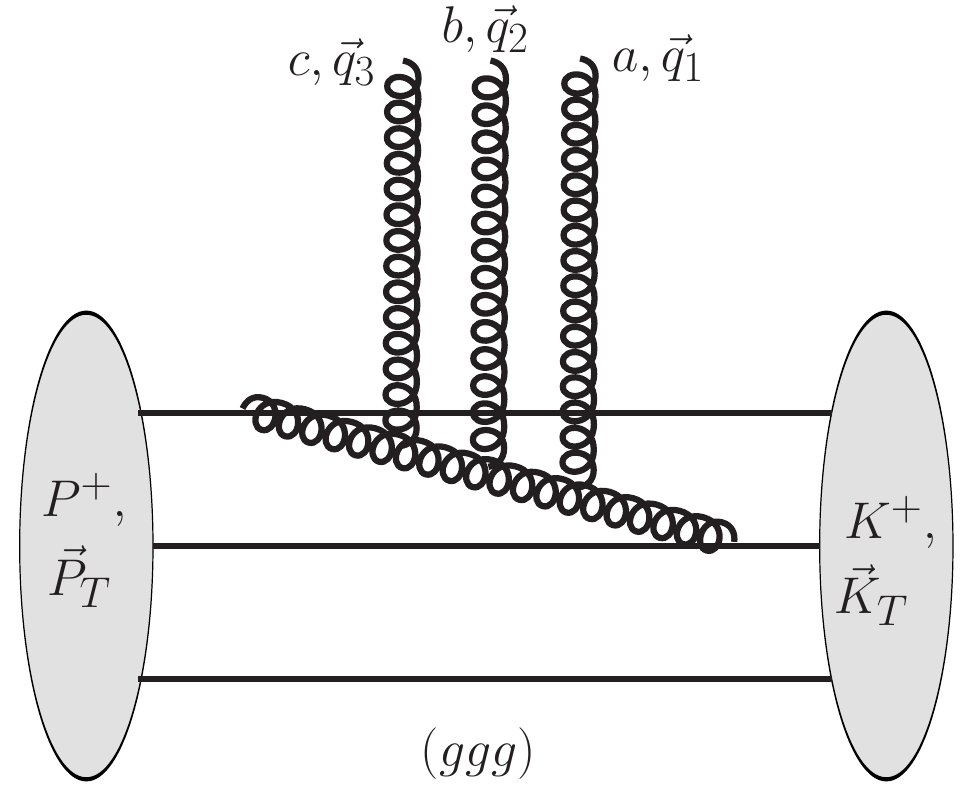}
    \hspace{0.1cm}
    \includegraphics[width=0.23\linewidth]{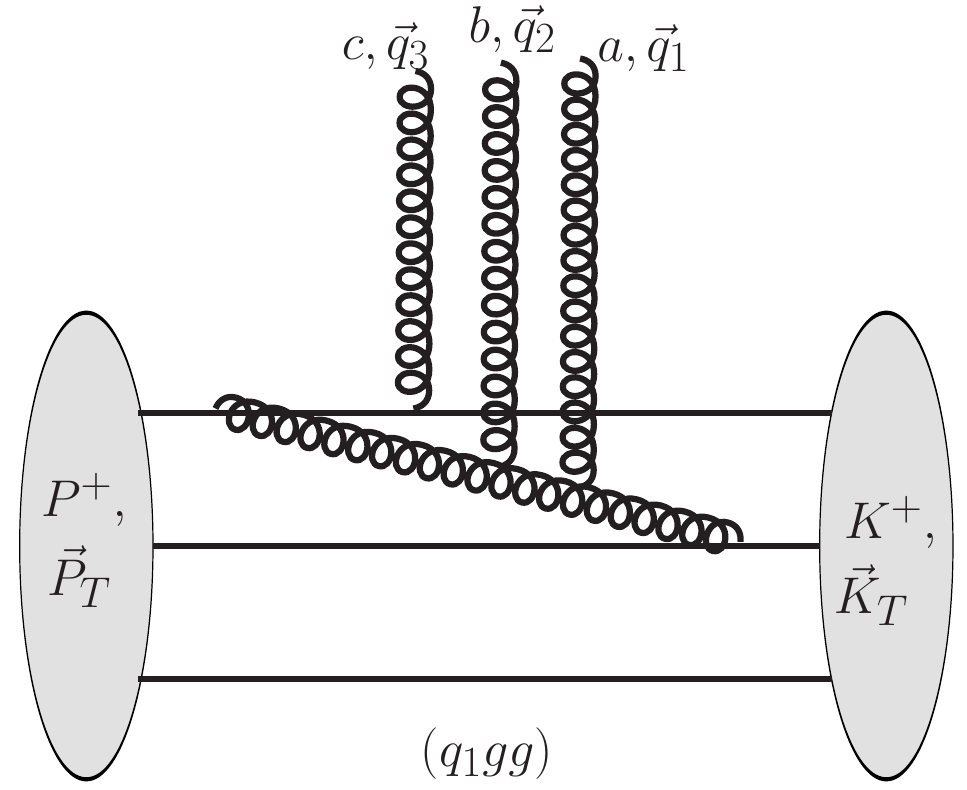}
    \hspace{0.1cm}
    \includegraphics[width=0.23\linewidth]{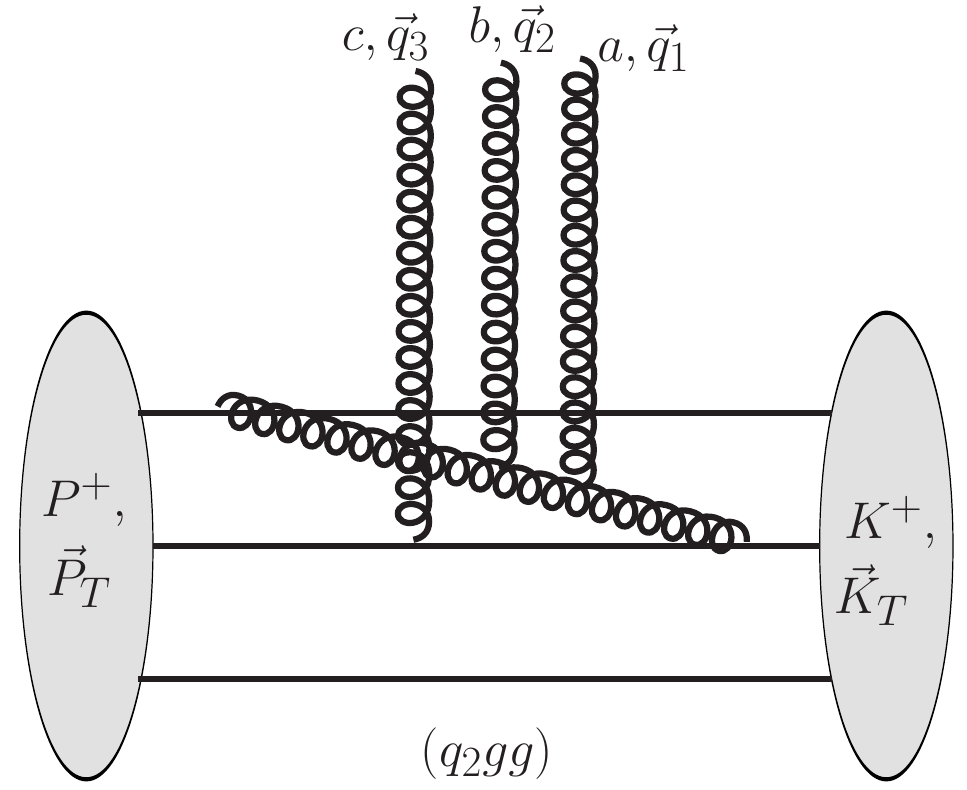}
    \hspace{0.1cm}
    \includegraphics[width=0.23\linewidth]{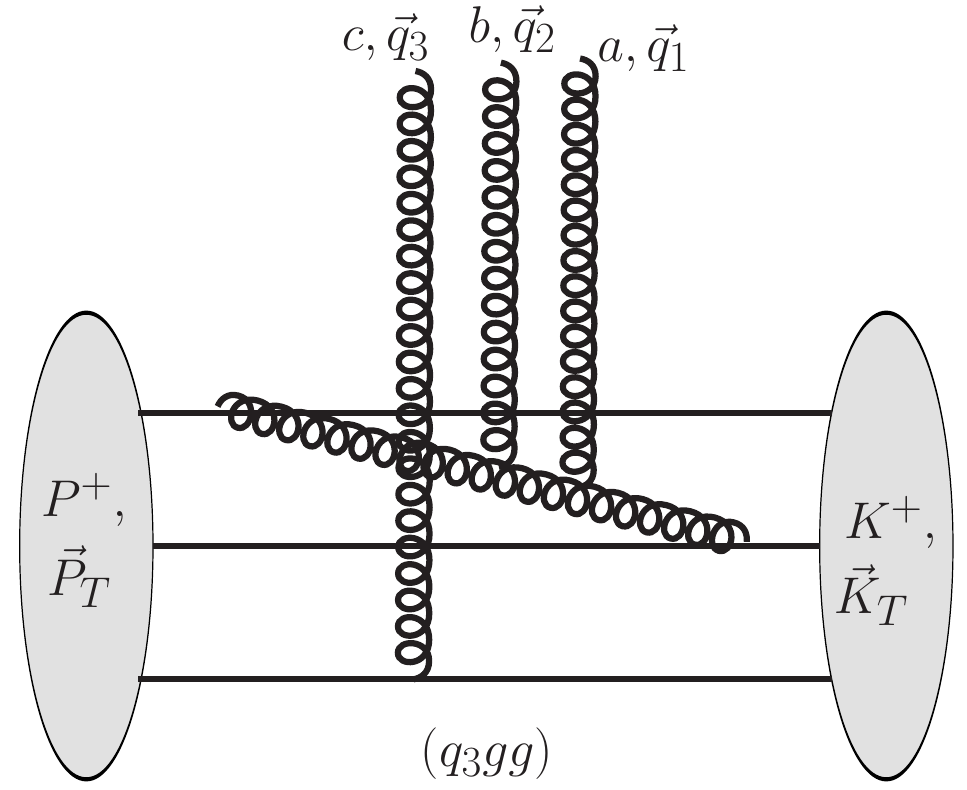}
  \end{minipage}
\caption{A sample of UV finite diagrams for $\langle \rho^a(\vec
  q_1)\, \rho^b(\vec q_2)\,\rho^c(\vec q_3)\rangle$ where at least two
  of the probes attach to the gluon in the proton.  The cut is located
  at the insertion of the three color charge operators.}
\label{fig:rho-rho-rho_FIN}
\end{figure}
We now proceed to the UV-finite contributions.  To write the following
expressions in more compact form we introduce the integral operator
\bea
\vec I &=& 
\frac{g^5}{3\cdot 16\pi^3} \int \left[\dd x_i\right]
  \int \left[\dd^2 k_i\right] \,
  \Psi_{qqq}(x_1,\vec k_1; x_2,\vec k_2; x_3,\vec k_3)\,
\nn\\
& &
\int\limits_x^{\mathrm{min}(x_1,1-x_2)}\frac{\dd x_g}{x_g}
\left(1-\frac{z_1+z_2}{2}+\frac{z_1 z_2}{6}\right)
\sqrt{\frac{x_1}{x_1-x_g}}\sqrt{\frac{x_2}{x_2+x_g}}
\int\dd^2 k_g
\frac{z_1\vec p_1-\vec k_g}
     {\left(z_1\vec p_1-\vec k_g \right)^2}~,
\eea
with $z_1 = x_g / x_1$ and $z_2 = x_g/(x_2+x_g)$.
We begin with the diagrams shown in fig.~\ref{fig:rho-rho-rho_FIN}
where two distinct quarks exchange a gluon across the operator
insertion, and where two of the probes attach to that gluon.\\

\bea
\mathrm{fig.}\, \ref{fig:rho-rho-rho_FIN}(ggg)
& = &
- \frac{1}{2}\, \tr T^a T^b T^c ~
\vec I \cdot
\frac{z_2\vec p_2 - (1-z_2)(\vec k_g -\vec q)}
  {\left(z_2\vec p_2 - (1-z_2)(\vec k_g -\vec q)\right)^2}
  \nn\\
  & &
  \Psi_{qqq}^*(x_1-x_g,\vec k_1+x_1\vec q - \vec k_g + x_g\vec K;
x_2+x_g,\vec k_2 -(1-x_2)\vec q + \vec k_g - x_g\vec K;
x_3,\vec k_3 +x_3\vec q)~,
     \label{eq:rho-rho-rho_FIN_ggg}
\eea
The symmetry factor
is 6 which includes a factor of 2 for interchanging the gluon emission
and absorption vertices between quarks 1 and 2 (so that in $|P\rangle$
the gluon couples to quark 2, and in $\langle K|$ it couples to quark
1).

\bea
\mathrm{fig.}\, \ref{fig:rho-rho-rho_FIN}(q_1gg) &=& -
\frac{1}{4}\, \tr (T^a T^b D^c - T^a T^b T^c )
~ \vec I \cdot
\frac{z_2\vec p_2 - (1-z_2)(\vec k_g -\vec q_{12})}
     {\left( z_2\vec p_2 - (1-z_2)(\vec k_g -\vec q_{12}) \right)^2}
\nn\\
& &  \Psi_{qqq}^*(x_1-x_g,\vec k_1+x_1\vec q - \vec q_3-\vec k_g + x_g\vec K;
x_2+x_g,\vec k_2 +x_2\vec q-\vec q_{12}+\vec k_g-x_g\vec K;
x_3,\vec k_3 +x_3\vec q)~.
     \label{eq:rho-rho-rho_FIN_q1gg}
\eea
The symmetry factor 6 includes a factor of 2 due to the
contribution from fig.~\ref{fig:rho-rho-rho_FIN}($q_2gg$) with the
gluon emission and absorption vertices between quarks 1 and 2
interchanged.
\bea
\mathrm{fig.}\, \ref{fig:rho-rho-rho_FIN}(gq_1g) &=& -
\frac{1}{4}\, \tr (T^a T^c D^b - T^a T^c T^b )
~ \vec I \cdot
\frac{z_2\vec p_2 - (1-z_2)(\vec k_g -\vec q_{13})}
     {\left( z_2\vec p_2 - (1-z_2)(\vec k_g -\vec q_{13}) \right)^2}
\nn\\
& &  \Psi_{qqq}^*(x_1-x_g,\vec k_1+x_1\vec q - \vec q_2-\vec k_g +x_g\vec K;
x_2+x_g,\vec k_2 +x_2\vec q-\vec q_{13}+\vec k_g-x_g\vec K;
x_3,\vec k_3 +x_3\vec q)~.
     \label{eq:rho-rho-rho_FIN_gq1g}
\eea
The symmetry factor of 6 includes a factor of 2 due to the
contribution of diagram fig.~\ref{fig:rho-rho-rho_FIN}($gq_2g$) (not
shown) with the gluon emission and absorption vertices between quarks
1 and 2 interchanged.
\bea
\mathrm{fig.}\, \ref{fig:rho-rho-rho_FIN}(ggq_1) &=& -
\frac{1}{4}\, \tr (T^b T^c D^a - T^a T^b T^c )
~ \vec I
\cdot
\frac{z_2\vec p_2 - (1-z_2)(\vec k_g -\vec q_{23})}
     {\left( z_2\vec p_2 - (1-z_2)(\vec k_g -\vec q_{23}) \right)^2}
\nn\\
& &  \Psi_{qqq}^*(x_1-x_g,\vec k_1+x_1\vec q - \vec q_1-\vec k_g+x_g\vec K;
x_2+x_g,\vec k_2 +x_2\vec q-\vec q_{23}+\vec k_g-x_g\vec K;
x_3,\vec k_3 +x_3\vec q)~.
     \label{eq:rho-rho-rho_FIN_ggq1}
\eea
Again, the symmetry factor for this diagram is 6 which includes the
contribution of diagram fig.~\ref{fig:rho-rho-rho_FIN}($ggq_2$) (not
shown) with the gluon emission and absorption vertices between quarks
1 and 2 interchanged.

\bea
\mathrm{fig.}\, \ref{fig:rho-rho-rho_FIN}(q_2gg) &=& -
\frac{1}{4}\, \tr (T^a T^b D^c - T^a T^b T^c )
~ \vec I
\cdot
\frac{z_2(\vec p_2-\vec q_3) - (1-z_2)(\vec k_g -\vec q_{12})}
     {\left(z_2(\vec p_2-\vec q_3) - (1-z_2)(\vec k_g -\vec q_{12}) \right)^2}
\nn\\
& &  \Psi_{qqq}^*(x_1-x_g,\vec k_1+x_1\vec q - \vec k_g + x_g\vec K;
x_2+x_g,\vec k_2 +x_2\vec q-\vec q+\vec k_g - x_g\vec K;
x_3,\vec k_3 +x_3\vec q)~.
     \label{eq:rho-rho-rho_FIN_q2gg}
\eea
The symmetry factor is 6; this includes the contribution from
fig.~\ref{fig:rho-rho-rho_FIN}($q_1gg$) with the gluon emission and
absorption vertices between quarks 1 and 2 interchanged.
\bea
\mathrm{fig.}\, \ref{fig:rho-rho-rho_FIN}(gq_2g) &=& -
\frac{1}{4}\, \tr (T^a T^c D^b - T^a T^c T^b )
~ \vec I
\cdot
\frac{z_2(\vec p_2-\vec q_2) - (1-z_2)(\vec k_g -\vec q_{13})}
     {\left( z_2(\vec p_2-\vec q_2) - (1-z_2)(\vec k_g -\vec q_{13}) \right)^2}
\nn\\
& &  \Psi_{qqq}^*(x_1-x_g,\vec k_1+x_1\vec q - \vec k_g + x_g\vec K;
x_2+x_g,\vec k_2 +x_2\vec q-\vec q+\vec k_g - x_g\vec K;
x_3,\vec k_3 +x_3\vec q)~.
     \label{eq:rho-rho-rho_FIN_gq2g}
\eea
The symmetry factor is 6.
\bea
\mathrm{fig.}\, \ref{fig:rho-rho-rho_FIN}(ggq_2) &=& -
\frac{1}{4}\, \tr (T^a T^b D^c - T^a T^b T^c )
~ \vec I
\cdot
\frac{z_2(\vec p_2-\vec q_1) - (1-z_2)(\vec k_g -\vec q_{23})}
     {\left( z_2(\vec p_2-\vec q_1) - (1-z_2)(\vec k_g -\vec q_{23}) \right)^2}
\nn\\
& &  \Psi_{qqq}^*(x_1-x_g,\vec k_1+x_1\vec q - \vec k_g + x_g\vec K;
x_2+x_g,\vec k_2 +x_2\vec q-\vec q+\vec k_g - x_g\vec K;
x_3,\vec k_3 +x_3\vec q)~.
     \label{eq:rho-rho-rho_FIN_ggq2}
\eea
The symmetry factor is 6.

\bea
\mathrm{fig.}\, \ref{fig:rho-rho-rho_FIN}(q_3gg) &=&
\frac{1}{2}\, \tr T^a T^b D^c
~ \vec I
\cdot
\frac{z_2\vec p_2 - (1-z_2)(\vec k_g -\vec q_{12})}
     {\left( z_2\vec p_2 - (1-z_2)(\vec k_g -\vec q_{12}) \right)^2}
\nn\\
& &  \Psi_{qqq}^*(x_1-x_g,\vec k_1+x_1\vec q - \vec k_g + x_g\vec K;
x_2+x_g,\vec k_2 +x_2\vec q-\vec q_{12}+\vec k_g - x_g\vec K;
x_3,\vec k_3 +x_3\vec q-\vec q_{3})~.
     \label{eq:rho-rho-rho_FIN_q3gg}
\eea
The symmetry factor is 6.
\bea
\mathrm{fig.}\, \ref{fig:rho-rho-rho_FIN}(gq_3g) &=&
\frac{1}{2}\, \tr T^a T^c D^b
~ \vec I
\cdot
\frac{z_2\vec p_2 - (1-z_2)(\vec k_g -\vec q_{13})}
     {\left( z_2\vec p_2 - (1-z_2)(\vec k_g -\vec q_{13}) \right)^2}
\nn\\
& &  \Psi_{qqq}^*(x_1-x_g,\vec k_1+x_1\vec q - \vec k_g + x_g\vec K;
x_2+x_g,\vec k_2 +x_2\vec q-\vec q_{13}+\vec k_g - x_g\vec K;
x_3,\vec k_3 +x_3\vec q-\vec q_{2})~.
     \label{eq:rho-rho-rho_FIN_gq3g}
\eea
The symmetry factor is 6.
\bea
\mathrm{fig.}\, \ref{fig:rho-rho-rho_FIN}(ggq_3) &=&
\frac{1}{2}\, \tr T^a T^b D^c
~ \vec I
\cdot
\frac{z_2\vec p_2 - (1-z_2)(\vec k_g -\vec q_{23})}
     {\left( z_2\vec p_2 - (1-z_2)(\vec k_g -\vec q_{23}) \right)^2}
\nn\\
& &  \Psi_{qqq}^*(x_1-x_g,\vec k_1+x_1\vec q - \vec k_g + x_g\vec K;
x_2+x_g,\vec k_2 +x_2\vec q-\vec q_{23}+\vec k_g - x_g\vec K;
x_3,\vec k_3 +x_3\vec q-\vec q_{1})~.
     \label{eq:rho-rho-rho_FIN_ggq3}
\eea
The symmetry factor is 6.
\\~~\\

\begin{figure}[htb]
  \centering
  \begin{minipage}[hb]{\linewidth}
    \includegraphics[width=0.23\linewidth]{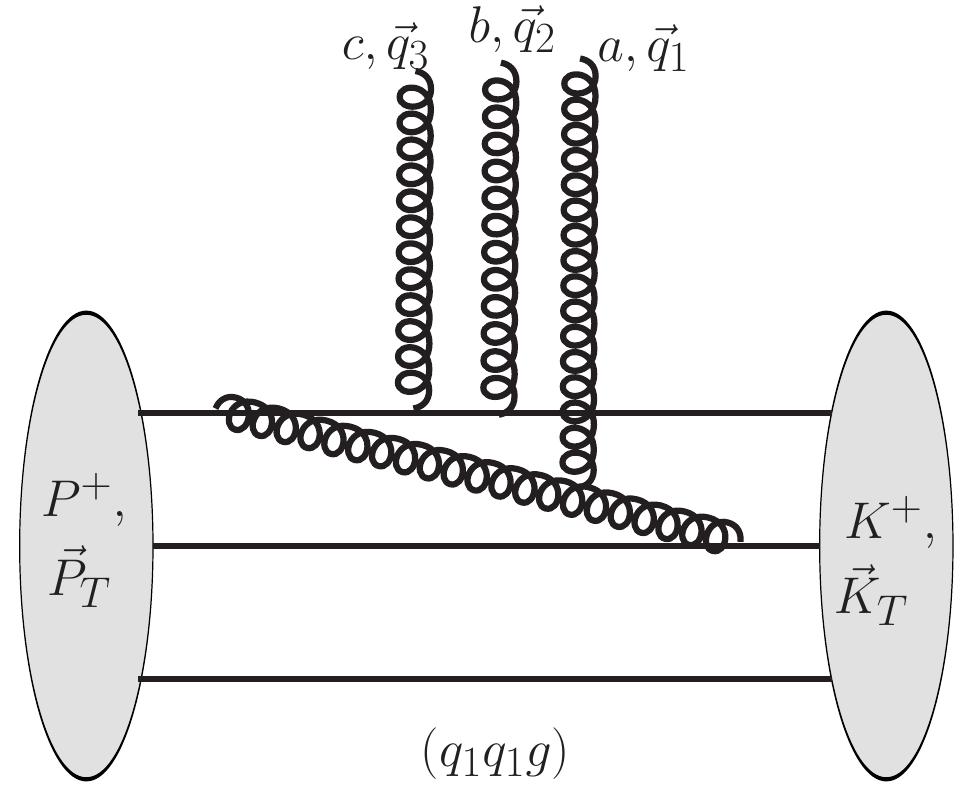}
    \hspace{0.1cm}
    \includegraphics[width=0.23\linewidth]{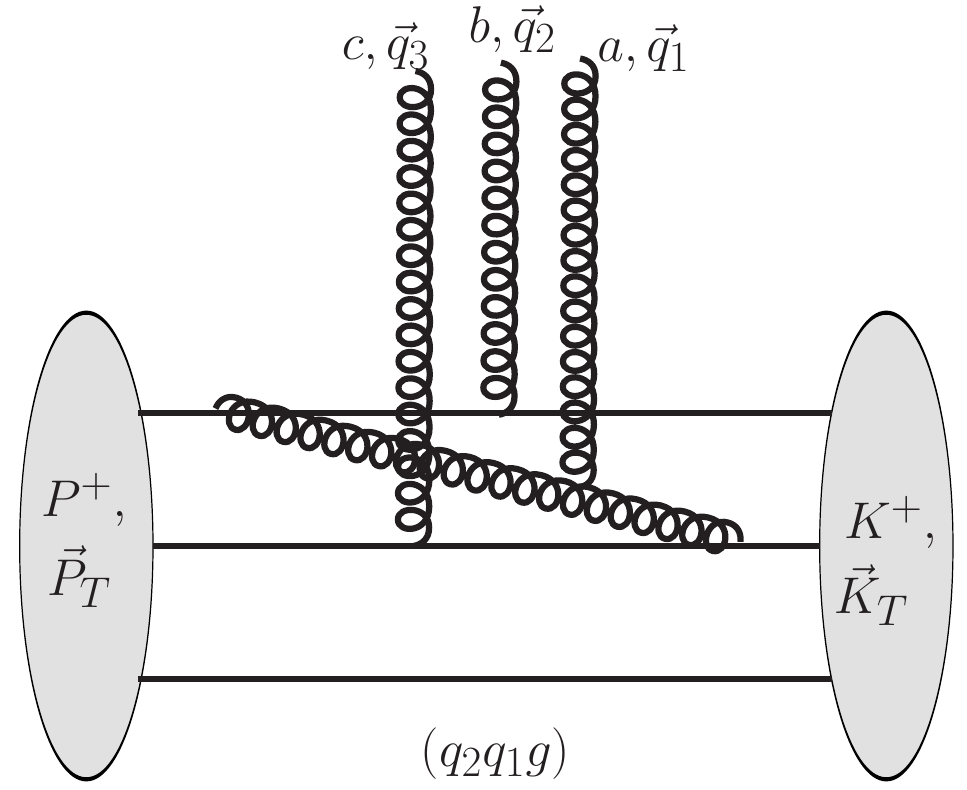}
    \hspace{0.1cm}
    \includegraphics[width=0.23\linewidth]{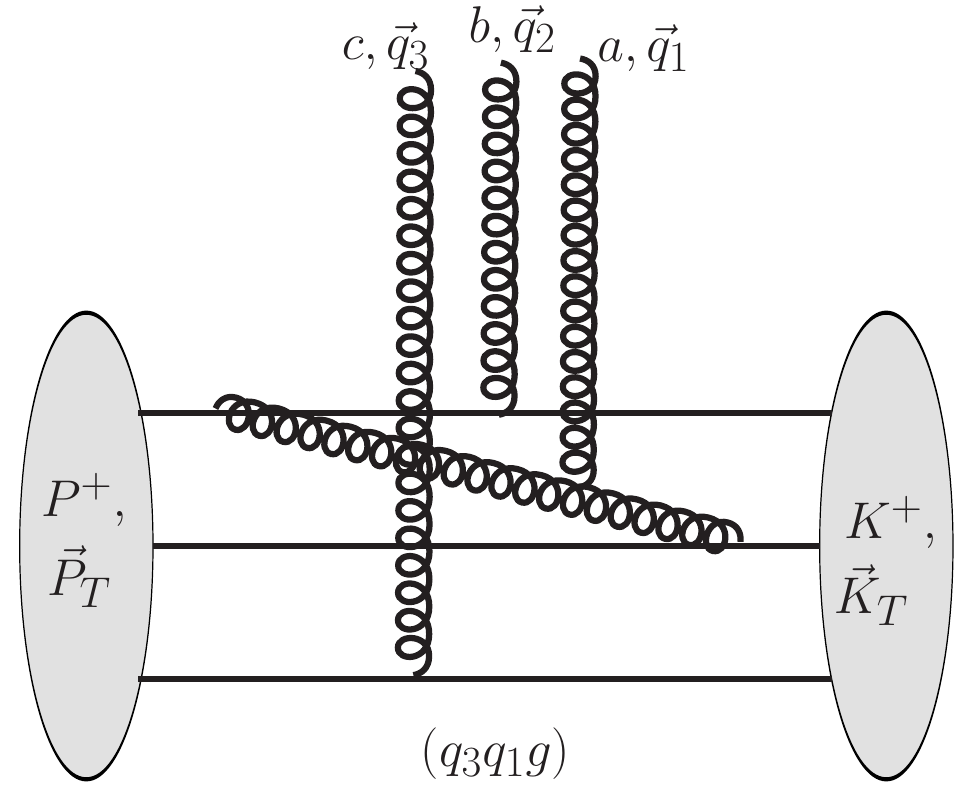}
    \hspace{0.1cm}
    \includegraphics[width=0.23\linewidth]{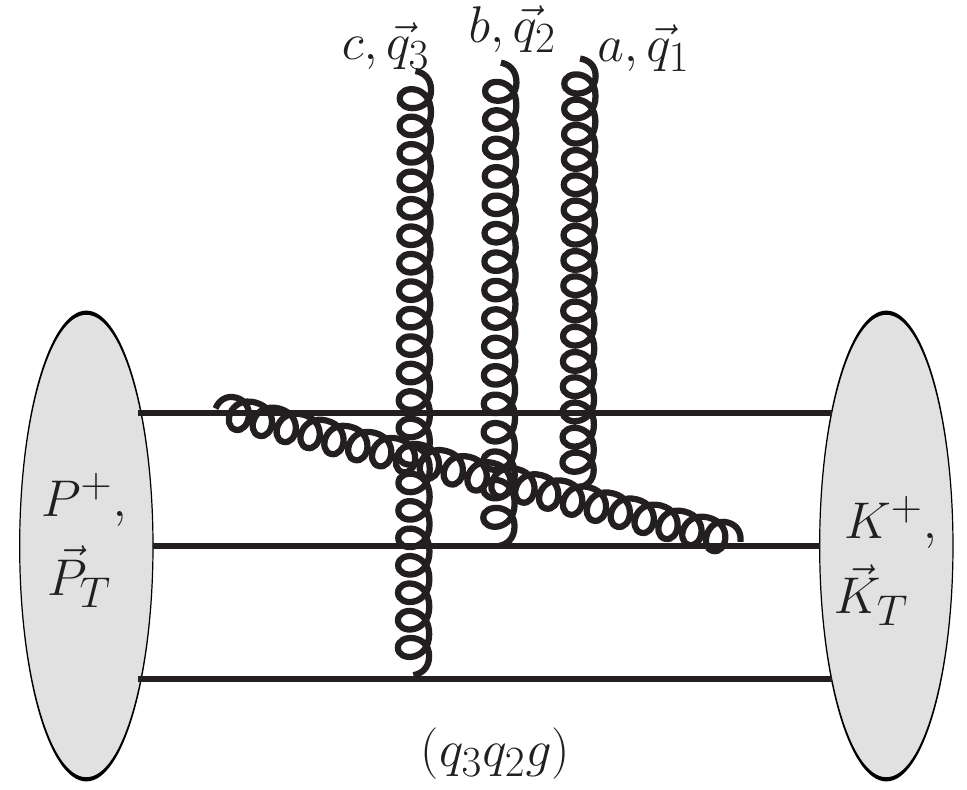}
  \end{minipage}
\caption{A sample of UV finite diagrams for $\langle \rho^a(\vec
  q_1)\, \rho^b(\vec q_2)\,\rho^c(\vec q_3)\rangle$ where one
  of the probes attaches to the gluon in the proton.  The cut is located
  at the insertion of the three color charge operators.}
\label{fig:rho-rho-rho_FIN-2}
\end{figure}
We continue with the diagrams shown in
fig.~\ref{fig:rho-rho-rho_FIN-2} where two distinct quarks exchange a
gluon across the operator insertion, and where one of the probes
attaches to that gluon.
\bea
\mathrm{fig.}\, \ref{fig:rho-rho-rho_FIN-2}(q_1 q_1 g) &=&
\frac{N_c}{2}\, \tr t^a t^b t^c
~ \vec I
\cdot
\frac{z_2\vec p_2 - (1-z_2)(\vec k_g -\vec q_{1})}
     {\left( z_2\vec p_2 - (1-z_2)(\vec k_g -\vec q_{1}) \right)^2}
\nn\\
& &  \Psi_{qqq}^*(x_1-x_g,\vec k_1+x_1\vec q - \vec k_g + x_g\vec K - \vec q_{23};
x_2+x_g,\vec k_2 +x_2\vec q-\vec q_{1}+\vec k_g - x_g\vec K;
x_3,\vec k_3 +x_3\vec q)~.
     \label{eq:rho-rho-rho_FIN-2_q1q1g}
\eea
The symmetry factor is 6 (this includes the contribution from diagram
fig.~\ref{fig:rho-rho-rho_FIN-2}($q_2 q_2 g$) (not shown) with the
gluon emission and absorption vertices between quarks 1 and 2
interchanged).
\bea
\mathrm{fig.}\, \ref{fig:rho-rho-rho_FIN-2}(q_1 g q_1) &=&
\frac{N_c}{2}\, \tr t^a t^c t^b
~ \vec I
\cdot
\frac{z_2\vec p_2 - (1-z_2)(\vec k_g -\vec q_{2})}
     {\left( z_2\vec p_2 - (1-z_2)(\vec k_g -\vec q_{2}) \right)^2}
\nn\\
& &  \Psi_{qqq}^*(x_1-x_g,\vec k_1+x_1\vec q - \vec k_g + x_g\vec K - \vec q_{13};
x_2+x_g,\vec k_2 +x_2\vec q-\vec q_{2}+\vec k_g - x_g\vec K;
x_3,\vec k_3 +x_3\vec q)~.
     \label{eq:rho-rho-rho_FIN-2_q1gq1}
\eea
Again, the symmetry factor is 6 (this includes the contribution from diagram
fig.~\ref{fig:rho-rho-rho_FIN-2}($q_2 g q_2$) (not shown) with the
gluon emission and absorption vertices between quarks 1 and 2
interchanged).
\bea
\mathrm{fig.}\, \ref{fig:rho-rho-rho_FIN-2}(g q_1 q_1) &=&
\frac{N_c}{2}\, \tr t^a t^b t^c
~ \vec I
\cdot
\frac{z_2\vec p_2 - (1-z_2)(\vec k_g -\vec q_{3})}
     {\left( z_2\vec p_2 - (1-z_2)(\vec k_g -\vec q_{3}) \right)^2}
\nn\\
& &  \Psi_{qqq}^*(x_1-x_g,\vec k_1+x_1\vec q - \vec k_g + x_g\vec K - \vec q_{12};
x_2+x_g,\vec k_2 +x_2\vec q-\vec q_{3}+\vec k_g - x_g\vec K;
x_3,\vec k_3 +x_3\vec q)~.
     \label{eq:rho-rho-rho_FIN-2_gq1q1}
\eea
Again, the symmetry factor is 6 (this includes the contribution from diagram
fig.~\ref{fig:rho-rho-rho_FIN-2}($g q_2 q_2$) (not shown) with the
gluon emission and absorption vertices between quarks 1 and 2
interchanged).
\bea
\mathrm{fig.}\, \ref{fig:rho-rho-rho_FIN-2}(q_2 q_1g) &=&
\left( -\frac{1}{4}(T^a)_{bc} +
\frac{N_c}{2}\, \tr t^a t^c t^b\right)
~ \vec I
\cdot
\frac{z_2(\vec p_2-\vec q_3) - (1-z_2)(\vec k_g -\vec q_{1})}
     {\left( z_2(\vec p_2-\vec q_3) - (1-z_2)(\vec k_g -\vec q_{1}) \right)^2}
\nn\\
& &  \Psi_{qqq}^*(x_1-x_g,\vec k_1+x_1\vec q - \vec k_g + x_g\vec K - \vec q_{2};
x_2+x_g,\vec k_2 +x_2\vec q-\vec q_{13}+\vec k_g - x_g\vec K;
x_3,\vec k_3 +x_3\vec q)~.
     \label{eq:rho-rho-rho_FIN-2_q2q1g}
\eea
The symmetry factor is 6, including the contribution from diagram
fig.~\ref{fig:rho-rho-rho_FIN-2}($q_1 q_2 g$) (not shown) with the
gluon emission and absorption vertices between quarks 1 and 2
interchanged.
\bea
\mathrm{fig.}\, \ref{fig:rho-rho-rho_FIN-2}(q_2 g q_1) &=&
\left( \frac{1}{4}(T^b)_{ca} +
\frac{N_c}{2}\, \tr t^a t^b t^c\right)
~ \vec I
\cdot
\frac{z_2(\vec p_2-\vec q_3) - (1-z_2)(\vec k_g -\vec q_{2})}
     {\left( z_2(\vec p_2-\vec q_3) - (1-z_2)(\vec k_g -\vec q_{2}) \right)^2}
\nn\\
& &  \Psi_{qqq}^*(x_1-x_g,\vec k_1+x_1\vec q - \vec k_g + x_g\vec K - \vec q_{1};
x_2+x_g,\vec k_2 +x_2\vec q-\vec q_{23}+\vec k_g - x_g\vec K;
x_3,\vec k_3 +x_3\vec q)~.
     \label{eq:rho-rho-rho_FIN-2_q2gq1}
\eea
The symmetry factor is 6, including the contribution from diagram
fig.~\ref{fig:rho-rho-rho_FIN-2}($q_1 g q_2$) (not shown) with the
gluon emission and absorption vertices between quarks 1 and 2
interchanged.
\bea
\mathrm{fig.}\, \ref{fig:rho-rho-rho_FIN-2}(g q_2 q_1) &=&
\left( \frac{1}{4}(T^c)_{ba} +
\frac{N_c}{2}\, \tr t^a t^c t^b\right)
~ \vec I
\cdot
\frac{z_2(\vec p_2-\vec q_2) - (1-z_2)(\vec k_g -\vec q_{3})}
     {\left( z_2(\vec p_2-\vec q_2) - (1-z_2)(\vec k_g -\vec q_{3}) \right)^2}
\nn\\
& &  \Psi_{qqq}^*(x_1-x_g,\vec k_1+x_1\vec q - \vec k_g + x_g\vec K - \vec q_{1};
x_2+x_g,\vec k_2 +x_2\vec q-\vec q_{23}+\vec k_g - x_g\vec K;
x_3,\vec k_3 +x_3\vec q)~.
     \label{eq:rho-rho-rho_FIN-2_gq2q1}
\eea
The symmetry factor is 6, including the contribution from diagram
fig.~\ref{fig:rho-rho-rho_FIN-2}($g q_1 q_2$) (not shown) with the
gluon emission and absorption vertices between quarks 1 and 2
interchanged.
\bea
\mathrm{fig.}\, \ref{fig:rho-rho-rho_FIN-2}(q_1 q_2 g) &=&
\left( \frac{1}{4}(T^a)_{bc} +
\frac{N_c}{2}\, \tr t^a t^b t^c\right)
~ \vec I
\cdot
\frac{z_2(\vec p_2-\vec q_2) - (1-z_2)(\vec k_g -\vec q_{1})}
     {\left( z_2(\vec p_2-\vec q_2) - (1-z_2)(\vec k_g -\vec q_{1}) \right)^2}
\nn\\
& &  \Psi_{qqq}^*(x_1-x_g,\vec k_1+x_1\vec q - \vec k_g + x_g\vec K - \vec q_{3};
x_2+x_g,\vec k_2 +x_2\vec q-\vec q_{12}+\vec k_g -x_g\vec K;
x_3,\vec k_3 +x_3\vec q)~.
     \label{eq:rho-rho-rho_FIN-2_q1q2g}
\eea
The symmetry factor is 6, including the contribution from diagram
fig.~\ref{fig:rho-rho-rho_FIN-2}($q_2 q_1 g$) with the gluon emission
and absorption vertices between quarks 1 and 2 interchanged.
\bea
\mathrm{fig.}\, \ref{fig:rho-rho-rho_FIN-2}(q_1 g q_2) &=&
\left( \frac{1}{4}(T^b)_{ac} +
\frac{N_c}{2}\, \tr t^a t^c t^b\right)
~ \vec I
\cdot
\frac{z_2(\vec p_2-\vec q_1) - (1-z_2)(\vec k_g -\vec q_{2})}
     {\left( z_2(\vec p_2-\vec q_1) - (1-z_2)(\vec k_g -\vec q_{2}) \right)^2}
\nn\\
& &  \Psi_{qqq}^*(x_1-x_g,\vec k_1+x_1\vec q - \vec k_g + x_g\vec K - \vec q_{3};
x_2+x_g,\vec k_2 +x_2\vec q-\vec q_{12}+\vec k_g - x_g\vec K;
x_3,\vec k_3 +x_3\vec q)~.
     \label{eq:rho-rho-rho_FIN-2_q1gq2}
\eea
The symmetry factor is 6, including the contribution from diagram
fig.~\ref{fig:rho-rho-rho_FIN-2}($q_2 g q_1$) with the gluon emission
and absorption vertices between quarks 1 and 2 interchanged.
\bea
\mathrm{fig.}\, \ref{fig:rho-rho-rho_FIN-2}(g q_1 q_2) &=&
\left( \frac{1}{4}(T^c)_{ab} +
\frac{N_c}{2}\, \tr t^a t^b t^c\right)
~ \vec I
\cdot
\frac{z_2(\vec p_2-\vec q_1) - (1-z_2)(\vec k_g -\vec q_{3})}
     {\left( z_2(\vec p_2-\vec q_1) - (1-z_2)(\vec k_g -\vec q_{3}) \right)^2}
\nn\\
& &  \Psi_{qqq}^*(x_1-x_g,\vec k_1+x_1\vec q - \vec k_g + x_g\vec K - \vec q_{2};
x_2+x_g,\vec k_2 +x_2\vec q-\vec q_{13}+\vec k_g - x_g\vec K;
x_3,\vec k_3 +x_3\vec q)~.
     \label{eq:rho-rho-rho_FIN-2_gq1q2}
\eea
The symmetry factor is 6, including the contribution from diagram
fig.~\ref{fig:rho-rho-rho_FIN-2}($g q_2 q_1$) with the gluon emission
and absorption vertices between quarks 1 and 2 interchanged.
\bea
\mathrm{fig.}\, \ref{fig:rho-rho-rho_FIN-2}(q_2 q_2 g) &=&
\frac{N_c}{2}\, \tr t^a t^b t^c
~ \vec I
\cdot
\frac{z_2(\vec p_2-\vec q_{23}) - (1-z_2)(\vec k_g -\vec q_{1})}
     {\left( z_2(\vec p_2-\vec q_{23}) - (1-z_2)(\vec k_g -\vec q_{1}) \right)^2}
\nn\\
& &  \Psi_{qqq}^*(x_1-x_g,\vec k_1+x_1\vec q - \vec k_g + x_g\vec K;
x_2+x_g,\vec k_2 +x_2\vec q-\vec q+\vec k_g - x_g\vec K;
x_3,\vec k_3 +x_3\vec q)~.
     \label{eq:rho-rho-rho_FIN-2_q2q2g}
\eea
The symmetry factor is 6, including the contribution from diagram
fig.~\ref{fig:rho-rho-rho_FIN-2}($q_1 q_1 g$) with the gluon emission
and absorption vertices between quarks 1 and 2 interchanged.
\bea
\mathrm{fig.}\, \ref{fig:rho-rho-rho_FIN-2}(q_2 g q_2) &=&
\frac{N_c}{2}\, \tr t^b t^a t^c
~ \vec I
\cdot
\frac{z_2(\vec p_2-\vec q_{13}) - (1-z_2)(\vec k_g -\vec q_{2})}
     {\left( z_2(\vec p_2-\vec q_{13}) - (1-z_2)(\vec k_g -\vec q_{2}) \right)^2}
\nn\\
& &  \Psi_{qqq}^*(x_1-x_g,\vec k_1+x_1\vec q - \vec k_g + x_g\vec K;
x_2+x_g,\vec k_2 +x_2\vec q-\vec q+\vec k_g - x_g\vec K;
x_3,\vec k_3 +x_3\vec q)~.
     \label{eq:rho-rho-rho_FIN-2_q2gq2}
\eea
The symmetry factor is 6, including the contribution from diagram
fig.~\ref{fig:rho-rho-rho_FIN-2}($q_1 g q_1$) with the gluon emission
and absorption vertices between quarks 1 and 2 interchanged.
\bea
\mathrm{fig.}\, \ref{fig:rho-rho-rho_FIN-2}(g q_2 q_2) &=&
\frac{N_c}{2}\, \tr t^a t^b t^c
~ \vec I
\cdot
\frac{z_2(\vec p_2-\vec q_{12}) - (1-z_2)(\vec k_g -\vec q_{3})}
     {\left( z_2(\vec p_2-\vec q_{12}) - (1-z_2)(\vec k_g -\vec q_{3}) \right)^2}
\nn\\
& &  \Psi_{qqq}^*(x_1-x_g,\vec k_1+x_1\vec q - \vec k_g + x_g\vec K;
x_2+x_g,\vec k_2 +x_2\vec q-\vec q+\vec k_g - x_g\vec K;
x_3,\vec k_3 +x_3\vec q)~.
     \label{eq:rho-rho-rho_FIN-2_gq2q2}
\eea
The symmetry factor is 6, including the contribution from diagram
fig.~\ref{fig:rho-rho-rho_FIN-2}($g q_1 q_1$) with the gluon emission
and absorption vertices between quarks 1 and 2 interchanged.
\bea
\mathrm{fig.}\, \ref{fig:rho-rho-rho_FIN-2}(q_3 q_1 g) &=&
\left( - \frac{1}{4}(T^a)_{cb} -
N_c \tr t^a t^c t^b + \frac{1}{4} \tr T^c T^a (D^b-T^b)\right)
~ \vec I
\cdot
\frac{z_2\vec p_2 - (1-z_2)(\vec k_g -\vec q_{1})}
     {\left( z_2\vec p_2 - (1-z_2)(\vec k_g -\vec q_{1}) \right)^2}
\nn\\
& &  \Psi_{qqq}^*(x_1-x_g,\vec k_1+x_1\vec q - \vec k_g + x_g\vec K - \vec q_{2};
x_2+x_g,\vec k_2 +x_2\vec q-\vec q_{1}+\vec k_g - x_g\vec K;
x_3,\vec k_3 +x_3\vec q-\vec q_{3})~.
     \label{eq:rho-rho-rho_FIN-2_q3q1g}
\eea
The symmetry factor of 6 includes the contribution from diagram
fig.~\ref{fig:rho-rho-rho_FIN-2}($q_3 q_2 g$) with the gluon emission
and absorption vertices between quarks 1 and 2 interchanged.
\bea
\mathrm{fig.}\, \ref{fig:rho-rho-rho_FIN-2}(q_3 g q_1) &=&
\left( -\frac{1}{4}(T^b)_{ca} -
N_c \tr t^a t^b t^c + \frac{1}{4}\tr T^c T^b (D^a-T^a)\right)
~ \vec I
\cdot
\frac{z_2\vec p_2 - (1-z_2)(\vec k_g -\vec q_{2})}
     {\left( z_2\vec p_2 - (1-z_2)(\vec k_g -\vec q_{2}) \right)^2}
\nn\\
& &  \Psi_{qqq}^*(x_1-x_g,\vec k_1+x_1\vec q - \vec k_g + x_g\vec K - \vec q_{1};
x_2+x_g,\vec k_2 +x_2\vec q-\vec q_{2}+\vec k_g - x_g\vec K;
x_3,\vec k_3 +x_3\vec q-\vec q_{3})~.
     \label{eq:rho-rho-rho_FIN-2_q3gq1}
\eea
The symmetry factor of 6 includes the contribution from diagram
fig.~\ref{fig:rho-rho-rho_FIN-2}($q_3 g q_2$) with the gluon emission
and absorption vertices between quarks 1 and 2 interchanged.
\bea
\mathrm{fig.}\, \ref{fig:rho-rho-rho_FIN-2}(g q_3 q_1) &=&
\left( -\frac{1}{4}(T^c)_{ba} -
N_c \tr t^a t^c t^b + \frac{1}{4}\tr T^b T^c (D^a-T^a)\right)
~ \vec I
\cdot
\frac{z_2\vec p_2 - (1-z_2)(\vec k_g -\vec q_{3})}
     {\left( z_2\vec p_2 - (1-z_2)(\vec k_g -\vec q_{3}) \right)^2}
\nn\\
& &  \Psi_{qqq}^*(x_1-x_g,\vec k_1+x_1\vec q - \vec k_g + x_g\vec K - \vec q_{1};
x_2+x_g,\vec k_2 +x_2\vec q-\vec q_{3}+\vec k_g - x_g\vec K;
x_3,\vec k_3 +x_3\vec q-\vec q_{2})~.
     \label{eq:rho-rho-rho_FIN-2_gq3q1}
\eea
The symmetry factor of 6 includes the contribution from diagram
fig.~\ref{fig:rho-rho-rho_FIN-2}($g q_3 q_2$) with the gluon emission
and absorption vertices between quarks 1 and 2 interchanged.
\bea
\mathrm{fig.}\, \ref{fig:rho-rho-rho_FIN-2}(q_1 q_3 g) &=&
\left( -\frac{1}{4}(T^a)_{bc} -
N_c \tr t^a t^b t^c + \frac{1}{4}\tr T^b T^a (D^c-T^c)\right)
~ \vec I
\cdot
\frac{z_2\vec p_2 - (1-z_2)(\vec k_g -\vec q_{1})}
     {\left( z_2\vec p_2 - (1-z_2)(\vec k_g -\vec q_{1}) \right)^2}
\nn\\
& &  \Psi_{qqq}^*(x_1-x_g,\vec k_1+x_1\vec q - \vec k_g + x_g\vec K - \vec q_{3};
x_2+x_g,\vec k_2 +x_2\vec q-\vec q_{1}+\vec k_g - x_g\vec K;
x_3,\vec k_3 +x_3\vec q-\vec q_{2})~.
     \label{eq:rho-rho-rho_FIN-2_q1q3g}
\eea
The symmetry factor of 6 includes the contribution from diagram
fig.~\ref{fig:rho-rho-rho_FIN-2}($q_2 q_3 g$) with the gluon emission
and absorption vertices between quarks 1 and 2 interchanged.
\bea
\mathrm{fig.}\, \ref{fig:rho-rho-rho_FIN-2}(q_1 g q_3) &=&
\left( -\frac{1}{4}(T^b)_{ac} -
N_c \tr t^b t^a t^c + \frac{1}{4}\tr T^a T^b (D^c-T^c)\right)
~ \vec I
\cdot
\frac{z_2\vec p_2 - (1-z_2)(\vec k_g -\vec q_{2})}
     {\left( z_2\vec p_2 - (1-z_2)(\vec k_g -\vec q_{2}) \right)^2}
\nn\\
& &  \Psi_{qqq}^*(x_1-x_g,\vec k_1+x_1\vec q - \vec k_g + x_g\vec K- \vec q_{3};
x_2+x_g,\vec k_2 +x_2\vec q-\vec q_{2}+\vec k_g - x_g\vec K;
x_3,\vec k_3 +x_3\vec q-\vec q_{1})~.
     \label{eq:rho-rho-rho_FIN-2_q1gq3}
\eea
The symmetry factor of 6 includes the contribution from diagram
fig.~\ref{fig:rho-rho-rho_FIN-2}($q_2 g q_3$) with the gluon emission
and absorption vertices between quarks 1 and 2 interchanged.
\bea
\mathrm{fig.}\, \ref{fig:rho-rho-rho_FIN-2}(g q_1 q_3) &=&
\left( -\frac{1}{4}(T^c)_{ab} -
N_c \tr t^a t^b t^c + \frac{1}{4}\tr T^a T^c (D^b-T^b)\right)
~ \vec I
\cdot
\frac{z_2\vec p_2 - (1-z_2)(\vec k_g -\vec q_{3})}
     {\left( z_2\vec p_2 - (1-z_2)(\vec k_g -\vec q_{3}) \right)^2}
\nn\\
& &  \Psi_{qqq}^*(x_1-x_g,\vec k_1+x_1\vec q - \vec k_g + x_g\vec K - \vec q_{2};
x_2+x_g,\vec k_2 +x_2\vec q-\vec q_{3}+\vec k_g - x_g\vec K;
x_3,\vec k_3 +x_3\vec q-\vec q_{1})~.
     \label{eq:rho-rho-rho_FIN-2_gq1q3}
\eea
The symmetry factor of 6 includes the contribution from diagram
fig.~\ref{fig:rho-rho-rho_FIN-2}($g q_2 q_3$) with the gluon emission
and absorption vertices between quarks 1 and 2 interchanged.
\bea
\mathrm{fig.}\, \ref{fig:rho-rho-rho_FIN-2}(q_3 q_2 g) &=&
\left( -\frac{1}{4}(T^a)_{bc} -
\frac{N_c}{4}(D^a)_{bc}  + \frac{1}{4}\tr T^a (D^b-T^b) T^c \right)
~ \vec I
\cdot
\frac{z_2(\vec p_2-\vec q_2) - (1-z_2)(\vec k_g -\vec q_{1})}
     {\left( z_2(\vec p_2-\vec q_2) - (1-z_2)(\vec k_g -\vec q_{1}) \right)^2}
\nn\\
& &  \Psi_{qqq}^*(x_1-x_g,\vec k_1+x_1\vec q - \vec k_g + x_g\vec K;
x_2+x_g,\vec k_2 +x_2\vec q-\vec q_{12}+\vec k_g - x_g\vec K;
x_3,\vec k_3 +x_3\vec q-\vec q_{3})~.
     \label{eq:rho-rho-rho_FIN-2_q3q2g}
\eea
The symmetry factor of 6 includes the contribution from diagram
fig.~\ref{fig:rho-rho-rho_FIN-2}($q_3 q_1 g$) with the gluon emission
and absorption vertices between quarks 1 and 2 interchanged.
\bea
\mathrm{fig.}\, \ref{fig:rho-rho-rho_FIN-2}(q_3 g q_2) &=&
\left( -\frac{1}{4}(T^b)_{ac} -
\frac{N_c}{4}(D^b)_{ac}  + \frac{1}{4}\tr T^b (D^a-T^a) T^c \right)
~ \vec I
\cdot
\frac{z_2(\vec p_2-\vec q_1) - (1-z_2)(\vec k_g -\vec q_{2})}
     {\left( z_2(\vec p_2-\vec q_1) - (1-z_2)(\vec k_g -\vec q_{2}) \right)^2}
\nn\\
& &  \Psi_{qqq}^*(x_1-x_g,\vec k_1+x_1\vec q - \vec k_g + x_g\vec K;
x_2+x_g,\vec k_2 +x_2\vec q-\vec q_{12}+\vec k_g - x_g\vec K;
x_3,\vec k_3 +x_3\vec q-\vec q_{3})~.
     \label{eq:rho-rho-rho_FIN-2_q3gq2}
\eea
The symmetry factor of 6 includes the contribution from diagram
fig.~\ref{fig:rho-rho-rho_FIN-2}($q_3 g q_1$) with the gluon emission
and absorption vertices between quarks 1 and 2 interchanged.
\bea
\mathrm{fig.}\, \ref{fig:rho-rho-rho_FIN-2}(gq_3 q_2) &=&
\left( -\frac{1}{4}(T^c)_{ab} -
\frac{N_c}{4}(D^c)_{ab}  + \frac{1}{4}\tr T^c (D^a-T^a) T^b \right)
~ \vec I
\cdot
\frac{z_2(\vec p_2-\vec q_1) - (1-z_2)(\vec k_g -\vec q_{3})}
     {\left( z_2(\vec p_2-\vec q_1) - (1-z_2)(\vec k_g -\vec q_{3}) \right)^2}
\nn\\
& &  \Psi_{qqq}^*(x_1-x_g,\vec k_1+x_1\vec q - \vec k_g + x_g\vec K;
x_2+x_g,\vec k_2 +x_2\vec q-\vec q_{13}+\vec k_g - x_g\vec K;
x_3,\vec k_3 +x_3\vec q-\vec q_{2})~.
     \label{eq:rho-rho-rho_FIN-2_gq3q2}
\eea
The symmetry factor of 6 includes the contribution from diagram
fig.~\ref{fig:rho-rho-rho_FIN-2}($g q_3 q_1$) with the gluon emission
and absorption vertices between quarks 1 and 2 interchanged.
\bea
\mathrm{fig.}\, \ref{fig:rho-rho-rho_FIN-2}(q_2 q_3 g) &=&
\left( \frac{1}{4}(T^a)_{bc} -
\frac{N_c}{4}(D^a)_{bc}  + \frac{1}{4}\tr T^a (D^c-T^c) T^b \right)
~ \vec I
\cdot
\frac{z_2(\vec p_2-\vec q_3) - (1-z_2)(\vec k_g -\vec q_{1})}
     {\left( z_2(\vec p_2-\vec q_3) - (1-z_2)(\vec k_g -\vec q_{1}) \right)^2}
\nn\\
& &  \Psi_{qqq}^*(x_1-x_g,\vec k_1+x_1\vec q - \vec k_g + x_g\vec K;
x_2+x_g,\vec k_2 +x_2\vec q-\vec q_{13}+\vec k_g - x_g\vec K;
x_3,\vec k_3 +x_3\vec q-\vec q_{2})~.
     \label{eq:rho-rho-rho_FIN-2_q2q3g}
\eea
The symmetry factor of 6 includes the contribution from diagram
fig.~\ref{fig:rho-rho-rho_FIN-2}($q_1 q_3 g$) with the gluon emission
and absorption vertices between quarks 1 and 2 interchanged.
\bea
\mathrm{fig.}\, \ref{fig:rho-rho-rho_FIN-2}(q_2 g q_3) &=&
\left( \frac{1}{4}(T^b)_{ac} -
\frac{N_c}{4}(D^b)_{ac}  + \frac{1}{4}\tr T^b (D^c-T^c) T^a \right)
~ \vec I
\cdot
\frac{z_2(\vec p_2-\vec q_3) - (1-z_2)(\vec k_g -\vec q_{2})}
     {\left( z_2(\vec p_2-\vec q_3) - (1-z_2)(\vec k_g -\vec q_{2}) \right)^2}
\nn\\
& &  \Psi_{qqq}^*(x_1-x_g,\vec k_1+x_1\vec q - \vec k_g + x_g\vec K;
x_2+x_g,\vec k_2 +x_2\vec q-\vec q_{23}+\vec k_g - x_g\vec K;
x_3,\vec k_3 +x_3\vec q-\vec q_{1})~.
     \label{eq:rho-rho-rho_FIN-2_q2gq3}
\eea
The symmetry factor of 6 includes the contribution from diagram
fig.~\ref{fig:rho-rho-rho_FIN-2}($q_1 gq_3$) with the gluon emission
and absorption vertices between quarks 1 and 2 interchanged.
\bea
\mathrm{fig.}\, \ref{fig:rho-rho-rho_FIN-2}(g q_2 q_3) &=&
\left( \frac{1}{4}(T^c)_{ab} -
\frac{N_c}{4}(D^c)_{ab}  + \frac{1}{4}\tr T^c (D^b-T^b) T^a \right)
~ \vec I
\cdot
\frac{z_2(\vec p_2-\vec q_2) - (1-z_2)(\vec k_g -\vec q_{3})}
     {\left( z_2(\vec p_2-\vec q_2) - (1-z_2)(\vec k_g -\vec q_{3}) \right)^2}
\nn\\
& &  \Psi_{qqq}^*(x_1-x_g,\vec k_1+x_1\vec q - \vec k_g + x_g\vec K;
x_2+x_g,\vec k_2 +x_2\vec q-\vec q_{23}+\vec k_g - x_g\vec K;
x_3,\vec k_3 +x_3\vec q-\vec q_{1})~.
     \label{eq:rho-rho-rho_FIN-2_gq2q3}
\eea
The symmetry factor of 6 includes the contribution from diagram
fig.~\ref{fig:rho-rho-rho_FIN-2}($g q_1 q_3$) with the gluon emission
and absorption vertices between quarks 1 and 2 interchanged.
\bea
\mathrm{fig.}\, \ref{fig:rho-rho-rho_FIN-2}(q_3 q_3 g) &=& 0 \\
\mathrm{fig.}\, \ref{fig:rho-rho-rho_FIN-2}(q_3 g q_3) &=& 0 \\
\mathrm{fig.}\, \ref{fig:rho-rho-rho_FIN-2}(g q_3 q_3) &=& 0~.
     \label{eq:rho-rho-rho_FIN-2_q3q3g}
\eea

\subsubsection{Coupling only to quarks}
\label{sec:finite-only-to-quarks}

\begin{figure}[htb]
  \centering
  \begin{minipage}[hb]{\linewidth}
    \includegraphics[width=0.23\linewidth]{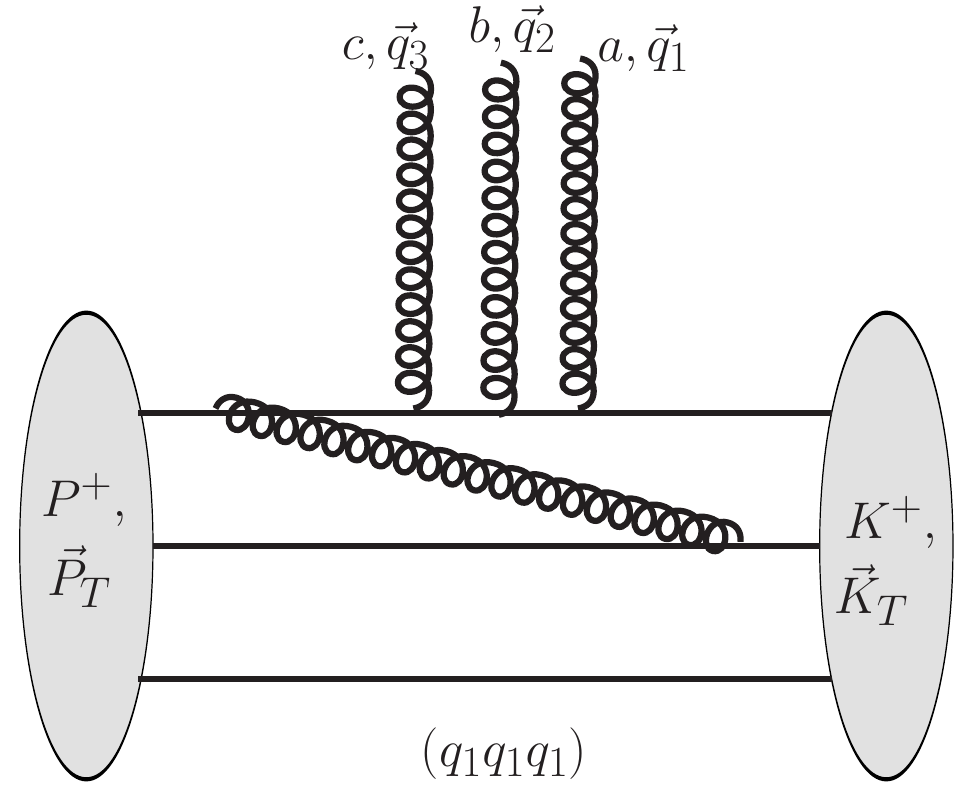}
    \hspace{0.1cm}
    \includegraphics[width=0.23\linewidth]{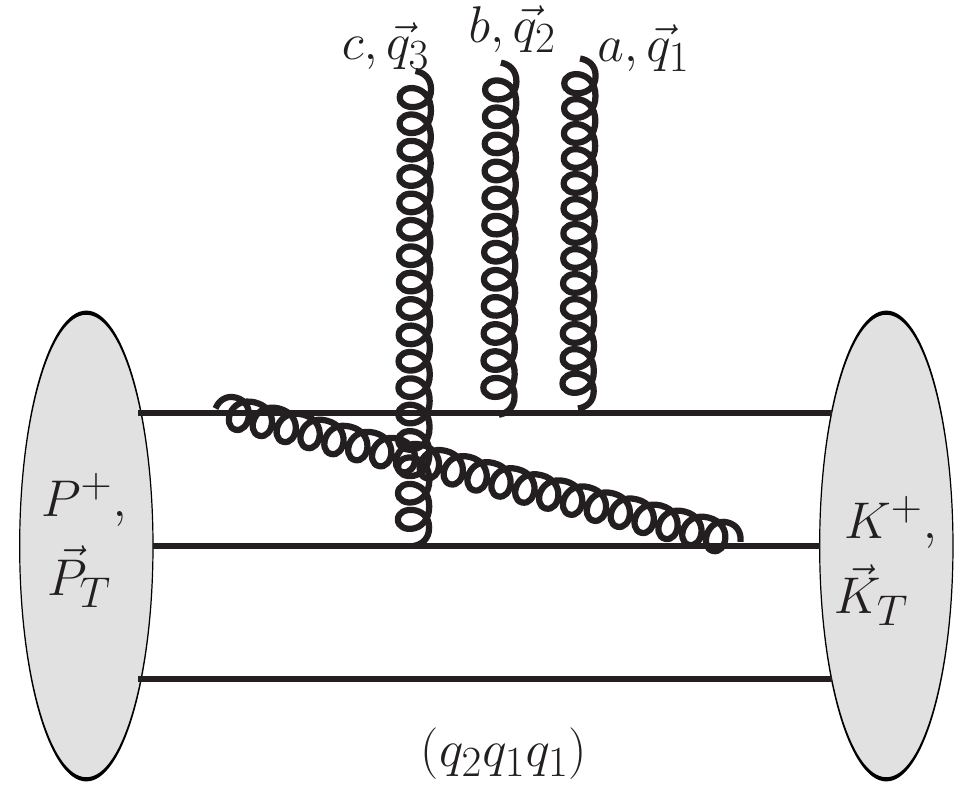}
    \hspace{0.1cm}
    \includegraphics[width=0.23\linewidth]{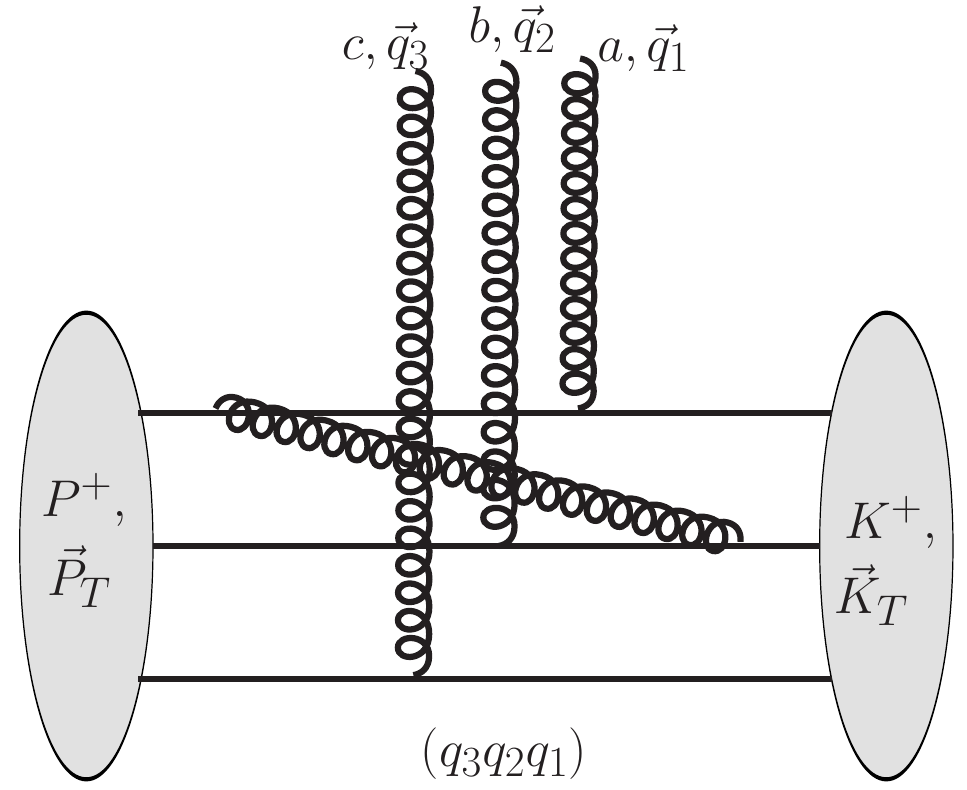}
    \hspace{0.1cm}
    \includegraphics[width=0.23\linewidth]{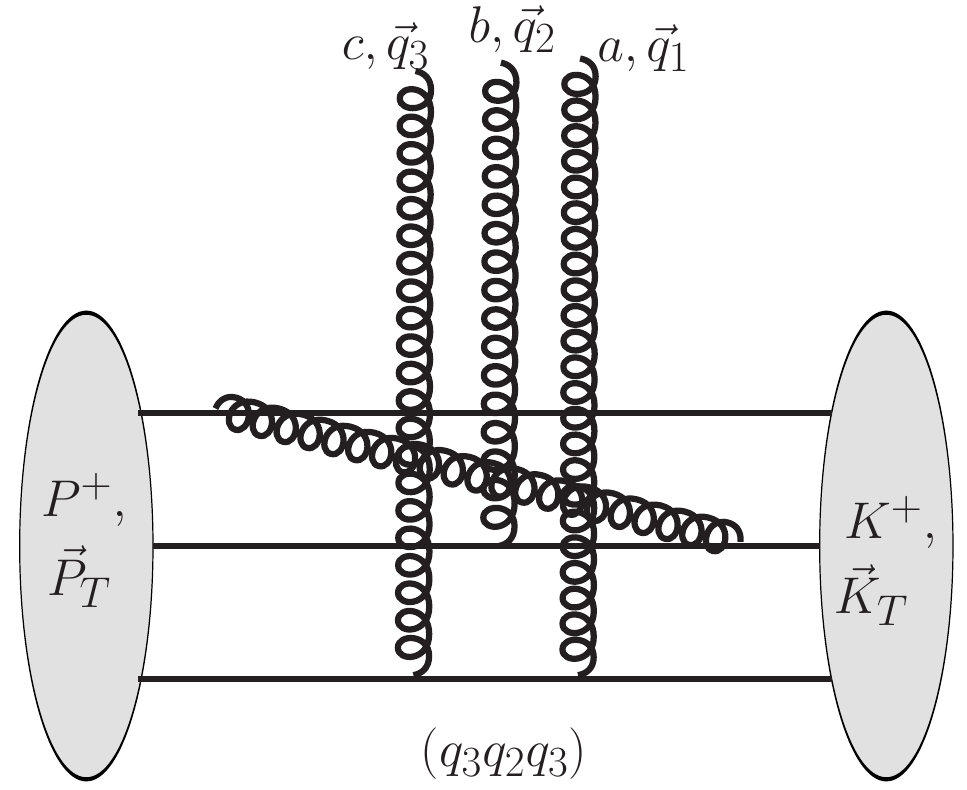}
  \end{minipage}
\caption{A sample of UV finite diagrams for $\langle \rho^a(\vec
  q_1)\, \rho^b(\vec q_2)\,\rho^c(\vec q_3)\rangle$ evaluated in the
  $|qqqg\rangle$ Fock state, where all three probes attach to quarks in
  the proton.  The cut is located at the insertion of the three color
  charge operators.}
\label{fig:rho-rho-rho_FIN-q.a}
\end{figure}
Here we consider the diagrams where all three gluon probes couple
exclusively to quarks in the proton. We begin with the matrix element
of $\rho^a(\vec q_1)\, \rho^b(\vec q_2)\,\rho^c(\vec q_3)$ in the
$|qqqg\rangle$ Fock state. A few examples are shown in
fig.~\ref{fig:rho-rho-rho_FIN-q.a}.

%
\bea
\mathrm{fig.}\, \ref{fig:rho-rho-rho_FIN-q.a}(q_1 q_1 q_1) &=&
- C_F \tr t^a t^b t^c\,
\vec I \cdot
\frac{z_2\vec p_2 - (1-z_2)\vec k_g }
     {\left( z_2\vec p_2 - (1-z_2)\vec k_g \right)^2}
\nn\\
& &  \Psi_{qqq}^*(x_1-x_g,\vec k_1-(1-x_1)\vec q - \vec k_g + x_g\vec K;
x_2+x_g,\vec k_2 +x_2\vec q +\vec k_g - x_g\vec K;
x_3,\vec k_3 +x_3\vec q)~.
     \label{eq:rho-rho-rho_FIN-q1q1q1}
\eea
The symmetry factor of 6 includes the contribution from diagram
fig.~\ref{fig:rho-rho-rho_FIN-q.a}($q_2 q_2 q_2$) with the gluon
emission and absorption vertices between quarks 1 and 2 interchanged.
\bea
\mathrm{fig.}\, \ref{fig:rho-rho-rho_FIN-q.a}(q_2 q_1 q_1) &=&
- (C_F-\frac{1}{2}) \tr t^a t^b t^c\,
\vec I \cdot
\frac{z_2(\vec p_2 -\vec q_3) - (1-z_2)\vec k_g }
     {\left( z_2(\vec p_2  -\vec q_3) - (1-z_2)\vec k_g \right)^2}
\nn\\
& &  \Psi_{qqq}^*(x_1-x_g,\vec k_1+x_1\vec q - \vec q_{12}- \vec k_g + x_g\vec K;
x_2+x_g,\vec k_2 +x_2\vec q  - \vec q_{3} +\vec k_g - x_g\vec K;
x_3,\vec k_3 +x_3\vec q)~.
     \label{eq:rho-rho-rho_FIN-q2q1q1}
\eea
The symmetry factor of 6 includes the contribution from diagram
fig.~\ref{fig:rho-rho-rho_FIN-q.a}($q_1 q_2 q_2$) with the gluon
emission and absorption vertices between quarks 1 and 2 interchanged.
\bea
\mathrm{fig.}\, \ref{fig:rho-rho-rho_FIN-q.a}(q_3 q_1 q_1) &=&
(C_F-\frac{1}{2}-\frac{1}{2N_c}) \tr t^a t^b t^c\,
\vec I \cdot
\frac{z_2\vec p_2 - (1-z_2)\vec k_g }
     {\left( z_2\vec p_2  - (1-z_2)\vec k_g \right)^2}
\nn\\
& &  \Psi_{qqq}^*(x_1-x_g,\vec k_1+x_1\vec q - \vec q_{12}- \vec k_g + x_g\vec K;
x_2+x_g,\vec k_2 +x_2\vec q +\vec k_g - x_g\vec K;
x_3,\vec k_3 +x_3\vec q - \vec q_{3})~.
     \label{eq:rho-rho-rho_FIN-q3q1q1}
\eea
The symmetry factor of 6 includes the contribution from diagram
fig.~\ref{fig:rho-rho-rho_FIN-q.a}($q_3 q_2 q_2$) with the gluon
emission and absorption vertices between quarks 1 and 2 interchanged.
\bea
\mathrm{fig.}\, \ref{fig:rho-rho-rho_FIN-q.a}(q_1 q_2 q_1) &=&
- (C_F-\frac{1}{2}) \tr t^a t^c t^b\,
\vec I \cdot
\frac{z_2(\vec p_2 -\vec q_2) - (1-z_2)\vec k_g }
     {\left( z_2(\vec p_2  -\vec q_2) - (1-z_2)\vec k_g \right)^2}
\nn\\
& &  \Psi_{qqq}^*(x_1-x_g,\vec k_1+x_1\vec q - \vec q_{13}- \vec k_g + x_g\vec K;
x_2+x_g,\vec k_2 +x_2\vec q  - \vec q_{2} +\vec k_g - x_g\vec K;
x_3,\vec k_3 +x_3\vec q)~.
     \label{eq:rho-rho-rho_FIN-q1q2q1}
\eea
The symmetry factor of 6 includes the contribution from diagram
fig.~\ref{fig:rho-rho-rho_FIN-q.a}($q_2 q_1 q_2$) with the gluon
emission and absorption vertices between quarks 1 and 2 interchanged.
\bea
\mathrm{fig.}\, \ref{fig:rho-rho-rho_FIN-q.a}(q_2 q_2 q_1) &=&
- (C_F-\frac{1}{2}) \tr t^a t^b t^c\,
\vec I \cdot
\frac{z_2(\vec p_2 -\vec q_{23}) - (1-z_2)\vec k_g }
     {\left( z_2(\vec p_2  -\vec q_{23}) - (1-z_2)\vec k_g \right)^2}
\nn\\
& &  \Psi_{qqq}^*(x_1-x_g,\vec k_1+x_1\vec q - \vec q_{1}- \vec k_g + x_g\vec K;
x_2+x_g,\vec k_2 +x_2\vec q  - \vec q_{23} +\vec k_g - x_g\vec K;
x_3,\vec k_3 +x_3\vec q)~.
     \label{eq:rho-rho-rho_FIN-q2q2q1}
\eea
The symmetry factor of 6 includes the contribution from diagram
fig.~\ref{fig:rho-rho-rho_FIN-q.a}($q_1 q_1 q_2$) with the gluon
emission and absorption vertices between quarks 1 and 2 interchanged.
\bea
\mathrm{fig.}\, \ref{fig:rho-rho-rho_FIN-q.a}(q_3 q_2 q_1) &=&
[(C_F-1) \tr t^a t^b t^c + (C_F-\frac{N_c}{2}) \tr t^a t^c t^b]\,
\vec I \cdot
\frac{z_2(\vec p_2 -\vec q_{2}) - (1-z_2)\vec k_g }
     {\left( z_2(\vec p_2  -\vec q_{2}) - (1-z_2)\vec k_g \right)^2}
\nn\\
& &  \Psi_{qqq}^*(x_1-x_g,\vec k_1+x_1\vec q - \vec q_{1}- \vec k_g + x_g\vec K;
x_2+x_g,\vec k_2 +x_2\vec q  - \vec q_{2} +\vec k_g - x_g\vec K;
x_3,\vec k_3 +x_3\vec q - \vec q_{3} )~.
     \label{eq:rho-rho-rho_FIN-q3q2q1}
\eea
The symmetry factor of 6 includes the contribution from diagram
fig.~\ref{fig:rho-rho-rho_FIN-q.a}($q_3 q_1 q_2$) with the gluon
emission and absorption vertices between quarks 1 and 2 interchanged.
\bea
\mathrm{fig.}\, \ref{fig:rho-rho-rho_FIN-q.a}(q_1 q_3 q_1) &=&
(2C_F-\frac{1}{2}-\frac{N_c}{2}) \tr t^a t^c t^b\,
\vec I \cdot
\frac{z_2\vec p_2 - (1-z_2)\vec k_g }
     {\left( z_2\vec p_2  - (1-z_2)\vec k_g \right)^2}
\nn\\
& &  \Psi_{qqq}^*(x_1-x_g,\vec k_1+x_1\vec q - \vec q_{13}- \vec k_g + x_g\vec K;
x_2+x_g,\vec k_2 +x_2\vec q  +\vec k_g - x_g\vec K;
x_3,\vec k_3 +x_3\vec q - \vec q_{2})~.
     \label{eq:rho-rho-rho_FIN-q1q3q1}
\eea
The symmetry factor of 6 includes the contribution from diagram
fig.~\ref{fig:rho-rho-rho_FIN-q.a}($q_2 q_3 q_2$) with the gluon
emission and absorption vertices between quarks 1 and 2 interchanged.
\bea
\mathrm{fig.}\, \ref{fig:rho-rho-rho_FIN-q.a}(q_2 q_3 q_1) &=&
[(C_F-1) \tr t^a t^c t^b + (C_F-\frac{N_c}{2}) \tr t^a t^b t^c]\,
\vec I \cdot
\frac{z_2(\vec p_2 -\vec q_3) - (1-z_2)\vec k_g }
     {\left( z_2(\vec p_2 -\vec q_3)  - (1-z_2)\vec k_g \right)^2}
\nn\\
& &  \Psi_{qqq}^*(x_1-x_g,\vec k_1+x_1\vec q - \vec q_{1}- \vec k_g + x_g\vec K;
x_2+x_g,\vec k_2 +x_2\vec q - \vec q_{3}  +\vec k_g - x_g\vec K;
x_3,\vec k_3 +x_3\vec q - \vec q_{2})~.
     \label{eq:rho-rho-rho_FIN-q2q3q1}
\eea
The symmetry factor of 6 includes the contribution from diagram
fig.~\ref{fig:rho-rho-rho_FIN-q.a}($q_1 q_3 q_2$) with the gluon
emission and absorption vertices between quarks 1 and 2 interchanged.
\bea
\mathrm{fig.}\, \ref{fig:rho-rho-rho_FIN-q.a}(q_3 q_3 q_1) &=&
       (2C_F-\frac{1}{2}-\frac{N_c}{2}) \tr t^a t^b t^c\,
\vec I \cdot
\frac{z_2\vec p_2  - (1-z_2)\vec k_g }
     {\left( z_2\vec p_2  - (1-z_2)\vec k_g \right)^2}
\nn\\
& &  \Psi_{qqq}^*(x_1-x_g,\vec k_1+x_1\vec q - \vec q_{1}- \vec k_g + x_g\vec K;
x_2+x_g,\vec k_2 +x_2\vec q  +\vec k_g - x_g\vec K;
x_3,\vec k_3 +x_3\vec q - \vec q_{23})~.
     \label{eq:rho-rho-rho_FIN-q3q3q1}
\eea
The symmetry factor of 6 includes the contribution from diagram
fig.~\ref{fig:rho-rho-rho_FIN-q.a}($q_3 q_3 q_2$) with the gluon
emission and absorption vertices between quarks 1 and 2 interchanged.


%
\bea
\mathrm{fig.}\, \ref{fig:rho-rho-rho_FIN-q.a}(q_1 q_1 q_2) &=&
- (C_F - \frac{1}{2}) \tr t^a t^b t^c\,
\vec I \cdot
\frac{z_2(\vec p_2 -\vec q_1) - (1-z_2)\vec k_g }
     {\left( z_2(\vec p_2 -\vec q_1) - (1-z_2)\vec k_g \right)^2}
\nn\\
& &  \Psi_{qqq}^*(x_1-x_g,\vec k_1+x_1\vec q - \vec q_{23} - \vec k_g + x_g\vec K;
x_2+x_g,\vec k_2 +x_2\vec q  - \vec q_{1} +\vec k_g - x_g\vec K;
x_3,\vec k_3 +x_3\vec q)~.
     \label{eq:rho-rho-rho_FIN-q1q1q2}
\eea
The symmetry factor of 6 includes the contribution from diagram
fig.~\ref{fig:rho-rho-rho_FIN-q.a}($q_2 q_2 q_1$) with the gluon
emission and absorption vertices between quarks 1 and 2 interchanged.
\bea
\mathrm{fig.}\, \ref{fig:rho-rho-rho_FIN-q.a}(q_2 q_1 q_2) &=&
- (C_F-\frac{1}{2}) \tr t^a t^c t^b\,
\vec I \cdot
\frac{z_2(\vec p_2 -\vec q_{13}) - (1-z_2)\vec k_g }
     {\left( z_2(\vec p_2  -\vec q_{13}) - (1-z_2)\vec k_g \right)^2}
\nn\\
& &  \Psi_{qqq}^*(x_1-x_g,\vec k_1+x_1\vec q - \vec q_{2}- \vec k_g + x_g\vec K;
x_2+x_g,\vec k_2 +x_2\vec q  - \vec q_{13} +\vec k_g - x_g\vec K;
x_3,\vec k_3 +x_3\vec q)~.
     \label{eq:rho-rho-rho_FIN-q2q1q2}
\eea
The symmetry factor of 6 includes the contribution from diagram
fig.~\ref{fig:rho-rho-rho_FIN-q.a}($q_1 q_2 q_1$) with the gluon
emission and absorption vertices between quarks 1 and 2 interchanged.
\bea
\mathrm{fig.}\, \ref{fig:rho-rho-rho_FIN-q.a}(q_3 q_1 q_2) &=&
[(C_F-\frac{N_c}{2}) \tr t^a t^b t^c + (C_F-1) \tr t^a t^c t^b]\,
\vec I \cdot
\frac{z_2(\vec p_2 - \vec q_1) - (1-z_2)\vec k_g }
     {\left( z_2(\vec p_2 - \vec q_1) - (1-z_2)\vec k_g \right)^2}
\nn\\
& &  \Psi_{qqq}^*(x_1-x_g,\vec k_1+x_1\vec q - \vec q_{2}- \vec k_g + x_g\vec K;
x_2+x_g,\vec k_2 +x_2\vec q  - \vec q_{1} +\vec k_g - x_g\vec K;
x_3,\vec k_3 +x_3\vec q - \vec q_{3})~.
     \label{eq:rho-rho-rho_FIN-q3q1q2}
\eea
The symmetry factor of 6 includes the contribution from diagram
fig.~\ref{fig:rho-rho-rho_FIN-q.a}($q_3 q_2 q_1$) with the gluon
emission and absorption vertices between quarks 1 and 2 interchanged.
\bea
\mathrm{fig.}\, \ref{fig:rho-rho-rho_FIN-q.a}(q_1 q_2 q_2) &=&
- (C_F-\frac{1}{2}) \tr t^a t^b t^c\,
\vec I \cdot
\frac{z_2(\vec p_2 -\vec q_{12}) - (1-z_2)\vec k_g }
     {\left( z_2(\vec p_2  -\vec q_{12}) - (1-z_2)\vec k_g \right)^2}
\nn\\
& &  \Psi_{qqq}^*(x_1-x_g,\vec k_1+x_1\vec q - \vec q_{3}- \vec k_g + x_g\vec K;
x_2+x_g,\vec k_2 +x_2\vec q  - \vec q_{12} +\vec k_g - x_g\vec K;
x_3,\vec k_3 +x_3\vec q)~.
     \label{eq:rho-rho-rho_FIN-q1q2q2}
\eea
The symmetry factor of 6 includes the contribution from diagram
fig.~\ref{fig:rho-rho-rho_FIN-q.a}($q_2 q_1 q_1$) with the gluon
emission and absorption vertices between quarks 1 and 2 interchanged.
\bea
\mathrm{fig.}\, \ref{fig:rho-rho-rho_FIN-q.a}(q_2 q_2 q_2) &=&
- C_F \tr t^a t^b t^c\,
\vec I \cdot
\frac{z_2(\vec p_2 -\vec q) - (1-z_2)\vec k_g }
     {\left( z_2(\vec p_2  -\vec q) - (1-z_2)\vec k_g \right)^2}
\nn\\
& &  \Psi_{qqq}^*(x_1-x_g,\vec k_1+x_1\vec q - \vec k_g + x_g\vec K;
x_2+x_g,\vec k_2 +x_2\vec q  - \vec q +\vec k_g - x_g\vec K;
x_3,\vec k_3 +x_3\vec q)~.
     \label{eq:rho-rho-rho_FIN-q2q2q2}
\eea
The symmetry factor of 6 includes the contribution from diagram
fig.~\ref{fig:rho-rho-rho_FIN-q.a}($q_1 q_1 q_1$) with the gluon
emission and absorption vertices between quarks 1 and 2 interchanged.
\bea
\mathrm{fig.}\, \ref{fig:rho-rho-rho_FIN-q.a}(q_3 q_2 q_2) &=&
(2C_F-\frac{1}{2}-\frac{N_c}{2}) \tr t^a t^b t^c \,
\vec I \cdot
\frac{z_2(\vec p_2 -\vec q_{12}) - (1-z_2)\vec k_g }
     {\left( z_2(\vec p_2  -\vec q_{12}) - (1-z_2)\vec k_g \right)^2}
\nn\\
& &  \Psi_{qqq}^*(x_1-x_g,\vec k_1+x_1\vec q - \vec k_g + x_g\vec K;
x_2+x_g,\vec k_2 +x_2\vec q  - \vec q_{12} +\vec k_g - x_g\vec K;
x_3,\vec k_3 +x_3\vec q - \vec q_{3} )~.
     \label{eq:rho-rho-rho_FIN-q3q2q2}
\eea
The symmetry factor of 6 includes the contribution from diagram
fig.~\ref{fig:rho-rho-rho_FIN-q.a}($q_3 q_1 q_1$) with the gluon
emission and absorption vertices between quarks 1 and 2 interchanged.
\bea
\mathrm{fig.}\, \ref{fig:rho-rho-rho_FIN-q.a}(q_1 q_3 q_2) &=&
       [(C_F-1)\tr t^a t^b t^c + (C_F-\frac{N_c}{2}) \tr t^a t^c t^b]\,
\vec I \cdot
\frac{z_2(\vec p_2 -\vec q_1) - (1-z_2)\vec k_g }
     {\left( z_2(\vec p_2 - \vec q_1) - (1-z_2)\vec k_g \right)^2}
\nn\\
& &  \Psi_{qqq}^*(x_1-x_g,\vec k_1+x_1\vec q - \vec q_{3}- \vec k_g + x_g\vec K;
x_2+x_g,\vec k_2 +x_2\vec q - \vec q_{1}  +\vec k_g - x_g\vec K;
x_3,\vec k_3 +x_3\vec q - \vec q_{2})~.
     \label{eq:rho-rho-rho_FIN-q1q3q2}
\eea
The symmetry factor of 6 includes the contribution from diagram
fig.~\ref{fig:rho-rho-rho_FIN-q.a}($q_2 q_3 q_1$) with the gluon
emission and absorption vertices between quarks 1 and 2 interchanged.
\bea
\mathrm{fig.}\, \ref{fig:rho-rho-rho_FIN-q.a}(q_2 q_3 q_2) &=&
(2C_F-\frac{1}{2} -\frac{N_c}{2}) \tr t^a t^c t^b\,
\vec I \cdot
\frac{z_2(\vec p_2 -\vec q_{13}) - (1-z_2)\vec k_g }
     {\left( z_2(\vec p_2 -\vec q_{13})  - (1-z_2)\vec k_g \right)^2}
\nn\\
& &  \Psi_{qqq}^*(x_1-x_g,\vec k_1+x_1\vec q - \vec k_g + x_g\vec K;
x_2+x_g,\vec k_2 +x_2\vec q - \vec q_{13}  +\vec k_g - x_g\vec K;
x_3,\vec k_3 +x_3\vec q - \vec q_{2})~.
     \label{eq:rho-rho-rho_FIN-q2q3q2}
\eea
The symmetry factor of 6 includes the contribution from diagram
fig.~\ref{fig:rho-rho-rho_FIN-q.a}($q_1 q_3 q_1$) with the gluon
emission and absorption vertices between quarks 1 and 2 interchanged.
\bea
\mathrm{fig.}\, \ref{fig:rho-rho-rho_FIN-q.a}(q_3 q_3 q_2) &=&
       (2C_F-\frac{1}{2}-\frac{N_c}{2}) \tr t^a t^b t^c\,
\vec I \cdot
\frac{z_2(\vec p_2 -\vec q_1) - (1-z_2)\vec k_g }
     {\left( z_2(\vec p_2 -\vec q_1) - (1-z_2)\vec k_g \right)^2}
\nn\\
& &  \Psi_{qqq}^*(x_1-x_g,\vec k_1+x_1\vec q - \vec k_g + x_g\vec K;
x_2+x_g,\vec k_2 +x_2\vec q -\vec q_1 +\vec k_g - x_g\vec K;
x_3,\vec k_3 +x_3\vec q - \vec q_{23})~.
     \label{eq:rho-rho-rho_FIN-q3q3q2}
\eea
The symmetry factor of 6 includes the contribution from diagram
fig.~\ref{fig:rho-rho-rho_FIN-q.a}($q_3 q_3 q_1$) with the gluon
emission and absorption vertices between quarks 1 and 2 interchanged.


%
\bea
\mathrm{fig.}\, \ref{fig:rho-rho-rho_FIN-q.a}(q_1 q_1 q_3) &=&
(2C_F - \frac{1}{2}- \frac{N_c}{2}) \tr t^a t^b t^c\,
\vec I \cdot
\frac{z_2\vec p_2 - (1-z_2)\vec k_g }
     {\left( z_2\vec p_2 - (1-z_2)\vec k_g \right)^2}
\nn\\
& &  \Psi_{qqq}^*(x_1-x_g,\vec k_1+x_1\vec q - \vec q_{23} - \vec k_g + x_g\vec K;
x_2+x_g,\vec k_2 +x_2\vec q  +\vec k_g - x_g\vec K;
x_3,\vec k_3 +x_3\vec q - \vec q_{1})~.
     \label{eq:rho-rho-rho_FIN-q1q1q3}
\eea
The symmetry factor of 6 includes the contribution from diagram
fig.~\ref{fig:rho-rho-rho_FIN-q.a}($q_2 q_2 q_3$) with the gluon
emission and absorption vertices between quarks 1 and 2 interchanged.
\bea
\mathrm{fig.}\, \ref{fig:rho-rho-rho_FIN-q.a}(q_2 q_1 q_3) &=&
[(C_F-1) \tr t^a t^b t^c + (C_F-\frac{N_c}{2}) \tr t^a t^c t^b] \,
\vec I \cdot
\frac{z_2(\vec p_2 -\vec q_3) - (1-z_2)\vec k_g }
     {\left( z_2(\vec p_2  -\vec q_3) - (1-z_2)\vec k_g \right)^2}
\nn\\
& &  \Psi_{qqq}^*(x_1-x_g,\vec k_1+x_1\vec q - \vec q_{2}- \vec k_g + x_g\vec K;
x_2+x_g,\vec k_2 +x_2\vec q  - \vec q_{3} +\vec k_g - x_g\vec K;
x_3,\vec k_3 +x_3\vec q  - \vec q_{1})~.
     \label{eq:rho-rho-rho_FIN-q2q1q3}
\eea
The symmetry factor of 6 includes the contribution from diagram
fig.~\ref{fig:rho-rho-rho_FIN-q.a}($q_1 q_2 q_3$) with the gluon
emission and absorption vertices between quarks 1 and 2 interchanged.
\bea
\mathrm{fig.}\, \ref{fig:rho-rho-rho_FIN-q.a}(q_3 q_1 q_3) &=&
(2C_F-\frac{1}{2}-\frac{N_c}{2}) \tr t^a t^c t^b\,
\vec I \cdot
\frac{z_2\vec p_2 - (1-z_2)\vec k_g }
     {\left( z_2\vec p_2  - (1-z_2)\vec k_g \right)^2}
\nn\\
& &  \Psi_{qqq}^*(x_1-x_g,\vec k_1+x_1\vec q - \vec q_{2}- \vec k_g + x_g\vec K;
x_2+x_g,\vec k_2 +x_2\vec q +\vec k_g - x_g\vec K;
x_3,\vec k_3 +x_3\vec q - \vec q_{13})~.
     \label{eq:rho-rho-rho_FIN-q3q1q3}
\eea
The symmetry factor of 6 includes the contribution from diagram
fig.~\ref{fig:rho-rho-rho_FIN-q.a}($q_3 q_2 q_3$) with the gluon
emission and absorption vertices between quarks 1 and 2 interchanged.
\bea
\mathrm{fig.}\, \ref{fig:rho-rho-rho_FIN-q.a}(q_1 q_2 q_3) &=&
[(C_F-1) \tr t^a t^c t^b + (C_F-\frac{N_c}{2}) \tr t^a t^b t^c]\,
\vec I \cdot
\frac{z_2(\vec p_2 -\vec q_2) - (1-z_2)\vec k_g }
     {\left( z_2(\vec p_2  -\vec q_2) - (1-z_2)\vec k_g \right)^2}
\nn\\
& &  \Psi_{qqq}^*(x_1-x_g,\vec k_1+x_1\vec q - \vec q_{3}- \vec k_g + x_g\vec K;
x_2+x_g,\vec k_2 +x_2\vec q  - \vec q_{2} +\vec k_g - x_g\vec K;
x_3,\vec k_3 +x_3\vec q - \vec q_{1})~.
     \label{eq:rho-rho-rho_FIN-q1q2q3}
\eea
The symmetry factor of 6 includes the contribution from diagram
fig.~\ref{fig:rho-rho-rho_FIN-q.a}($q_2 q_1 q_3$) with the gluon
emission and absorption vertices between quarks 1 and 2 interchanged.
\bea
\mathrm{fig.}\, \ref{fig:rho-rho-rho_FIN-q.a}(q_2 q_2 q_3) &=&
 (2C_F-\frac{1}{2}-\frac{N_c}{2}) \tr t^a t^b t^c\,
\vec I \cdot
\frac{z_2(\vec p_2 -\vec q_{23}) - (1-z_2)\vec k_g }
     {\left( z_2(\vec p_2  -\vec q_{23}) - (1-z_2)\vec k_g \right)^2}
\nn\\
& &  \Psi_{qqq}^*(x_1-x_g,\vec k_1+x_1\vec q - \vec k_g + x_g\vec K;
x_2+x_g,\vec k_2 +x_2\vec q  - \vec q_{23} +\vec k_g - x_g\vec K;
x_3,\vec k_3 +x_3\vec q - \vec q_{1})~.
     \label{eq:rho-rho-rho_FIN-q2q2q3}
\eea
The symmetry factor of 6 includes the contribution from diagram
fig.~\ref{fig:rho-rho-rho_FIN-q.a}($q_1 q_1 q_3$) with the gluon
emission and absorption vertices between quarks 1 and 2 interchanged.
\bea
\mathrm{fig.}\, \ref{fig:rho-rho-rho_FIN-q.a}(q_3 q_2 q_3) &=&
(2C_F-\frac{1}{2}-\frac{N_c}{2}) \tr t^a t^c t^b\,
\vec I \cdot
\frac{z_2(\vec p_2 -\vec q_{2}) - (1-z_2)\vec k_g }
     {\left( z_2(\vec p_2  -\vec q_{2}) - (1-z_2)\vec k_g \right)^2}
\nn\\
& &  \Psi_{qqq}^*(x_1-x_g,\vec k_1+x_1\vec q - \vec k_g + x_g\vec K;
x_2+x_g,\vec k_2 +x_2\vec q  - \vec q_{2} +\vec k_g - x_g\vec K;
x_3,\vec k_3 +x_3\vec q - \vec q_{13} )~.
     \label{eq:rho-rho-rho_FIN-q3q2q3}
\eea
The symmetry factor of 6 includes the contribution from diagram
fig.~\ref{fig:rho-rho-rho_FIN-q.a}($q_3 q_1 q_3$) with the gluon
emission and absorption vertices between quarks 1 and 2 interchanged.
\bea
\mathrm{fig.}\, \ref{fig:rho-rho-rho_FIN-q.a}(q_1 q_3 q_3) &=&
(2C_F-\frac{1}{2}-\frac{N_c}{2}) \tr t^a t^b t^c\,
\vec I \cdot
\frac{z_2\vec p_2 - (1-z_2)\vec k_g }
     {\left( z_2\vec p_2  - (1-z_2)\vec k_g \right)^2}
\nn\\
& &  \Psi_{qqq}^*(x_1-x_g,\vec k_1+x_1\vec q - \vec q_{3}- \vec k_g + x_g\vec K;
x_2+x_g,\vec k_2 +x_2\vec q  +\vec k_g - x_g\vec K;
x_3,\vec k_3 +x_3\vec q - \vec q_{12})~.
     \label{eq:rho-rho-rho_FIN-q1q3q3}
\eea
The symmetry factor of 6 includes the contribution from diagram
fig.~\ref{fig:rho-rho-rho_FIN-q.a}($q_2 q_3 q_3$) with the gluon
emission and absorption vertices between quarks 1 and 2 interchanged.
\bea
\mathrm{fig.}\, \ref{fig:rho-rho-rho_FIN-q.a}(q_2 q_3 q_3) &=&
(2C_F-\frac{1}{2}-\frac{N_c}{2}) \tr t^a t^b t^c \,
\vec I \cdot
\frac{z_2(\vec p_2 -\vec q_3) - (1-z_2)\vec k_g }
     {\left( z_2(\vec p_2 -\vec q_3)  - (1-z_2)\vec k_g \right)^2}
\nn\\
& &  \Psi_{qqq}^*(x_1-x_g,\vec k_1+x_1\vec q - \vec k_g + x_g\vec K;
x_2+x_g,\vec k_2 +x_2\vec q - \vec q_{3}  +\vec k_g - x_g\vec K;
x_3,\vec k_3 +x_3\vec q - \vec q_{12})~.
     \label{eq:rho-rho-rho_FIN-q2q3q3}
\eea
The symmetry factor of 6 includes the contribution from diagram
fig.~\ref{fig:rho-rho-rho_FIN-q.a}($q_1 q_3 q_3$) with the gluon
emission and absorption vertices between quarks 1 and 2 interchanged.
\bea
\mathrm{fig.}\, \ref{fig:rho-rho-rho_FIN-q.a}(q_3 q_3 q_3) &=&
       C_F (2-N_c) \tr t^a t^b t^c\,
\vec I \cdot
\frac{z_2\vec p_2  - (1-z_2)\vec k_g }
     {\left( z_2\vec p_2  - (1-z_2)\vec k_g \right)^2}
\nn\\
& &  \Psi_{qqq}^*(x_1-x_g,\vec k_1+x_1\vec q - \vec k_g + x_g\vec K;
x_2+x_g,\vec k_2 +x_2\vec q  +\vec k_g - x_g\vec K;
x_3,\vec k_3 +x_3\vec q - \vec q)~.
     \label{eq:rho-rho-rho_FIN-q3q3q3}
\eea
The symmetry factor of 6 includes the contribution from diagram
fig.~\ref{fig:rho-rho-rho_FIN-q.a}($q_3 q_3 q_3$) with the gluon
emission and absorption vertices between quarks 1 and 2 interchanged.
\\

\begin{figure}[htb]
  \centering
  \begin{minipage}[hb]{\linewidth}
    \includegraphics[width=0.23\linewidth]{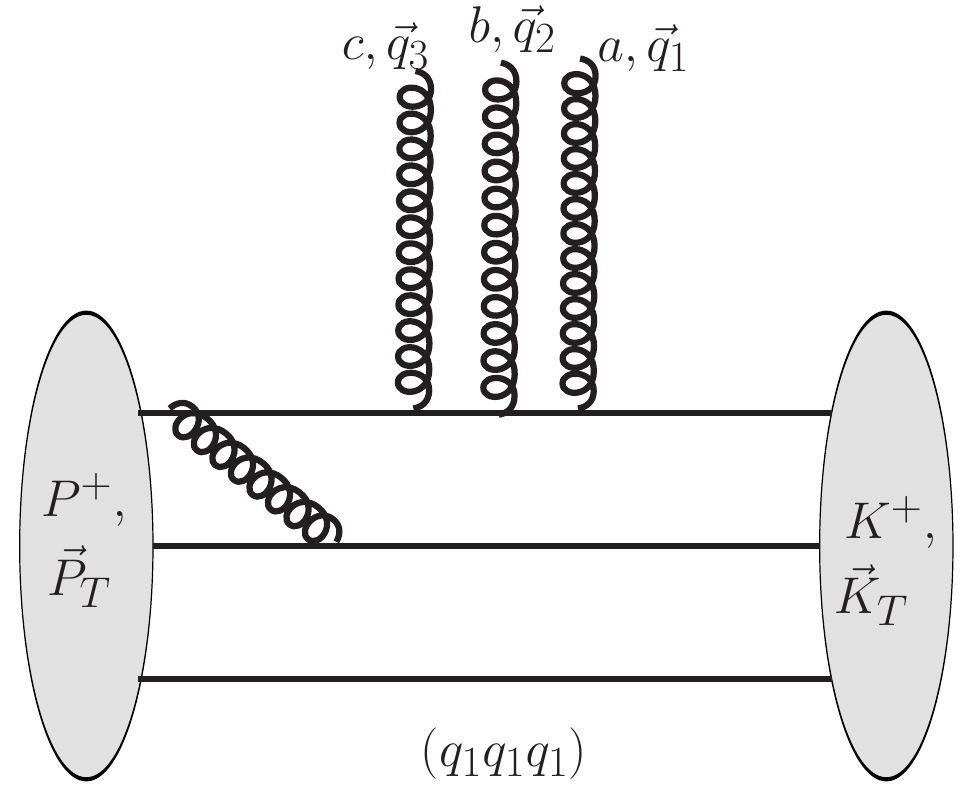}
    \hspace{0.1cm}
    \includegraphics[width=0.23\linewidth]{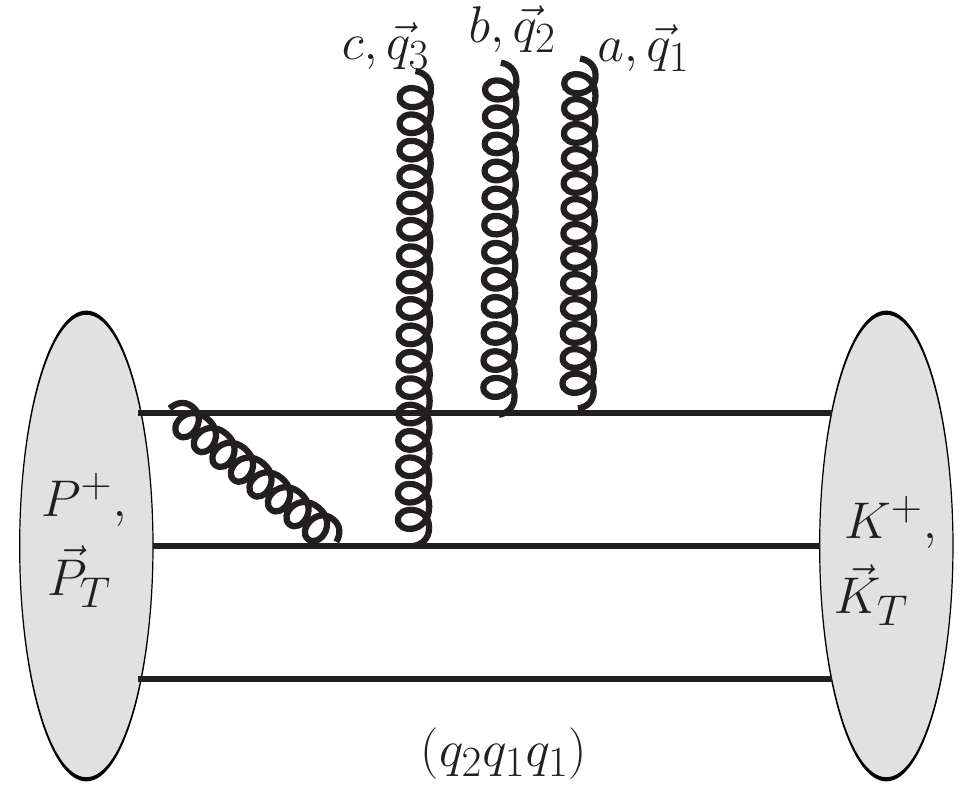}
    \hspace{0.1cm}
    \includegraphics[width=0.23\linewidth]{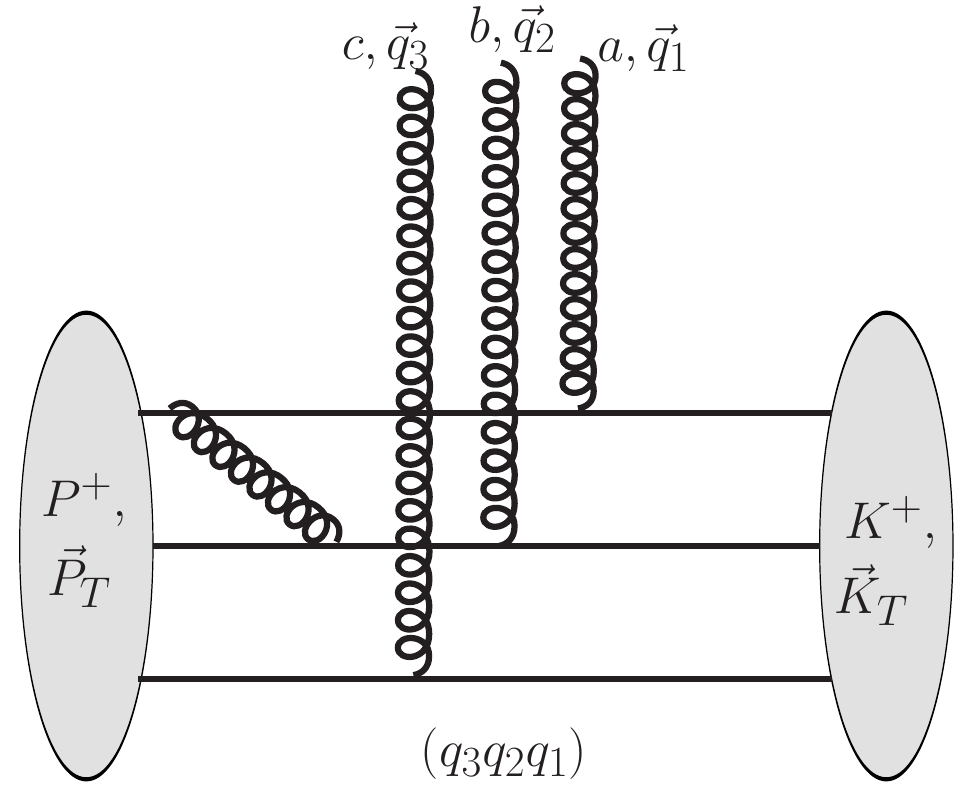}
    \hspace{0.1cm}
    \includegraphics[width=0.23\linewidth]{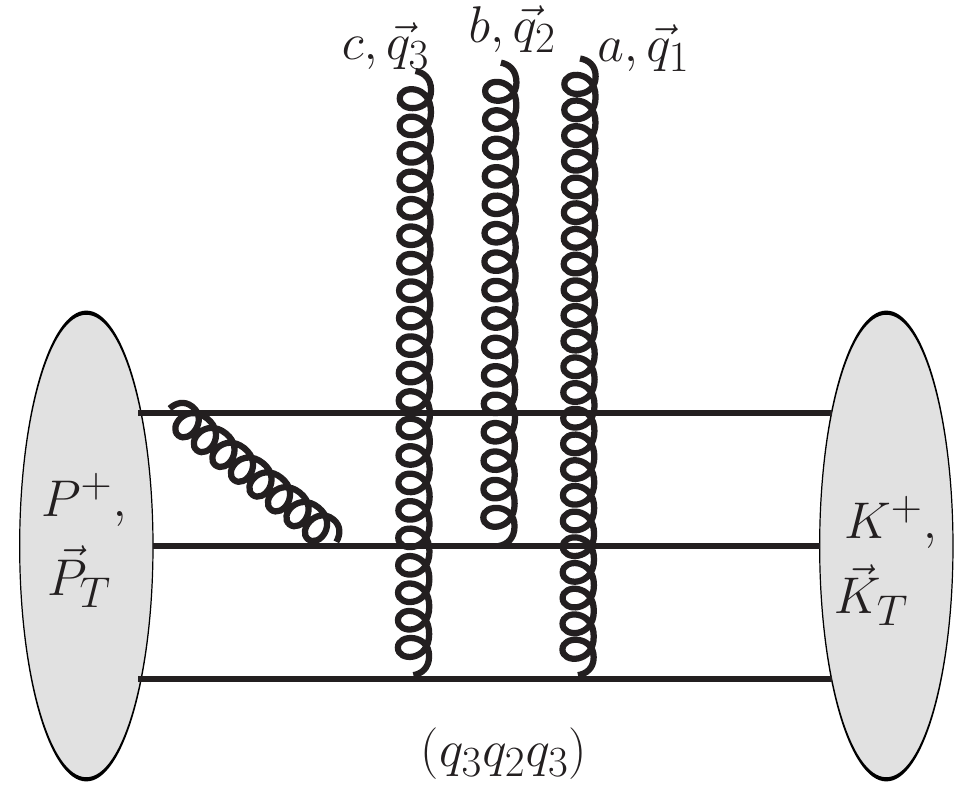}
  \end{minipage}
\caption{A sample of UV finite diagrams for $\langle \rho^a(\vec
  q_1)\, \rho^b(\vec q_2)\,\rho^c(\vec q_3)\rangle$ evaluated in the
  $|qqq\rangle$ Fock state incl.\ ${\cal O}(g^2)$ virtual correction.
  The cut is located at the insertion of the three color charge
  operators.}
\label{fig:rho-rho-rho_FIN-q.b}
\end{figure}
The final set of diagrams corresponds to the virtual corrections where
two quarks exchange a gluon on either side of the insertion of the
$\rho^a(\vec q_1)\, \rho^b(\vec q_2)\,\rho^c(\vec q_3)$
operator. All the symmetry factors are $6\times \frac{1}{2}=3$;
the factor of $\frac{1}{2}$ corrects for double counting of
internal gluon exchanges as mentioned at the end of appendix~\ref{sec:LFwf_qqqg}.

Fig.~\ref{fig:rho-rho-rho_FIN-q.b} shows a small subset of
these diagrams. To write them in compact form we introduce the integral
operator
\bea
J &=& 
\frac{g^5}{3\cdot 16\pi^3} \int \left[\dd x_i\right]
  \int \left[\dd^2 k_i\right] \,
  \Psi_{qqq}(x_1,\vec k_1; x_2,\vec k_2; x_3,\vec k_3)\,
\nn\\
& &
\int\limits_x^{\mathrm{min}(x_1,1-x_2)}\frac{\dd x_g}{x_g}
\left(1-\frac{z_1+z_2}{2}+\frac{z_1 z_2}{6}\right)
\sqrt{\frac{x_1}{x_1-x_g}}\sqrt{\frac{x_2}{x_2+x_g}}
\int\dd^2 k_g~.
\eea
We shall also include right away the contribution from the diagram where
the gluon exchange occurs on the other side of the insertion from quark 2 to
quark 1. With this,
\bea
\mathrm{fig.}\, \ref{fig:rho-rho-rho_FIN-q.b}(q_1 q_1 q_1) &=&
       C_F \tr t^a t^b t^c\,
J \, \frac{z_2\vec p_2-(1-z_2)\vec k_g}
     {\left(z_2\vec p_2-(1-z_2)\vec k_g \right)^2} \cdot
\left(\frac{z_1\vec p_1  - \vec k_g }
     {\left( z_1\vec p_1  - \vec k_g \right)^2} +
     \frac{z_1(\vec p_1-\vec q)  - \vec k_g }
          {\left( z_1(\vec p_1-\vec q)  - \vec k_g \right)^2}
\right)
\nn\\
& &  \Psi_{qqq}^*(x_1-x_g,\vec k_1+x_1\vec q -\vec q - \vec k_g + x_g\vec K;
x_2+x_g,\vec k_2 +x_2\vec q  +\vec k_g - x_g\vec K;
x_3,\vec k_3 +x_3\vec q)~.
     \label{eq:rho-rho-rho_FINv-q1q1q1}
\eea
\bea
\mathrm{fig.}\, \ref{fig:rho-rho-rho_FIN-q.b}(q_2 q_1 q_1) &=&
       (C_F-\frac{N_c+1}{2}) \tr t^a t^b t^c\,
J \, \left(
     \frac{z_2\vec p_2-(1-z_2)\vec k_g}
     {\left(z_2\vec p_2-(1-z_2)\vec k_g \right)^2} \cdot
     \frac{z_1\vec p_1  - \vec k_g }
     {\left( z_1\vec p_1  - \vec k_g \right)^2} \right. \nn\\
     & & \left. \hspace{4cm}     
     +
     \frac{z_2(\vec p_2-\vec q_3)-(1-z_2)\vec k_g}
     {\left(z_2(\vec p_2-\vec q_3)-(1-z_2)\vec k_g \right)^2} \cdot
     \frac{z_1(\vec p_1-\vec q_{12})  - \vec k_g }
     {\left( z_1(\vec p_1-\vec q_{12})  - \vec k_g \right)^2}
\right)
\nn\\
& &  \Psi_{qqq}^*(x_1-x_g,\vec k_1+x_1\vec q -\vec q_{12} - \vec k_g + x_g\vec K;
x_2+x_g,\vec k_2 +x_2\vec q-\vec q_3  +\vec k_g - x_g\vec K;
x_3,\vec k_3 +x_3\vec q)~.
     \label{eq:rho-rho-rho_FINv-q2q1q1}
\eea
\bea
\mathrm{fig.}\, \ref{fig:rho-rho-rho_FIN-q.b}(q_3 q_1 q_1) &=&
     -(2C_F -\frac{N_c+1}{2}) \tr t^a t^b t^c\,
J \, \frac{z_2\vec p_2-(1-z_2)\vec k_g}
     {\left(z_2\vec p_2-(1-z_2)\vec k_g \right)^2} \cdot
\left(\frac{z_1\vec p_1  - \vec k_g }
     {\left( z_1\vec p_1  - \vec k_g \right)^2} +
     \frac{z_1(\vec p_1-\vec q_{12})  - \vec k_g }
          {\left( z_1(\vec p_1-\vec q_{12})  - \vec k_g \right)^2}
\right)
\nn\\
& &  \Psi_{qqq}^*(x_1-x_g,\vec k_1+x_1\vec q -\vec q_{12} - \vec k_g + x_g\vec K;
x_2+x_g,\vec k_2 +x_2\vec q  +\vec k_g - x_g\vec K;
x_3,\vec k_3 +x_3\vec q - \vec q_3)~.
     \label{eq:rho-rho-rho_FINv-q3q1q1}
\eea
\bea
\mathrm{fig.}\, \ref{fig:rho-rho-rho_FIN-q.b}(q_1 q_2 q_1) &=&
       (C_F-\frac{N_c+1}{2}) \tr t^a t^c t^b\,
J \, \left(
     \frac{z_2\vec p_2-(1-z_2)\vec k_g}
     {\left(z_2\vec p_2-(1-z_2)\vec k_g \right)^2} \cdot
     \frac{z_1\vec p_1  - \vec k_g }
     {\left( z_1\vec p_1  - \vec k_g \right)^2} \right. \nn\\
     & & \left. \hspace{4cm}     
     +
     \frac{z_2(\vec p_2-\vec q_2)-(1-z_2)\vec k_g}
     {\left(z_2(\vec p_2-\vec q_2)-(1-z_2)\vec k_g \right)^2} \cdot
     \frac{z_1(\vec p_1-\vec q_{13})  - \vec k_g }
     {\left( z_1(\vec p_1-\vec q_{13})  - \vec k_g \right)^2}
\right)
\nn\\
& &  \Psi_{qqq}^*(x_1-x_g,\vec k_1+x_1\vec q -\vec q_{13} - \vec k_g + x_g\vec K;
x_2+x_g,\vec k_2 +x_2\vec q-\vec q_2  +\vec k_g - x_g\vec K;
x_3,\vec k_3 +x_3\vec q)~.
     \label{eq:rho-rho-rho_FINv-q1q2q1}
\eea
\bea
\mathrm{fig.}\, \ref{fig:rho-rho-rho_FIN-q.b}(q_2 q_2 q_1) &=&
       (C_F-\frac{N_c+1}{2}) \tr t^a t^b t^c\,
J \, \left(
     \frac{z_2\vec p_2-(1-z_2)\vec k_g}
     {\left(z_2\vec p_2-(1-z_2)\vec k_g \right)^2} \cdot
     \frac{z_1\vec p_1  - \vec k_g }
     {\left( z_1\vec p_1  - \vec k_g \right)^2} \right. \nn\\
     & & \left. \hspace{4cm}     
     +
     \frac{z_2(\vec p_2-\vec q_{23})-(1-z_2)\vec k_g}
     {\left(z_2(\vec p_2-\vec q_{23})-(1-z_2)\vec k_g \right)^2} \cdot
     \frac{z_1(\vec p_1-\vec q_{1})  - \vec k_g }
     {\left( z_1(\vec p_1-\vec q_{1})  - \vec k_g \right)^2}
\right)
\nn\\
& &  \Psi_{qqq}^*(x_1-x_g,\vec k_1+x_1\vec q -\vec q_{1} - \vec k_g + x_g\vec K;
x_2+x_g,\vec k_2 +x_2\vec q-\vec q_{23}  +\vec k_g - x_g\vec K;
x_3,\vec k_3 +x_3\vec q)~.
     \label{eq:rho-rho-rho_FINv-q2q2q1}
\eea
\bea
\mathrm{fig.}\, \ref{fig:rho-rho-rho_FIN-q.b}(q_3 q_2 q_1) &=&
- (C_F-\frac{N_c+1}{2}) (\tr t^a t^b t^c+\tr t^a t^c t^b) \,
J \, \left(
     \frac{z_2\vec p_2-(1-z_2)\vec k_g}
     {\left(z_2\vec p_2-(1-z_2)\vec k_g \right)^2} \cdot
     \frac{z_1\vec p_1  - \vec k_g }
     {\left( z_1\vec p_1  - \vec k_g \right)^2} \right. \nn\\
     & & \left. \hspace{4cm}     
     +
     \frac{z_2(\vec p_2-\vec q_{2})-(1-z_2)\vec k_g}
     {\left(z_2(\vec p_2-\vec q_{2})-(1-z_2)\vec k_g \right)^2} \cdot
     \frac{z_1(\vec p_1-\vec q_{1})  - \vec k_g }
     {\left( z_1(\vec p_1-\vec q_{1})  - \vec k_g \right)^2}
\right)
\nn\\
& &  \Psi_{qqq}^*(x_1-x_g,\vec k_1+x_1\vec q -\vec q_{1} - \vec k_g + x_g\vec K;
x_2+x_g,\vec k_2 +x_2\vec q-\vec q_{2}  +\vec k_g - x_g\vec K;
x_3,\vec k_3 +x_3\vec q-\vec q_{3})~.
     \label{eq:rho-rho-rho_FINv-q3q2q1}
\eea
\bea
\mathrm{fig.}\, \ref{fig:rho-rho-rho_FIN-q.b}(q_1 q_3 q_1) &=&
- (2C_F-\frac{N_c+1}{2}) \tr t^a t^c t^b\,
J \,      \frac{z_2\vec p_2-(1-z_2)\vec k_g}
{\left(z_2\vec p_2-(1-z_2)\vec k_g \right)^2} \cdot
\left(
     \frac{z_1\vec p_1  - \vec k_g }
     {\left( z_1\vec p_1  - \vec k_g \right)^2}
     +
     \frac{z_1(\vec p_1-\vec q_{13})  - \vec k_g }
     {\left( z_1(\vec p_1-\vec q_{13})  - \vec k_g \right)^2}
\right)
\nn\\
& &  \Psi_{qqq}^*(x_1-x_g,\vec k_1+x_1\vec q -\vec q_{13} - \vec k_g + x_g\vec K;
x_2+x_g,\vec k_2 +x_2\vec q  +\vec k_g - x_g\vec K;
x_3,\vec k_3 +x_3\vec q -\vec q_2)~.
     \label{eq:rho-rho-rho_FINv-q1q3q1}
\eea
\bea
\mathrm{fig.}\, \ref{fig:rho-rho-rho_FIN-q.b}(q_2 q_3 q_1) &=&
- (C_F-\frac{N_c+1}{2}) (\tr t^a t^b t^c+\tr t^a t^c t^b) \,
J \, \left(
     \frac{z_2\vec p_2-(1-z_2)\vec k_g}
     {\left(z_2\vec p_2-(1-z_2)\vec k_g \right)^2} \cdot
     \frac{z_1\vec p_1  - \vec k_g }
     {\left( z_1\vec p_1  - \vec k_g \right)^2} \right. \nn\\
     & & \left. \hspace{4cm}     
     +
     \frac{z_2(\vec p_2-\vec q_{3})-(1-z_2)\vec k_g}
     {\left(z_2(\vec p_2-\vec q_{3})-(1-z_2)\vec k_g \right)^2} \cdot
     \frac{z_1(\vec p_1-\vec q_{1})  - \vec k_g }
     {\left( z_1(\vec p_1-\vec q_{1})  - \vec k_g \right)^2}
\right)
\nn\\
& &  \Psi_{qqq}^*(x_1-x_g,\vec k_1+x_1\vec q -\vec q_{1} - \vec k_g + x_g\vec K;
x_2+x_g,\vec k_2 +x_2\vec q-\vec q_{3}  +\vec k_g - x_g\vec K;
x_3,\vec k_3 +x_3\vec q -\vec q_{2})~.
     \label{eq:rho-rho-rho_FINv-q2q3q1}
\eea
\bea
\mathrm{fig.}\, \ref{fig:rho-rho-rho_FIN-q.b}(q_3 q_3 q_1) &=&
 - (2C_F-\frac{N_c+1}{2}) \tr t^a t^b t^c \,
J \,      \frac{z_2\vec p_2-(1-z_2)\vec k_g}
     {\left(z_2\vec p_2-(1-z_2)\vec k_g \right)^2} \cdot \left(
     \frac{z_1\vec p_1  - \vec k_g }
     {\left( z_1\vec p_1  - \vec k_g \right)^2}
     +
     \frac{z_1(\vec p_1-\vec q_{1})  - \vec k_g }
     {\left( z_1(\vec p_1-\vec q_{1})  - \vec k_g \right)^2}
\right)
\nn\\
& &  \Psi_{qqq}^*(x_1-x_g,\vec k_1+x_1\vec q -\vec q_{1} - \vec k_g + x_g\vec K;
x_2+x_g,\vec k_2 +x_2\vec q +\vec k_g - x_g\vec K;
x_3,\vec k_3 +x_3\vec q-\vec q_{23})~.
     \label{eq:rho-rho-rho_FINv-q3q3q1}
\eea
%

%
\bea
\mathrm{fig.}\, \ref{fig:rho-rho-rho_FIN-q.b}(q_1 q_1 q_2) &=&
      (C_F - \frac{N_c+1}{2}) \tr t^a t^b t^c\,
J \, \left(
     \frac{z_2\vec p_2-(1-z_2)\vec k_g}
     {\left(z_2\vec p_2-(1-z_2)\vec k_g \right)^2} \cdot
     \frac{z_1\vec p_1  - \vec k_g }
          {\left( z_1\vec p_1  - \vec k_g \right)^2} \right. \nn\\
          & & \left. \hspace{4cm} +
     \frac{z_2(\vec p_2-\vec q_1)-(1-z_2)\vec k_g}
     {\left(z_2(\vec p_2-\vec q_1)-(1-z_2)\vec k_g \right)^2} \cdot
     \frac{z_1(\vec p_1-\vec q_{23})  - \vec k_g }
          {\left( z_1(\vec p_1-\vec q_{23})  - \vec k_g \right)^2}
\right)
\nn\\
& &  \Psi_{qqq}^*(x_1-x_g,\vec k_1+x_1\vec q -\vec q_{23} - \vec k_g + x_g\vec K;
x_2+x_g,\vec k_2 +x_2\vec q -\vec q_1 +\vec k_g - x_g\vec K;
x_3,\vec k_3 +x_3\vec q)~.
     \label{eq:rho-rho-rho_FINv-q1q1q2}
\eea
\bea
\mathrm{fig.}\, \ref{fig:rho-rho-rho_FIN-q.b}(q_2 q_1 q_2) &=&
       (C_F-\frac{N_c+1}{2}) \tr t^a t^c t^b\,
J \, \left(
     \frac{z_2\vec p_2-(1-z_2)\vec k_g}
     {\left(z_2\vec p_2-(1-z_2)\vec k_g \right)^2} \cdot
     \frac{z_1\vec p_1  - \vec k_g }
     {\left( z_1\vec p_1  - \vec k_g \right)^2} \right. \nn\\
     & & \left. \hspace{4cm}     
     +
     \frac{z_2(\vec p_2-\vec q_{13})-(1-z_2)\vec k_g}
     {\left(z_2(\vec p_2-\vec q_{13})-(1-z_2)\vec k_g \right)^2} \cdot
     \frac{z_1(\vec p_1-\vec q_{2})  - \vec k_g }
     {\left( z_1(\vec p_1-\vec q_{2})  - \vec k_g \right)^2}
\right)
\nn\\
& &  \Psi_{qqq}^*(x_1-x_g,\vec k_1+x_1\vec q -\vec q_{2} - \vec k_g + x_g\vec K;
x_2+x_g,\vec k_2 +x_2\vec q-\vec q_{13}  +\vec k_g - x_g\vec K;
x_3,\vec k_3 +x_3\vec q)~.
     \label{eq:rho-rho-rho_FINv-q2q1q2}
\eea
\bea
\mathrm{fig.}\, \ref{fig:rho-rho-rho_FIN-q.b}(q_3 q_1 q_2) &=&
     -(C_F -\frac{N_c+1}{2}) (\tr t^a t^b t^c+\tr t^a t^c t^b) \,
J \, \left(
     \frac{z_2\vec p_2-(1-z_2)\vec k_g}
     {\left(z_2\vec p_2-(1-z_2)\vec k_g \right)^2} \cdot
     \frac{z_1\vec p_1  - \vec k_g }
     {\left( z_1\vec p_1  - \vec k_g \right)^2} \right. \nn\\
 & & \left. \hspace{4cm}+
     \frac{z_2(\vec p_2-\vec q_1)-(1-z_2)\vec k_g}
     {\left(z_2(\vec p_2-\vec q_1)-(1-z_2)\vec k_g \right)^2} \cdot
     \frac{z_1(\vec p_1-\vec q_{2})  - \vec k_g }
     {\left( z_1(\vec p_1-\vec q_{2})  - \vec k_g \right)^2}
\right)
\nn\\
& &  \Psi_{qqq}^*(x_1-x_g,\vec k_1+x_1\vec q -\vec q_{2} - \vec k_g + x_g\vec K;
x_2+x_g,\vec k_2 +x_2\vec q  -\vec q_{1} +\vec k_g - x_g\vec K;
x_3,\vec k_3 +x_3\vec q - \vec q_3)~.
     \label{eq:rho-rho-rho_FINv-q3q1q2}
\eea
\bea
\mathrm{fig.}\, \ref{fig:rho-rho-rho_FIN-q.b}(q_1 q_2 q_2) &=&
       (C_F-\frac{N_c+1}{2}) \tr t^a t^b t^c\,
J \, \left(
     \frac{z_2\vec p_2-(1-z_2)\vec k_g}
     {\left(z_2\vec p_2-(1-z_2)\vec k_g \right)^2} \cdot
     \frac{z_1\vec p_1  - \vec k_g }
     {\left( z_1\vec p_1  - \vec k_g \right)^2} \right. \nn\\
     & & \left. \hspace{4cm}     
     +
     \frac{z_2(\vec p_2-\vec q_{12})-(1-z_2)\vec k_g}
     {\left(z_2(\vec p_2-\vec q_{12})-(1-z_2)\vec k_g \right)^2} \cdot
     \frac{z_1(\vec p_1-\vec q_{3})  - \vec k_g }
     {\left( z_1(\vec p_1-\vec q_{3})  - \vec k_g \right)^2}
\right)
\nn\\
& &  \Psi_{qqq}^*(x_1-x_g,\vec k_1+x_1\vec q -\vec q_{3} - \vec k_g + x_g\vec K;
x_2+x_g,\vec k_2 +x_2\vec q-\vec q_{12}  +\vec k_g - x_g\vec K;
x_3,\vec k_3 +x_3\vec q)~.
     \label{eq:rho-rho-rho_FINv-q1q2q2}
\eea
\bea
\mathrm{fig.}\, \ref{fig:rho-rho-rho_FIN-q.b}(q_2 q_2 q_2) &=&
       C_F \tr t^a t^b t^c\,
J \, \left(
     \frac{z_2\vec p_2-(1-z_2)\vec k_g}
     {\left(z_2\vec p_2-(1-z_2)\vec k_g \right)^2}
     +
     \frac{z_2(\vec p_2-\vec q)-(1-z_2)\vec k_g}
     {\left(z_2(\vec p_2-\vec q)-(1-z_2)\vec k_g \right)^2}
\right)\cdot \frac{z_1\vec p_1  - \vec k_g }
     {\left( z_1\vec p_1  - \vec k_g \right)^2}
\nn\\
& &  \Psi_{qqq}^*(x_1-x_g,\vec k_1+x_1\vec q - \vec k_g + x_g\vec K;
x_2+x_g,\vec k_2 +x_2\vec q-\vec q  +\vec k_g - x_g\vec K;
x_3,\vec k_3 +x_3\vec q)~.
     \label{eq:rho-rho-rho_FINv-q2q2q2}
\eea
\bea
\mathrm{fig.}\, \ref{fig:rho-rho-rho_FIN-q.b}(q_3 q_2 q_2) &=&
- (2C_F-\frac{N_c+1}{2}) \tr t^a t^b t^c \,
J \, \left(
     \frac{z_2\vec p_2-(1-z_2)\vec k_g}
     {\left(z_2\vec p_2-(1-z_2)\vec k_g \right)^2}
     +
     \frac{z_2(\vec p_2-\vec q_{12})-(1-z_2)\vec k_g}
     {\left(z_2(\vec p_2-\vec q_{12})-(1-z_2)\vec k_g \right)^2}
\right) \cdot \frac{z_1\vec p_1  - \vec k_g }
     {\left( z_1\vec p_1  - \vec k_g \right)^2}
\nn\\
& &  \Psi_{qqq}^*(x_1-x_g,\vec k_1+x_1\vec q - \vec k_g + x_g\vec K;
x_2+x_g,\vec k_2 +x_2\vec q-\vec q_{12}  +\vec k_g - x_g\vec K;
x_3,\vec k_3 +x_3\vec q-\vec q_{3})~.
     \label{eq:rho-rho-rho_FINv-q3q2q2}
\eea
\bea
\mathrm{fig.}\, \ref{fig:rho-rho-rho_FIN-q.b}(q_1 q_3 q_2) &=&
- (C_F-\frac{N_c+1}{2}) (\tr t^a t^c t^b+\tr t^a t^b t^c)\,
J \, \left(
      \frac{z_2\vec p_2-(1-z_2)\vec k_g}
      {\left(z_2\vec p_2-(1-z_2)\vec k_g \right)^2} \cdot
      \frac{z_1\vec p_1  - \vec k_g }
     {\left( z_1\vec p_1  - \vec k_g \right)^2} \right. \nn\\
& & \left. \hspace{4cm}
     +
      \frac{z_2(\vec p_2-\vec q_1) -(1-z_2)\vec k_g}
      {\left(z_2(\vec p_2-\vec q_1) -(1-z_2)\vec k_g \right)^2} \cdot
      \frac{z_1(\vec p_1-\vec q_{3})  - \vec k_g }
     {\left( z_1(\vec p_1-\vec q_{3})  - \vec k_g \right)^2}
\right)
\nn\\
& &  \Psi_{qqq}^*(x_1-x_g,\vec k_1+x_1\vec q -\vec q_{3} - \vec k_g + x_g\vec K;
x_2+x_g,\vec k_2 +x_2\vec q  -\vec q_{1} +\vec k_g - x_g\vec K;
x_3,\vec k_3 +x_3\vec q -\vec q_2)~.
     \label{eq:rho-rho-rho_FINv-q1q3q2}
\eea
\bea
\mathrm{fig.}\, \ref{fig:rho-rho-rho_FIN-q.b}(q_2 q_3 q_2) &=&
- (2C_F-\frac{N_c+1}{2}) \tr t^a t^c t^b \,
J \, \left(
     \frac{z_2\vec p_2-(1-z_2)\vec k_g}
     {\left(z_2\vec p_2-(1-z_2)\vec k_g \right)^2}
     +
     \frac{z_2(\vec p_2-\vec q_{13})-(1-z_2)\vec k_g}
     {\left(z_2(\vec p_2-\vec q_{13})-(1-z_2)\vec k_g \right)^2} \cdot
\right)\cdot \frac{z_1\vec p_1  - \vec k_g }
     {\left( z_1\vec p_1  - \vec k_g \right)^2}
\nn\\
& &  \Psi_{qqq}^*(x_1-x_g,\vec k_1+x_1\vec q - \vec k_g + x_g\vec K;
x_2+x_g,\vec k_2 +x_2\vec q-\vec q_{13}  +\vec k_g - x_g\vec K;
x_3,\vec k_3 +x_3\vec q -\vec q_{2})~.
     \label{eq:rho-rho-rho_FINv-q2q3q2}
\eea
\bea
\mathrm{fig.}\, \ref{fig:rho-rho-rho_FIN-q.b}(q_3 q_3 q_2) &=&
 - (2C_F-\frac{N_c+1}{2}) \tr t^a t^b t^c \,
J \, \left(
     \frac{z_2\vec p_2-(1-z_2)\vec k_g}
     {\left(z_2\vec p_2-(1-z_2)\vec k_g \right)^2}
     +
     \frac{z_2(\vec p_2-\vec q_1)  -(1-z_2)\vec k_g}
     {\left(z_2(\vec p_2-\vec q_1) -(1-z_2)\vec k_g \right)^2}
\right)\cdot \frac{z_1\vec p_1  - \vec k_g }
     {\left( z_1\vec p_1  - \vec k_g \right)^2}
\nn\\
& &  \Psi_{qqq}^*(x_1-x_g,\vec k_1+x_1\vec q - \vec k_g + x_g\vec K;
x_2+x_g,\vec k_2 +x_2\vec q -\vec q_1 +\vec k_g - x_g\vec K;
x_3,\vec k_3 +x_3\vec q-\vec q_{23})~.
     \label{eq:rho-rho-rho_FINv-q3q3q2}
\eea
%

%
\bea
\mathrm{fig.}\, \ref{fig:rho-rho-rho_FIN-q.b}(q_1 q_1 q_3) &=&
     - (2C_F - \frac{N_c+1}{2}) \tr t^a t^b t^c\,
J \, \frac{z_2\vec p_2-(1-z_2)\vec k_g}
     {\left(z_2\vec p_2-(1-z_2)\vec k_g \right)^2} \cdot
\left(
     \frac{z_1\vec p_1  - \vec k_g }
          {\left( z_1\vec p_1  - \vec k_g \right)^2} +
     \frac{z_1(\vec p_1-\vec q_{23})  - \vec k_g }
          {\left( z_1(\vec p_1-\vec q_{23})  - \vec k_g \right)^2}
\right)
\nn\\
& &  \Psi_{qqq}^*(x_1-x_g,\vec k_1+x_1\vec q -\vec q_{23} - \vec k_g + x_g\vec K;
x_2+x_g,\vec k_2 +x_2\vec q +\vec k_g - x_g\vec K;
x_3,\vec k_3 +x_3\vec q  -\vec q_1)~.
     \label{eq:rho-rho-rho_FINv-q1q1q3}
\eea
\bea
\mathrm{fig.}\, \ref{fig:rho-rho-rho_FIN-q.b}(q_2 q_1 q_3) &=&
   -  (C_F-\frac{N_c+1}{2}) (\tr t^a t^c t^b+\tr t^a t^b t^c) \,
J \, \left(
     \frac{z_2\vec p_2-(1-z_2)\vec k_g}
     {\left(z_2\vec p_2-(1-z_2)\vec k_g \right)^2} \cdot
     \frac{z_1\vec p_1  - \vec k_g }
     {\left( z_1\vec p_1  - \vec k_g \right)^2} \right. \nn\\
     & & \left. \hspace{4cm}     
     +
     \frac{z_2(\vec p_2-\vec q_{3})-(1-z_2)\vec k_g}
     {\left(z_2(\vec p_2-\vec q_{3})-(1-z_2)\vec k_g \right)^2} \cdot
     \frac{z_1(\vec p_1-\vec q_{2})  - \vec k_g }
     {\left( z_1(\vec p_1-\vec q_{2})  - \vec k_g \right)^2}
\right)
\nn\\
& &  \Psi_{qqq}^*(x_1-x_g,\vec k_1+x_1\vec q -\vec q_{2} - \vec k_g + x_g\vec K;
x_2+x_g,\vec k_2 +x_2\vec q -\vec q_{3}  +\vec k_g - x_g\vec K;
x_3,\vec k_3 +x_3\vec q -\vec q_{1})~.
     \label{eq:rho-rho-rho_FINv-q2q1q3}
\eea
\bea
\mathrm{fig.}\, \ref{fig:rho-rho-rho_FIN-q.b}(q_3 q_1 q_3) &=&
     -(2C_F -\frac{N_c+1}{2}) \tr t^a t^c t^b \,
J \,      \frac{z_2\vec p_2-(1-z_2)\vec k_g}
     {\left(z_2\vec p_2-(1-z_2)\vec k_g \right)^2} \cdot
\left(
     \frac{z_1\vec p_1  - \vec k_g }
     {\left( z_1\vec p_1  - \vec k_g \right)^2}
 +
     \frac{z_1(\vec p_1-\vec q_{2})  - \vec k_g }
     {\left( z_1(\vec p_1-\vec q_{2})  - \vec k_g \right)^2}
\right)
\nn\\
& &  \Psi_{qqq}^*(x_1-x_g,\vec k_1+x_1\vec q -\vec q_{2} - \vec k_g + x_g\vec K;
x_2+x_g,\vec k_2 +x_2\vec q  +\vec k_g - x_g\vec K;
x_3,\vec k_3 +x_3\vec q - \vec q_{13})~.
     \label{eq:rho-rho-rho_FINv-q3q1q3}
\eea
\bea
\mathrm{fig.}\, \ref{fig:rho-rho-rho_FIN-q.b}(q_1 q_2 q_3) &=&
    -   (C_F-\frac{N_c+1}{2}) (\tr t^a t^b t^c+\tr t^a t^c t^b)\,
J \, \left(
     \frac{z_2\vec p_2-(1-z_2)\vec k_g}
     {\left(z_2\vec p_2-(1-z_2)\vec k_g \right)^2} \cdot
     \frac{z_1\vec p_1  - \vec k_g }
     {\left( z_1\vec p_1  - \vec k_g \right)^2} \right. \nn\\
     & & \left. \hspace{4cm}     
     +
     \frac{z_2(\vec p_2-\vec q_{2})-(1-z_2)\vec k_g}
     {\left(z_2(\vec p_2-\vec q_{2})-(1-z_2)\vec k_g \right)^2} \cdot
     \frac{z_1(\vec p_1-\vec q_{3})  - \vec k_g }
     {\left( z_1(\vec p_1-\vec q_{3})  - \vec k_g \right)^2}
\right)
\nn\\
& &  \Psi_{qqq}^*(x_1-x_g,\vec k_1+x_1\vec q -\vec q_{3} - \vec k_g + x_g\vec K;
x_2+x_g,\vec k_2 +x_2\vec q-\vec q_{2}  +\vec k_g - x_g\vec K;
x_3,\vec k_3 +x_3\vec q -\vec q_{1})~.
     \label{eq:rho-rho-rho_FINv-q1q2q3}
\eea
\bea
\mathrm{fig.}\, \ref{fig:rho-rho-rho_FIN-q.b}(q_2 q_2 q_3) &=&
     -(2C_F - \frac{N_c+1}{2}) \tr t^a t^b t^c\,
J \, \left(
     \frac{z_2\vec p_2-(1-z_2)\vec k_g}
     {\left(z_2\vec p_2-(1-z_2)\vec k_g \right)^2}
     +
     \frac{z_2(\vec p_2-\vec q_{23})-(1-z_2)\vec k_g}
     {\left(z_2(\vec p_2-\vec q_{23})-(1-z_2)\vec k_g \right)^2}
\right)\cdot \frac{z_1\vec p_1  - \vec k_g }
     {\left( z_1\vec p_1  - \vec k_g \right)^2}
\nn\\
& &  \Psi_{qqq}^*(x_1-x_g,\vec k_1+x_1\vec q - \vec k_g + x_g\vec K;
x_2+x_g,\vec k_2 +x_2\vec q-\vec q_{23}  +\vec k_g - x_g\vec K;
x_3,\vec k_3 +x_3\vec q -\vec q_{1})~.
     \label{eq:rho-rho-rho_FINv-q2q2q3}
\eea
\bea
\mathrm{fig.}\, \ref{fig:rho-rho-rho_FIN-q.b}(q_3 q_2 q_3) &=&
- (2C_F-\frac{N_c+1}{2}) \tr t^a t^c t^b \,
J \, \left(
     \frac{z_2\vec p_2-(1-z_2)\vec k_g}
     {\left(z_2\vec p_2-(1-z_2)\vec k_g \right)^2}
     +
     \frac{z_2(\vec p_2-\vec q_{2})-(1-z_2)\vec k_g}
     {\left(z_2(\vec p_2-\vec q_{2})-(1-z_2)\vec k_g \right)^2}
\right) \cdot \frac{z_1\vec p_1  - \vec k_g }
     {\left( z_1\vec p_1  - \vec k_g \right)^2}
\nn\\
& &  \Psi_{qqq}^*(x_1-x_g,\vec k_1+x_1\vec q - \vec k_g + x_g\vec K;
x_2+x_g,\vec k_2 +x_2\vec q-\vec q_{2}  +\vec k_g - x_g\vec K;
x_3,\vec k_3 +x_3\vec q-\vec q_{13})~.
     \label{eq:rho-rho-rho_FINv-q3q2q3}
\eea
\bea
\mathrm{fig.}\, \ref{fig:rho-rho-rho_FIN-q.b}(q_1 q_3 q_3) &=&
- (2C_F-\frac{N_c+1}{2}) \tr t^a t^b t^c\,
J \,  \frac{z_2\vec p_2-(1-z_2)\vec k_g}
      {\left(z_2\vec p_2-(1-z_2)\vec k_g \right)^2} \cdot
\left(
      \frac{z_1\vec p_1  - \vec k_g }
     {\left( z_1\vec p_1  - \vec k_g \right)^2}
     +
      \frac{z_1(\vec p_1-\vec q_{3})  - \vec k_g }
     {\left( z_1(\vec p_1-\vec q_{3})  - \vec k_g \right)^2}
\right)
\nn\\
& &  \Psi_{qqq}^*(x_1-x_g,\vec k_1+x_1\vec q -\vec q_{3} - \vec k_g + x_g\vec K;
x_2+x_g,\vec k_2 +x_2\vec q +\vec k_g - x_g\vec K;
x_3,\vec k_3 +x_3\vec q -\vec q_{12})~.
     \label{eq:rho-rho-rho_FINv-q1q3q3}
\eea
\bea
\mathrm{fig.}\, \ref{fig:rho-rho-rho_FIN-q.b}(q_2 q_3 q_3) &=&
- (2C_F-\frac{N_c+1}{2}) \tr t^a t^b t^c \,
J \, \left(
     \frac{z_2\vec p_2-(1-z_2)\vec k_g}
     {\left(z_2\vec p_2-(1-z_2)\vec k_g \right)^2}
     +
     \frac{z_2(\vec p_2-\vec q_{3})-(1-z_2)\vec k_g}
     {\left(z_2(\vec p_2-\vec q_{3})-(1-z_2)\vec k_g \right)^2} \cdot
\right)\cdot \frac{z_1\vec p_1  - \vec k_g }
     {\left( z_1\vec p_1  - \vec k_g \right)^2}
\nn\\
& &  \Psi_{qqq}^*(x_1-x_g,\vec k_1+x_1\vec q - \vec k_g + x_g\vec K;
x_2+x_g,\vec k_2 +x_2\vec q-\vec q_{3}  +\vec k_g - x_g\vec K;
x_3,\vec k_3 +x_3\vec q -\vec q_{12})~.
     \label{eq:rho-rho-rho_FINv-q2q3q3}
\eea
\bea
\mathrm{fig.}\, \ref{fig:rho-rho-rho_FIN-q.b}(q_3 q_3 q_3) &=&
 - 2C_F(2-N_c) \tr t^a t^b t^c \,
J \,
     \frac{z_2\vec p_2-(1-z_2)\vec k_g}
     {\left(z_2\vec p_2-(1-z_2)\vec k_g \right)^2} \cdot
     \frac{z_1\vec p_1  - \vec k_g }
     {\left( z_1\vec p_1  - \vec k_g \right)^2}
\nn\\
& &  \Psi_{qqq}^*(x_1-x_g,\vec k_1+x_1\vec q - \vec k_g + x_g\vec K;
x_2+x_g,\vec k_2 +x_2\vec q +\vec k_g - x_g\vec K;
x_3,\vec k_3 +x_3\vec q-\vec q)~.
     \label{eq:rho-rho-rho_FINv-q3q3q3}
\eea
\\

In closing this section we note that the subset of diagrams where all
three gluons attach to the same quark line is proportional to the UV
finite, ${\cal O}(g^2)$ correction to the electromagnetic form factor
of the proton, i.e.\ to the matrix element
$\langle\rho_\mathrm{em}(\vec q)\rangle$.  Hence, for $\vec q^2 \to 0$
this must vanish. Indeed, in this limit the sum of
eqs.~(\ref{eq:rho-rho-rho_FIN-q1q1q1},
\ref{eq:rho-rho-rho_FIN-q2q2q2}, \ref{eq:rho-rho-rho_FIN-q3q3q3},
\ref{eq:rho-rho-rho_FINv-q1q1q1}, \ref{eq:rho-rho-rho_FINv-q2q2q2},
\ref{eq:rho-rho-rho_FINv-q3q3q3}), multiplied by their respective
symmetry factors, is zero.  [The finite parts of the UV divergent
  diagrams eqs.~(\ref{eq:rho-rho-rho_UV_q1q1q1},
  \ref{eq:rho-rho-rho_UV_q2q2q2}) vanish individually when $\vec q^2
  \to 0$, see appendix~\ref{sec:UVcancel}.]

\section{The correlator in impact parameter space}
\label{eq:sec_G3-_b-space}

The vanishing of $\langle\rho^a(\vec q_1)\, \rho^b(\vec q_2)\,
\rho^c(\vec q_3)\rangle$ when $\vec q_1=0$ or $\vec q_3=0$ leads to a
sum rule in impact parameter space. Let us first separate $C$-odd and
even contributions via
\bea
G_3^-(\vec q_1,\vec q_2,\vec q_3) \sim d^{abc}\, \langle\rho^a(\vec q_1)\,
\rho^b(\vec q_2)\, \rho^c(\vec q_3)\rangle,~~~~~~~~~~
G_3^+(\vec q_1,\vec q_2,\vec q_3) \sim if^{abc}\, \langle\rho^a(\vec q_1)\,
\rho^b(\vec q_2)\, \rho^c(\vec q_3)\rangle~.
\eea
Introducing the total momentum transfer $\vec K = - (\vec q_1+\vec
q_2+ \vec q_3)$, where we assume $\vec P=0$ for the incoming proton,
and the relative momenta $\vec \Delta_{12}=\vec q_1-\vec q_2$, $\vec
\Delta_{23}=\vec q_2-\vec q_3$, we can Fourier transform these
correlators to impact parameter space:
\bea
G_3^\pm(\vec b, \vec \Delta_{12}, \vec \Delta_{23}) =
\int\frac{\dd^2 K}{(2\pi)^2}\, e^{-i\vec b \cdot \vec K}\,
G_3^\pm\left(
\frac{2\vec \Delta_{12}+\vec \Delta_{23}-\vec K}{3},
\frac{-\vec \Delta_{12}+\vec \Delta_{23}-\vec K}{3},
-\frac{\vec \Delta_{12}+2\vec \Delta_{23}+\vec K}{3}
\right)~.
\eea
These functions satisfy the sum rule
\be
\int \dd^2b\,\, G_3^\pm(\vec b, \vec \Delta_{12}, \vec \Delta_{23}) =0
\ee
when $2\vec \Delta_{12}=-\vec \Delta_{23}$ or $\vec \Delta_{12}=-2\vec
\Delta_{23}$ or $\vec \Delta_{12}=\vec \Delta_{23}=0$.\\

We proceed to show a numerical estimate for $G_3^-(b)$ for $\vec
\Delta_{12} = \vec \Delta_{23} = 0$, normalized according to
$G_3^-(\vec q_1,\vec q_2,\vec q_3) = 4 d^{abc}\, \langle\rho^a(\vec
q_1)\, \rho^b(\vec q_2)\, \rho^c(\vec q_3)\rangle / g^3$.
Here,
$\langle\rho^a(\vec q_1)\, \rho^b(\vec q_2)\, \rho^c(\vec q_3)\rangle$
is given by the sum of {\em all} diagrams computed above. For the
numerical results we employ the ``harmonic oscillator'' three-quark
model wave function $\Psi_{\mathrm{qqq}}(x_i,\vec k_i)$ by Brodsky and
Schlumpf~\cite{Schlumpf:1992vq,Brodsky:1994fz} used also in Ref.~\cite{Dumitru:2021tvw},
 which assumes a
Gaussian momentum distribution of quarks in two transverse dimensions,
with a specific $x$-dependent width. Also, the magnitude of the NLO
correction depends on the value of the coupling for which we use
$\alpha_s=0.2$; and on the collinear regulator (see
app.~\ref{sec:UVcancel}) which we take as 0.2~GeV. \\

\begin{figure}[htb]
  \centering
  \begin{minipage}[hb]{\linewidth}
    \includegraphics[width=0.75\linewidth]{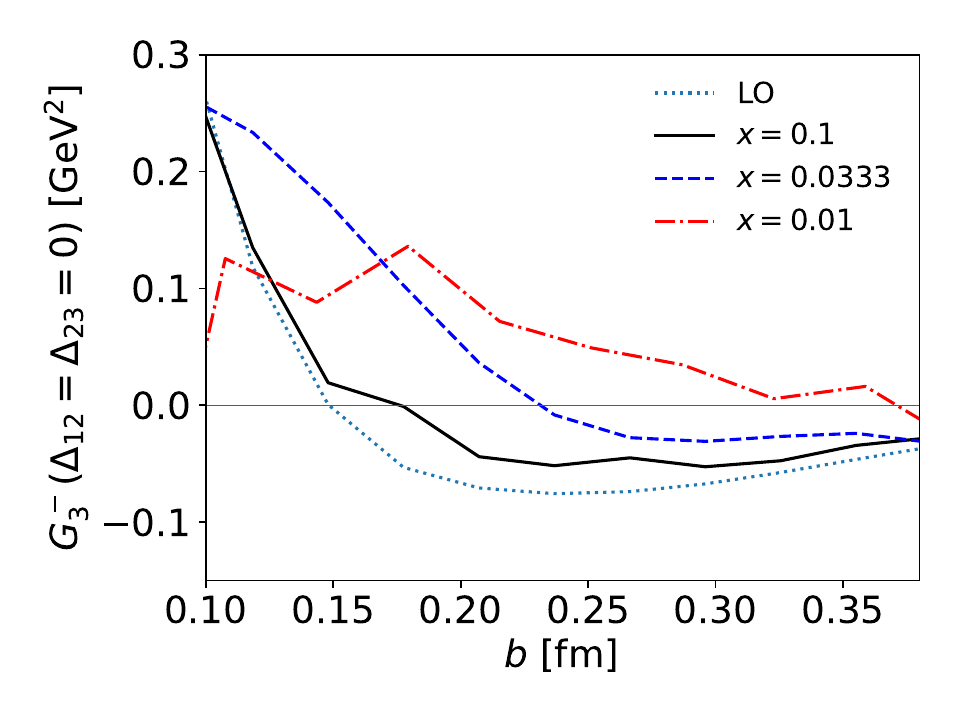}
  \end{minipage}
  \caption{Numerical model estimate of the impact parameter
    dependence of the $C$-odd part of the correlator of three
    color charge operators in the proton.}
\label{fig:G3-_vs_b}
\end{figure}

Our result is shown in fig.~\ref{fig:G3-_vs_b}. At $x=0.1$ the
correction to the LO result~\cite{Dumitru:2020fdh} is numerically
small; similar behavior was observed for the correlator of two color
charge operators in ref.~\cite{Dumitru:2021tvw}. This is an important
check of the perturbative expansion about a three-quark Fock state. With
decreasing $x$ the NLO correction grows substantially.\\

Fig.~\ref{fig:G3-_vs_b} also shows that the $b$-dependence of $G_3^-$
does not follow a positive definite 1-body ``parton density
distribution'', or a proton thickness function,
respectively, at small $b$. Rather, $n$-body quantum correlations of
color charge depend non-trivially on impact parameter, reflecting in a
change of sign of $G_3^-$ at $b \simeq 0.15$ fm. 
Lastly, we note that the generic magnitude of $G_3^-$, for impact
parameters in the perturbative region, is similar to that of the
correlator of two color charge operators $G_2$ shown in
ref~\cite{Dumitru:2021tvw}. Hence, for realistic values of the
coupling there are substantial corrections to Gaussian color
charge fluctuations in the proton at moderately small $x$.


\section{Summary}  \label{sec:summary}

We have computed the diagrams for the light cone gauge color charge
correlator $\langle\rho^a(\vec q_1)\, \rho^b(\vec q_2)\, \rho^c(\vec
q_3)\rangle$ in a proton made of three quarks and a perturbative gluon
which is not required to carry a small light-cone momentum. This
correlator provides the leading correction to Gaussian color charge
fluctuations in the proton.  It is independent of the renormalization
scale since UV divergences cancel, but like the correlator of two color
charge operators~\cite{Dumitru:2020gla} it also exhibits logarithmic
collinear and soft singularities. \\

These results may be used to obtain a more realistic picture of
correlations in the proton at moderate $x \gsim 0.01$. In particular,
there exist contributions where one can not ``pair up'' the transverse
momenta of two of the three exchanged gluons, i.e.\ where the three
probes hit the target proton at three different impact parameters.
Furthermore, our explicit expressions could be used as initial
conditions (in the weak field / dilute regime) for small-$x$
evolution, in particular for impact parameter dependent
evolution~\cite{GolecBiernat:2003ym, Berger:2010sh, Cepila:2018faq,Mantysaari:2018zdd,
  Bendova:2019psy} {\em with} a contribution that is odd under $C$ and
$P$. \\

The expressions for $\langle\rho^a(\vec q_1)\, \rho^b(\vec q_2)\,
\rho^c(\vec q_3)\rangle$ have a form similar to those for
$\langle\rho^a(\vec q_1)\, \rho^b(\vec q_2)\rangle$ obtained
previously~\cite{Dumitru:2020gla}. Therefore, they can be evaluated
numerically using the same code developed for the two-point charge
correlator~\cite{Dumitru:2021tvw}.  We have provided first numerical
estimates for the $C$-odd part $G_3^-$ of $\langle\rho^a(\vec q_1)\,
\rho^b(\vec q_2)\, \rho^c(\vec q_3)\rangle$ in
sec.~\ref{eq:sec_G3-_b-space}. We find that the NLO correction due to
the $|qqqg\rangle$ Fock state of the proton is numerically small at
$x=0.1$ but that it increases rapidly as $x\to 0.01$. Also, the
dependence of the three-charge correlator on impact parameter $b$ is
rather non-trivial, and its sign changes indicate $b$-dependent transitions
in the nature of the quantum correlations of color charge in the proton.
  This result,
together with published results on the $\langle\rho^a\, \rho^b\rangle$
correlator~\cite{Dumitru:2021tvw}, provides guidance for
phenomenological models of color charge correlations in the proton
(at $x \sim 0.01 - 0.1$).

\section*{Acknowledgements}

The figures have been prepared with Jaxodraw~\cite{Binosi:2008ig}. A.D.\ thanks the US Department of Energy, Office of Nuclear Physics, for support via Grant DE-SC0002307; and The City University of New York for PSC-CUNY Research grant 64025-00~52. R.P.\  is  supported  by  the Academy of Finland, project 1322507 and by the European  Research  Council,  grant  no.\ 725369. H.M.\ is supported by the Academy of Finland projects 338263 and 346567, and by the European Research Council project STRONG-2020 (grant agreement no.\ 824093).  The content of this article does not reflect the official opinion of the European Union and responsibility for the information and views expressed therein lies entirely with the authors. Computing resources from CSC – IT Center for Science in Espoo, Finland and from the Finnish Grid and Cloud Infrastructure (persistent identifier \texttt{urn:nbn:fi:research-infras-2016072533}) were used in this work.


\appendix

\section{The proton on the light front} \label{sec:P_LF}
For completeness we briefly review in this appendix the Fock state
description of the proton on the light front as used in the main text.

\subsection{Three quark Fock state of the proton} \label{sec:LFwf}

The Fock space description of the eigenstates of the QCD Hamiltonian
on the light cone has been discussed many times in the literature, see
e.g.\ refs.~\cite{Lepage:1980fj,Brodsky:1997de,Brodsky:2000ii}. Any
such state, such as the proton, will involve a superposition of $n$
parton Fock states, for arbitrary $n$. However, it is well known from
phenomenology that at $x \gsim 0.1$ the proton can be understood to
first approximation in terms of a state of three massive quarks (with
the appropriate flavors and spins, and in an anti-symmetric color
singlet state), see e.g.\ refs.~\cite{Schlumpf:1992vq,
Brodsky:1994fz}. Therefore, we consider such a phenomenological state
of three massive quarks, a ``light-cone constituent quark model'', a
reasonable point of expansion, to which we add a perturbative
$|qqqg\rangle$ correction\footnote{Let us also note that the two-point
color charge correlator $\langle\rho^a(\vec q_1)\, \rho^b(\vec
q_2)\rangle$ at $x=0.1$ receives fairly small corrections from the
perturbative gluon emission~\cite{Dumitru:2021tvw}. We take this as
another confirmation for the validity (at such $x$) of expanding from
a state of three massive quarks.}.\\

We write the light cone state of an on-shell proton with four-momentum
$P^\mu = (P^+, P^-,\vec{P})$ composed of three quarks as
\bea
|P\rangle &=& \frac{1}{\sqrt{N_c !}}
\int \left[\dd x_i\right]
\int \left[\dd^2 k_i\right]\,
\Psi_{qqq}(x_1, \vec k_1; x_2, \vec k_2; x_3, \vec k_3)
\sum_{i_1, i_2, i_3}\epsilon_{i_1 i_2 i_3}\,
|p_1,i_1; \, p_2,i_2; \, p_3,i_3\rangle \,
|S\rangle~.
\label{eq:valence-proton}
\eea
$N_c=3$ is the number of colors while $|S\rangle$ is the helicity wave
function of the proton,
normalized to $\langle S | S\rangle=1$. We have used the following
compact notation:
\be \left[\dd x_i\right] \equiv \frac{\dd x_1 \dd x_2 \dd x_3}{8 x_1 x_2 x_3}
\delta(1-x_1-x_2-x_3),\quad\quad
\left[\dd^2 k_i\right] \equiv
\frac{\dd^2 k_1 \dd^2 k_2 \dd^2 k_3}{(2\pi)^6}\,
\delta(\vec{k}_1+\vec{k}_2+\vec{k}_3)~.
\label{eq:[dxi][dki]}
\ee
The three on-shell quark momenta are specified by their lightcone
momentum components $p_i^+ = x_i P^+$ and their transverse components
$\vec{p}_{i} = x_i \vec{P} + \vec{k}_i$.  The quark colors are denoted
as $i_{1,2,3}$.  $\Psi_{qqq}$ is the three-quark wave function.  It is
symmetric under exchange of any two of the quarks: $\Psi_{qqq}(x_1,
\vec k_1; x_2, \vec k_2; x_3, \vec k_3) = \Psi_{qqq}(x_2, \vec k_2;
x_1, \vec k_1; x_3, \vec k_3)$ etc.

For simplicity, we will assume that the momentum space wave function
$\Psi_{qqq}$ does not depend on the helicities $h_i=\pm 1$ of the
quarks, i.e.\ that the helicity wave function factorizes from the
color-momentum wave function. Helicity matrix elements are given by
$\langle h_i\rangle =0$ and $\langle h_i h_{j\ne i}\rangle = -
\frac{1}{3}$, see ref.~\cite{Dumitru:2020gla}.
\\

We neglect plus momentum transfer so that $\xi = (K^+ - P^+)/P^+ \to
0$.  The proton state is then normalized according to
\bea
\langle K | P\rangle &=& 16\pi^3 \, P^+ \delta(P^+ - K^+)
\, \delta(\vec{P} - \vec{K}) \label{eq:ProtonNorm1}
~.
\eea

The one-particle quark states introduced above are created by the
action of the quark creation operator on the vacuum $|0\rangle$:
\be
|p,i,h\rangle =   b^\dagger_{ih}(p) |0\rangle~.
\ee
The quark creation and annihilation operators satisfy the
anti-commutation relation
\be
\{ b_{jh}(k) , b^\dagger_{ih'}(p) \} = \delta^{ji}_{h h'}
  \,\, 16\pi^3\, k^+\delta(k^+-p^+)\, \delta(\vec k - \vec
p)~,\label{anticomm} 
\ee
so that
\be
\langle k, j, h\,|\, p,i,h'\rangle = \delta^{ji}_{h h'}
  \,\, 16\pi^3\, k^+\delta(k^+-p^+)\, \delta(\vec k - \vec
p)~.
\ee
Similarly, the operators which create or destroy a gluon satisfy
the commutation relations
\be
[ a_{a\lambda}(k) , a^\dagger_{b\rho}(p) ] =
  \delta^{ab}_{\lambda\rho}\,\, 16\pi^3\, k^+\delta(k^+-p^+)\,
  \delta(\vec k - \vec p)~.\label{eq:acomm}
\ee

The normalization of the valence quark wave function is given by
\be \label{eq:Norm_psi3}
\frac{1}{2}
\int \left[\dd x_i\right]
  \int \left[\dd^2 k_i\right]\,
  |\Psi_{qqq}(x_1, \vec k_1; x_2, \vec k_2; x_3, \vec k_3)|^2 = 1~.
\ee
%

\subsection{Three quarks and a gluon} \label{sec:LFwf_qqqg}

In this section we outline the computation in light-cone perturbation
theory of the three quark plus one gluon wave function, and of the
virtual corrections~\cite{Dumitru:2020gla,Hanninen:2017ddy,Lappi:2016oup}.

The physical incoming one-particle quark state can be written as a
simultaneous perturbative and Fock state decomposition in terms of the
bare states
\be
\begin{split}
\label{eq:qpertexp}
\vert q(p, h, i)\rangle = Z^{1/2}_q(p^+) \biggl
(\vert q(p, h, i)\rangle_0 +
\sum_{h',\sigma,j,a}\int \dk_q
\dk_g(2\pi)^{3}&\delta(p^+ - k_q^+ - k_g^+)\delta(\pvec - \kvec_q - \kvec_g)\psi_{q\to
  qg}(p; k_q, k_g)\\
& \times \vert q(k_q, h', j) \,
g(k_g,\sigma,a)\rangle_0 + \dots \biggr )~.
\end{split}
\ee
Here, the LCwf for $q \to q g$ splitting is denoted as $\psi_{q\to
  qg}$ and the Lorentz invariant measures $\dk_q$ and $\dk_g$ are
defined as
\be
\begin{split}
  \int \dk & \equiv \int \frac{\ud k^+}{2k^+}\frac{\ud^{2}k}{(2\pi)^{3}}
  \to (\mu^2)^{2-D/2}\int \frac{\ud k^+}{(2\pi)\, 2k^+}
  \frac{\ud^{D-2}k}{(2\pi)^{D-2}}
~.
\end{split}
\ee
The latter form is used to regularize ultraviolet (UV) divergences by
integrating over the momenta of all particles in $D$ dimensions. Here,
an arbitrary scale $\mu^2$ has been introduced so that the transverse
integrals preserve their natural dimensions.  The quark wave function
renormalization coefficient $Z_q$ can be calculated from the
normalization requirement
\be
\label{eq:1qnorm}
\langle q(p, h, i)\vert q(p, h, i)\rangle =
        {}_{0}\langle q(p, h, i)\vert q(p, h, i)\rangle_0 =
        2p^+ (2\pi)^{3}\delta^{(3)}(0)~.
\ee
\\

Now we replace each quark state in \eq\nr{eq:valence-proton} by
the perturbative expansion in \eq\nr{eq:qpertexp}. This yields
\be
\label{eq:qqqg-A-proton}
\begin{split}
&\vert q(p_1,h_1,i_1)\, q(p_2,h_2,i_2)\,
  q(p_3,h_3,i_3)\rangle\, \vert S\rangle
   = \\
  &\biggl [\left(1
  - \frac{C_q(p^+_1)}{2}\right)\vert q(p_1,h_1,i_1)\rangle_0
  + \sum_{h',\sigma,j ,a}2g(t^{a})_{j i_1}
  \int \frac{\dk_g}{2(p_1^+ - k_g^+)}\hat \psi_{q\to qg}(p_1; p_1- k_g, k_g)\\
  &\hspace{5cm} \times
  \vert q(p_1-k_g,h',j )\, g(k_g,\sigma,a)\rangle_0
  + \cdots \biggr ]\\
  \otimes & \biggl [\left(1
  - \frac{C_q(p^+_2)}{2}\right)\vert q(p_2,h_2,i_2)\rangle_0
  + \sum_{h',\sigma,j ,a}2g(t^{a})_{j i_2}
  \int \frac{\dk_g}{2(p_2^+ - k_g^+)}\hat \psi_{q\to qg}(p_2; p_2-k_g, k_g)\\
  &\hspace{5cm} \times
  \vert q(p_2-k_g,h',j )\, g(k_g,\sigma,a)\rangle_0
  + \cdots   \biggr ]\\
  \otimes & \biggl [\left(1
  - \frac{C_q(p^+_3)}{2}\right)\vert q(p_3,h_3,i_3)\rangle_0
  + \sum_{h',\sigma,j ,a}2g(t^{a})_{j i_3}
  \int \frac{\dk_g}{2(p_3^+ - k_g^+)}\hat \psi_{q\to qg}(p_3; p_3-k_g, k_g)\\
  &\hspace{5cm} \times
  \vert q(p_3-k_g,h',j )\, g(k_g,\sigma,a)\rangle_0
  + \cdots   \biggr ]\, \vert S\rangle~.
\end{split}
\ee
We have extracted the common factor $2g(t^a)_{j i_i}$ from $\psi_{q\to
  qg}$ by defining $\psi_{q\to qg}(p_i; p_i- k_g, k_g) \equiv
2g(t^a)_{j i_i}\hat \psi_{q\to qg}(p_i; p_i- k_g, k_g)$. The latter
involves the quark helicities and the gluon polarization.  Also
note that $C_q(p^+_i) \sim \mathcal{O}(g^2)$ while $\psi_{q \to qg}
\sim \mathcal{O}(g)$, and that terms of order $\mathcal{O}(g^3)$ and
higher must be dropped. Finally, the integration over the plus
momentum of the gluon extends up to the plus momentum of the parent
quark; for example, $k_g^+ < p_1^+$ in the first line, and so on.\\

We also need to add to the r.h.s.\ of eq.~(\ref{eq:qqqg-A-proton}) the
${\cal O}(g^2)$ contributions where one quark emits a gluon which is
then absorbed by a second (distinct) quark. For example, if the first
quark emits and the second quark absorbs the gluon, that gives the
contribution
\be
\begin{split}
  & \sum_{h_1',h_2',\sigma,j ,n,a}
  4g^2 (t^a)_{j i_1} (t^a)_{n i_2} \\
  & \int
  \frac{\dk_g}{2(p_1^+ - k_g^+)}
  \hat \psi_{q\to qg}(p_1; p_1-k_g, k_g)\,
  \frac{1}{2(p_2^+ + k_g^+)}
  \hat \psi_{qg\to q}(p_2, k_g; p_2 + k_g)\\
  &
  \vert q(p_1- k_g,h_1',j)\,
  q(p_2+k_g,h_2',n)\rangle_0 \otimes
  \vert q(p_3,h_3,i_3)\rangle_0\, \vert S\rangle
  ~.
\end{split}
\ee
Here, the integration over $k_g^+$ extends up to $\mathrm{min}(p_1^+,
P^+-p_2^+)$. There are analogous contributions corresponding to gluon
emission from quark 2 and absorption by quark 1 as well as from other
pairings.  Since we sum over all permutations of emitter and absorber,
to avoid double counting, we should either multiply the above
expression by $\frac{1}{2}$ or else include this factor in the
symmetry factors of the corresponding diagrams. We choose the latter
option so that all symmetry factors for the
fig.~\ref{fig:rho-rho-rho_FIN-q.b} type diagrams in
sec.~\ref{sec:finite-only-to-quarks} are 3.

\section{Cancellation of UV divergences}
\label{sec:UVcancel}

Here, we verify the cancellation of UV divergences. We quote the following
result from the appendix of ref.~\cite{Dumitru:2020gla}:
\bea
& & \frac{(2\pi)^{D-1}}{2p_1^{+}} \int \frac{\dk_g}{2(p_1^+ -
  k_g^+)} \, \langle S\vert \hat\psi_{q \to qg}(\pvec_1 ;\pvec_1 - \kvec_g,
\kvec_g)
\, \hat\psi^{\ast}_{q \to qg}(\pvec_1 - \lvec, \pvec_1 - \kvec_g
- \lvec_1, \kvec_g - \lvec + \lvec_1) \vert S\rangle \nn\\
&=& 2\pi^3
\frac{C_q(x_1)}{g^2\cf} + 2\pi^3 F(\lvec, \lvec_1;\frac{x}{x_1},m^2)~,
\eea
where $\lvec, \lvec_1$ are 2d transverse momenta ($l^+=l_1^+=0$),
$C_q(x)=1-Z_q(x)$ is the ${\cal O}(g^2)$ correction to the quark wave
function renormalization factor, and $F$ is a UV finite function
satisfying $F(0,0;x/x_1,m^2) \to 0$.  Note that the UV divergent part
does not depend on the momenta $\lvec, \lvec_1$. The
parameter  $m^2$ is an infrared regulator
for the DGLAP collinear singularity~\cite{Gribov:1972ri,
  Gribov:1972rt, Altarelli:1977zs, Dokshitzer:1977sg}, and $x$ is a
cutoff for the soft singularity.\\

We begin by collecting the terms proportional to
\be \label{eq:UV_prefac1}
\frac{g^3}{8 C_F} \int \left[\dd x_i\right]
  \int \left[\dd^2 k_i\right]\,
  \Psi_{qqq}(x_1,\vec k_1; x_2,\vec k_2; x_3,\vec k_3)\,
  \Psi_{qqq}^*(x_1,\vec k_1-(1-x_1)\vec q;
x_2,\vec k_2 +x_2\vec q;
x_3,\vec k_3 +x_3\vec q)\, C_q(x_1)~.
\ee
These are eqs.~(\ref{eq:rho-rho-rho_UV_ggg},
\ref{eq:rho-rho-rho_UV_q1gg}, \ref{eq:rho-rho-rho_UV_gq1g},
\ref{eq:rho-rho-rho_UV_ggq1}, \ref{eq:rho-rho-rho_UV_q1q1g},
\ref{eq:rho-rho-rho_UV_q1gq1}, \ref{eq:rho-rho-rho_UV_gq1q1}), the
first term in eq.~(\ref{eq:rho3_qqq}) times $-C_q(x_1)$, and
eq.~(\ref{eq:rho-rho-rho_UV_q1q1q1}):
\bea
& & 2\frac{3}{2} if^{abc} \nn\\
&+& 2\frac{3}{4} (d^{abc} - i f^{abc}) \nn\\
&+& 2\frac{3}{4} (d^{abc} + i f^{abc}) \nn\\
&+& 2\frac{3}{4} (d^{abc} - i f^{abc}) \nn\\
&-& 6 \frac{1}{4}  (d^{abc} + i f^{abc}) \nn\\
&-& 6 \frac{1}{4}  (d^{abc} - i f^{abc}) \nn\\
&-& 6 \frac{1}{4}  (d^{abc} + i f^{abc}) \nn\\
&-& \frac{8C_F}{2} \frac{1}{4}  (d^{abc} + i f^{abc}) \nn\\
&+& 4C_F\,\frac{1}{4}  (d^{abc} + i f^{abc}) \nn\\
&=& 0~.
\eea

Next, consider the diagrams which involve $\Psi_{qqq}^*(x_1,\vec
k_1+x_1\vec q; x_2,\vec k_2 +x_2\vec q-\vec q; x_3,\vec k_3 +x_3\vec q)$
in eq.~(\ref{eq:UV_prefac1}). These are: the
first term in eq.~(\ref{eq:rho3_qqq}) times $-C_q(x_2)$ and with
quarks 1 and 2 interchanged; and half of eq.~(\ref{eq:rho-rho-rho_UV_q2q2q2}):
\bea
&-& \frac{8C_F}{2} \, \frac{1}{4}  (d^{abc} + i f^{abc}) \nn\\
&+& 2\, C_F \cdot 2 \, \frac{1}{4}  (d^{abc} + i f^{abc}) = 0~.
\eea
Similarly, the diagrams which involve $\Psi_{qqq}^*(x_1,\vec
k_1+x_1\vec q; x_2,\vec k_2 +x_2\vec q; x_3,\vec k_3 +x_3\vec q-\vec q)$
in eq.~(\ref{eq:UV_prefac1}) are: the
first term in eq.~(\ref{eq:rho3_qqq}) times $-C_q(x_3)$ and with
quarks 1 and 3 interchanged; and the remaining
half of eq.~(\ref{eq:rho-rho-rho_UV_q2q2q2}) with quarks 2 and 3 interchanged:
\bea
&-& \frac{8C_F}{2} \, \frac{1}{4}  (d^{abc} + i f^{abc}) \nn\\
&+& 2\, C_F \cdot 2 \, \frac{1}{4}  (d^{abc} + i f^{abc}) = 0~.
\eea
\\

The above were diagrams where the transverse momentum of all three
probe gluons flowed into one single quark line. We now consider the divergent
diagrams where one of the momenta flows into a different quark line.

We begin with terms which involve $\Psi_{qqq}^*(x_1,\vec k_1+x_1\vec q
- \vec q_{12}; x_2,\vec k_2 +x_2\vec q - \vec q_3; x_3,\vec k_3
+x_3\vec q)$. These are eq.~(\ref{eq:rho-rho-rho_UV_q2gg}), the fourth
line in eq.~(\ref{eq:rho3_qqq}) times $-C_q(x_1)$, and
eq.~(\ref{eq:rho-rho-rho_UV_q2q1q1}):
\bea
&-& 4\, \frac{1}{4}\, \frac{3}{2}(d^{abc} + i f^{abc}) \nn\\
&+& \frac{8 C_F}{2}\, \frac{1}{4}  (d^{abc} + i f^{abc}) \nn\\
&+& 4\, \frac{1}{6}\,  \frac{1}{4}(d^{abc} + i f^{abc}) = 0~.
\eea
Now terms which involve $\Psi_{qqq}^*(x_1,\vec k_1+x_1\vec q - \vec
q_{13}; x_2,\vec k_2 +x_2\vec q - \vec q_2; x_3,\vec k_3 +x_3\vec q)$:
eq.~(\ref{eq:rho-rho-rho_UV_gq2g}), the third line in
eq.~(\ref{eq:rho3_qqq}) times $-C_q(x_1)$, and
eq.~(\ref{eq:rho-rho-rho_UV_q1q2q1}):
\bea
&-& 4\, \frac{1}{4}\, \frac{3}{2}\, (d^{abc} - i f^{abc}) \nn\\
&+& \frac{8 C_F}{2}\, \frac{1}{4}  (d^{abc} - i f^{abc}) \nn\\
&+& 4\, \frac{1}{6}\,  \frac{1}{4}(d^{abc} - i f^{abc}) = 0~.
\eea
Next are terms which involve $\Psi_{qqq}^*(x_1,\vec k_1+x_1\vec q -
\vec q_{23}; x_2,\vec k_2 +x_2\vec q - \vec q_1; x_3,\vec k_3 +x_3\vec
q)$: eq.~(\ref{eq:rho-rho-rho_UV_ggq2}), the second line in
eq.~(\ref{eq:rho3_qqq}) times $-C_q(x_2)$ and with quarks 1 and 2
interchanged, and eq.~(\ref{eq:rho-rho-rho_UV_q1q1q2}):
\bea
&-& 4\, \frac{1}{4}\, \frac{3}{2}\, (d^{abc} + i f^{abc}) \nn\\
&+& \frac{8 C_F}{2}\, \frac{1}{4}  (d^{abc} + i f^{abc}) \nn\\
&+& 4\, \frac{1}{6}\,  \frac{1}{4}(d^{abc} + i f^{abc}) = 0~.
\eea
Terms which involve $\Psi_{qqq}^*(x_1,\vec k_1+x_1\vec q - \vec q_{1};
x_2,\vec k_2 +x_2\vec q - \vec q_{23}; x_3,\vec k_3 +x_3\vec q)$:
eq.~(\ref{eq:rho-rho-rho_UV_q2q2g}), the second line in
eq.~(\ref{eq:rho3_qqq}) times $-C_q(x_1)$, and
eq.~(\ref{eq:rho-rho-rho_UV_q2q2q1}):
\bea
&-& 4\, \frac{3}{2}\, \frac{1}{4}(d^{abc} + i f^{abc}) \nn\\
&+& \frac{8 C_F}{2}\, \frac{1}{4}  (d^{abc} + i f^{abc}) \nn\\
&+& 4\, \frac{1}{6}\,  \frac{1}{4}(d^{abc} + i f^{abc}) = 0.
\eea
Terms which involve $\Psi_{qqq}^*(x_1,\vec k_1+x_1\vec q - \vec q_{2};
x_2,\vec k_2 +x_2\vec q - \vec q_{13}; x_3,\vec k_3 +x_3\vec q)$:
eq.~(\ref{eq:rho-rho-rho_UV_q2gq2}), the third line in
eq.~(\ref{eq:rho3_qqq}) times $-C_q(x_2)$ with quarks 1 and 2
interchanged, and eq.~(\ref{eq:rho-rho-rho_UV_q2q1q2}):
\bea
&-& 4\, \frac{3}{2}\, \frac{1}{4}(d^{abc} - i f^{abc}) \nn\\
&+& \frac{8 C_F}{2}\, \frac{1}{4}  (d^{abc} - i f^{abc}) \nn\\
&+& 4\, \frac{1}{6}\,  \frac{1}{4}(d^{abc} - i f^{abc}) = 0.
\eea
Terms which involve $\Psi_{qqq}^*(x_1,\vec k_1+x_1\vec q -
\vec q_{3}; x_2,\vec k_2 +x_2\vec q - \vec q_{12}; x_3,\vec k_3
+x_3\vec q)$: eq.~(\ref{eq:rho-rho-rho_UV_gq2q2}), the fourth line in
eq.~(\ref{eq:rho3_qqq}) times $-C_q(x_2)$ with quarks 1 and 2
interchanged, and eq.~(\ref{eq:rho-rho-rho_UV_q1q2q2}):
\bea
&-& 4\, \frac{3}{2}\, \frac{1}{4}(d^{abc} + i f^{abc}) \nn\\
&+& \frac{8 C_F}{2}\, \frac{1}{4}  (d^{abc} + i f^{abc}) \nn\\
&+& 4\, \frac{1}{6}\,  \frac{1}{4}(d^{abc} + i f^{abc}) = 0.
\eea
Terms which involve $\Psi_{qqq}^*(x_1,\vec k_1+x_1\vec q; x_2,\vec k_2
+x_2\vec q - \vec q_{12}; x_3,\vec k_3 +x_3\vec q - \vec q_{3})$:
eq.~(\ref{eq:rho-rho-rho_UV_q3q2q2}), and the fourth term in
eq.~(\ref{eq:rho3_qqq}) times $-C_q(x_3)$ with quarks 1 and 3
interchanged (followed by a second interchange of quarks 2 and 3):
\bea
&-& 4\, C_F \frac{1}{4}(d^{abc} + i f^{abc}) \nn\\
&+& \frac{8 C_F}{2}\, \frac{1}{4}  (d^{abc} + i f^{abc}) = 0 ~.
\eea
Terms which involve $\Psi_{qqq}^*(x_1,\vec k_1+x_1\vec q; x_2,\vec k_2
+x_2\vec q - \vec q_{13}; x_3,\vec k_3 +x_3\vec q - \vec q_{2})$:
eq.~(\ref{eq:rho-rho-rho_UV_q2q3q2}), and the third term in
eq.~(\ref{eq:rho3_qqq}) times $-C_q(x_3)$ with quarks 1 and 3
interchanged (followed by a second interchange of quarks 2 and 3):
\bea
&-& 4\, C_F \frac{1}{4}(d^{abc} - i f^{abc}) \nn\\
&+& \frac{8 C_F}{2}\, \frac{1}{4}  (d^{abc} - i f^{abc}) = 0 ~.
\eea
Terms which involve $\Psi_{qqq}^*(x_1,\vec k_1+x_1\vec q; x_2,\vec k_2
+x_2\vec q - \vec q_{1}; x_3,\vec k_3 +x_3\vec q - \vec q_{23})$:
eq.~(\ref{eq:rho-rho-rho_UV_q3q3q2}), and the second term in
eq.~(\ref{eq:rho3_qqq}) times $-C_q(x_3)$ with quarks 1 and 3
interchanged (followed by a second interchange of quarks 2 and 3):
\bea
&-& 4\, C_F \frac{1}{4}(d^{abc} + i f^{abc}) \nn\\
&+& \frac{8 C_F}{2}\, \frac{1}{4}  (d^{abc} + i f^{abc}) = 0 ~.
\eea
\\

The last structure to be checked corresponds to the diagrams where
each $\vec q_i$ flows into a different quark line. Terms which involve
$\Psi_{qqq}^*(x_1,\vec k_1+x_1\vec q - \vec q_{1}; x_2,\vec k_2
+x_2\vec q - \vec q_{2}; x_3,\vec k_3 +x_3\vec q - \vec q_{3})$:
eq.~(\ref{eq:rho-rho-rho_UV_q3q2g}), the fifth line in
eq.~(\ref{eq:rho3_qqq}) times $-C_q(x_1)$, and
eq.~(\ref{eq:rho-rho-rho_UV_q3q2q1}):
\bea
& & 4\, \frac{3}{2}\, \frac{1}{2} \, d^{abc} \nn\\
&-& \frac{8 C_F}{2}\, \frac{1}{2}\,  d^{abc} \nn\\
&-& 4\, \frac{1}{6}\, \frac{1}{2}\,  d^{abc} = 0~.
\eea
Terms which involve $\Psi_{qqq}^*(x_1,\vec k_1+x_1\vec q - \vec q_{2};
x_2,\vec k_2 +x_2\vec q - \vec q_{1}; x_3,\vec k_3 +x_3\vec q - \vec
q_{3})$: eq.~(\ref{eq:rho-rho-rho_UV_q3gq2}), the fifth line in
eq.~(\ref{eq:rho3_qqq}) times $-C_q(x_2)$ with quarks 1 and 2
interchanged, and eq.~(\ref{eq:rho-rho-rho_UV_q3q1q2}):
\bea
& & 4\, \frac{3}{2}\, \frac{1}{2} \, d^{abc} \nn\\
&-& \frac{8 C_F}{2}\, \frac{1}{2}\,  d^{abc} \nn\\
&-& 4\, \frac{1}{6}\, \frac{1}{2}\,  d^{abc} = 0~.
\eea
Terms which involve $\Psi_{qqq}^*(x_1,\vec k_1+x_1\vec q - \vec q_{3};
x_2,\vec k_2 +x_2\vec q - \vec q_{1}; x_3,\vec k_3 +x_3\vec q - \vec
q_{2})$: eq.~(\ref{eq:rho-rho-rho_UV_gq3q2}), the fifth line in
eq.~(\ref{eq:rho3_qqq}) times $-C_q(x_3)$ with quarks 1 and 3
interchanged, and eq.~(\ref{eq:rho-rho-rho_UV_q1q3q2}):
\bea
& & 4\, \frac{3}{2}\, \frac{1}{2} \, d^{abc} \nn\\
&-& \frac{8 C_F}{2}\, \frac{1}{2}\,  d^{abc} \nn\\
&-& 4\, \frac{1}{6}\, \frac{1}{2}\,  d^{abc} = 0~.
\eea
%

\section{Ward identity}
\label{sec:Ward}

In this section we verify that $\langle\rho^a(\vec q_1)\,\rho^b(\vec
q_2)\,\rho^c(\vec q_3)\, \rangle$ vanishes when $\vec q_1 \to 0$ or
$\vec q_3 \to 0$~\cite{Bartels:1999aw, Ewerz:2001fb}; the first and
last of the three gluon probes couple to a color singlet proton state
and must not have infinite transverse wave length. On the other hand,
this correlator need not vanish when $\vec q_2 \to 0$ at finite $\vec
q_{1,3}$ since the second gluon probe in that case does not couple to
a color singlet. It is straightforward to confirm that the expectation
value of $\rho^a(\vec q_1)\,\rho^b(\vec q_2)\,\rho^c(\vec q_3)$
between three-quark states at ${\cal O}(g^3)$ written in
eq.~(\ref{eq:rho3_qqq}) does indeed vanish when either $\vec q_1, \vec
q_3 \to 0$ (see, also, refs.~\cite{Czyzewski:1996bv, Engel:1997cga,
  Dumitru:2020fdh}). In what follows we consider the correction at
${\cal O}(g^5)$.  We first show that the finite part of the UV
divergent diagrams cancel when either $\vec q_1, \vec q_3 \to
0$. After this we repeat the same exercise and verify that the sum of
the remaining finite diagrams also cancel when either $\vec q_1, \vec
q_3 \to 0$.

\subsection{UV divergent diagrams}

First we verify that the subset of UV divergent diagrams satisfies the
Ward identity, i.e.\ that their finite parts cancel when $\vec q_1 \to
0$ or $\vec q_3 \to 0$. We only demonstrate here the case $\vec q_3
\to 0$ but we have checked the symmetry of these diagrams under $\vec
q_1 \leftrightarrow \vec q_3$. For the purpose of more compact
expressions we will split off the pre-''factor''
\begin{equation}
\frac{2g^5}{3\cdot 16\pi^3} \cdot 2\pi^3 \int [\ud x_i] \int [\ud^2 k_i]\Psi_{qqq}(x_1,\vec k_1;x_2,\vec k_2;x_3,\vec k_3)
\end{equation}
from the following expressions.\\

First, we collect all the terms from the UV divergent diagrams which
involve the structure $\Psi_{qqq}(x_1,\vec k_1 - (1-x_1)\vec
q_{12};x_2, \vec k _2 + x_2\vec q_{12};x_3, \vec k_3 + x_3 \vec
q_{12})$. These are given by eqs.~(\ref{eq:rho-rho-rho_UV_ggg},
\ref{eq:rho-rho-rho_UV_q1gg}, \ref{eq:rho-rho-rho_UV_gq1g},
\ref{eq:rho-rho-rho_UV_ggq1}, \ref{eq:rho-rho-rho_UV_q2gg},
\ref{eq:rho-rho-rho_UV_q1q1g}, \ref{eq:rho-rho-rho_UV_q1gq1},
\ref{eq:rho-rho-rho_UV_gq1q1}, \ref{eq:rho-rho-rho_UV_q1q1q1},
\ref{eq:rho-rho-rho_UV_q2q1q1}):
\begin{equation}
\begin{split}
& +3 \cdot \frac{N_c}{2}if^{abc} \cdot  F(\vec q_{12},0;\alpha_1,m^2)\\
& +3 \cdot \frac{N_c}{4}\left (d^{abc} - if^{abc} \right ) \cdot F(\vec q_{12},0;\alpha_1,m^2)\\
& +3 \cdot \frac{N_c}{4}\left (d^{abc} + if^{abc} \right ) \cdot F(\vec q_{12},\vec q_2;\alpha_1,m^2)\\
& +3 \cdot \frac{N_c}{4}\left (d^{abc} - if^{abc} \right ) \cdot F(\vec q_{12},\vec q_1;\alpha_1,m^2)\\ 
& -6 \cdot \frac{N_c}{2\cdot 4}\left (d^{abc} + if^{abc} \right ) \cdot F(\vec q_{12},0;\alpha_1,m^2)\\
& -3  \cdot \frac{N_c}{4}\left (d^{abc} + if^{abc} \right ) \cdot F(\vec q_{12},\vec q_2;\alpha_1,m^2)\\
& -3  \cdot \frac{N_c}{4}\left (d^{abc} - if^{abc} \right ) \cdot F(\vec q_{12},\vec q_1;\alpha_1,m^2)\\
& -3  \cdot \frac{N_c}{4}\left (d^{abc} + if^{abc} \right ) \cdot F(\vec q_{12},\vec q_{12};\alpha_1,m^2)\\
& +3  \cdot \frac{2\cf}{4}\left (d^{abc} + if^{abc} \right ) \cdot F(\vec q_{12},\vec q_{12};\alpha_1,m^2)\\
& +6  \cdot \frac{1}{2N_c \cdot 4}\left (d^{abc} + if^{abc} \right ) \cdot F(\vec q_{12},\vec q_{12};\alpha_1,m^2)\\
& = 0.
\end{split}
\end{equation}
Next, we collect terms which involve the structure $\Psi_{qqq}(x_1,\vec k_1 + x_1\vec q_{12} - \vec q_1;x_2, \vec k
_2 + x_2\vec q_{12} - \vec q_2;x_3, \vec k_3 + x_3 \vec q_{12})$. These are given by eqs.~(\ref{eq:rho-rho-rho_UV_gq2g},
\ref{eq:rho-rho-rho_UV_q2q2g}, \ref{eq:rho-rho-rho_UV_q3q2g},
\ref{eq:rho-rho-rho_UV_q1q2q1}, \ref{eq:rho-rho-rho_UV_q2q2q1}, \ref{eq:rho-rho-rho_UV_q3q2q1}):
\begin{equation}
\begin{split}
& -6 \cdot \frac{N_c}{2\cdot 4}\left (d^{abc} - if^{abc} \right ) \cdot F(\vec q_{1},0;\alpha_1,m^2)\\
& -6 \cdot \frac{N_c}{2\cdot 4}\left (d^{abc} + if^{abc} \right ) \cdot F(\vec q_{1},0;\alpha_1,m^2)\\
& -6 \cdot \frac{N_c}{2\cdot 4}\biggl [\left (d^{abc} + if^{abc} \right ) + \left (d^{abc} - if^{abc} \right ) \biggr ] \cdot F(\vec q_{1},0;\alpha_1,m^2)\\
& +6 \cdot \frac{1}{2N_c\cdot 4}\left (d^{abc} - if^{abc} \right ) \cdot F(\vec q_{1},\vec q_{1};\alpha_1,m^2)\\
& +6 \cdot \frac{1}{2N_c\cdot 4}\left (d^{abc} + if^{abc} \right ) \cdot F(\vec q_{1},\vec q_{1};\alpha_1,m^2)\\
& -6 \cdot \frac{1}{2N_c\cdot 4}\biggl [\left (d^{abc} + if^{abc} \right ) + \left (d^{abc} - if^{abc} \right ) \biggr ] \cdot F(\vec q_{1},\vec q_{1};\alpha_1,m^2)\\
& = 0.
\end{split}
\end{equation}
Similarly, the diagrams which involve the structure $\Psi_{qqq}(x_1,\vec k_1 + x_1\vec q_{12} - \vec q_2;x_2, \vec k
_2 + x_2\vec q_{12} - \vec q_1;x_3, \vec k_3 + x_3 \vec q_{12})$. These are given by eqs.~(\ref{eq:rho-rho-rho_UV_ggq2},
\ref{eq:rho-rho-rho_UV_q2gq2}, \ref{eq:rho-rho-rho_UV_q3gq2},
\ref{eq:rho-rho-rho_UV_q1q1q2},  \ref{eq:rho-rho-rho_UV_q2q1q2}, \ref{eq:rho-rho-rho_UV_q3q1q2}):
\begin{equation}
\begin{split}
& -6 \cdot \frac{N_c}{2\cdot 4}\left (d^{abc} + if^{abc} \right ) \cdot F(\vec q_{2},0;\alpha_1,m^2)\\
& -6 \cdot \frac{N_c}{2\cdot 4}\left (d^{abc} - if^{abc} \right ) \cdot F(\vec q_{2},0;\alpha_1,m^2)\\
& +6 \cdot \frac{N_c}{2\cdot 4}\biggl [\left (d^{abc} + if^{abc} \right ) + \left (d^{abc} - if^{abc} \right ) \biggr ] \cdot F(\vec q_{2},0;\alpha_1,m^2)\\
& +6 \cdot \frac{1}{2N_c\cdot 4}\left (d^{abc} + if^{abc} \right ) \cdot F(\vec q_{2},\vec q_{2};\alpha_1,m^2)\\
& +6 \cdot \frac{1}{2N_c\cdot 4}\left (d^{abc} - if^{abc} \right ) \cdot F(\vec q_{2},\vec q_{2};\alpha_1,m^2)\\
& -6 \cdot \frac{1}{2N_c\cdot 4}\biggl [\left (d^{abc} + if^{abc} \right ) + \left (d^{abc} - if^{abc} \right ) \biggr ] \cdot F(\vec q_{2},\vec q_{2};\alpha_1,m^2)\\
& = 0.
\end{split}
\end{equation}
Finally, all the other UV divergent diagrams (in the $\vec q_3 \to 0$
case) are proportional to the finite function $F(0,0;x/x_1,m^2)$,
which is zero.

\subsection{Finite diagrams}

In this section we verify that all the UV finite diagrams
satisfies the Ward identity. As in the UV divergent case, the sum of
all finite diagams should cancel when $\vec q_1 \rightarrow 0$ or
$\vec q_3 \rightarrow 0$. Here we only show details for the $\vec
q_3 \rightarrow 0$ case, but we have also checked that the case $\vec
q_1 \rightarrow 0$ satisfies the Ward identity.
  
First, we collect all the terms from the UV finite diagrams which involve the structure 
\begin{equation}
\begin{split}
\vec I \cdot & \frac{z_2 \vec p_2 - (1-z_2)(\vec k_g - \vec q_{12})}{\left (z_2 \vec p_2 - (1-z_2)(\vec k_g - \vec q_{12})\right )^2}  \\
& \times  \Psi^{\ast}_{qqq}(x_1 - x_g, \vec k_1 + x_1\vec q_{12} - \vec k_g + x_g \vec K; x_2 + x_g, \vec k_2 - (1-x_2)\vec q_{12} + \vec k_g - x_g\vec K; x_3, \vec k_3 + x_3\vec q_{12}).
\end{split}
\end{equation}
These are given by eqs.~(\ref{eq:rho-rho-rho_FIN_ggg}, \ref{eq:rho-rho-rho_FIN_q1gg}, \ref{eq:rho-rho-rho_FIN_q2gg},  \ref{eq:rho-rho-rho_FIN_q3gg}):
\begin{equation}
\begin{split}
& -6 \cdot \frac{1}{2} \tr T^a T^b T^c\\
& -6 \cdot \frac{1}{4} \tr (T^a T^b D^c - T^a T^b T^c)\\
& -6 \cdot \frac{1}{4} \tr (T^a T^b D^c - T^a T^b T^c)\\
& +6 \cdot \frac{1}{2} \tr T^a T^b D^c\\
& = 0.
\end{split}
\end{equation}
Next, we collect terms which involve the structure: 
\begin{equation}
\label{eq:str1}
\begin{split}
\vec I \cdot & \frac{z_2 \vec p_2 - (1-z_2)(\vec k_g - \vec q_{1})}{\left (z_2 \vec p_2 - (1-z_2)(\vec k_g - \vec q_{1})\right )^2} \\
& \times  \Psi^{\ast}_{qqq}(x_1 - x_g, \vec k_1 + x_1\vec q_{12} - \vec q_2 - \vec k_g + x_g \vec K; x_2 + x_g, \vec k_2 + x_2 \vec q_{12} - \vec q_1 + \vec k_g - x_g\vec K; x_3, \vec k_3 + x_3\vec q_{12})
\end{split}
\end{equation}
or 
\begin{equation}
\label{eq:str2}
\begin{split}
\vec I \cdot & \frac{z_2 \vec p_2 - (1-z_2)(\vec k_g - \vec q_{2})}{\left (z_2 \vec p_2 - (1-z_2)(\vec k_g - \vec q_{2})\right )^2} \\
& \times  \Psi^{\ast}_{qqq}(x_1 - x_g, \vec k_1 + x_1\vec q_{12} - \vec q_1 - \vec k_g + x_g \vec K; x_2 + x_g, \vec k_2 + x_2 \vec q_{12} - \vec q_2 + \vec k_g - x_g\vec K; x_3, \vec k_3 + x_3\vec q_{12}).
\end{split}
\end{equation}
Terms that involve the structure \eqref{eq:str1} are given by eqs.~(\ref{eq:rho-rho-rho_FIN_gq1g}, \ref{eq:rho-rho-rho_FIN-2_q1q1g}, \ref{eq:rho-rho-rho_FIN-2_q2q1g}, \ref{eq:rho-rho-rho_FIN-2_q3q1g}):
\begin{equation}
\begin{split}
& -6 \cdot \frac{1}{4} \tr (T^a T^c D^b - T^a T^c T^b)\\
& +6 \cdot \frac{N_c}{2} \tr t^a t^b t^c\\
& +6 \cdot  \left (-\frac{1}{4}(T^a)_{bc} + \frac{N_c}{2} \tr t^a t^c t^b \right ) \\
& +6 \cdot \left (-\frac{1}{4}(T^a)_{cb} - N_c \tr t^a t^c t^b + \frac{1}{4}\tr T^c T^a (D^b - T^b) \right ).\\
\end{split}
\end{equation}
Here the color factors can be simplified by noting that 
\begin{equation}
\tr T^a T^b D^c = \frac{N_c}{2}d^{abc}, \quad \tr T^a T^b T^c = \frac{iN_c}{2}f^{abc}, \quad \tr t^at^bt^c = \frac{1}{4}\left (d^{abc} + if^{abc}\right ),
\end{equation}
where $\nc = 3$ and the strucutre constants $d^{abc}$ and $f^{abc}$ are totally symmetric and antisymmetric, respectively. This gives
\begin{equation}
\begin{split}
& -6 \cdot \frac{1}{4} \left (\frac{N_c}{2}d^{abc} + \frac{iN_c}{2}f^{abc} \right )\\
& +6 \cdot \frac{1}{4} \left (\frac{N_c}{2}d^{abc} + \frac{iN_c}{2}f^{abc} \right )\\
& +6 \cdot  \frac{1}{4} \left (-(T^a)_{bc} + \left (\frac{N_c}{2}d^{abc} - \frac{iN_c}{2}f^{abc} \right )\right ) \\
& +6 \cdot \frac{1}{4}\left (+(T^a)_{bc} - \left (\frac{N_c}{2}d^{abc} - \frac{iN_c}{2}f^{abc} \right ) \right )\\
& = 0.
\end{split}
\end{equation}
Similarly, terms which involve the structure \eqref{eq:str2} are given by eqs.~(\ref{eq:rho-rho-rho_FIN_ggq1}, \ref{eq:rho-rho-rho_FIN-2_q1gq1}, \ref{eq:rho-rho-rho_FIN-2_q2gq1}, \ref{eq:rho-rho-rho_FIN-2_q3gq1}):
\begin{equation}
\begin{split}
& -6 \cdot \frac{1}{4} \tr (T^b T^c D^a - T^a T^b T^c)\\
& +6 \cdot \frac{N_c}{2} \tr t^a t^c t^b\\
& +6 \cdot  \left (+\frac{1}{4}(T^b)_{ca} + \frac{N_c}{2} \tr t^a t^b t^c \right ) \\
& +6 \cdot \left (-\frac{1}{4}(T^b)_{ca} - N_c \tr t^a t^b t^c + \frac{1}{4}\tr T^c T^b (D^a - T^a) \right )\\
& = 0.
\end{split}
\end{equation}

Let us then collect all the terms which involve the structure 
\begin{equation}
\label{eq:str11}
\begin{split}
\vec I \cdot & \frac{z_2 (\vec p_2 - \vec q_2) - (1-z_2)(\vec k_g - \vec q_{1})}{\left (z_2 (\vec p_2 - \vec q_2) - (1-z_2)(\vec k_g - \vec q_{1})\right )^2} \\
& \times  \Psi^{\ast}_{qqq}(x_1 - x_g, \vec k_1 + x_1\vec q_{12} - \vec k_g + x_g \vec K; x_2 + x_g, \vec k_2 -(1 - x_2) \vec q_{12}  + \vec k_g - x_g\vec K; x_3, \vec k_3 + x_3\vec q_{12})
\end{split}
\end{equation}
or 
\begin{equation}
\label{eq:str22}
\begin{split}
\vec I \cdot & \frac{z_2 (\vec p_2 - \vec q_1) - (1-z_2)(\vec k_g - \vec q_{2})}{\left (z_2 (\vec p_2 - \vec q_1) - (1-z_2)(\vec k_g - \vec q_{2})\right )^2} \\
& \times  \Psi^{\ast}_{qqq}(x_1 - x_g, \vec k_1 + x_1\vec q_{12} - \vec k_g + x_g \vec K; x_2 + x_g, \vec k_2 - (1 - x_2) \vec q_{12}  + \vec k_g - x_g\vec K; x_3, \vec k_3 + x_3\vec q_{12}).
\end{split}
\end{equation}
Terms that involve the structure in \eqref{eq:str11} are given by eqs.~(\ref{eq:rho-rho-rho_FIN_gq2g}, \ref{eq:rho-rho-rho_FIN-2_q1q2g}, \ref{eq:rho-rho-rho_FIN-2_q2q2g}, \ref{eq:rho-rho-rho_FIN-2_q3q2g}):
\begin{equation}
\begin{split}
& -6 \cdot \frac{1}{4} \tr (T^a T^c D^b - T^a T^c T^b)\\
& +6 \cdot \left (+\frac{1}{4}(T^a)_{bc} + \frac{N_c}{2} \tr t^a t^b t^c \right )\\
& +6 \cdot \frac{N_c}{2} \tr t^a t^b t^c \\
& +6 \cdot \left (-\frac{1}{4}(T^a)_{bc} - \frac{N_c}{4}(D^a)_{bc} + \frac{1}{4}\tr T^a (D^b - T^b)T^c \right )\\
& = 0.
\end{split}
\end{equation}
Terms that involve the structure in \eqref{eq:str22} are given by eqs.~(\ref{eq:rho-rho-rho_FIN_ggq2}, \ref{eq:rho-rho-rho_FIN-2_q1gq2}, \ref{eq:rho-rho-rho_FIN-2_q2gq2}, \ref{eq:rho-rho-rho_FIN-2_q3gq2}):
\begin{equation}
\begin{split}
& -6 \cdot \frac{1}{4} \tr (T^a T^b D^c - T^a T^b T^c)\\
& +6 \cdot \left (+\frac{1}{4}(T^b)_{ac} + \frac{N_c}{2} \tr t^a t^c t^b \right )\\
& +6 \cdot \frac{N_c}{2} \tr t^b t^a t^c \\
& +6 \cdot \left (-\frac{1}{4}(T^b)_{ac} - \frac{N_c}{4}(D^b)_{ac} + \frac{1}{4}\tr T^b (D^a - T^a)T^c \right )\\
& = 0.
\end{split}
\end{equation}
We also have diagrams which involve the sturcture
\begin{equation}
\begin{split}
\vec I \cdot & \frac{z_2 (\vec p_2 - \vec q_{12}) - (1-z_2)\vec k_g }{\left (z_2 (\vec p_2 - \vec q_{12}) - (1-z_2)\vec k_g\right )^2} \\
& \times  \Psi^{\ast}_{qqq}(x_1 - x_g, \vec k_1 + x_1\vec q_{12} - \vec k_g + x_g \vec K; x_2 + x_g, \vec k_2 -(1 - x_2) \vec q_{12}  + \vec k_g - x_g\vec K; x_3, \vec k_3 + x_3\vec q_{12}).
\end{split}
\end{equation}
These diagrams are given by eqs.~(\ref{eq:rho-rho-rho_FIN-2_gq2q2}, \ref{eq:rho-rho-rho_FIN-q1q2q2}, \ref{eq:rho-rho-rho_FIN-q2q2q2}, \ref{eq:rho-rho-rho_FIN-q3q2q2}):
\begin{equation}
\begin{split}
& +6 \cdot \frac{\nc}{2} \tr t^a t^b t^c \\
& -6 \cdot \left (\cf - \frac{1}{2}\right ) \tr t^at^bt^c\\
& -6 \cdot \cf \tr t^a t^b t^c \\
& +6 \cdot \left (2\cf - \frac{1}{2} - \frac{\nc}{2} \right )\tr t^a t^b t^c\\
& = 0.
\end{split}
\end{equation}

We then consider terms that involve the structure 
\begin{equation}
\label{eq:str3}
\begin{split}
\vec I \cdot & \frac{z_2 \vec p_2 - (1-z_2)(\vec k_g - \vec q_{1})}{\left (z_2 \vec p_2 - (1-z_2)(\vec k_g - \vec q_{1})\right )^2} \\
& \times  \Psi^{\ast}_{qqq}(x_1 - x_g, \vec k_1 + x_1\vec q_{12} - \vec k_g + x_g \vec K; x_2 + x_g, \vec k_2 + x_2 \vec q_{12} - \vec q_1 + \vec k_g - x_g\vec K; x_3, \vec k_3 + x_3\vec q_{12} - \vec q_2)
\end{split}
\end{equation}
or 
\begin{equation}
\label{eq:str4}
\begin{split}
\vec I \cdot & \frac{z_2 \vec p_2 - (1-z_2)(\vec k_g - \vec q_{2})}{\left (z_2 \vec p_2 - (1-z_2)(\vec k_g - \vec q_{2})\right )^2} \\
& \times  \Psi^{\ast}_{qqq}(x_1 - x_g, \vec k_1 + x_1\vec q_{12} - \vec k_g + x_g \vec K; x_2 + x_g, \vec k_2 + x_2 \vec q_{12} - \vec q_2 + \vec k_g - x_g\vec K; x_3, \vec k_3 + x_3\vec q_{12}-\vec q_1).
\end{split}
\end{equation}
Terms that involve the structure \eqref{eq:str3} are given by eqs.~(\ref{eq:rho-rho-rho_FIN_gq3g}, \ref{eq:rho-rho-rho_FIN-2_q1q3g}, \ref{eq:rho-rho-rho_FIN-2_q2q3g}):
\begin{equation}
\begin{split}
&  +6 \cdot \frac{1}{2} \tr T^a T^c D^b \\
&  +6 \cdot \left (-\frac{1}{4}(T^a)_{bc} - \nc \tr t^at^bt^c + \frac{1}{4}\tr T^b T^a (D^c - T^c) \right ) \\
& +6 \cdot \left (+\frac{1}{4}(T^a)_{bc} - \frac{\nc}{4}(D^a)_{bc} + \frac{1}{4}\tr T^a (D^c - T^c) T^b \right ) \\
& = 0.
\end{split}
\end{equation}
Similarly, terms that involve the structure \eqref{eq:str4} are given by eqs.~(\ref{eq:rho-rho-rho_FIN_ggq3}, \ref{eq:rho-rho-rho_FIN-2_q1gq3}, \ref{eq:rho-rho-rho_FIN-2_q2gq3}):
\begin{equation}
\begin{split}
&  +6 \cdot \frac{1}{2} \tr T^a T^b D^c \\
&  +6 \cdot \left (-\frac{1}{4}(T^b)_{ac} - \nc \tr t^b t^a t^c + \frac{1}{4}\tr T^a T^b (D^c - T^c) \right ) \\
& +6 \cdot \left (+\frac{1}{4}(T^b)_{ac} - \frac{\nc}{4}(D^b)_{ac} + \frac{1}{4}\tr T^b (D^c - T^c) T^a \right ) \\
& = 0.
\end{split}
\end{equation}
A similar set of diagrams are given by terms that involve the structure 
\begin{equation}
\label{eq:str5}
\begin{split}
\vec I \cdot & \frac{z_2 (\vec p_2 - \vec q_1) - (1-z_2) \vec k_g }{\left (z_2 (\vec p_2 - \vec q_1) - (1-z_2)\vec k_g\right )^2} \\
& \times  \Psi^{\ast}_{qqq}(x_1 - x_g, \vec k_1 + x_1\vec q_{12} - \vec k_g + x_g \vec K - \vec q_2; x_2 + x_g, \vec k_2 + x_2 \vec q_{12} - \vec q_1 + \vec k_g - x_g\vec K; x_3, \vec k_3 + x_3\vec q_{12})
\end{split}
\end{equation}
or 
\begin{equation}
\label{eq:str6}
\begin{split}
\vec I \cdot & \frac{z_2 (\vec p_2 - \vec q_2) - (1-z_2) \vec k_g }{\left (z_2 (\vec p_2 - \vec q_2) - (1-z_2)\vec k_g\right )^2} \\
& \times  \Psi^{\ast}_{qqq}(x_1 - x_g, \vec k_1 + x_1\vec q_{12} - \vec k_g + x_g \vec K - \vec q_1; x_2 + x_g, \vec k_2 + x_2 \vec q_{12} - \vec q_2 + \vec k_g - x_g\vec K; x_3, \vec k_3 + x_3\vec q_{12}).
\end{split}
\end{equation}
Terms that involve the structure \eqref{eq:str5} are given by eqs.~(\ref{eq:rho-rho-rho_FIN-2_gq1q2}, \ref{eq:rho-rho-rho_FIN-q1q1q2}, \ref{eq:rho-rho-rho_FIN-q2q1q2}, \ref{eq:rho-rho-rho_FIN-q3q1q2}):
\begin{equation}
\begin{split}
&  +6 \cdot \left (\frac{1}{4}(T^c)_{ab} + \frac{\nc}{2} \tr t^a t^b t^c \right ) \\
&  -6 \cdot \left (\cf - \frac{1}{2} \right )\tr t^a t^b t^c \\
&  -6 \cdot \left (\cf - \frac{1}{2} \right )\tr t^a t^c t^b \\
& + 6 \cdot \left ((\cf - \frac{\nc}{2})\tr t^a t^b t^c + (\cf - 1)\tr t^a t^c t^b \right )\\
& = 0.
\end{split}
\end{equation}
Terms that involve the structure \eqref{eq:str6} are given by eqs.~(\ref{eq:rho-rho-rho_FIN-2_gq2q1}, \ref{eq:rho-rho-rho_FIN-q1q2q1}, \ref{eq:rho-rho-rho_FIN-q2q2q1}, \ref{eq:rho-rho-rho_FIN-q3q2q1}):
\begin{equation}
\begin{split}
&  +6 \cdot \left (\frac{1}{4}(T^c)_{ba} + \frac{\nc}{2} \tr t^a t^c t^b \right ) \\
&  -6 \cdot \left (\cf - \frac{1}{2} \right )\tr t^a t^c t^b \\
&  -6 \cdot \left (\cf - \frac{1}{2} \right )\tr t^a t^b t^c \\
& + 6 \cdot \left ((\cf - \frac{\nc}{2})\tr t^a t^c t^b + (\cf - 1)\tr t^a t^b t^c \right )\\
& = 0.
\end{split}
\end{equation}

We then consider the structure 
\begin{equation}
\label{eq:str7}
\begin{split}
\vec I \cdot & \frac{z_2 \vec p_2  - (1-z_2) \vec k_g }{\left (z_2 \vec p_2 - (1-z_2)\vec k_g\right )^2}. 
\end{split}
\end{equation}
For the terms shown in eqs.~(\ref{eq:rho-rho-rho_FIN-2_gq1q1}, \ref{eq:rho-rho-rho_FIN-q1q1q1}, \ref{eq:rho-rho-rho_FIN-q2q1q1}, \ref{eq:rho-rho-rho_FIN-q3q1q1}) the structure above is multiplied with the three-quark wave function
\begin{equation}
\Psi^{\ast}_{qqq}(x_1 - x_g, \vec k_1 + x_1\vec q_{12} - \vec k_g + x_g \vec K - \vec q_{12}; x_2 + x_g, \vec k_2 + x_2 \vec q_{12} + \vec k_g - x_g\vec K; x_3, \vec k_3 + x_3\vec q_{12}),
\end{equation}
and the sum of the color factors:
\begin{equation}
\begin{split}
& +6 \cdot \frac{\nc}{2} \tr t^a t^b t^c\\
& -6 \cdot \cf  \tr t^a t^b t^c\\
& -6 \cdot \left (\cf - \frac{1}{2} \right ) \tr t^a t^b t^c\\
& +6 \cdot \left (\cf - \frac{1}{2} - \frac{1}{2\nc} \right )\tr t^a t^b t^c\\
& = 0.
\end{split}
\end{equation}
For the terms (\ref{eq:rho-rho-rho_FIN-2_gq3q1}, \ref{eq:rho-rho-rho_FIN-q1q3q1}, \ref{eq:rho-rho-rho_FIN-q2q3q1}, \ref{eq:rho-rho-rho_FIN-q3q3q1}) or (\ref{eq:rho-rho-rho_FIN-2_gq1q3}, \ref{eq:rho-rho-rho_FIN-q1q1q3}, \ref{eq:rho-rho-rho_FIN-q2q1q3}, \ref{eq:rho-rho-rho_FIN-q3q1q3}) the structure in \eqref{eq:str7} is multiplied with the three-quark wave function
\begin{equation}
\Psi^{\ast}_{qqq}(x_1 - x_g, \vec k_1 + x_1\vec q_{12} - \vec k_g + x_g \vec K - \vec q_{1}; x_2 + x_g, \vec k_2 + x_2 \vec q_{12} + \vec k_g - x_g\vec K; x_3, \vec k_3 + x_3\vec q_{12} - \vec q_2),
\end{equation}
or 
\begin{equation}
\Psi^{\ast}_{qqq}(x_1 - x_g, \vec k_1 + x_1\vec q_{12} - \vec k_g + x_g \vec K - \vec q_{2}; x_2 + x_g, \vec k_2 + x_2 \vec q_{12} + \vec k_g - x_g\vec K; x_3, \vec k_3 + x_3\vec q_{12} - \vec q_1),
\end{equation}
respectively. For the first set of terms, the sum of the color factors yields
\begin{equation}
\begin{split}
& +6 \cdot \left (-\frac{1}{4}(T^c)_{ba} - \nc \tr t^a t^c t^b + \frac{1}{4} \tr T^b T^c (D^a - T^a) \right )\\
& +6 \cdot  \left (2\cf - \frac{1}{2} - \frac{\nc}{2} \right )\tr t^a t^c t^b\\
& +6 \cdot \left ((\cf - \frac{\nc}{2})\tr t^a t^b t^c + (\cf - 1)\tr t^a t^c t^b \right )\\
& +6 \cdot \left (2\cf - \frac{1}{2} - \frac{\nc}{2} \right )\tr t^a t^b t^c\\
& = 0.
\end{split}
\end{equation}
Similarly, for the second set of terms
\begin{equation}
\begin{split}
& +6 \cdot \left (-\frac{1}{4}(T^c)_{ab} - \nc \tr t^a t^b t^c + \frac{1}{4} \tr T^a T^c (D^b - T^b) \right )\\
& +6 \cdot  \left (2\cf - \frac{1}{2} - \frac{\nc}{2} \right )\tr t^a t^b t^c\\
& +6 \cdot \left ((\cf - \frac{\nc}{2})\tr t^a t^c t^b + (\cf - 1)\tr t^a t^b t^c \right )\\
& +6 \cdot \left (2\cf - \frac{1}{2} - \frac{\nc}{2} \right )\tr t^a t^c t^b\\
& = 0.
\end{split}
\end{equation}
Finally, we consider the set of terms (\ref{eq:rho-rho-rho_FIN-q1q3q3}, \ref{eq:rho-rho-rho_FIN-q2q3q3}, \ref{eq:rho-rho-rho_FIN-q3q3q3}) that contain the structure in \eqref{eq:str7} multiplied with the three-quark wave function 
\begin{equation}
\Psi^{\ast}_{qqq}(x_1 - x_g, \vec k_1 + x_1\vec q_{12} - \vec k_g + x_g \vec K - \vec q_{12}; x_2 + x_g, \vec k_2 + x_2 \vec q_{12} + \vec k_g - x_g\vec K; x_3, \vec k_3 -(1-x_3)\vec q_{12}).
\end{equation}
For these terms, the sum of the color factors:
\begin{equation}
\begin{split}
& +6 \cdot \left (2\cf - \frac{1}{2} - \frac{\nc}{2} \right ) \tr t^a t^b t^c\\
& +6 \cdot \left (2\cf - \frac{1}{2} - \frac{\nc}{2} \right ) \tr t^a t^b t^c\\
& +6 \cdot \left (2\cf - \nc\cf \right ) \tr t^a t^b t^c\\
& = 0.
\end{split}
\end{equation}

We are then left with two more contributions that contain the operator $\vec I$. These two contributions involve the structure 
\begin{equation}
\label{eq:str8}
\begin{split}
\vec I \cdot & \frac{z_2 (\vec p_2 - \vec q_1) - (1-z_2)\vec k_g}{\left (z_2 (\vec p_2 - \vec q_1) - (1-z_2)\vec k_g\right )^2} \\
& \times  \Psi^{\ast}_{qqq}(x_1 - x_g, \vec k_1 + x_1\vec q_{12} - \vec k_g + x_g \vec K; x_2 + x_g, \vec k_2 + x_2 \vec q_{12} - \vec q_1 + \vec k_g - x_g\vec K; x_3, \vec k_3 + x_3\vec q_{12} - \vec q_2)
\end{split}
\end{equation}
and
\begin{equation}
\label{eq:str9}
\begin{split}
\vec I \cdot & \frac{z_2 (\vec p_2 - \vec q_2) - (1-z_2)\vec k_g}{\left (z_2 (\vec p_2  - \vec q_2) - (1-z_2)\vec k_g\right )^2} \\
& \times  \Psi^{\ast}_{qqq}(x_1 - x_g, \vec k_1 + x_1\vec q_{12} - \vec k_g + x_g \vec K; x_2 + x_g, \vec k_2 + x_2 \vec q_{12} - \vec q_2 + \vec k_g - x_g\vec K; x_3, \vec k_3 + x_3\vec q_{12}-\vec q_1).
\end{split}
\end{equation}
Terms that involve the structure \eqref{eq:str8} are given by eqs.~(\ref{eq:rho-rho-rho_FIN-2_gq3q2}, \ref{eq:rho-rho-rho_FIN-q1q3q2}, \ref{eq:rho-rho-rho_FIN-q2q3q2}, \ref{eq:rho-rho-rho_FIN-q3q3q2}), and the sum of the color factors: 
\begin{equation}
\begin{split}
& +6 \cdot \left (-\frac{1}{4}(T^c)_{ab} - \frac{\nc}{4}(D^c)_{ab} + \frac{1}{4}\tr T^c (D^a - T^a)T^b\right )\\
& +6 \cdot \left ((\cf - \frac{\nc}{2})\tr t^a t^c t^b + (\cf - 1)\tr t^a t^b t^c \right )\\
& +6 \cdot \left (2\cf - \frac{1}{2} - \frac{\nc}{2} \right ) \tr t^a t^c t^b\\
& +6 \cdot \left (2\cf - \frac{1}{2} - \frac{\nc}{2} \right ) \tr t^a t^b t^c\\
& = 0.
\end{split}
\end{equation}
Terms that involve the structure \eqref{eq:str9} are given by eqs.~(\ref{eq:rho-rho-rho_FIN-2_gq2q3}, \ref{eq:rho-rho-rho_FIN-q1q2q3}, \ref{eq:rho-rho-rho_FIN-q2q2q3}, \ref{eq:rho-rho-rho_FIN-q3q2q3}), and the sum of the color factors: 
\begin{equation}
\begin{split}
& +6 \cdot \left (+\frac{1}{4}(T^c)_{ab} - \frac{\nc}{4}(D^c)_{ab} + \frac{1}{4}\tr T^c (D^b - T^b)T^a\right )\\
& +6 \cdot \left ((\cf - \frac{\nc}{2})\tr t^a t^b t^c + (\cf - 1)\tr t^a t^c t^b \right )\\
& +6 \cdot \left (2\cf - \frac{1}{2} - \frac{\nc}{2} \right ) \tr t^a t^b t^c\\
& +6 \cdot \left (2\cf - \frac{1}{2} - \frac{\nc}{2} \right ) \tr t^a t^c t^b\\
& = 0.
\end{split}
\end{equation}

We then move on and consider contributions which involve the integral operator $J$. Let us start by considering terms which have the structure 
\begin{equation}
\begin{split}	
J & \frac{z_2 \vec p_2 - (1-z_2)\vec k_g}{(z_2 \vec p_2 - (1-z_2)\vec k_g)^2} \cdot \left (\frac{z_1 \vec p_1 - \vec k_g}{(z_1 \vec p_1 - \vec k_g)^2} + \frac{z_1(\vec p_1 - \vec q_{12}) - \vec k_g}{(z_1(\vec p_1 - \vec q_{12}) - \vec k_g)^2} \right ) \\
& \times \Psi^\ast_{qqq}\left (x_1 - x_g, \vec k_1 - (1-x_1)\vec q_{12} - \vec k_g + x_g\vec K; x_2 + x_g, \vec k_2 + x_2 \vec q_{12} + \vec k_g - x_g\vec K; x_3, \vec k_3 + x_3\vec q_{12} \right ).
\end{split}
\end{equation}
These terms are given by eqs.~(\ref{eq:rho-rho-rho_FINv-q1q1q1}, \ref{eq:rho-rho-rho_FINv-q2q1q1}, \ref{eq:rho-rho-rho_FINv-q3q1q1}), and the sum of the color factors:
\begin{equation}
\begin{split}
& +3 \cdot \cf \tr t^a t^b t^c\\
& +3 \cdot \left (\cf - \frac{\nc + 1}{2} \right ) \tr t^a t^b t^c\\
& -3 \cdot \left (2\cf - \frac{\nc+1}{2} \right ) \tr t^a t^b t^c\\
& = 0.
\end{split}
\end{equation}

We continue by considering the contributions which involve the structure 
\begin{equation}
\begin{split}	
J & \biggl (\frac{z_2 \vec p_2 - (1-z_2)\vec k_g}{(z_2 \vec p_2 - (1-z_2)\vec k_g)^2} \cdot \frac{z_1 \vec p_1 - \vec k_g}{(z_1 \vec p_1 - \vec k_g)^2} + \frac{z_2 (\vec p_2 - \vec q_2) - (1-z_2)\vec k_g}{(z_2 (\vec p_2 - \vec q_2) - (1-z_2)\vec k_g)^2} \cdot \frac{z_1(\vec p_1 - \vec q_{1}) - \vec k_g}{(z_1(\vec p_1 - \vec q_{1}) - \vec k_g)^2}\biggr  )\\
& \times \Psi^\ast_{qqq}\left (x_1 - x_g, \vec k_1 + x_1\vec q_{12} - \vec q_1  - \vec k_g + x_g\vec K; x_2 + x_g, \vec k_2 + x_2 \vec q_{12} - \vec q_2 + \vec k_g - x_g\vec K; x_3, \vec k_3 + x_3\vec q_{12} \right ).
\end{split}
\end{equation}
Terms that have this structure are given by eqs.~(\ref{eq:rho-rho-rho_FINv-q1q2q1}, \ref{eq:rho-rho-rho_FINv-q2q2q1}, \ref{eq:rho-rho-rho_FINv-q3q2q1}), and the sum of the color factors:
\begin{equation}
\begin{split}
& +3 \cdot \left (\cf - \frac{\nc + 1}{2} \right )\tr t^a t^c t^b\\
& +3 \cdot \left (\cf - \frac{\nc + 1}{2} \right ) \tr t^a t^b t^c\\
& -3 \cdot \left (\cf - \frac{\nc+1}{2} \right ) \tr \left (t^a t^b t^c + t^a t^c t^b \right )\\
& = 0.
\end{split}
\end{equation}

The contributions that contain the structure
\begin{equation}
\begin{split}	
J & \frac{z_2 \vec p_2 - (1-z_2)\vec k_g}{(z_2 \vec p_2 - (1-z_2)\vec k_g)^2} \cdot \left (\frac{z_1 \vec p_1 - \vec k_g}{(z_1 \vec p_1 - \vec k_g)^2} + \frac{z_1(\vec p_1 - \vec q_{1}) - \vec k_g}{(z_1(\vec p_1 - \vec q_{1}) - \vec k_g)^2} \right ) \\
& \times \Psi^\ast_{qqq}\left (x_1 - x_g, \vec k_1 - (1-x_1)\vec q_{12} - \vec q_1 - \vec k_g + x_g\vec K; x_2 + x_g, \vec k_2 + x_2 \vec q_{12} + \vec k_g - x_g\vec K; x_3, \vec k_3 + x_3\vec q_{12} - \vec q_2 \right )
\end{split}
\end{equation}
are given by eqs.~(\ref{eq:rho-rho-rho_FINv-q1q3q1}, \ref{eq:rho-rho-rho_FINv-q2q3q1}, \ref{eq:rho-rho-rho_FINv-q3q3q1}). The sum of the color factors:
\begin{equation}
\begin{split}
& -3 \cdot \left (2\cf - \frac{\nc + 1}{2} \right ) \tr t^a t^c t^b\\
& -3 \cdot \left (\cf - \frac{\nc + 1}{2} \right ) \tr \left (t^a t^b t^c + t^a t^c t^b \right )\\
& -3 \cdot \left (2\cf - \frac{\nc+1}{2} \right ) \tr t^a t^b t^c\\
& = 0.
\end{split}
\end{equation}

The contributions that contain the structure
\begin{equation}
\begin{split}	
J & \biggl (\frac{z_2 \vec p_2 - (1-z_2)\vec k_g}{(z_2 \vec p_2 - (1-z_2)\vec k_g)^2} \cdot \frac{z_1 \vec p_1 - \vec k_g}{(z_1 \vec p_1 - \vec k_g)^2} + \frac{z_2 (\vec p_2 - \vec q_1) - (1-z_2)\vec k_g}{(z_2 (\vec p_2 - \vec q_1) - (1-z_2)\vec k_g)^2} \cdot \frac{z_1(\vec p_1 - \vec q_{2}) - \vec k_g}{(z_1(\vec p_1 - \vec q_{2}) - \vec k_g)^2}\biggr  ) \\
& \times \Psi^\ast_{qqq}\left (x_1 - x_g, \vec k_1 + x_1\vec q_{12} - \vec q_2  - \vec k_g + x_g\vec K; x_2 + x_g, \vec k_2 + x_2 \vec q_{12} - \vec q_1 + \vec k_g - x_g\vec K; x_3, \vec k_3 + x_3\vec q_{12} \right )
\end{split}
\end{equation}
are given by eqs.~(\ref{eq:rho-rho-rho_FINv-q1q1q2}, \ref{eq:rho-rho-rho_FINv-q2q1q2}, \ref{eq:rho-rho-rho_FINv-q3q1q2}). The sum of the color factors:
\begin{equation}
\begin{split}
& +3 \cdot \left (\cf - \frac{\nc + 1}{2} \right ) \tr t^a t^b t^c\\
& +3 \cdot \left (\cf - \frac{\nc+1}{2} \right ) \tr t^a t^c t^b\\
& -3 \cdot \left (\cf - \frac{\nc + 1}{2} \right ) \tr \left (t^a t^b t^c + t^a t^c t^b \right )\\
& = 0.
\end{split}
\end{equation}

Furthermore, the contributions that contain the structure
\begin{equation}
\begin{split}	
J & \cdot \frac{z_1 \vec p_1 - \vec k_g}{(z_1 \vec p_1 - \vec k_g)^2}\biggl (\frac{z_2 \vec p_2 - (1-z_2)\vec k_g}{(z_2 \vec p_2 - (1-z_2)\vec k_g)^2}  + \frac{z_2 (\vec p_2 - \vec q_{12}) - (1-z_2)\vec k_g}{(z_2 (\vec p_2 - \vec q_{12}) - (1-z_2)\vec k_g)^2} \biggr  ) \\
& \times \Psi^\ast_{qqq}\left (x_1 - x_g, \vec k_1 + x_1\vec q_{12}  - \vec k_g + x_g\vec K; x_2 + x_g, \vec k_2 -(1-x_2) \vec q_{12} + \vec k_g - x_g\vec K; x_3, \vec k_3 + x_3\vec q_{12} \right )
\end{split}
\end{equation}
and
\begin{equation}
\begin{split}	
J & \cdot \frac{z_1 \vec p_1 - \vec k_g}{(z_1 \vec p_1 - \vec k_g)^2}\biggl (\frac{z_2 \vec p_2 - (1-z_2)\vec k_g}{(z_2 \vec p_2 - (1-z_2)\vec k_g)^2}  + \frac{z_2 (\vec p_2 - \vec q_1) - (1-z_2)\vec k_g}{(z_2 (\vec p_2 - \vec q_1) - (1-z_2)\vec k_g)^2} \biggr  ) \\
& \times \Psi^\ast_{qqq}\left (x_1 - x_g, \vec k_1 + x_1\vec q_{12}  - \vec k_g + x_g\vec K; x_2 + x_g, \vec k_2 + x_2 \vec q_{12} - \vec q_1 + \vec k_g - x_g\vec K; x_3, \vec k_3 + x_3\vec q_{12} - \vec q_2\right )
\end{split}
\end{equation}
are given by eqs.~(\ref{eq:rho-rho-rho_FINv-q1q2q2}, \ref{eq:rho-rho-rho_FINv-q2q2q2}, \ref{eq:rho-rho-rho_FINv-q3q2q2}) and eqs.~(\ref{eq:rho-rho-rho_FINv-q1q3q2}, \ref{eq:rho-rho-rho_FINv-q2q3q2}, \ref{eq:rho-rho-rho_FINv-q3q3q2}), respectively. Correspondingly, the sum of the color factors
\begin{equation}
\begin{split}
& +3 \cdot \left (\cf - \frac{\nc + 1}{2} \right ) \tr t^a t^b t^c\\
& +3 \cdot \cf \tr t^a t^b t^c\\
& -3 \cdot \left (2\cf - \frac{\nc + 1}{2} \right ) \tr t^a t^b t^c \\
& = 0
\end{split}
\end{equation}
and
\begin{equation}
\begin{split}
& -3 \cdot \left (\cf - \frac{\nc + 1}{2} \right ) \tr \left (t^a t^c t^b + \tr t^a t^b t^c\right )\\
& -3 \cdot \left (2\cf - \frac{\nc+1}{2} \right ) \tr t^a t^c t^b\\
& -3 \cdot \left (2\cf - \frac{\nc + 1}{2} \right ) \tr t^a t^b t^c\\
& = 0.
\end{split}
\end{equation}

The final set of contributions are given by eqs.~(\ref{eq:rho-rho-rho_FINv-q1q1q3}, \ref{eq:rho-rho-rho_FINv-q2q1q3}, \ref{eq:rho-rho-rho_FINv-q3q1q3}), eqs.~(\ref{eq:rho-rho-rho_FINv-q1q2q3}, \ref{eq:rho-rho-rho_FINv-q2q2q3}, \ref{eq:rho-rho-rho_FINv-q3q2q3}) and eqs.~(\ref{eq:rho-rho-rho_FINv-q1q3q3}, \ref{eq:rho-rho-rho_FINv-q2q3q3}, \ref{eq:rho-rho-rho_FINv-q3q3q3}). These terms contain three independent structures
\begin{equation}
\begin{split}	
J & \frac{z_2 \vec p_2 - (1-z_2)\vec k_g}{(z_2 \vec p_2 - (1-z_2)\vec k_g)^2} \cdot  \biggl (\frac{z_1 \vec p_1 - \vec k_g}{(z_1 \vec p_1 - \vec k_g)^2}  + \frac{z_1 (\vec p_1 - \vec q_{2}) - (1-z_2)\vec k_g}{(z_1 (\vec p_1 - \vec q_{2}) - (1-z_2)\vec k_g)^2} \biggr  ) \\
& \times \Psi^\ast_{qqq}\left (x_1 - x_g, \vec k_1 + x_1\vec q_{12} - \vec q_2  - \vec k_g + x_g\vec K; x_2 + x_g, \vec k_2 + x_2 \vec q_{12} + \vec k_g - x_g\vec K; x_3, \vec k_3 + x_3\vec q_{12} - \vec q_1 \right )
\end{split}
\end{equation}
\begin{equation}
\begin{split}	
J &  \frac{z_1 \vec p_1 - \vec k_g}{(z_1 \vec p_1 - \vec k_g)^2}\cdot \biggl (\frac{z_2 \vec p_2 - (1-z_2)\vec k_g}{(z_2 \vec p_2 - (1-z_2)\vec k_g)^2}  + \frac{z_2 (\vec p_2 - \vec q_2) - (1-z_2)\vec k_g}{(z_2 (\vec p_2 - \vec q_2) - (1-z_2)\vec k_g)^2} \biggr  ) \\
& \times \Psi^\ast_{qqq}\left (x_1 - x_g, \vec k_1 + x_1\vec q_{12}  - \vec k_g + x_g\vec K; x_2 + x_g, \vec k_2 + x_2 \vec q_{12} - \vec q_2 + \vec k_g - x_g\vec K; x_3, \vec k_3 + x_3\vec q_{12} - \vec q_1\right )
\end{split}
\end{equation}
and
\begin{equation}
\begin{split}	
J &  \frac{z_1 \vec p_1 - \vec k_g}{(z_1 \vec p_1 - \vec k_g)^2}\cdot \frac{z_2 \vec p_2 - (1-z_2)\vec k_g}{(z_2 \vec p_2 - (1-z_2)\vec k_g)^2} \\
& \times \Psi^\ast_{qqq}\left (x_1 - x_g, \vec k_1 + x_1\vec q_{12}  - \vec k_g + x_g\vec K; x_2 + x_g, \vec k_2 + x_2 \vec q_{12} + \vec k_g - x_g\vec K; x_3, \vec k_3 - (1- x_3)\vec q_{12} \right ),
\end{split}
\end{equation}
respectively. Correspondingly, the sum of the color factors 
\begin{equation}
\begin{split}
& -3 \cdot \left (2\cf - \frac{\nc + 1}{2} \right ) \tr t^a t^b t^c\\
& -3 \cdot \left (\cf - \frac{\nc+1}{2} \right ) \tr \left (t^a t^b t^c + t^a t^c t^b \right )\\
& -3 \cdot \left (2\cf - \frac{\nc+1}{2}\right ) \tr t^a t^c t^b\\
& = 0.
\end{split}
\end{equation}
\begin{equation}
\begin{split}
& -3 \cdot \left (\cf - \frac{\nc+1}{2} \right ) \tr \left (t^a t^b t^c + t^a t^c t^b \right )\\
& -3 \cdot \left (2\cf - \frac{\nc + 1}{2} \right ) \tr t^a t^b t^c\\
& -3 \cdot \left (2\cf - \frac{\nc+1}{2}\right ) \tr t^a t^c t^b\\
& = 0.
\end{split}
\end{equation}
and
\begin{equation}
\begin{split}
& -3 \cdot 2\left (2\cf - \frac{\nc + 1}{2} \right ) \tr t^a t^b t^c\\
& -3 \cdot 2\left (2\cf - \frac{\nc+1}{2} \right ) \tr t^a t^b t^c\\
& -3 \cdot 2\cf\left (2 - \nc\right ) \tr t^a t^b t^c\\
& = 0.
\end{split}
\end{equation}

This concludes our check that the sum of all UV finite diagrams cancel when either $\vec q_1, \vec q_3 \rightarrow 0$.

\bibliography{spires}

\begin{thebibliography}{73}
\expandafter\ifx\csname natexlab\endcsname\relax\def\natexlab#1{#1}\fi
\expandafter\ifx\csname bibnamefont\endcsname\relax
  \def\bibnamefont#1{#1}\fi
\expandafter\ifx\csname bibfnamefont\endcsname\relax
  \def\bibfnamefont#1{#1}\fi
\expandafter\ifx\csname citenamefont\endcsname\relax
  \def\citenamefont#1{#1}\fi
\expandafter\ifx\csname url\endcsname\relax
  \def\url#1{\texttt{#1}}\fi
\expandafter\ifx\csname urlprefix\endcsname\relax\def\urlprefix{URL }\fi
\providecommand{\bibinfo}[2]{#2}
\providecommand{\eprint}[2][]{\url{#2}}

\bibitem[{\citenamefont{Dumitru and Paatelainen}(2021)}]{Dumitru:2020gla}
\bibinfo{author}{\bibfnamefont{A.}~\bibnamefont{Dumitru}} \bibnamefont{and}
  \bibinfo{author}{\bibfnamefont{R.}~\bibnamefont{Paatelainen}},
  \bibinfo{journal}{Phys. Rev. D} \textbf{\bibinfo{volume}{103}},
  \bibinfo{pages}{034026} (\bibinfo{year}{2021}), \eprint{2010.11245}.

\bibitem[{\citenamefont{Dumitru
  et~al.}(2021{\natexlab{a}})\citenamefont{Dumitru, M\"antysaari, and
  Paatelainen}}]{Dumitru:2021tvw}
\bibinfo{author}{\bibfnamefont{A.}~\bibnamefont{Dumitru}},
  \bibinfo{author}{\bibfnamefont{H.}~\bibnamefont{M\"antysaari}},
  \bibnamefont{and}
  \bibinfo{author}{\bibfnamefont{R.}~\bibnamefont{Paatelainen}},
  \bibinfo{journal}{Phys. Lett. B} \textbf{\bibinfo{volume}{820}},
  \bibinfo{pages}{136560} (\bibinfo{year}{2021}{\natexlab{a}}),
  \eprint{2103.11682}.

\bibitem[{\citenamefont{Dumitru
  et~al.}(2021{\natexlab{b}})\citenamefont{Dumitru, M\"antysaari, Paatelainen,
  Roy, Salazar, and Schenke}}]{Dumitru:2021mab}
\bibinfo{author}{\bibfnamefont{A.}~\bibnamefont{Dumitru}},
  \bibinfo{author}{\bibfnamefont{H.}~\bibnamefont{M\"antysaari}},
  \bibinfo{author}{\bibfnamefont{R.}~\bibnamefont{Paatelainen}},
  \bibinfo{author}{\bibfnamefont{K.}~\bibnamefont{Roy}},
  \bibinfo{author}{\bibfnamefont{F.}~\bibnamefont{Salazar}}, \bibnamefont{and}
  \bibinfo{author}{\bibfnamefont{B.}~\bibnamefont{Schenke}}, in
  \emph{\bibinfo{booktitle}{{28th International Workshop on Deep Inelastic
  Scattering and Related Subjects}}} (\bibinfo{year}{2021}{\natexlab{b}}),
  \eprint{2105.10144}.

\bibitem[{\citenamefont{Mueller}(2001)}]{Mueller:2001fv}
\bibinfo{author}{\bibfnamefont{A.~H.} \bibnamefont{Mueller}}, in
  \emph{\bibinfo{booktitle}{Cargese 2001, {QCD} perspectives on hot and dense
  matter}} (\bibinfo{year}{2001}), pp. \bibinfo{pages}{45--72},
  \eprint{hep-ph/0111244}.

\bibitem[{\citenamefont{Kovchegov and Levin}(2012)}]{Kovchegov:2012mbw}
\bibinfo{author}{\bibfnamefont{Y.~V.} \bibnamefont{Kovchegov}}
  \bibnamefont{and} \bibinfo{author}{\bibfnamefont{E.}~\bibnamefont{Levin}},
  \emph{\bibinfo{title}{{Quantum chromodynamics at high energy}}},
  vol.~\bibinfo{volume}{33} (\bibinfo{publisher}{Cambridge University Press},
  \bibinfo{year}{2012}), ISBN \bibinfo{isbn}{978-0-521-11257-4,
  978-1-139-55768-9}.

\bibitem[{\citenamefont{Bartels}(1980)}]{Bartels:1980pe}
\bibinfo{author}{\bibfnamefont{J.}~\bibnamefont{Bartels}},
  \bibinfo{journal}{Nucl. Phys. B} \textbf{\bibinfo{volume}{175}},
  \bibinfo{pages}{365} (\bibinfo{year}{1980}).

\bibitem[{\citenamefont{Jaroszewicz}(1980)}]{Jaroszewicz:1980mq}
\bibinfo{author}{\bibfnamefont{T.}~\bibnamefont{Jaroszewicz}},
  \bibinfo{journal}{Acta Phys. Polon. B} \textbf{\bibinfo{volume}{11}},
  \bibinfo{pages}{965} (\bibinfo{year}{1980}).

\bibitem[{\citenamefont{Kwiecinski and Praszalowicz}(1980)}]{Kwiecinski:1980wb}
\bibinfo{author}{\bibfnamefont{J.}~\bibnamefont{Kwiecinski}} \bibnamefont{and}
  \bibinfo{author}{\bibfnamefont{M.}~\bibnamefont{Praszalowicz}},
  \bibinfo{journal}{Phys. Lett. B} \textbf{\bibinfo{volume}{94}},
  \bibinfo{pages}{413} (\bibinfo{year}{1980}).

\bibitem[{\citenamefont{Braun}(1998)}]{Braun:1998fs}
\bibinfo{author}{\bibfnamefont{M.~A.} \bibnamefont{Braun}}
  (\bibinfo{year}{1998}), \eprint{hep-ph/9805394}.

\bibitem[{\citenamefont{Ewerz}(2003)}]{Ewerz:2003xi}
\bibinfo{author}{\bibfnamefont{C.}~\bibnamefont{Ewerz}} (\bibinfo{year}{2003}),
  \eprint{hep-ph/0306137}.

\bibitem[{\citenamefont{Kovner and Lublinsky}(2007)}]{Kovner:2005qj}
\bibinfo{author}{\bibfnamefont{A.}~\bibnamefont{Kovner}} \bibnamefont{and}
  \bibinfo{author}{\bibfnamefont{M.}~\bibnamefont{Lublinsky}},
  \bibinfo{journal}{JHEP} \textbf{\bibinfo{volume}{02}}, \bibinfo{pages}{058}
  (\bibinfo{year}{2007}), \eprint{hep-ph/0512316}.

\bibitem[{\citenamefont{Abazov et~al.}(2020)}]{Abazov:2020rus}
\bibinfo{author}{\bibfnamefont{V.~M.} \bibnamefont{Abazov}}
  \bibnamefont{et~al.} (\bibinfo{collaboration}{D0, TOTEM})
  (\bibinfo{year}{2020}), \eprint{2012.03981}.

\bibitem[{\citenamefont{Antchev et~al.}(2019)}]{Antchev:2017yns}
\bibinfo{author}{\bibfnamefont{G.}~\bibnamefont{Antchev}} \bibnamefont{et~al.}
  (\bibinfo{collaboration}{TOTEM}), \bibinfo{journal}{Eur. Phys. J. C}
  \textbf{\bibinfo{volume}{79}}, \bibinfo{pages}{785} (\bibinfo{year}{2019}),
  \eprint{1812.04732}.

\bibitem[{\citenamefont{Martynov and Nicolescu}(2019)}]{Martynov:2018sga}
\bibinfo{author}{\bibfnamefont{E.}~\bibnamefont{Martynov}} \bibnamefont{and}
  \bibinfo{author}{\bibfnamefont{B.}~\bibnamefont{Nicolescu}},
  \bibinfo{journal}{Eur. Phys. J. C} \textbf{\bibinfo{volume}{79}},
  \bibinfo{pages}{461} (\bibinfo{year}{2019}), \eprint{1808.08580}.

\bibitem[{\citenamefont{Balitsky}(1996)}]{Balitsky:1995ub}
\bibinfo{author}{\bibfnamefont{I.}~\bibnamefont{Balitsky}},
  \bibinfo{journal}{Nucl. Phys.} \textbf{\bibinfo{volume}{B463}},
  \bibinfo{pages}{99} (\bibinfo{year}{1996}), \eprint{hep-ph/9509348}.

\bibitem[{\citenamefont{Balitsky}(1999)}]{Balitsky:1998ya}
\bibinfo{author}{\bibfnamefont{I.}~\bibnamefont{Balitsky}},
  \bibinfo{journal}{Phys. Rev.} \textbf{\bibinfo{volume}{D60}},
  \bibinfo{pages}{014020} (\bibinfo{year}{1999}), \eprint{hep-ph/9812311}.

\bibitem[{\citenamefont{Balitsky}(2001)}]{Balitsky:2001re}
\bibinfo{author}{\bibfnamefont{I.}~\bibnamefont{Balitsky}},
  \bibinfo{journal}{Phys. Lett.} \textbf{\bibinfo{volume}{B518}},
  \bibinfo{pages}{235} (\bibinfo{year}{2001}), \eprint{hep-ph/0105334}.

\bibitem[{\citenamefont{Jalilian-Marian
  et~al.}(1997)\citenamefont{Jalilian-Marian, Kovner, Leonidov, and
  Weigert}}]{Jalilian-Marian:1997jx}
\bibinfo{author}{\bibfnamefont{J.}~\bibnamefont{Jalilian-Marian}},
  \bibinfo{author}{\bibfnamefont{A.}~\bibnamefont{Kovner}},
  \bibinfo{author}{\bibfnamefont{A.}~\bibnamefont{Leonidov}}, \bibnamefont{and}
  \bibinfo{author}{\bibfnamefont{H.}~\bibnamefont{Weigert}},
  \bibinfo{journal}{Nucl. Phys.} \textbf{\bibinfo{volume}{B504}},
  \bibinfo{pages}{415} (\bibinfo{year}{1997}), \eprint{hep-ph/9701284}.

\bibitem[{\citenamefont{Jalilian-Marian
  et~al.}(1999{\natexlab{a}})\citenamefont{Jalilian-Marian, Kovner, Leonidov,
  and Weigert}}]{Jalilian-Marian:1997gr}
\bibinfo{author}{\bibfnamefont{J.}~\bibnamefont{Jalilian-Marian}},
  \bibinfo{author}{\bibfnamefont{A.}~\bibnamefont{Kovner}},
  \bibinfo{author}{\bibfnamefont{A.}~\bibnamefont{Leonidov}}, \bibnamefont{and}
  \bibinfo{author}{\bibfnamefont{H.}~\bibnamefont{Weigert}},
  \bibinfo{journal}{Phys. Rev.} \textbf{\bibinfo{volume}{D59}},
  \bibinfo{pages}{014014} (\bibinfo{year}{1999}{\natexlab{a}}),
  \eprint{hep-ph/9706377}.

\bibitem[{\citenamefont{Jalilian-Marian
  et~al.}(1999{\natexlab{b}})\citenamefont{Jalilian-Marian, Kovner, and
  Weigert}}]{JalilianMarian:1997dw}
\bibinfo{author}{\bibfnamefont{J.}~\bibnamefont{Jalilian-Marian}},
  \bibinfo{author}{\bibfnamefont{A.}~\bibnamefont{Kovner}}, \bibnamefont{and}
  \bibinfo{author}{\bibfnamefont{H.}~\bibnamefont{Weigert}},
  \bibinfo{journal}{Phys. Rev.} \textbf{\bibinfo{volume}{D59}},
  \bibinfo{pages}{014015} (\bibinfo{year}{1999}{\natexlab{b}}),
  \eprint{hep-ph/9709432}.

\bibitem[{\citenamefont{Iancu et~al.}(2001{\natexlab{a}})\citenamefont{Iancu,
  Leonidov, and McLerran}}]{Iancu:2001ad}
\bibinfo{author}{\bibfnamefont{E.}~\bibnamefont{Iancu}},
  \bibinfo{author}{\bibfnamefont{A.}~\bibnamefont{Leonidov}}, \bibnamefont{and}
  \bibinfo{author}{\bibfnamefont{L.~D.} \bibnamefont{McLerran}},
  \bibinfo{journal}{Phys. Lett.} \textbf{\bibinfo{volume}{B510}},
  \bibinfo{pages}{133} (\bibinfo{year}{2001}{\natexlab{a}}),
  \eprint{hep-ph/0102009}.

\bibitem[{\citenamefont{Iancu et~al.}(2001{\natexlab{b}})\citenamefont{Iancu,
  Leonidov, and McLerran}}]{Iancu:2000hn}
\bibinfo{author}{\bibfnamefont{E.}~\bibnamefont{Iancu}},
  \bibinfo{author}{\bibfnamefont{A.}~\bibnamefont{Leonidov}}, \bibnamefont{and}
  \bibinfo{author}{\bibfnamefont{L.~D.} \bibnamefont{McLerran}},
  \bibinfo{journal}{Nucl. Phys.} \textbf{\bibinfo{volume}{A692}},
  \bibinfo{pages}{583} (\bibinfo{year}{2001}{\natexlab{b}}),
  \eprint{hep-ph/0011241}.

\bibitem[{\citenamefont{Ferreiro et~al.}(2002)\citenamefont{Ferreiro, Iancu,
  Leonidov, and McLerran}}]{Ferreiro:2001qy}
\bibinfo{author}{\bibfnamefont{E.}~\bibnamefont{Ferreiro}},
  \bibinfo{author}{\bibfnamefont{E.}~\bibnamefont{Iancu}},
  \bibinfo{author}{\bibfnamefont{A.}~\bibnamefont{Leonidov}}, \bibnamefont{and}
  \bibinfo{author}{\bibfnamefont{L.}~\bibnamefont{McLerran}},
  \bibinfo{journal}{Nucl. Phys.} \textbf{\bibinfo{volume}{A703}},
  \bibinfo{pages}{489} (\bibinfo{year}{2002}), \eprint{hep-ph/0109115}.

\bibitem[{\citenamefont{Weigert}(2002)}]{Weigert:2000gi}
\bibinfo{author}{\bibfnamefont{H.}~\bibnamefont{Weigert}},
  \bibinfo{journal}{Nucl. Phys.} \textbf{\bibinfo{volume}{A703}},
  \bibinfo{pages}{823} (\bibinfo{year}{2002}), \eprint{hep-ph/0004044}.

\bibitem[{\citenamefont{Kovchegov}(1999)}]{Kovchegov:1999yj}
\bibinfo{author}{\bibfnamefont{Y.~V.} \bibnamefont{Kovchegov}},
  \bibinfo{journal}{Phys. Rev.} \textbf{\bibinfo{volume}{D60}},
  \bibinfo{pages}{034008} (\bibinfo{year}{1999}), \eprint{hep-ph/9901281}.

\bibitem[{\citenamefont{Kovchegov et~al.}(2004)\citenamefont{Kovchegov,
  Szymanowski, and Wallon}}]{Kovchegov:2003dm}
\bibinfo{author}{\bibfnamefont{Y.~V.} \bibnamefont{Kovchegov}},
  \bibinfo{author}{\bibfnamefont{L.}~\bibnamefont{Szymanowski}},
  \bibnamefont{and} \bibinfo{author}{\bibfnamefont{S.}~\bibnamefont{Wallon}},
  \bibinfo{journal}{Phys. Lett. B} \textbf{\bibinfo{volume}{586}},
  \bibinfo{pages}{267} (\bibinfo{year}{2004}), \eprint{hep-ph/0309281}.

\bibitem[{\citenamefont{Hatta et~al.}(2005)\citenamefont{Hatta, Iancu, Itakura,
  and McLerran}}]{Hatta:2005as}
\bibinfo{author}{\bibfnamefont{Y.}~\bibnamefont{Hatta}},
  \bibinfo{author}{\bibfnamefont{E.}~\bibnamefont{Iancu}},
  \bibinfo{author}{\bibfnamefont{K.}~\bibnamefont{Itakura}}, \bibnamefont{and}
  \bibinfo{author}{\bibfnamefont{L.}~\bibnamefont{McLerran}},
  \bibinfo{journal}{Nucl. Phys. A} \textbf{\bibinfo{volume}{760}},
  \bibinfo{pages}{172} (\bibinfo{year}{2005}), \eprint{hep-ph/0501171}.

\bibitem[{\citenamefont{Lappi et~al.}(2016)\citenamefont{Lappi, Ramnath,
  Rummukainen, and Weigert}}]{Lappi:2016gqe}
\bibinfo{author}{\bibfnamefont{T.}~\bibnamefont{Lappi}},
  \bibinfo{author}{\bibfnamefont{A.}~\bibnamefont{Ramnath}},
  \bibinfo{author}{\bibfnamefont{K.}~\bibnamefont{Rummukainen}},
  \bibnamefont{and} \bibinfo{author}{\bibfnamefont{H.}~\bibnamefont{Weigert}},
  \bibinfo{journal}{Phys. Rev. D} \textbf{\bibinfo{volume}{94}},
  \bibinfo{pages}{054014} (\bibinfo{year}{2016}), \eprint{1606.00551}.

\bibitem[{\citenamefont{Zhou}(2014)}]{Zhou:2013gsa}
\bibinfo{author}{\bibfnamefont{J.}~\bibnamefont{Zhou}}, \bibinfo{journal}{Phys.
  Rev. D} \textbf{\bibinfo{volume}{89}}, \bibinfo{pages}{074050}
  (\bibinfo{year}{2014}), \eprint{1308.5912}.

\bibitem[{\citenamefont{Boer et~al.}(2016)\citenamefont{Boer, Echevarria,
  Mulders, and Zhou}}]{Boer:2015pni}
\bibinfo{author}{\bibfnamefont{D.}~\bibnamefont{Boer}},
  \bibinfo{author}{\bibfnamefont{M.~G.} \bibnamefont{Echevarria}},
  \bibinfo{author}{\bibfnamefont{P.}~\bibnamefont{Mulders}}, \bibnamefont{and}
  \bibinfo{author}{\bibfnamefont{J.}~\bibnamefont{Zhou}},
  \bibinfo{journal}{Phys. Rev. Lett.} \textbf{\bibinfo{volume}{116}},
  \bibinfo{pages}{122001} (\bibinfo{year}{2016}), \eprint{1511.03485}.

\bibitem[{\citenamefont{Yao et~al.}(2019)\citenamefont{Yao, Hagiwara, and
  Hatta}}]{Yao:2018vcg}
\bibinfo{author}{\bibfnamefont{X.}~\bibnamefont{Yao}},
  \bibinfo{author}{\bibfnamefont{Y.}~\bibnamefont{Hagiwara}}, \bibnamefont{and}
  \bibinfo{author}{\bibfnamefont{Y.}~\bibnamefont{Hatta}},
  \bibinfo{journal}{Phys. Lett. B} \textbf{\bibinfo{volume}{790}},
  \bibinfo{pages}{361} (\bibinfo{year}{2019}), \eprint{1812.03959}.

\bibitem[{\citenamefont{H\"agler
  et~al.}(2002{\natexlab{a}})\citenamefont{H\"agler, Pire, Szymanowski, and
  Teryaev}}]{Hagler:2002nh}
\bibinfo{author}{\bibfnamefont{P.}~\bibnamefont{H\"agler}},
  \bibinfo{author}{\bibfnamefont{B.}~\bibnamefont{Pire}},
  \bibinfo{author}{\bibfnamefont{L.}~\bibnamefont{Szymanowski}},
  \bibnamefont{and} \bibinfo{author}{\bibfnamefont{O.}~\bibnamefont{Teryaev}},
  \bibinfo{journal}{Phys. Lett. B} \textbf{\bibinfo{volume}{535}},
  \bibinfo{pages}{117} (\bibinfo{year}{2002}{\natexlab{a}}),
  \bibinfo{note}{[Erratum: Phys.Lett.B 540, 324--325 (2002)]},
  \eprint{hep-ph/0202231}.

\bibitem[{\citenamefont{H\"agler
  et~al.}(2002{\natexlab{b}})\citenamefont{H\"agler, Pire, Szymanowski, and
  Teryaev}}]{Hagler:2002nf}
\bibinfo{author}{\bibfnamefont{P.}~\bibnamefont{H\"agler}},
  \bibinfo{author}{\bibfnamefont{B.}~\bibnamefont{Pire}},
  \bibinfo{author}{\bibfnamefont{L.}~\bibnamefont{Szymanowski}},
  \bibnamefont{and} \bibinfo{author}{\bibfnamefont{O.}~\bibnamefont{Teryaev}},
  \bibinfo{journal}{Eur. Phys. J. C} \textbf{\bibinfo{volume}{26}},
  \bibinfo{pages}{261} (\bibinfo{year}{2002}{\natexlab{b}}),
  \eprint{hep-ph/0207224}.

\bibitem[{\citenamefont{Dumitru and Stebel}(2019)}]{Dumitru:2019qec}
\bibinfo{author}{\bibfnamefont{A.}~\bibnamefont{Dumitru}} \bibnamefont{and}
  \bibinfo{author}{\bibfnamefont{T.}~\bibnamefont{Stebel}},
  \bibinfo{journal}{Phys. Rev. D} \textbf{\bibinfo{volume}{99}},
  \bibinfo{pages}{094038} (\bibinfo{year}{2019}), \eprint{1903.07660}.

\bibitem[{\citenamefont{Czyzewski et~al.}(1997)\citenamefont{Czyzewski,
  Kwiecinski, Motyka, and Sadzikowski}}]{Czyzewski:1996bv}
\bibinfo{author}{\bibfnamefont{J.}~\bibnamefont{Czyzewski}},
  \bibinfo{author}{\bibfnamefont{J.}~\bibnamefont{Kwiecinski}},
  \bibinfo{author}{\bibfnamefont{L.}~\bibnamefont{Motyka}}, \bibnamefont{and}
  \bibinfo{author}{\bibfnamefont{M.}~\bibnamefont{Sadzikowski}},
  \bibinfo{journal}{Phys. Lett. B} \textbf{\bibinfo{volume}{398}},
  \bibinfo{pages}{400} (\bibinfo{year}{1997}), \bibinfo{note}{[Erratum:
  Phys.Lett.B 411, 402 (1997)]}, \eprint{hep-ph/9611225}.

\bibitem[{\citenamefont{Engel et~al.}(1998)\citenamefont{Engel, Ivanov,
  Kirschner, and Szymanowski}}]{Engel:1997cga}
\bibinfo{author}{\bibfnamefont{R.}~\bibnamefont{Engel}},
  \bibinfo{author}{\bibfnamefont{D.}~\bibnamefont{Ivanov}},
  \bibinfo{author}{\bibfnamefont{R.}~\bibnamefont{Kirschner}},
  \bibnamefont{and}
  \bibinfo{author}{\bibfnamefont{L.}~\bibnamefont{Szymanowski}},
  \bibinfo{journal}{Eur. Phys. J. C} \textbf{\bibinfo{volume}{4}},
  \bibinfo{pages}{93} (\bibinfo{year}{1998}), \eprint{hep-ph/9707362}.

\bibitem[{\citenamefont{Kilian and Nachtmann}(1998)}]{Kilian:1997ew}
\bibinfo{author}{\bibfnamefont{W.}~\bibnamefont{Kilian}} \bibnamefont{and}
  \bibinfo{author}{\bibfnamefont{O.}~\bibnamefont{Nachtmann}},
  \bibinfo{journal}{Eur. Phys. J. C} \textbf{\bibinfo{volume}{5}},
  \bibinfo{pages}{317} (\bibinfo{year}{1998}), \eprint{hep-ph/9712371}.

\bibitem[{\citenamefont{Rueter et~al.}(1999)\citenamefont{Rueter, Dosch, and
  Nachtmann}}]{Rueter:1998gj}
\bibinfo{author}{\bibfnamefont{M.}~\bibnamefont{Rueter}},
  \bibinfo{author}{\bibfnamefont{H.~G.} \bibnamefont{Dosch}}, \bibnamefont{and}
  \bibinfo{author}{\bibfnamefont{O.}~\bibnamefont{Nachtmann}},
  \bibinfo{journal}{Phys. Rev. D} \textbf{\bibinfo{volume}{59}},
  \bibinfo{pages}{014018} (\bibinfo{year}{1999}), \eprint{hep-ph/9806342}.

\bibitem[{\citenamefont{Albacete and Soto-Ontoso}(2017)}]{Albacete:2016pmp}
\bibinfo{author}{\bibfnamefont{J.~L.} \bibnamefont{Albacete}} \bibnamefont{and}
  \bibinfo{author}{\bibfnamefont{A.}~\bibnamefont{Soto-Ontoso}},
  \bibinfo{journal}{Phys. Lett. B} \textbf{\bibinfo{volume}{770}},
  \bibinfo{pages}{149} (\bibinfo{year}{2017}), \eprint{1605.09176}.

\bibitem[{\citenamefont{Albacete et~al.}(2017)\citenamefont{Albacete, Petersen,
  and Soto-Ontoso}}]{Albacete:2016gxu}
\bibinfo{author}{\bibfnamefont{J.~L.} \bibnamefont{Albacete}},
  \bibinfo{author}{\bibfnamefont{H.}~\bibnamefont{Petersen}}, \bibnamefont{and}
  \bibinfo{author}{\bibfnamefont{A.}~\bibnamefont{Soto-Ontoso}},
  \bibinfo{journal}{Phys. Rev. C} \textbf{\bibinfo{volume}{95}},
  \bibinfo{pages}{064909} (\bibinfo{year}{2017}), \eprint{1612.06274}.

\bibitem[{\citenamefont{Albacete et~al.}(2018)\citenamefont{Albacete, Petersen,
  and Soto-Ontoso}}]{Albacete:2017ajt}
\bibinfo{author}{\bibfnamefont{J.~L.} \bibnamefont{Albacete}},
  \bibinfo{author}{\bibfnamefont{H.}~\bibnamefont{Petersen}}, \bibnamefont{and}
  \bibinfo{author}{\bibfnamefont{A.}~\bibnamefont{Soto-Ontoso}},
  \bibinfo{journal}{Phys. Lett. B} \textbf{\bibinfo{volume}{778}},
  \bibinfo{pages}{128} (\bibinfo{year}{2018}), \eprint{1707.05592}.

\bibitem[{\citenamefont{Cs\"org\H{o} et~al.}(2020)\citenamefont{Cs\"org\H{o},
  Pasechnik, and Ster}}]{Csorgo:2019egs}
\bibinfo{author}{\bibfnamefont{T.}~\bibnamefont{Cs\"org\H{o}}},
  \bibinfo{author}{\bibfnamefont{R.}~\bibnamefont{Pasechnik}},
  \bibnamefont{and} \bibinfo{author}{\bibfnamefont{A.}~\bibnamefont{Ster}},
  \bibinfo{journal}{Eur. Phys. J. C} \textbf{\bibinfo{volume}{80}},
  \bibinfo{pages}{126} (\bibinfo{year}{2020}), \eprint{1910.08817}.

\bibitem[{\citenamefont{M\"antysaari}(2020)}]{Mantysaari:2020axf}
\bibinfo{author}{\bibfnamefont{H.}~\bibnamefont{M\"antysaari}},
  \bibinfo{journal}{Rept. Prog. Phys.} \textbf{\bibinfo{volume}{83}},
  \bibinfo{pages}{082201} (\bibinfo{year}{2020}), \eprint{2001.10705}.

\bibitem[{\citenamefont{Boer et~al.}(2011)}]{Boer:2011fh}
\bibinfo{author}{\bibfnamefont{D.}~\bibnamefont{Boer}} \bibnamefont{et~al.}
  (\bibinfo{year}{2011}), \eprint{1108.1713}.

\bibitem[{\citenamefont{Accardi et~al.}(2016)}]{Accardi:2012qut}
\bibinfo{author}{\bibfnamefont{A.}~\bibnamefont{Accardi}} \bibnamefont{et~al.},
  \bibinfo{journal}{Eur. Phys. J. A} \textbf{\bibinfo{volume}{52}},
  \bibinfo{pages}{268} (\bibinfo{year}{2016}), \eprint{1212.1701}.

\bibitem[{\citenamefont{Aschenauer et~al.}(2019)\citenamefont{Aschenauer,
  Fazio, Lee, Mantysaari, Page, Schenke, Ullrich, Venugopalan, and
  Zurita}}]{Aschenauer:2017jsk}
\bibinfo{author}{\bibfnamefont{E.}~\bibnamefont{Aschenauer}},
  \bibinfo{author}{\bibfnamefont{S.}~\bibnamefont{Fazio}},
  \bibinfo{author}{\bibfnamefont{J.}~\bibnamefont{Lee}},
  \bibinfo{author}{\bibfnamefont{H.}~\bibnamefont{Mantysaari}},
  \bibinfo{author}{\bibfnamefont{B.}~\bibnamefont{Page}},
  \bibinfo{author}{\bibfnamefont{B.}~\bibnamefont{Schenke}},
  \bibinfo{author}{\bibfnamefont{T.}~\bibnamefont{Ullrich}},
  \bibinfo{author}{\bibfnamefont{R.}~\bibnamefont{Venugopalan}},
  \bibnamefont{and} \bibinfo{author}{\bibfnamefont{P.}~\bibnamefont{Zurita}},
  \bibinfo{journal}{Rept. Prog. Phys.} \textbf{\bibinfo{volume}{82}},
  \bibinfo{pages}{024301} (\bibinfo{year}{2019}), \eprint{1708.01527}.

\bibitem[{\citenamefont{Abdul~Khalek et~al.}(2021)}]{AbdulKhalek:2021gbh}
\bibinfo{author}{\bibfnamefont{R.}~\bibnamefont{Abdul~Khalek}}
  \bibnamefont{et~al.} (\bibinfo{year}{2021}), \eprint{2103.05419}.

\bibitem[{\citenamefont{Dumitru et~al.}(2018)\citenamefont{Dumitru, Miller, and
  Venugopalan}}]{Dumitru:2018vpr}
\bibinfo{author}{\bibfnamefont{A.}~\bibnamefont{Dumitru}},
  \bibinfo{author}{\bibfnamefont{G.~A.} \bibnamefont{Miller}},
  \bibnamefont{and}
  \bibinfo{author}{\bibfnamefont{R.}~\bibnamefont{Venugopalan}},
  \bibinfo{journal}{Phys. Rev. D} \textbf{\bibinfo{volume}{98}},
  \bibinfo{pages}{094004} (\bibinfo{year}{2018}), \eprint{1808.02501}.

\bibitem[{\citenamefont{Dumitru et~al.}(2020)\citenamefont{Dumitru, Skokov, and
  Stebel}}]{Dumitru:2020fdh}
\bibinfo{author}{\bibfnamefont{A.}~\bibnamefont{Dumitru}},
  \bibinfo{author}{\bibfnamefont{V.}~\bibnamefont{Skokov}}, \bibnamefont{and}
  \bibinfo{author}{\bibfnamefont{T.}~\bibnamefont{Stebel}},
  \bibinfo{journal}{Phys. Rev. D} \textbf{\bibinfo{volume}{101}},
  \bibinfo{pages}{054004} (\bibinfo{year}{2020}), \eprint{2001.04516}.

\bibitem[{\citenamefont{Bartels and Motyka}(2008)}]{Bartels:2007aa}
\bibinfo{author}{\bibfnamefont{J.}~\bibnamefont{Bartels}} \bibnamefont{and}
  \bibinfo{author}{\bibfnamefont{L.}~\bibnamefont{Motyka}},
  \bibinfo{journal}{Eur. Phys. J. C} \textbf{\bibinfo{volume}{55}},
  \bibinfo{pages}{65} (\bibinfo{year}{2008}), \eprint{0711.2196}.

\bibitem[{\citenamefont{Beuf}(2014)}]{Beuf:2014uia}
\bibinfo{author}{\bibfnamefont{G.}~\bibnamefont{Beuf}}, \bibinfo{journal}{Phys.
  Rev.} \textbf{\bibinfo{volume}{D89}}, \bibinfo{pages}{074039}
  (\bibinfo{year}{2014}), \eprint{1401.0313}.

\bibitem[{\citenamefont{Duclou\'e et~al.}(2019)\citenamefont{Duclou\'e, Iancu,
  Mueller, Soyez, and Triantafyllopoulos}}]{Ducloue:2019ezk}
\bibinfo{author}{\bibfnamefont{B.}~\bibnamefont{Duclou\'e}},
  \bibinfo{author}{\bibfnamefont{E.}~\bibnamefont{Iancu}},
  \bibinfo{author}{\bibfnamefont{A.}~\bibnamefont{Mueller}},
  \bibinfo{author}{\bibfnamefont{G.}~\bibnamefont{Soyez}}, \bibnamefont{and}
  \bibinfo{author}{\bibfnamefont{D.}~\bibnamefont{Triantafyllopoulos}},
  \bibinfo{journal}{JHEP} \textbf{\bibinfo{volume}{04}}, \bibinfo{pages}{081}
  (\bibinfo{year}{2019}), \eprint{1902.06637}.

\bibitem[{\citenamefont{Boussarie and Mehtar-Tani}(2020)}]{Boussarie:2020fpb}
\bibinfo{author}{\bibfnamefont{R.}~\bibnamefont{Boussarie}} \bibnamefont{and}
  \bibinfo{author}{\bibfnamefont{Y.}~\bibnamefont{Mehtar-Tani}}
  (\bibinfo{year}{2020}), \eprint{2006.14569}.

\bibitem[{\citenamefont{Kovchegov and Sievert}(2012)}]{Kovchegov:2012ga}
\bibinfo{author}{\bibfnamefont{Y.~V.} \bibnamefont{Kovchegov}}
  \bibnamefont{and} \bibinfo{author}{\bibfnamefont{M.~D.}
  \bibnamefont{Sievert}}, \bibinfo{journal}{Phys. Rev. D}
  \textbf{\bibinfo{volume}{86}}, \bibinfo{pages}{034028}
  (\bibinfo{year}{2012}), \bibinfo{note}{[Erratum: Phys.Rev.D 86, 079906
  (2012)]}, \eprint{1201.5890}.

\bibitem[{\citenamefont{Schlumpf}(1993)}]{Schlumpf:1992vq}
\bibinfo{author}{\bibfnamefont{F.}~\bibnamefont{Schlumpf}},
  \bibinfo{journal}{Phys. Rev. D} \textbf{\bibinfo{volume}{47}},
  \bibinfo{pages}{4114} (\bibinfo{year}{1993}), \bibinfo{note}{[Erratum:
  Phys.Rev.D 49, 6246 (1994)]}, \eprint{hep-ph/9212250}.

\bibitem[{\citenamefont{Brodsky and Schlumpf}(1994)}]{Brodsky:1994fz}
\bibinfo{author}{\bibfnamefont{S.~J.} \bibnamefont{Brodsky}} \bibnamefont{and}
  \bibinfo{author}{\bibfnamefont{F.}~\bibnamefont{Schlumpf}},
  \bibinfo{journal}{Phys. Lett. B} \textbf{\bibinfo{volume}{329}},
  \bibinfo{pages}{111} (\bibinfo{year}{1994}), \eprint{hep-ph/9402214}.

\bibitem[{\citenamefont{Golec-Biernat and Stasto}(2003)}]{GolecBiernat:2003ym}
\bibinfo{author}{\bibfnamefont{K.~J.} \bibnamefont{Golec-Biernat}}
  \bibnamefont{and} \bibinfo{author}{\bibfnamefont{A.~M.}
  \bibnamefont{Stasto}}, \bibinfo{journal}{Nucl. Phys.}
  \textbf{\bibinfo{volume}{B668}}, \bibinfo{pages}{345} (\bibinfo{year}{2003}),
  \eprint{hep-ph/0306279}.

\bibitem[{\citenamefont{Berger and Stasto}(2011)}]{Berger:2010sh}
\bibinfo{author}{\bibfnamefont{J.}~\bibnamefont{Berger}} \bibnamefont{and}
  \bibinfo{author}{\bibfnamefont{A.}~\bibnamefont{Stasto}},
  \bibinfo{journal}{Phys. Rev.} \textbf{\bibinfo{volume}{D83}},
  \bibinfo{pages}{034015} (\bibinfo{year}{2011}), \eprint{1010.0671}.

\bibitem[{\citenamefont{Cepila et~al.}(2019)\citenamefont{Cepila, Contreras,
  and Matas}}]{Cepila:2018faq}
\bibinfo{author}{\bibfnamefont{J.}~\bibnamefont{Cepila}},
  \bibinfo{author}{\bibfnamefont{J.}~\bibnamefont{Contreras}},
  \bibnamefont{and} \bibinfo{author}{\bibfnamefont{M.}~\bibnamefont{Matas}},
  \bibinfo{journal}{Phys. Rev. D} \textbf{\bibinfo{volume}{99}},
  \bibinfo{pages}{051502} (\bibinfo{year}{2019}), \eprint{1812.02548}.

\bibitem[{\citenamefont{M\"antysaari and Schenke}(2018)}]{Mantysaari:2018zdd}
\bibinfo{author}{\bibfnamefont{H.}~\bibnamefont{M\"antysaari}}
  \bibnamefont{and} \bibinfo{author}{\bibfnamefont{B.}~\bibnamefont{Schenke}},
  \bibinfo{journal}{Phys. Rev. D} \textbf{\bibinfo{volume}{98}},
  \bibinfo{pages}{034013} (\bibinfo{year}{2018}), \eprint{1806.06783}.

\bibitem[{\citenamefont{Bendova et~al.}(2019)\citenamefont{Bendova, Cepila,
  Contreras, and Matas}}]{Bendova:2019psy}
\bibinfo{author}{\bibfnamefont{D.}~\bibnamefont{Bendova}},
  \bibinfo{author}{\bibfnamefont{J.}~\bibnamefont{Cepila}},
  \bibinfo{author}{\bibfnamefont{J.}~\bibnamefont{Contreras}},
  \bibnamefont{and} \bibinfo{author}{\bibfnamefont{M.}~\bibnamefont{Matas}},
  \bibinfo{journal}{Phys. Rev. D} \textbf{\bibinfo{volume}{100}},
  \bibinfo{pages}{054015} (\bibinfo{year}{2019}), \eprint{1907.12123}.

\bibitem[{\citenamefont{Binosi et~al.}(2009)\citenamefont{Binosi, Collins,
  Kaufhold, and Theussl}}]{Binosi:2008ig}
\bibinfo{author}{\bibfnamefont{D.}~\bibnamefont{Binosi}},
  \bibinfo{author}{\bibfnamefont{J.}~\bibnamefont{Collins}},
  \bibinfo{author}{\bibfnamefont{C.}~\bibnamefont{Kaufhold}}, \bibnamefont{and}
  \bibinfo{author}{\bibfnamefont{L.}~\bibnamefont{Theussl}},
  \bibinfo{journal}{Comput. Phys. Commun.} \textbf{\bibinfo{volume}{180}},
  \bibinfo{pages}{1709} (\bibinfo{year}{2009}), \eprint{0811.4113}.

\bibitem[{\citenamefont{Lepage and Brodsky}(1980)}]{Lepage:1980fj}
\bibinfo{author}{\bibfnamefont{G.}~\bibnamefont{Lepage}} \bibnamefont{and}
  \bibinfo{author}{\bibfnamefont{S.~J.} \bibnamefont{Brodsky}},
  \bibinfo{journal}{Phys. Rev. D} \textbf{\bibinfo{volume}{22}},
  \bibinfo{pages}{2157} (\bibinfo{year}{1980}).

\bibitem[{\citenamefont{Brodsky et~al.}(1998)\citenamefont{Brodsky, Pauli, and
  Pinsky}}]{Brodsky:1997de}
\bibinfo{author}{\bibfnamefont{S.~J.} \bibnamefont{Brodsky}},
  \bibinfo{author}{\bibfnamefont{H.-C.} \bibnamefont{Pauli}}, \bibnamefont{and}
  \bibinfo{author}{\bibfnamefont{S.~S.} \bibnamefont{Pinsky}},
  \bibinfo{journal}{Phys. Rept.} \textbf{\bibinfo{volume}{301}},
  \bibinfo{pages}{299} (\bibinfo{year}{1998}), \eprint{hep-ph/9705477}.

\bibitem[{\citenamefont{Brodsky et~al.}(2001)\citenamefont{Brodsky, Hwang, Ma,
  and Schmidt}}]{Brodsky:2000ii}
\bibinfo{author}{\bibfnamefont{S.~J.} \bibnamefont{Brodsky}},
  \bibinfo{author}{\bibfnamefont{D.~S.} \bibnamefont{Hwang}},
  \bibinfo{author}{\bibfnamefont{B.-Q.} \bibnamefont{Ma}}, \bibnamefont{and}
  \bibinfo{author}{\bibfnamefont{I.}~\bibnamefont{Schmidt}},
  \bibinfo{journal}{Nucl. Phys. B} \textbf{\bibinfo{volume}{593}},
  \bibinfo{pages}{311} (\bibinfo{year}{2001}), \eprint{hep-th/0003082}.

\bibitem[{\citenamefont{Hanninen et~al.}(2018)\citenamefont{Hanninen, Lappi,
  and Paatelainen}}]{Hanninen:2017ddy}
\bibinfo{author}{\bibfnamefont{H.}~\bibnamefont{Hanninen}},
  \bibinfo{author}{\bibfnamefont{T.}~\bibnamefont{Lappi}}, \bibnamefont{and}
  \bibinfo{author}{\bibfnamefont{R.}~\bibnamefont{Paatelainen}},
  \bibinfo{journal}{Annals Phys.} \textbf{\bibinfo{volume}{393}},
  \bibinfo{pages}{358} (\bibinfo{year}{2018}), \eprint{1711.08207}.

\bibitem[{\citenamefont{Lappi and Paatelainen}(2017)}]{Lappi:2016oup}
\bibinfo{author}{\bibfnamefont{T.}~\bibnamefont{Lappi}} \bibnamefont{and}
  \bibinfo{author}{\bibfnamefont{R.}~\bibnamefont{Paatelainen}},
  \bibinfo{journal}{Annals Phys.} \textbf{\bibinfo{volume}{379}},
  \bibinfo{pages}{34} (\bibinfo{year}{2017}), \eprint{1611.00497}.

\bibitem[{\citenamefont{Gribov and
  Lipatov}(1972{\natexlab{a}})}]{Gribov:1972ri}
\bibinfo{author}{\bibfnamefont{V.}~\bibnamefont{Gribov}} \bibnamefont{and}
  \bibinfo{author}{\bibfnamefont{L.}~\bibnamefont{Lipatov}},
  \bibinfo{journal}{Sov. J. Nucl. Phys.} \textbf{\bibinfo{volume}{15}},
  \bibinfo{pages}{438} (\bibinfo{year}{1972}{\natexlab{a}}).

\bibitem[{\citenamefont{Gribov and
  Lipatov}(1972{\natexlab{b}})}]{Gribov:1972rt}
\bibinfo{author}{\bibfnamefont{V.}~\bibnamefont{Gribov}} \bibnamefont{and}
  \bibinfo{author}{\bibfnamefont{L.}~\bibnamefont{Lipatov}},
  \bibinfo{journal}{Sov. J. Nucl. Phys.} \textbf{\bibinfo{volume}{15}},
  \bibinfo{pages}{675} (\bibinfo{year}{1972}{\natexlab{b}}).

\bibitem[{\citenamefont{Altarelli and Parisi}(1977)}]{Altarelli:1977zs}
\bibinfo{author}{\bibfnamefont{G.}~\bibnamefont{Altarelli}} \bibnamefont{and}
  \bibinfo{author}{\bibfnamefont{G.}~\bibnamefont{Parisi}},
  \bibinfo{journal}{Nucl. Phys. B} \textbf{\bibinfo{volume}{126}},
  \bibinfo{pages}{298} (\bibinfo{year}{1977}).

\bibitem[{\citenamefont{Dokshitzer}(1977)}]{Dokshitzer:1977sg}
\bibinfo{author}{\bibfnamefont{Y.~L.} \bibnamefont{Dokshitzer}},
  \bibinfo{journal}{Sov. Phys. JETP} \textbf{\bibinfo{volume}{46}},
  \bibinfo{pages}{641} (\bibinfo{year}{1977}).

\bibitem[{\citenamefont{Bartels and Ewerz}(1999)}]{Bartels:1999aw}
\bibinfo{author}{\bibfnamefont{J.}~\bibnamefont{Bartels}} \bibnamefont{and}
  \bibinfo{author}{\bibfnamefont{C.}~\bibnamefont{Ewerz}},
  \bibinfo{journal}{JHEP} \textbf{\bibinfo{volume}{09}}, \bibinfo{pages}{026}
  (\bibinfo{year}{1999}), \eprint{hep-ph/9908454}.

\bibitem[{\citenamefont{Ewerz}(2001)}]{Ewerz:2001fb}
\bibinfo{author}{\bibfnamefont{C.}~\bibnamefont{Ewerz}},
  \bibinfo{journal}{JHEP} \textbf{\bibinfo{volume}{04}}, \bibinfo{pages}{031}
  (\bibinfo{year}{2001}), \eprint{hep-ph/0103260}.

\end{thebibliography}

\end{document}